\documentclass{iopjournal}

\usepackage{graphicx,bm,natbib,upgreek,amsmath, amssymb, mathrsfs,accents}
\usepackage[toc,page,titletoc]{appendix}
\usepackage[capitalise,nameinlink]{cleveref}
\crefname{supp}{Supplement}{Supplements}
\usepackage{subcaption}
\usepackage{listings}
\usepackage{booktabs}
\usepackage{braket}
\usepackage{tabularx}
\usepackage[ruled, vlined,linesnumbered]{algorithm2e}
\usepackage[textsize=tiny]{todonotes}
\usepackage{multirow}

\reversemarginpar

\newcommand{\ogsA}{$\text{OGS}_\text{A}$} \newcommand{\ogsB}{$\text{OGS}_\text{B}$}
\RestyleAlgo{ruled}
\renewcommand{\articletype}[1]{{\vspace*{-8mm}\noindent \Large \sf Journal Name}}
\renewcommand{\articletype}[1]{}

\begin{document}

\articletype{Paper} 

\title{Single-Satellite Quantum Repeater Performance Analysis}

\author{
Cameron Paterson$^{1,*}$, \orcid{0009-0004-7028-7569},
Jasminder S. Sidhu$^{1,2,3}$\orcid{0000-0002-6167-8224},
Thomas Brougham$^1$\orcid{0000-0002-9066-1771},
Sarah E. McCarthy$^{1,4}$, \orcid{0009-0003-5566-9668} and
and Daniel K. L. Oi$^{1}$\orcid{0000-0003-0965-9509}
}

\affil{$^1$SUPA Department of Physics, University of Strathclyde, Glasgow, G4 0NG, United Kingdom}

\affil{$^2$School of Mathematical and Physical Sciences, University of Sheffield, Sheffield, S3 7RH, UK}

\affil{$^3$Dahlem Center for Complex Quantum Systems, Freie Universit{\"a}t Berlin, 14195 Berlin, Germany}

\affil{$^4$Fraunhofer Centre for Applied Photonics, Glasgow, G1 1RD, United Kingdom}

\affil{$^*$Author to whom any correspondence should be addressed.}

\email{cameron.paterson@strath.ac.uk}

\keywords{satellite, quantum communications, quantum entanglement, quantum memory, quantum repeater}

\begin{abstract}
Satellite optical links have the potential to extend the range of quantum communication beyond that which is possible through optical fibres, which are constrained by exponential loss scaling. Quantum repeaters are a proposed technology to mitigate the effects of high channel losses in both fibre and satellite links. Although a satellite quantum repeater could improve the entanglement distribution efficiency, the rate at which entanglement distribution can be attempted is constrained by classical communication latency in the entanglement swapping process. Direct entanglement distribution does not suffer from this latency and can therefore attempt to compensate for high dual-downlink losses by increasing the source rate. A systematic comparative analysis of repeater-assisted and direct entanglement distribution rates using a satellite is absent from the literature. In addition to entanglement distribution rates, a detailed analysis of the fidelity of entanglement distributed by repeater-assisted satellite links has yet to be performed.  Here, we develop an in-depth framework for describing a space quantum link consisting of a single satellite and two ground stations. We compare the entanglement distribution rates of direct and repeater-assisted satellite entanglement distribution, taking into account variations in overpass geometry and quantum memory capacity. The long-term entanglement distribution performance to different pairs of major cities is also considered, and the altitudinal dependence is  explored. Furthermore, we develop a Monte Carlo model of the repeater satellite to explore the effect that different memory management policies and capacities have on the waiting time statistics of memory qubits. From these statistics we calculate the fidelity distributions of the entangled pairs established between the ground stations. These results will inform mission design for future space-borne quantum repeater nodes, as well as requirements on space-based memory platforms.
\end{abstract}

\section{Introduction}
Emerging quantum technologies offer the ability to establish provably secure cryptographic keys~\cite{BennettQuantumCryptography,ShorPreskillSecurityProof,ScaraniSecuirty.81.1301}, improve sensing capabilities~\cite{GiovannettiMetrology,DegenQuantumSensing}, enable precise clock synchronisation~\cite{ShiClockSynchronisation}, and computational speed-up for certain problems~\cite{ShorFactoring}. Robust and reliable distribution of quantum entanglement is essential for the implementation of these technologies. Optical fibre is limited by exponential photonic loss scaling with distance~\cite{PirandolaAdvancesInQuantumCryptography}, limiting its viability for direct entanglement distribution to a few hundred kilometres. This range can be extended using quantum repeaters~\cite{BriegelQRs}, but long-range fibre links still face implementation-related challenges due to the requirements for close spacing, number, and placement of repeater stations between end-nodes. 

Following landmark experiments by the Micius mission~\cite{miciusExperiemntsInSpace} there has been much interest in space quantum communications to extend the range of entanglement distribution~\cite{Sidhu_2021AdvancsIn,belenchia2022quantum, koudia2024spacebasedquantuminternetentanglement,goswami2023satellite}. Satellite links exploit the more benign inverse square diffraction loss scaling and vacuum propagation for the majority of a space-Earth link and have advantages of access to moving platforms and remote areas where fibre links may be impractical.
Micius was successful in distributing entanglement between ground stations over 1000~km apart, though direct pair distribution losses led to a received rate of $\sim1$ pairs s$^{-1}$ from $5.9\times 10^6$ pairs s$^{-1}$ sent.

Satellite quantum repeaters with quantum memories are a promising platform for  distributing entanglement at large scales~\cite{gundoganProposalSpaceMemories2020,fittipaldiEntSwappingInOrbit2024, LiorniQRsInSapce,gundoganJasBeyondLOS,dominguez2026satellite}. Quantum repeaters greatly increase the efficiency of distributing entanglement at the cost of introducing communication latency which impacts the distribution rate. Direct transmission does not incur such latencies, hence their overall entanglement distribution rate can be increased through ``brute-force'' by simply using a higher source rate. Single-satellite direct dual-downlink (DDDL) entanglement distribution (a la Micius) is the most feasible to implement in the near term, but there are growing efforts to develop space quantum memories and repeaters as a path to a longer-term solution. Thus it is important to understand their respective resource requirements, performance characteristics, and the cross-over point between the two approaches. In this work, we focus on a link consisting of a single satellite and two optical ground stations (OGSs), providing a systematic exploration of the performance of these dual downlinks for general overpass geometries. 

The paper is organised as follows. Section~\ref{sec:Preliminaries} provides: background and context to our work; describes the dual-downlink channel model and entanglement distribution protocols considered; and outlines the Monte Carlo (MC) simulation of repeater waiting time and entanglement fidelity statistics. Section~\ref{Results} presents results of: per-overpass entanglement distribution volume with and without quantum memories under varying overpass geometries; expected annual total number of distributed entangled pairs; distributions of waiting times and entanglement fidelities for representative overpasses. Section~\ref{sec:conclusion} concludes by identifying potential follow-on research.

\begin{figure}[!t]
	\centering
	\includegraphics[width=0.75\linewidth]{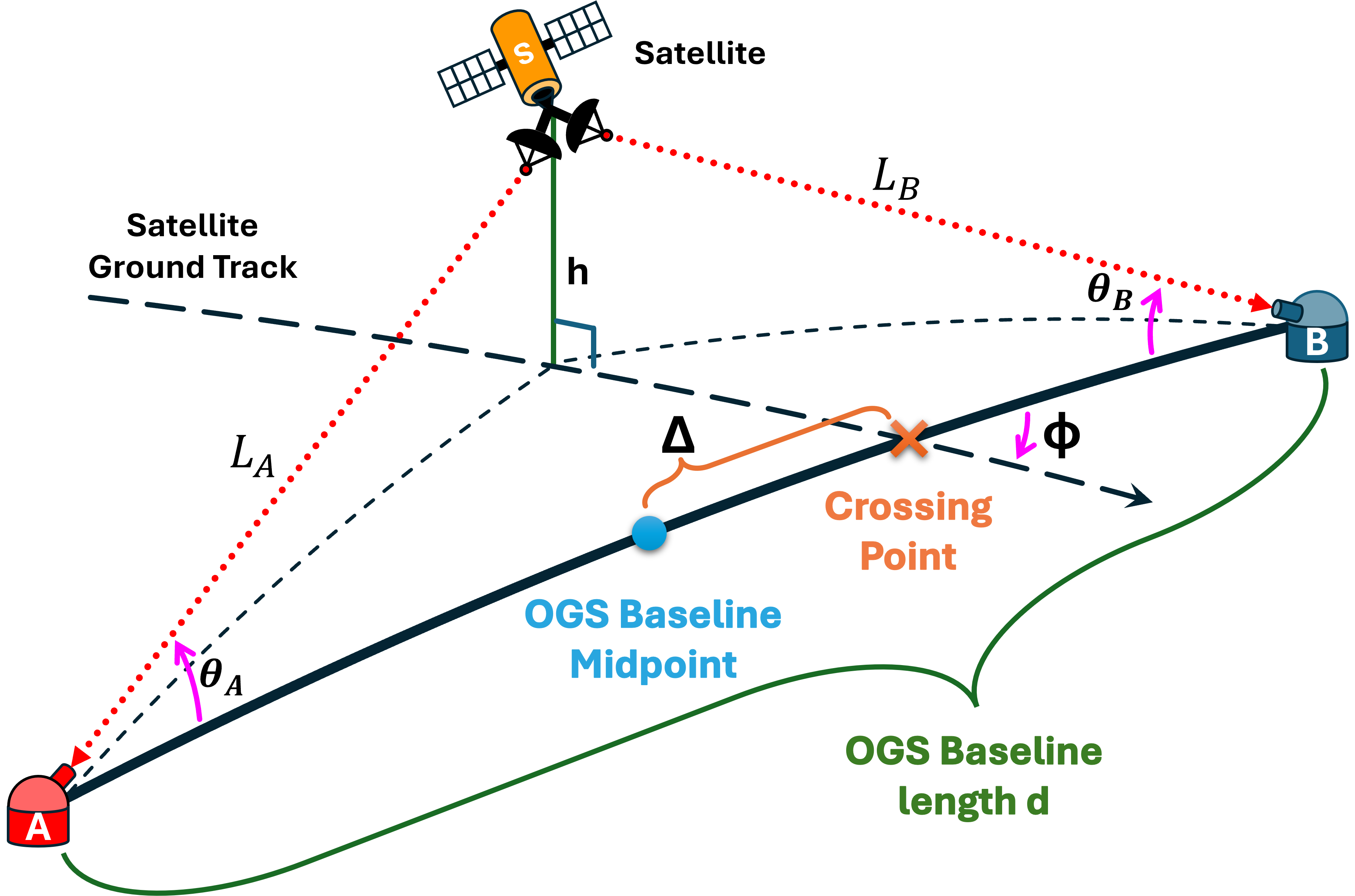}
	\caption{Single Satellite -- Two OGS Overpass Geometry. A single satellite simultaneously communicates with Alice (\ogsA) and Bob (\ogsB) in a single satellite quantum repeater (SSQR) configuration. We consider a satellite in a circular orbit of altitude $h$. The great-circle separating A and B of arc-length $d$ is denoted the OGS baseline. We choose to parameterise the overpass geometry by the intersection of the satellite ground track with the OGS baseline, defined by the crossing angle ($\phi$) and offset ($\Delta$) relative to the midpoint position (an alternative choice would be to parameterise each single downlink by the minimum ground track offset to each OGS, $\{d_{min}^A,d_{min}^B\}$). The slant distances to the OGS's $\{L_A(t),L_B(t)\}$, and the local elevations $\{\theta_A(t),\theta_B(t)\}$ vary during the overpass. Entanglement distribution is possible when the satellite is above the  minimum elevation $\theta_\text{min}$ for both A and B at the same time. An overpass is defined as the period for which this dual-visibility criterion is met. In the direct dual downlink (DDDL) configuration, the satellite transmits pairs of entangled photons simultaneously, one photon to each OGS. For a single satellite repeater (SSQR), the satellite has two independent sets of an entanglement source and quantum memory to establish an entangled link with each OGS, together with an entanglement swapping mechanism.
	}
	\label{fig:twoOGSLinkDiagram}
\end{figure}

\section{Preliminaries and Methods}
\label{sec:Preliminaries}

We consider a single satellite communicating with 2 OGSs through dual-downlink quantum optical channels and define a parameterisation of general overpass geometries. The two methods of entanglement distribution studied are introduced, together with the estimation procedure of the short and long-term performance. We finally cover the MC of memory qubit waiting times used to estimate the entanglement fidelity of pairs distribution to the ground stations, for different memory management policies. In this preliminary investigation, we will not consider weather effects or time varying background light alongside other simplifications of the problem such as identical OGSs and night operations.

\subsection{Dual-Downlink Overpass Channel Loss}

The single-satellite 2-OGS configuration is the simplest space-based entanglement distribution topology. The most general dual-downlink overpass geometry is shown in Fig.~\ref{fig:twoOGSLinkDiagram} and can be parameterised by the pair of variables $\{\Delta,\phi\}$~\cite{sidhu2026operational}.
We are interested in determining the effect of overpass geometry on the ability of the satellite to distribute entanglement, to inform future mission and constellation design considerations.
The entanglement distribution performance will be determined by the optical losses in each satellite-OGS channel. We model each link as for a single-satellite one OGS configuration~\cite{sidhu2021finitekeyeffectssatellite, sidhuFiniteResource} (See Appendix~\ref{appendix:singleOGSLinks}). The slant distances $\{L_A(t),L_B(t)\}$ and elevations $\{\theta_A(t),\theta_B(t)\}$ determine the transmittances $\eta_A (t)$ and $\eta_B (t)$ of the \ogsA and \ogsB  downlinks, respectively (see Appendix~\ref{appendix:linkBudget} for details). For the DDDL configuration, the effective dual channel transmittance is $\eta_{AB}=\eta_A \eta_B$, whereas for SSQR it scales as $\sim\min\{\eta_A,\eta_B\}$.

\subsection{Satellite Entanglement Distribution}

We describe the two entanglement distribution protocols that we consider corresponding to: DDDL in which the satellite is equipped only with an entangled pair source (EPS) and no quantum memory and SSQR using the Sender-Receiver protocol~\cite{Jones_2016}. This is not the only possible choice of repeater protocol, for example~\cite{azuma2015all}, but we defer analysis of these to future work. 

\subsubsection{Direct Dual-Downlink.}

In DDDL, the satellite is only equipped with an EPS. The satellite generates entangled photon pairs and sends one photon of each pair to Alice and Bob. If the EPS produces entangled photons at a rate $R_\text{EPS}$, then the instantaneous pair distribution rate (PDR) is given by
\begin{equation}
    R^\text{(DDDL)}(t) = \eta_A(t) \eta_B(t) R_\text{EPS}.
\label{equ:DDDLentDistRate}
\end{equation}
For simplicity, we consider a deterministic source~\cite{pan2026quantum}. In practice, EPSs for satellite deployment (e.g.~\cite{villar2020entanglement,mccarthy2025compact}) are currently based on spontaneous parametric down conversion (SPDC), either pulsed or continuously pumped, where the production of entangled pairs becomes probabilistic and the performance is inferior to a deterministic source of the same average pair production rate. Hence our model provides an upper bound to the performance of DDDL.
The pair distribution volume (PDV) for DDDL can be calculated by integrating Eq.~\ref{equ:DDDLentDistRate} over the duration of the overpass. DDDL quantum key distribution has been studied in~\cite{wang2026scalability,sidhu2026operational}.

\subsubsection{Satellite Quantum Repeater Entanglement Distribution.}

\begin{figure}[!tbh]
    \centering
    \includegraphics[width=0.75\linewidth]{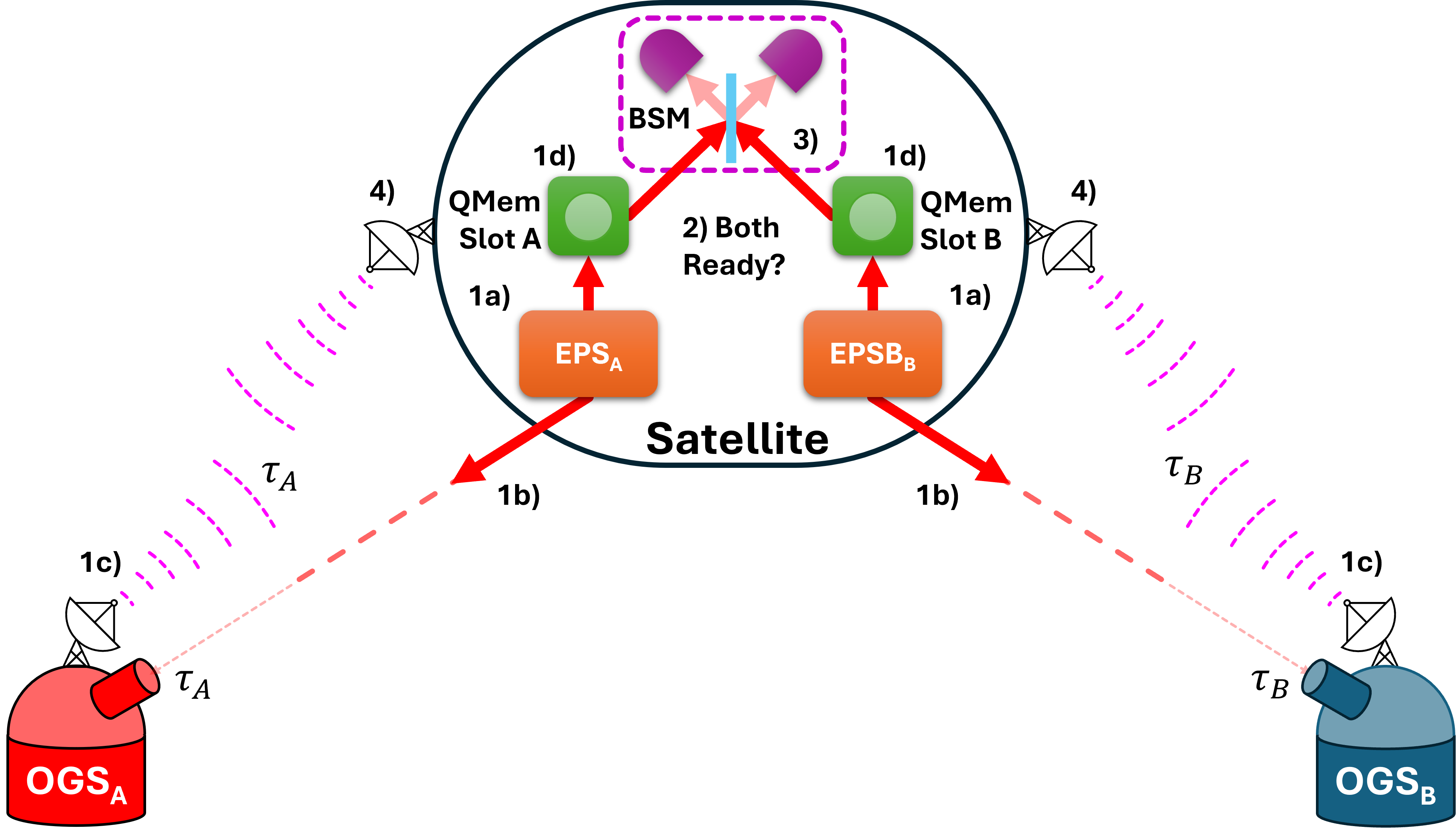}
    \caption{Single Satellite Quantum Repeater. 1) Each downlink independently establishes an entangled pair the satellite quantum memory and an OGS: 1a) An EPS generates an entangled pair of photons, one photon is stored in a satellite quantum memory mode slot (alternatively, a DLCZ-type-entanglement source~\cite{duan2001long} can be used);
    1b) The other photon is sent to the OGS with time-of-flight $\tau_{A,B}$;
    1c) The OGS notifies the satellite of transmission success/failure, with return signal time-of-flight $\tau_{A,B}$;
    1d) In the case of transmission failure, the memory is reset and a new entanglement distribution attempt is made. If the photon was received by the OGS, the protocol proceeds to the next stage.
    2) Once a link has been successful in establishing an entangled pair, it either waits for the other link to succeed, if it has not already been successful.
    3) Once both links are ready, the satellite performs entanglement swapping to link the two OGSs. We assume a success probability of 50\% for a linear optical Bell State Measurement (BSM).
    4) The swap result is conveyed to the OGSs allowing them to perform corrective operations or discard the attempt in the case of failure. The latency of this stage has no impact upon the operational rate.
    Each link can be parallelised using multiply memory mode slots, to provide a linear increase in the PDR.}
    \label{fig:SSR_protocol}
\end{figure}

Quantum repeaters have been proposed to address the challenge of distributing entanglement over high loss channels typical of long distance links~\cite{BriegelQRs}. This is achieved by splitting a large link into multiple smaller sublinks over which quantum transmission is possible and connecting them using quantum memories, classical communications and entanglement swapping~\cite{azumaRepeatersReview2023, munroInsideRepeaters2015} in order to overcome the direct transmission limit. Although satellite links exhibit much more favourable loss scaling than fiber links they still present a high-loss environment. For example, the Micius dual-downlink experiments reported losses of the order of $27\ \text{dB}$ or greater per downlink, or of the order of $>55 \text{dB}$ for the combined DDDL link loss. Satellite quantum memory repeaters have been proposed to improve the performance of the satellite for quantum communications~\cite{gundoganProposalSpaceMemories2020}\footnote{Ground-based quantum repeaters with satellite-borne EPSs was proposed in~\cite{boone2015entanglement} and considered in~\cite{ji2026global,dawar2026feasibility}.}.

The basic operation of a repeater satellite is shown in Fig.~\ref{fig:SSR_protocol}. Repeater protocols require classical communication between the satellite and each OGS, introducing an unavoidable communication latency leading to two important practical considerations. First, there is an immediate constraint on the minimum lifetime of a candidate memory platform. These latency times are determined by the two-way photon transit times which are on the order of $\text{ms}$ for a satellite in low Earth orbit (LEO). Second, the latency also imposes a bottleneck on the rate at which entanglement distribution can be attempted. This reduction to the PDR is compounded by the fact that the Bell state measurement (BSM) required to perform entanglement swapping will typically only succeed with a probability $p_\text{BSM}$~\cite{BSMRefdoi:10.1126/sciadv.adf4080}.
Although future technology may allow for deterministic BSMs, here we assume a conventional BSM with $p_\text{BSM} = 50\%$. 

\subsubsection{Initial Comparison of DDDL and SSQR.}

To illustrate the potential benefit a quantum memory could offer entanglement distribution, consider the case of static symmetric links with transmittance $\eta=10^{-3}$ to each of Alice and Bob. If the satellite has an onboard entangled pair source (EPS) with the Micius source rate of $5.9 \times 10^{6}\ \text{ pairs s}^{-1}$, the DDDL distribution rate to Alice and Bob is, $10^{-3}\cdot10^{-3} \cdot 5.9 \times 10^6 = 5.9\ \text{pairs s}^{-1}$.
With an unconstrained number of memory slots and using the same entanglement source rate, the satellite establishes pairs with Alice and Bob individually at a rate, $R_\text{A,B} = 10^{-3} \cdot 5.9 \times 10^6 = 5900\ \text{pairs s}^{-1}$.
If the BSM succeeds with probability $p_\text{BSM}$, then the SSQR rate is, $p_\text{BSM} \times  5900\ \text{pairs s}^{-1}$,
which is $3$ orders of magnitude higher than the rate without memories, up to $p_\text{BSM}$. This advantage arises because having memories on board removes the need for simultaneous success in dual-downlink transmission, i.e. splitting the simultaneous dual downlink into two independent single downlinks. As a result, the distribution rates scales as $\eta$ for SSQR compared to $\eta^2$ for DDDL.

\subsubsection{Memory Capacity Impact on SSQR Rate.}

The above assumes that the memory was not subject to any constraints in terms of its capacity and lifetime. In practice, quantum memories are limited in the total number of memory modes or slots. This affects the rate at which entanglement distribution can be attempted. For a single mode, since distribution can only be reattempted following classical signalling from the ground stations, the rate at which entanglement distribution can be attempted is,
\begin{equation}
    R_\text{single-mode}^\text{(SSQR)} = \frac{1}{2\tau_{A,B}} = \frac{c}{2L_{A,B}}.
    \label{eq:latencylimitedrate}
\end{equation}
For a low-Earth orbit (LEO) altitude $h=500\ \text{km}$, this round trip time has a minimum value of $\sim 3.3\ \text{ms}$ when the satellite is at zenith (i.e. directly above an OGS), leading to a maximum attempt rate of $R_\text{EPS}^\text{SSQR} \lesssim 300\ \text{s}^{-1}$.
For $\eta=10^{-3}$, the rate at which entanglement can be established on a single link would then only be $0.3 \text{ pairs s}^{-1}$. This process can be multiplexed with multi satellite memory modes for a linear improvement in rate. Going forward, we impose a finite capacity $N_\text{sat}$. We do not impose any limit on the memory capacity at the OGGs. We note that the EPS rate only needs to be the same as the satellite-OGS link attempt rate, potentially relaxing the requirements for photon and quantum memory bandwidth by a considerable amount.

We consider the Sender-Receiver protocol~\cite{Jones_2016} where the satellite assigns $N_A$ and $N_B$ modes of its memory to each of Alice and Bob ($N_A+N_B=N_\text{sat}$), respectively. The  average rate ($R_{A,B}$) of entanglement generation to each OGS and the overall entanglement distribution rate after entanglement swapping ($R^\text{SSQR}$) are,
\begin{equation}
    R_{A,B} =\frac{1}{2 \tau_{A,B}} \sum_{k = 1}^{N_{A,B}} \binom{N_{A,B}}{k}\eta_{A,B}^k (1-\eta_{A,B})^{N_{A,B}-k} k, \quad R^\text{(SSQR)} = p_\text{BSM} \text{ min}\{R_A, R_B\}.
    \label{equ:SRRate}
\end{equation}
Note that we do not include the effect of any processing time required on the satellite or ground station in Eq.~\ref{equ:SRRate}, including the time to latch photons into the ground station memory and perform the BSM. This approximation is valid when the photon transit times are dominant. Furthermore, distribution rates calculated with Eq.~\ref{equ:SRRate} assume that the satellite's memory is not limited by its lifetime. 

From the instantaneous PDR, the single overpass pair distribution volume (PDV) is,
\begin{equation}
    N_\text{PDV}^\text{(DDDL,SSQR)}=\int_{t_\text{start}}^{t_\text{end}}\ R^\text{(DDDL,SSQR)}(t)\ dt,
\end{equation}
where $t_\text{start}$ and $t_\text{end}$ denote the times between which the satellite is in simultaneous view above the minimum elevation at each OGS. In principle, a SSQR could begin to load up the memory slots of the first OGS in view prior to simultaneous visibility, but we neglect this possibility to simplify the model. The extreme case would be the use of long-duration quantum memories to courier entanglement between distant OGSs in a time-delayed quantum repeater protocol~\cite{gundoganJasBeyondLOS}.

A natural question would be how best to assign the memory slots to each of the Alice and Bob downlinks. The most general scheme would be where these allocations dynamically vary with time during an overpass and are adjusted based on link parameters, i.e. $N_A = N_A(t)$ and $N_B(t) = N_\text{tot} - N_A(t)$, as considered in~\cite{fittipaldiEntSwappingInOrbit2024}. Since dynamical memory allocation between downlinks may introduce experimental complexity, and to simplify the analysis, we consider a static memory allocation for the duration of an overpass, though this can vary on an overpass to overpass basis. The simplest choice is $N_A = N_B = N_\text{tot}/2$, though this may not be optimal for a given overpass geometry. For example, if an overpass passes directly overhead Alice and not Bob, meaning that $\eta_A(t)>\eta_B(t)$ throughout, then the PDV can be improved by assigning the $OGS_B$ link more onboard memory to compensate for its higher link loss. This can be understood from Eq.~\ref{equ:SRRate} where the overall $R^\text{SSQR}$ is maximised when the individual satellite-Alice $R_A$ and satellite-Bob $R_B$ rates are equalised\footnote{An extension would be to have a general pool of memory slots, each of which would be allocated to either OGS link on a per attempt basis, determined on which link had exhausted its stored entanglement. A more comprehensive exploration of memory allocation and swapping policies is left to future work.}. We therefore consider the case where the satellite employs an optimal static allocation of its memory, chosen to maximise the per-pass PDV. This would offer a balance between improving the distribution rates compared to an equal split without introducing significant extra experimental complexity. 

The communications latency that repeater protocols suffer from is reflected in the $1/2\tau$ term in Eq.~\ref{eq:latencylimitedrate}, which DDDL does not suffer. A memory equipped satellite will therefore only outperform DDDL when there is sufficiently high memory capacity to compensate for this latency. We define the crossover capacity, $N_c$, as the memory capacity required for SSQR to match DDDL performance. One could consider the instantaneous $N_c (t)$ corresponding to setting $R_\text{DDDL}(t)=R_\text{SSQR}(t)$, though here we consider $N_c$ on a per-overpass basis where $N_\text{PDV}^\text{DDDL}=N_\text{PDV}^\text{SSQR}$. The crossover capacity has direct relevance for requirements on space-borne memory platforms and in gauging the feasibility of candidate platforms. 

\subsection{Long-term PDV}

\begin{figure}[!tbp]
    \centering
    \includegraphics[width=0.6\linewidth]{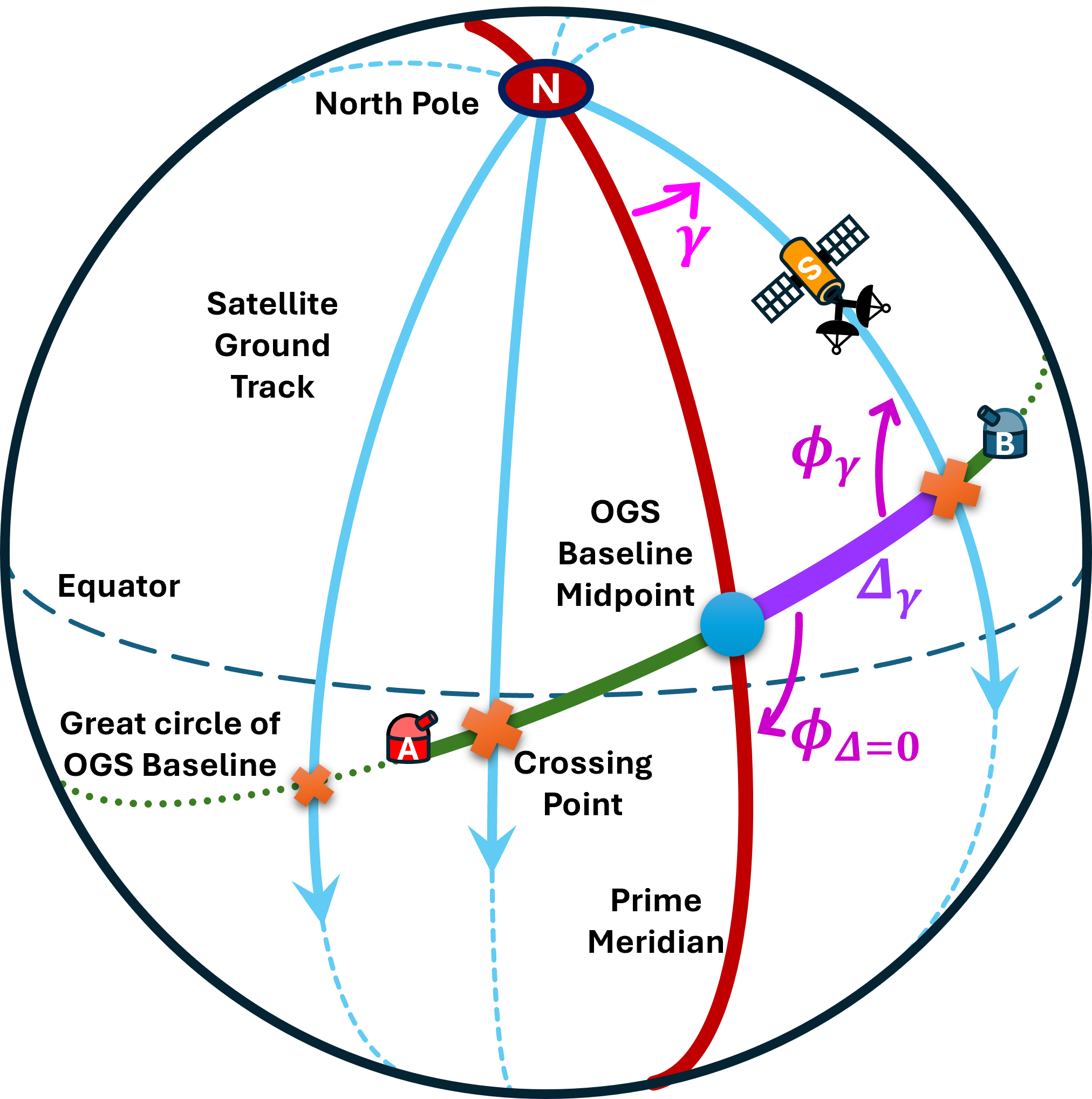}
    \caption{Expected Annual PDV. A polar orbit satellite, in the absence of Earth synchronism, will cross the OGS baseline great circle at randomly distributed longitudes. We only consider the North to South crossings, assuming these occur at local midnight for a noon-midnight SSO for which the polar orbit approximates. For convenience, we can take the prime meridian to be the longitude of the OGS baseline midpoint, the orbital angle is denote $\gamma$. At each crossing point, the overpass geometry is $\{\Delta_\gamma,\phi_\gamma\}$, hence PDV is then a function of $\gamma$. Over the course of year, the expected average can be determined from the integral of $N_{PDV}(\gamma)$ with respect to $\gamma$, suitably normalised and proportional to the number of orbits. Note that $N_{PDV}$ can be positive for crossing points outside of the OGS baseline (e.g. left most crossing point) if there is sufficient line of sight.}
    \label{fig:annualPDVDiagram_revised}
\end{figure}

The per-overpass PDV can depend on the overpass geometry that varies from overpass to overpass. Long-term term entanglement distribution performance is a more representative metric. We consider the expected annual PDV for a Noon-Midnight Sun synchronous orbit (SSO), assuming night-time only operation. We will approximate an SSO by a polar orbit, valid for OGS locations away from the North and South poles~\cite{sidhu2021finitekeyeffectssatellite}. For a circular polar satellite (Fig.~\ref{fig:annualPDVDiagram_revised}), overpasses occur along a meridian of fixed longitude $\gamma$. The prime meridian can be taken as the OGS baseline midpoint. The crossing point of the satellite ground track will define an overpass geometry $\Delta_\gamma$ and $\phi_\gamma$, and a corresponding per-pass PDV, $N_{PDV}(\Delta_\gamma, \phi_\gamma)=N_{PDV}(\gamma)$ pairs.

Over the course of the year, the set of $\gamma$ for each orbit can considered to be uniformly and randomly distributed, assuming that there is no Earth synchronism (some constellation designs do employ simultaneous Sun and Earth synchronism~\cite{mazzarella2020quarc}). We note that many orbits with longitude $\gamma$ will result in $N_{PDV}(\gamma)=0$ if the ground track does not approach close enough to both OGSs simultaneously. The average PDV per orbit, $\bar{N}_{\gamma}$, can be found by averaging over $\gamma$,
\begin{equation}
    \bar{N}_{\gamma} =  \frac{1}{2\pi} \int_0^{2\pi} N_{PDV}(\gamma)\ d\gamma.
    \label{equ:avgOverpassPairs}
\end{equation}
Scaling $\bar{N}_{\gamma}$ by the number of overpasses in a year ($N_\text{annual-overpasses}$) then gives an estimate of the total average annual PDV, $\bar{N}_\text{year}$.
As illustrated in Fig.~\ref{fig:annualPDVDiagram_revised}, overpasses can be characterised as having either a North-South or South-North trajectory. For an SSO satellite, which can be approximated as being polar, only one of these  will occur at night. The assumption that entangled pair distribution only occurs on the midnight overpass of a Noon-Midnight SSO can be incorporated into the model by only considering overpasses with $0\le\phi_\gamma\le\pi$, for example. This restriction enforces that only either North-South or South-North overpasses are counted.

\subsection{Fidelity of Distributed Entanglement}
\label{sec:entFidelityAndMemNoise}
In addition to the number of entangled pairs is it also important to consider the quality of the distributed entanglement, for example as quantified by entanglement measures~\cite{10.5555/2011706.2011707, Horodecki_2009QuantumEntanglement}. Entanglement measures are typically defined in the asymptotic limit and quantify how well a state can be used for a certain operational task. However, measures defined in this way are typically computationally intractable and are less useful for the characterisation of a small numbers of states that will be typical for near-term satellite quantum networks.

To quantify the quality of the entangled pairs distributed by the satellite we choose the single-copy entanglement fidelity of a 2-qubit state $\hat{\rho}_{AB}$,
\begin{equation}
    F =  _{AB}\bra{\Phi}\hat{\rho}_{AB}\ket{\Phi}_{AB},
\end{equation}
where $\ket{\Phi^+}_{AB}=\frac{1}{\sqrt{2}}\left(\ket{00}+\ket{11}\right)_{AB}$. In principle any maximally entangled state could be used in place of $\ket{\Phi^+}$, since all such states can be related by local transformations. The fidelity is a suitable metric to use in the context of finite block sizes. It allows us to evaluate the quality of the distributed pairs individually without appealing to an asymptotic limit. 
A fidelity of $1$ represents perfect entanglement with no noise, and it can be shown that if a two-qubit state has $F > 1/2$ then this implies that it is entangled~\cite{horoSingleFractionReference}. Furthermore, different quantum technology applications typically have fidelity thresholds (see e.g. Table $1$ in~\cite{de_Forges_de_Parny_2023}). 

To simplify the analysis, we assumed that the satellite produces perfect Bell pairs on board and that any fidelity loss due to transmission through the atmosphere and subsequent reception and storage on the ground is negligible (ignoring background light). Under these assumptions the only source of fidelity degradation is decoherence of photon stored in the satellite quantum memory whilst awaiting the ``success'' signal from the OGSs. Different quantum memory types have varying storage and retrieval characteristics, here we will assume that storage and retrieval efficiency is ideal but introduce a non-zero dephasing rate of a stored qubit state $\hat{\varrho}$ modelled as~\cite{Walln_fer_2022}, 
\begin{equation}
\label{equ:memoryNoiseChannel}
    \hat{\varrho} \mapsto (1 - \lambda(t)) \hat{\varrho} + \lambda(t) \hat{Z} \hat{\varrho} \hat{Z},\qquad \lambda(t) = \frac{1 - \exp(-t/\tau_\text{mem})}{2}
\end{equation}
after storage time $t$, $\tau_{mem}$ is the $1/e$ dephasing time of the memory, and $\hat{Z}$ is the Pauli Z-operator.

The repeater satellite performs entanglement swapping on qubits stored in its memory after Alice and Bob each confirm successful photon transmission. If the satellite-Alice and satellite-Bob memory qubits have been stored for times $t_A$ and $t_B$ respectively, then the Alice-Bob pair after swapping can be shown (Appendix~\ref{appendix:MemoryNoiseModel}) to be a Bell-diagonal state with fidelity
\begin{equation}
\label{equ:finalStateFidelity}
    F(t_A, t_B) = \lambda(t_A)\lambda(t_B)+(1-\lambda(t_A))(1-\lambda(t_B)).
\end{equation}

The development of memories that combine long lifetimes with on-demand photon emission and high levels of multimodality is of critical importance for the development of scalable repeater applications. Storage times on the order of microseconds have been reported~\cite{mustafaMicrosecondLifetime}, as well as times of $1\  \text{ms}$, through the use of clock states robust against decoherence caused by residual magnetic fields~\cite{ZhaoClockStates}. Dynamical decoupling techniques have allowed for storage times of six hours in rare-earth memories~\cite{ZhongSixHourCoherence}. Storage of $50$ and $100$ simultaneous modes has been demonstrated in atomic frequency comb memories~\cite{AFC50100Modes}, with capacities of $59$~\cite{RE59cap} and $1650$~\cite{AFC1650} demonstrated in rare-earth memories. However, no platform that combines on-demand emission, long storage times, high degrees of multiplexing and the low size, weight, and power (SWaP) requirements that would be necessary for satellite deployment have yet been demonstrated.

\subsection{Monte Carlo Model of Quantum Repeater Operation}
\label{sec:MCMethods}

The fidelity in Eq.~\ref{equ:finalStateFidelity} is not linear in the waiting time. To fully characterise the fidelity distribution it is therefore necessary to access the waiting time statics of memory qubits, motivating our use of MC methods. These distributions will inform future work on strategies involving entanglement purification. Although computationally expensive, MC methods are well suited to modelling complex probabilistic systems. They are a natural choice for modelling quantum repeaters and have been used in previous research (e.g.~\cite{Walln_fer_2024, niu2023all, Brand2020WaitingTimeFidelity}). The primary contribution of this work is extending such analysis to dynamic satellite links.

In the MC model, we keep track of the internal state of the satellite's onboard quantum memory throughout the overpass. The loading and swapping time stamps of each of the stored photons on board the satellite are recorded. From the stored waiting times the final fidelity of the distributed pairs between Alice and Bob are calculated using Eq.~\ref{equ:finalStateFidelity}.

\subsubsection{Memory Management and Swapping Policies.}

\begin{figure}[!tbp]
\centering
\newcommand{\figwidth}{0.48\textwidth}
\newcommand{\figheight}{0.15\textheight}
\begin{subfigure}{\figwidth}
\centering
\includegraphics[height=\figheight]{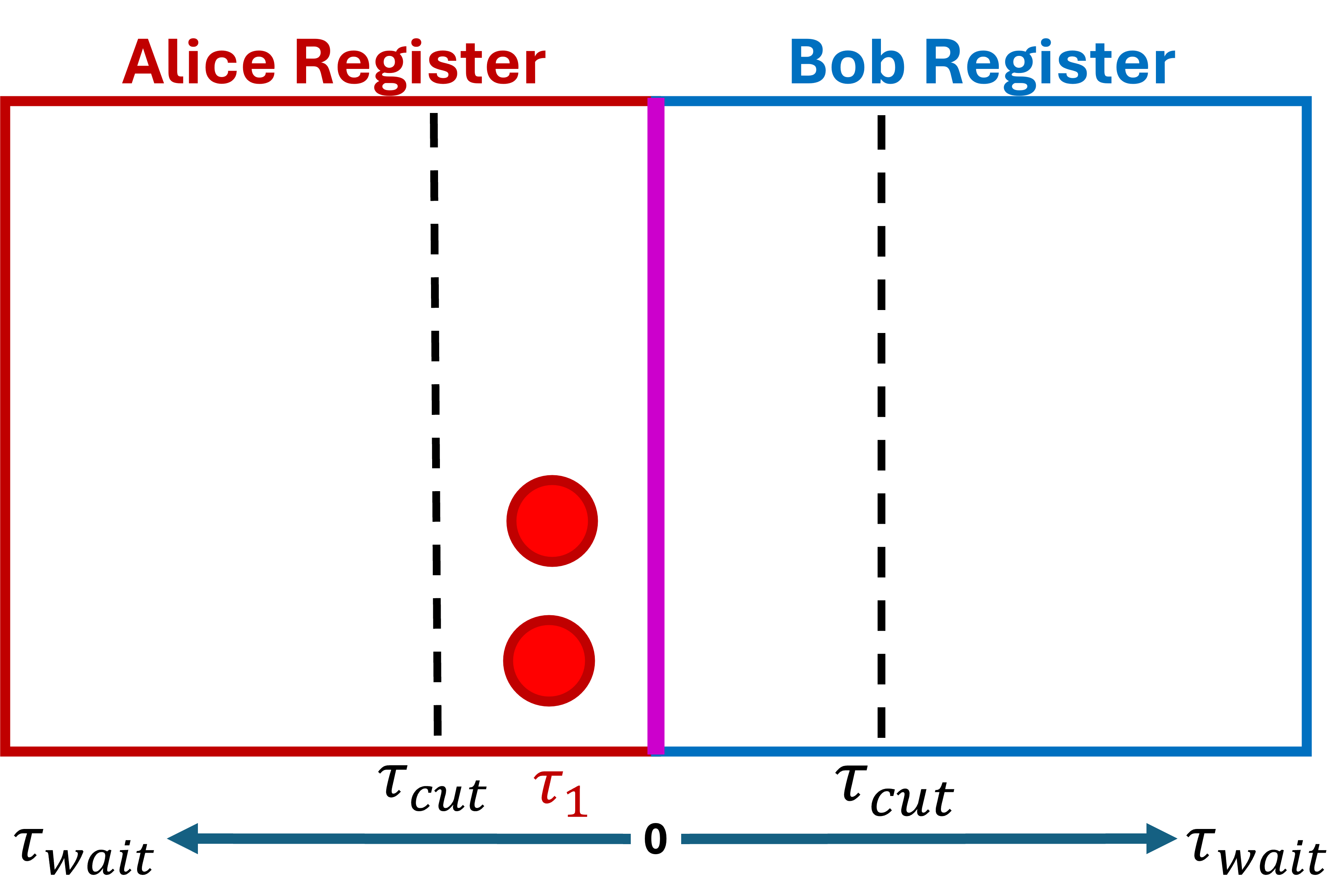
}
\caption{At the start of a round, the satellite may have memory qubits leftover from a previous time step. For this example, these are in the Alice register with current waiting time $\tau_1$. The satellite attempts transmission to the ground stations with any vacant memory slots. }
\label{fig:step1}
\end{subfigure}
\ 
\begin{subfigure}{\figwidth}
\centering
\includegraphics[height=\figheight]{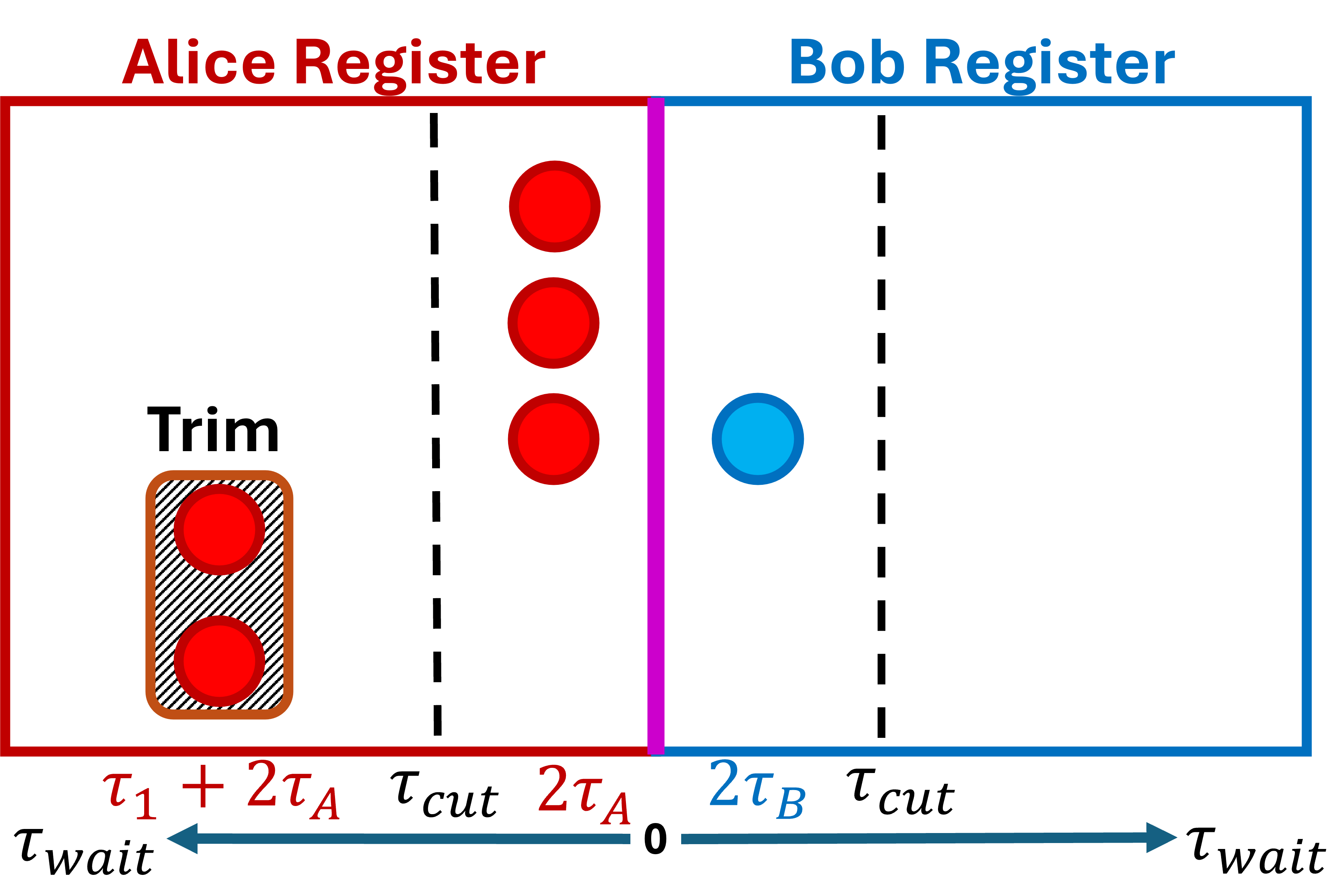
}
\caption{Entanglement swapping is performed when the ground stations confirm successful qubit transmission. Before performing swapping the satellite discards any pairs that have been stored longer than $\tau_\text{cut}$.}
\label{fig:step2}
\end{subfigure}\\
\begin{subfigure}{\figwidth}
\centering
\includegraphics[height=\figheight]{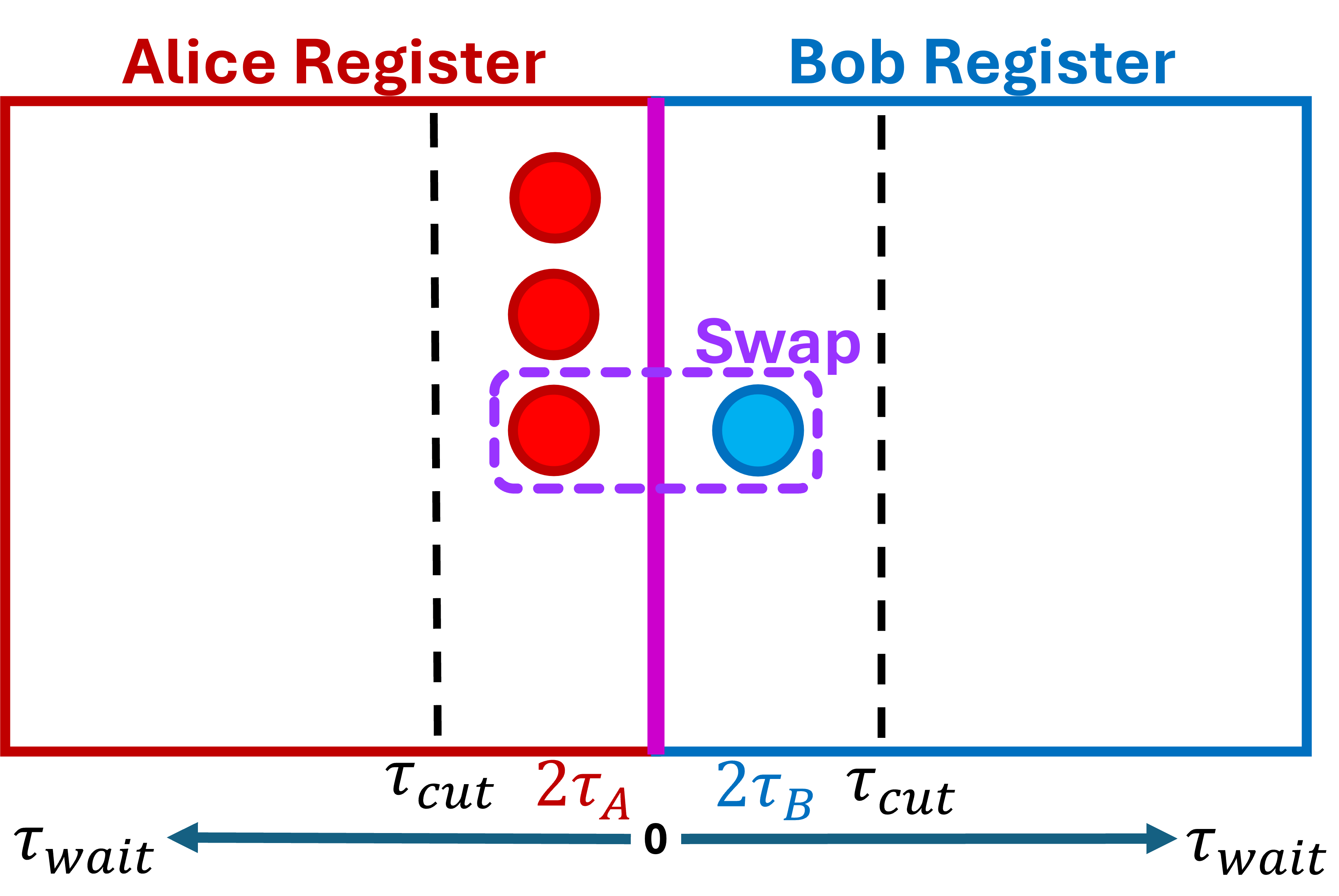
}
\caption{Following trimming the satellite performs swapping, using the youngest qubits.}
\label{fig:step3}
\end{subfigure}
\ 
\begin{subfigure}{\figwidth}
\centering
\includegraphics[height=\figheight]{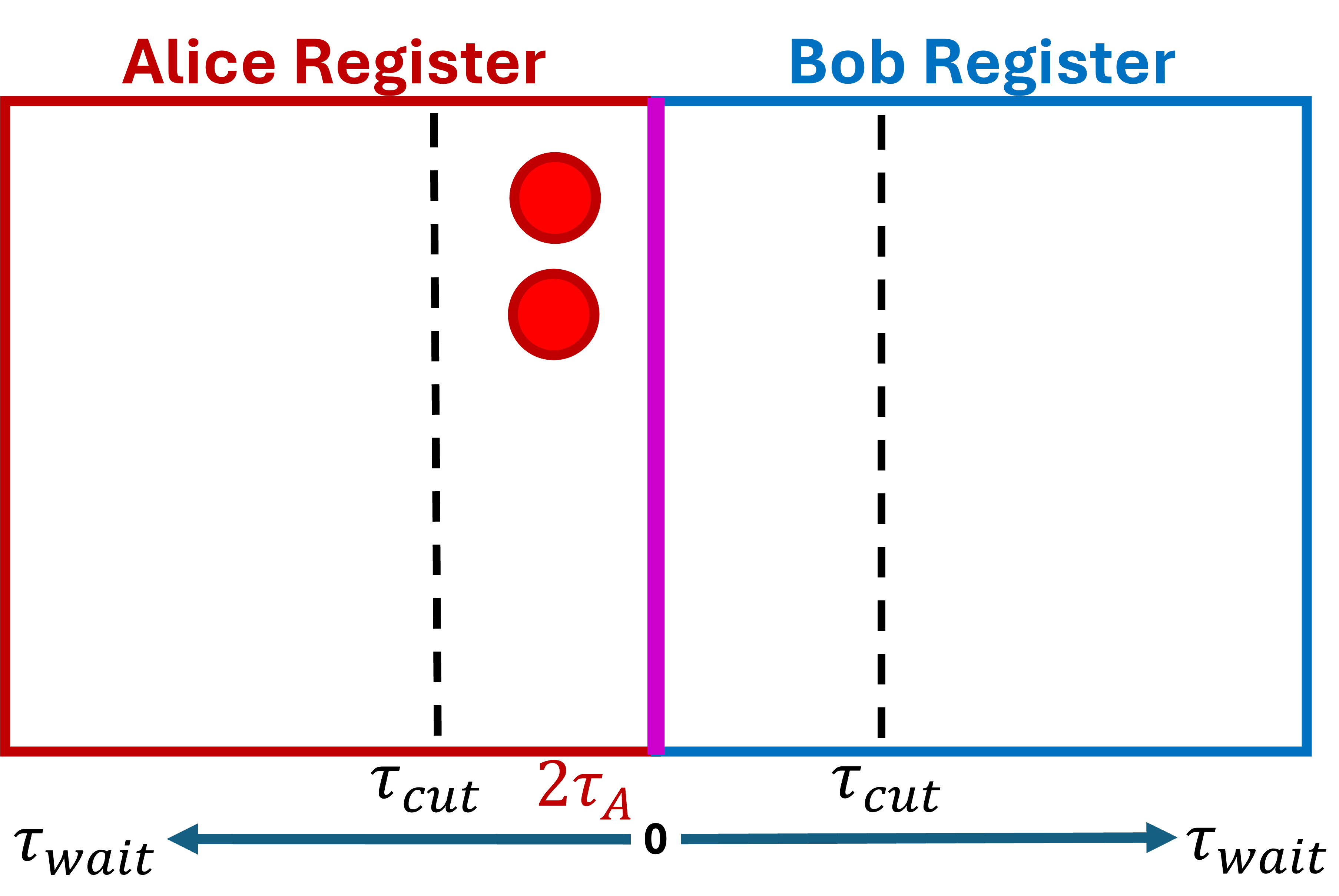
}
\caption{Following trimming and swapping the process is repeated.}
\label{fig:step4}
\end{subfigure}
\caption{Memory management strategy. The satellite memory is split into separate registers for Alice (Red) and Bob (Blue). The horizontal axis indicates the current waiting of memory qubits and increases moving away from the centre, with the vertical dotted line indicating the cut-off time $\tau_\text{cut}$. Alternatively, it is possible to trim memory slots based on the number of stored pairs. In this case step (b) instead involve discarding until a fixed number remains (a `buffer-size' memory management policy). The choice of whether to trim before or after swapping offers a trade-off between final state fidelity and the total number of pairs. }
\label{fig:memoryStrategy}
\end{figure}

There is freedom in how to manage the memory slots on the satellite, namely what to do with unpaired stored photons and the choice of which photons to swap once both links have successes. If a qubit has been stored for too long and its entanglement with the OGS has degraded, it may be preferable to discard it and attempt to refresh it. If the satellite has multiple qubits that are candidates for swapping, it must decide which ones to use (Fig.~\ref{fig:memoryStrategy}.).

To address these considerations, we assume that the satellite swaps pairs as soon as possible and always chooses the youngest (i.e. most recently confirmed) memory qubits to use in a swapping operation (Fig.~\ref{fig:memoryStrategy}c), corresponding to a `newest qubit first' policy \cite{PerezCastro2024Simulation}. This choice is motivated by the fact that the fidelity of the post-swapping pair decreases as both the Alice and Bob storage times increase. Choosing the most recent memory qubits to use in swapping therefore maximises the fidelity of the resulting pair. Before swapping the satellite trims its memory by discarding any qubits whose storage times have exceeded a cut-off time $\tau_\text{cut}$ (Fig.~\ref{fig:memoryStrategy}b). We have chosen to trim memory based on storage times rather than the number of stored qubits\footnote{It is also possible to trim memory qubits based on the number of stored pairs. In such a `buffer-size' memory management policy the satellite instead discards pairs until only a predetermined number remain, rather than trimming based on storage time. See Appendix~\ref{appen:bufferBasedMemoryManagement}.}. Trimming is performed before swapping to maximise the fidelity of the final pair. The use of cut-off times has been considered in previous studies on repeaters operating with finite lifetimes~\cite{cutoffs_I_esta_2023, Walln_fer_2022, cutoff_Rozpedek_2018}. By employing a cut-off, the satellite discards qubits that have been too degraded by memory noise, allowing these emptied slots to be re-employed in fresh attempts at entanglement distribution with its corresponding OGS.

When a memory qubit is used in swapping its storage time is recorded. These storage times are then used to construct the qubit waiting distributions. For a given overpass, this depends on $N_\text{sat}$ and $\tau_\text{cut}$. Once the distribution in waiting times has been obtained the fidelity distribution is constructed from it using Eq.~\ref{equ:finalStateFidelity}. 

\section{Results and Discussion}
\label{Results}

We concentrate on how varying overpass geometries affects the entanglement distribution performance of DDDL and SSQR. We calculate the per-pass PDV (Section~\ref{sec:perPassPDV}) of a satellite for different overpass geometries, which allows for an estimation of the memory resources required for a repeater satellite to match DDDL. Different optimal overpass geometries for DDDL and repeater satellites are established. The effect of changing baseline system loss values is also considered, as well as a comparison of a repeater satellite adopting an optimal static and equal split of its memory.

The long-term performance of the satellite is evaluate in terms of the average annual PDV (Section~\ref{sec:annualPDV}). This is done for a DDDL satellite with Micius source rate and a repeater satellite with $N_\text{sat} = 200$, with a baseline system loss of $\eta_\text{loss}^\text{sys}=25.9\ \text{dB}$. Higher optimal altitudes for the repeater satellite compared to DDDL are shown. The relative benefit of adopting an optimal static memory allocation compared to an equal one is shown to depend on the inclination of the OGS baseline with respect to the equator.

Although the PDV is a useful metric it does not quantify the quality of the distributed pairs. To address this, we impose finite memory lifetimes on the satellite and consider dephasing noise in the quantum memory. Evaluating the quality of the pairs then requires accessing the waiting time statistics of the qubits for which we use a round-based MC model of the repeater satellite. The waiting times for different overpasses and their dependence on $N_\text{sat}$ and memory management are investigated. We then give the fidelities of the distributed pairs as a function of overpass time, as well as their spread in the final ensemble established to OGSs at the end of the overpass.

\subsection{Per-Overpass PDV}
\label{sec:perPassPDV}

\begin{figure}[!t]
\centering
        \includegraphics[width=\linewidth]{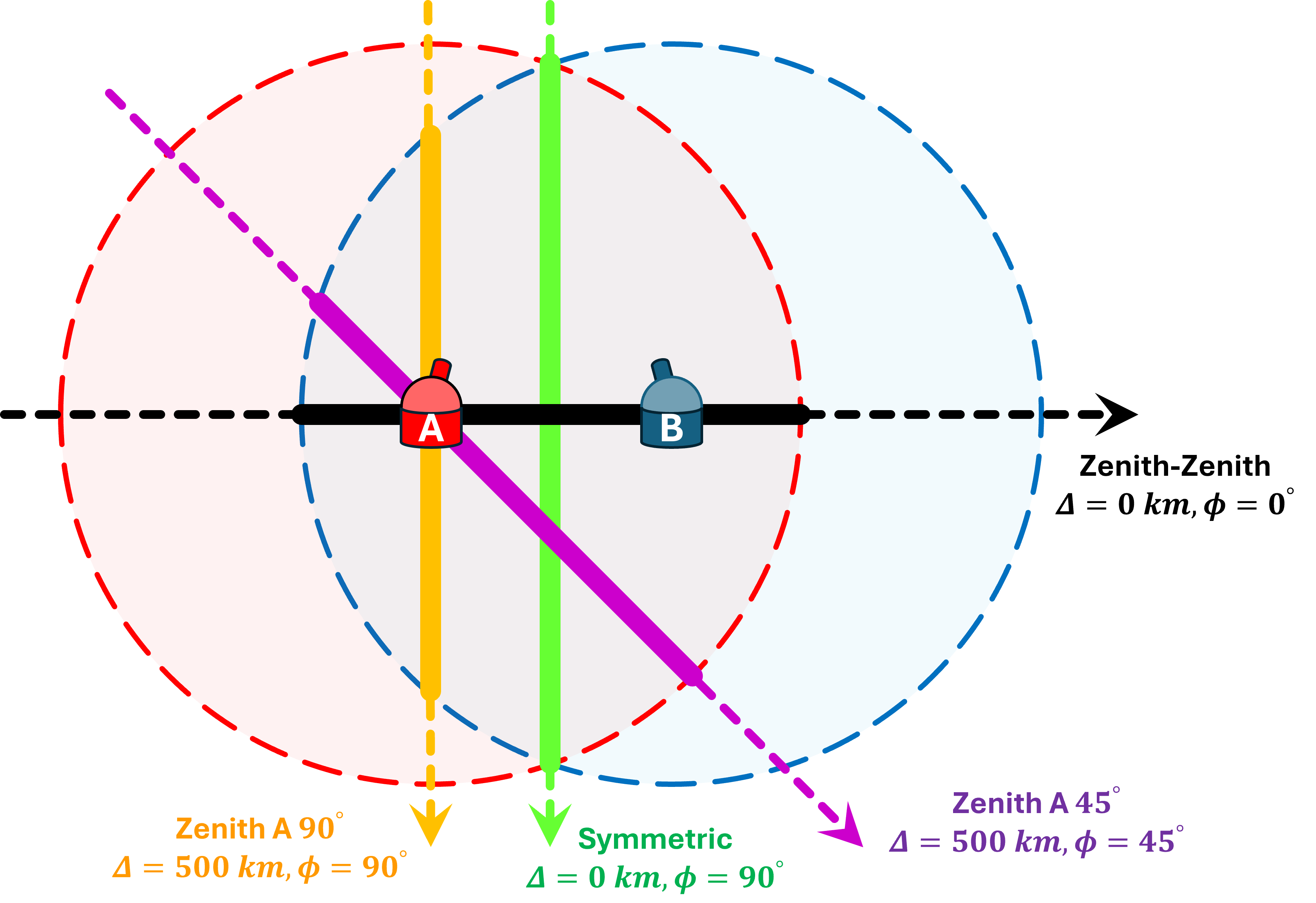}
        \caption{Representative overpass geometries. The dashed circles centred on Alice and Bob show their individual visibility regions. Their overlap defines the dual-visibility region, in which the satellite can attempt to distribute entanglement to the ground stations. The dashed arrows show the ground tracks of the satellite trajectories. Bolded segments are where dual-visibility is satisfied and it is the time intervals corresponding to these that defines an overpass.}
        \label{fig:4passesDiagram}
\end{figure}

\begin{figure}[!tbp]
\centering
\newcommand{\figwidth}{0.45\linewidth}
\newcommand{\figheight}{0.2\textheight}
\begin{subfigure}{\figwidth}
\centering
\includegraphics[width=\linewidth]{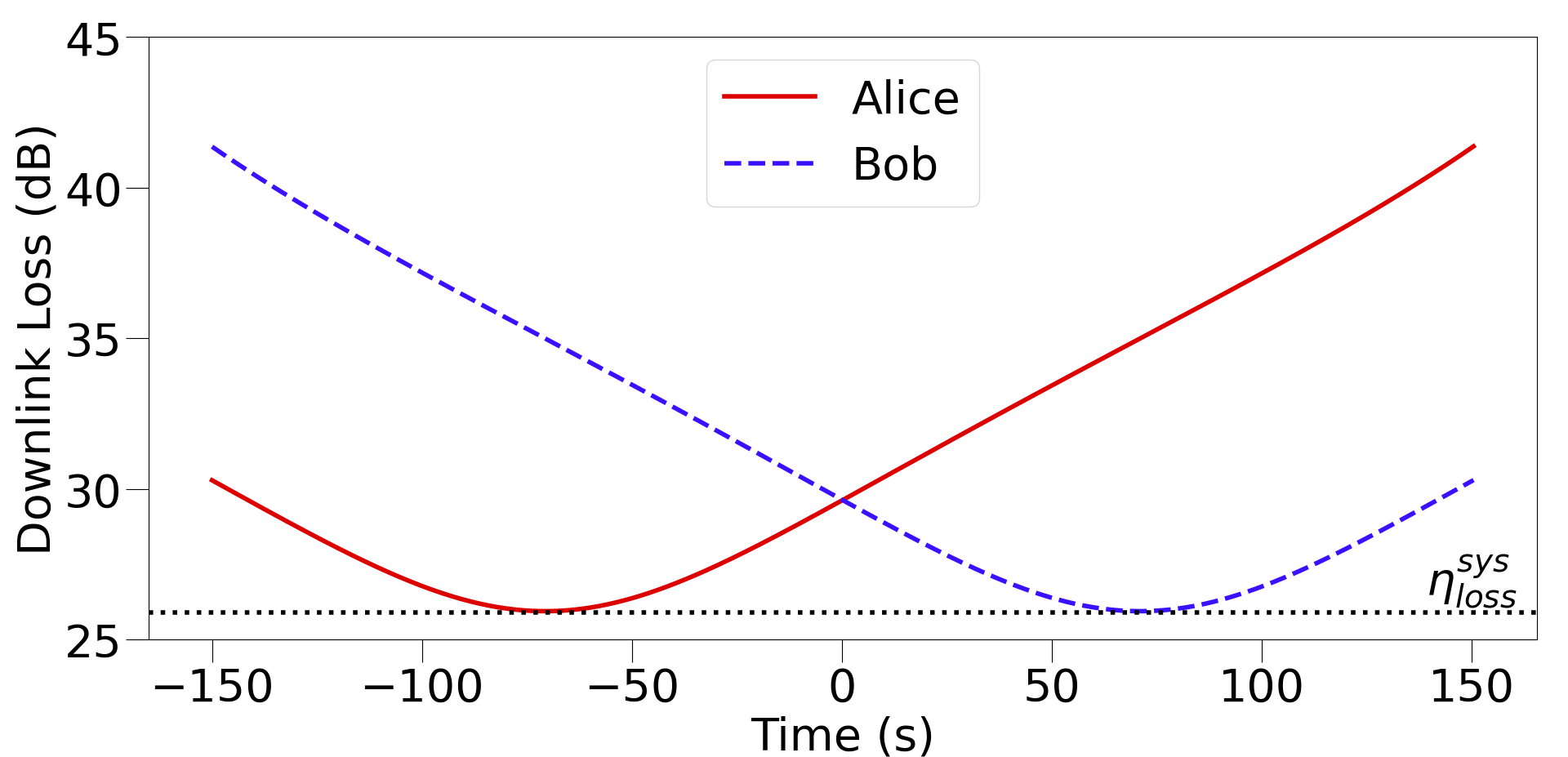}
\caption{Zenith-zenith ($\Delta = 0\ \text{km},\phi = 0^\circ$).}
\end{subfigure}
\hfill
\begin{subfigure}{\figwidth}
\centering
\includegraphics[width=\linewidth]{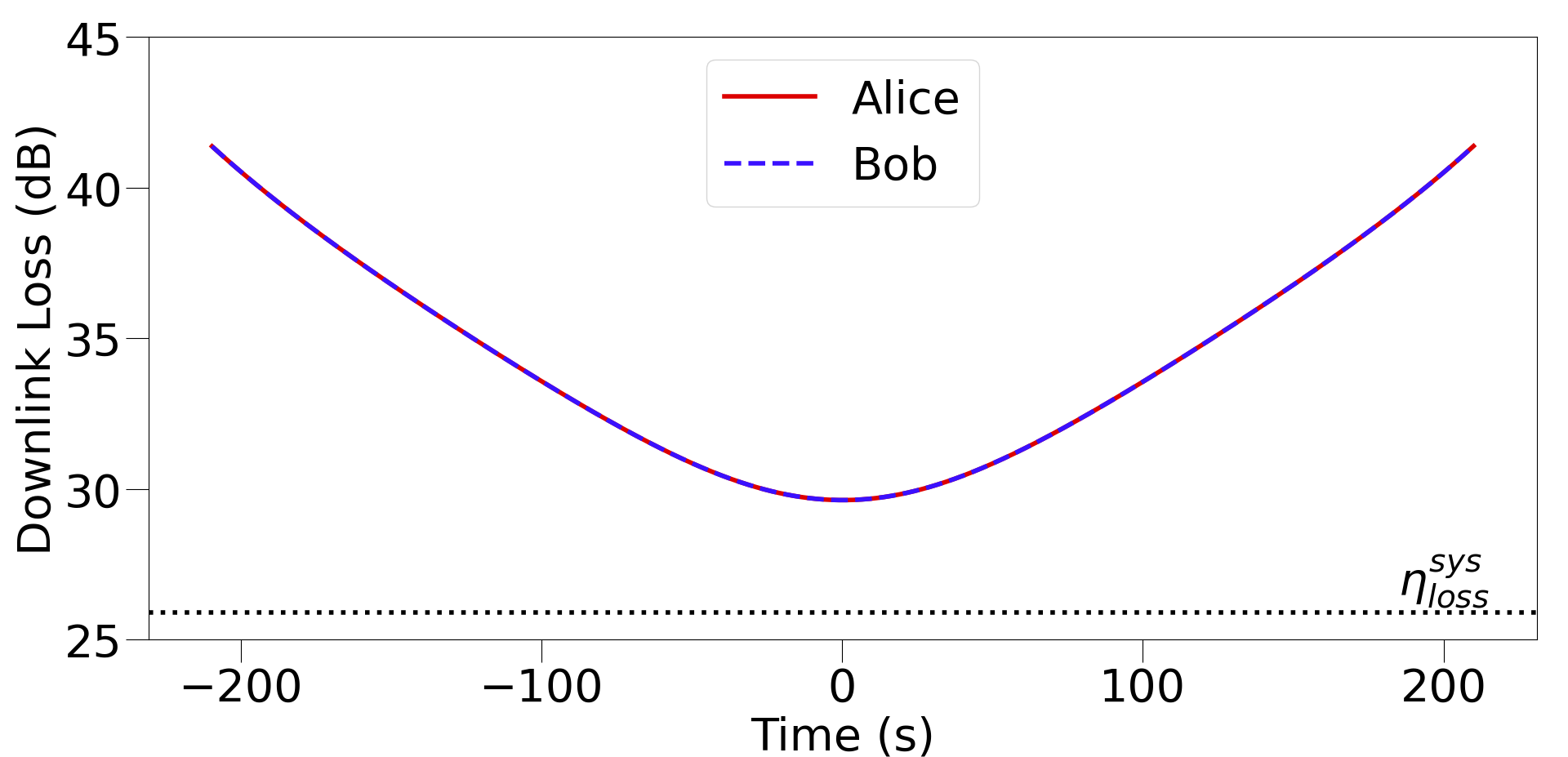}
\caption{Symmetric ($\Delta = 0\ \text{ km}, \phi = 90^\circ$).}
\end{subfigure}
\begin{subfigure}{\figwidth}
\centering
\vspace{12pt}
\includegraphics[width=\linewidth]{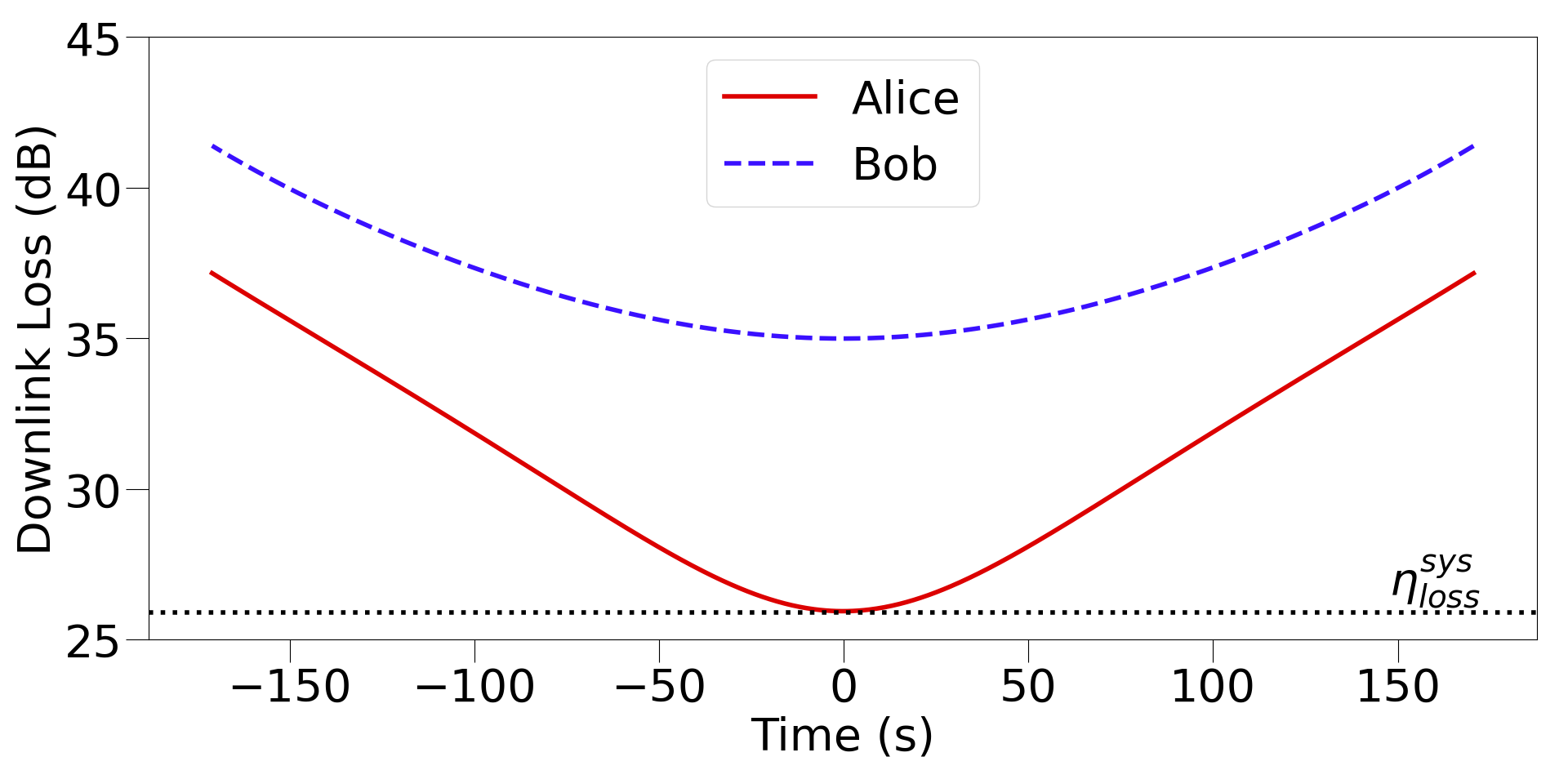}
\caption{Zenith A $90^\circ$ ($\Delta = 500 \text{ km}, \phi = 90^\circ$).}
\end{subfigure}
\hfill
\begin{subfigure}{\figwidth}
\centering
\vspace{12pt}
\includegraphics[width=\linewidth]{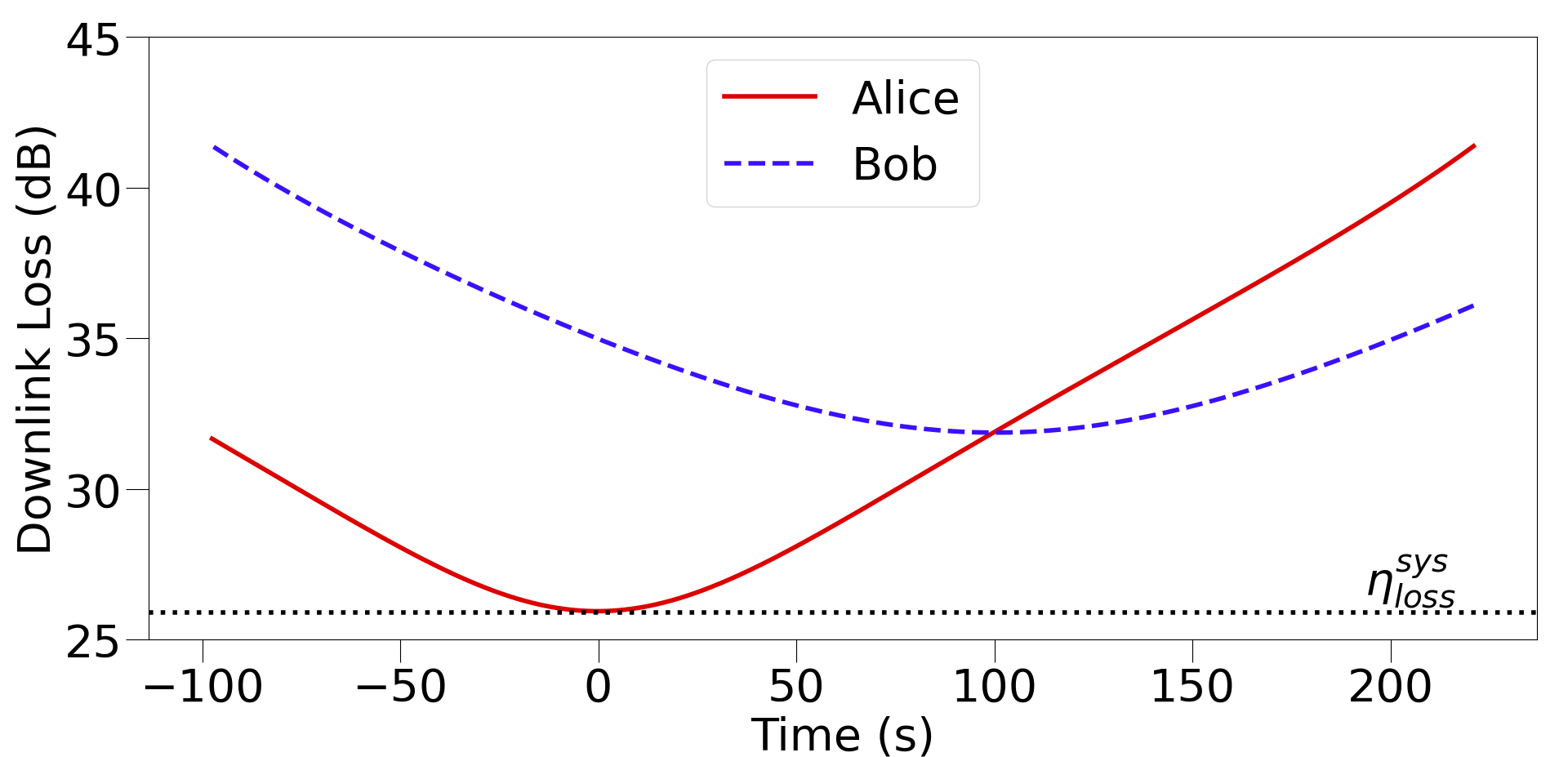}
\caption{Zenith A $45^\circ$ ($\Delta = 500 \text{ km}, \phi = 45^\circ$).}
\end{subfigure}
\caption{Individual channel losses for the representative overpasses: (a) Zenith-zenith, (b) Symmetric, (c) Zenith A $90 ^\circ$ and (d) Zenith A $45^\circ$. The $t = 0$ s point is when the satellite's ground track intersects the OGS baseline or the mid-point for the Zenith-zenith overpass. The dotted line indicates $\eta^\text{sys}_\text{loss}$ where the single-OGS link loss is at its minimum possible value. }
\label{fig:representativeLosses}
\end{figure}

\begin{table}[!tbp]
\centering
\begin{tabular}{lcl}
\toprule
Description & Parameter & Value \\ 
\midrule
OGS baseline length & $d$ & $1000$ km \\ 
Orbital altitude & $h$ & $500$ km\\ 
Lifetime of satellite memories & $\tau^\text{sat}_\text{mem}$ & $\infty$ \\ 
Minimum elevation angle for visibility & $\theta_\text{min}$ & $10^\circ$  \\ 
Bell state measurement success probability & $p_\text{BSM}$ & $\frac{1}{2}$  \\ 
DDDL entangled pair source rate & $R_\text{EPS}^\text{DDDL}$ & $5.9 \times 10^6$ pairs s$^{-1}$\\  
Source wavelength & $\lambda$ & $780$ nm \\
Atmospheric transmittance at zenith & $\eta_{atm}$ & 0.79\\ 
Transmitter Aperture diameter $T_X$& $D_t$ & $100\ \text{mm}$ \\
Transmitter Gaussian beam waist & $w_0$ & $45\ \text{mm}$ \\
Receiver aperture diameter $R_X$& $D_r$ & $1000\ \text{mm}$ \\
Intrinsic loss & $\eta_{int}^\text{dB}$ & $10\ \text{dB}$ \\
System loss metric & $\eta^\text{sys}_\text{loss}$ & $25.9\ \text{dB}$\\
\bottomrule
\end{tabular}
\caption{Baseline system parameters. The $R_\text{EPS}^\text{DDDL}$ rate matches Micius, the source rate for SSQR varies depending on the time-of-flight latency and $N_{A,B}$. The intrinsic loss incorporates imperfect beam quality, beam wander, and internal transmitter and receiver losses. Beam truncation (clipping) of the Gaussian beam has been included in the diffraction loss. Deterministic entangled photon pair sources and perfect photon detection at the OGSs are assumed.}
\label{table:IIIAParameters}
\end{table}


To evaluate the effect of overpass geometry on the PDV, we begin by examining four representative overpass geometries (Fig.~\ref{fig:4passesDiagram}): Zenith-Zenith; Symmetric; Zenith A $90^\circ$; and Zenith A $45^\circ$. Their corresponding downlink losses are shown in Fig.~\ref{fig:representativeLosses} with the system parameters as in Table~\ref{table:IIIAParameters}. For each overpass we choose the $t = 0$ point to correspond to the intersection of the satellite ground track and OGS baseline, or the midpoint for the zenith-zenith overpass. The per-pass PDV as a function of $N_\text{sat}$ and the effect of changing channel efficiencies are also considered. We note that all distribution rates and PDVs in this section are obtained under the assumption that the satellite memory is not constrained in its lifetime, only its capacity. 

\begin{figure}[!tbp]
\centering
\newcommand{\figwidth}{0.45\linewidth}
\newcommand{\figheight}{0.2\textheight}
\begin{subfigure}{\figwidth}
\centering
\includegraphics[width=\linewidth]{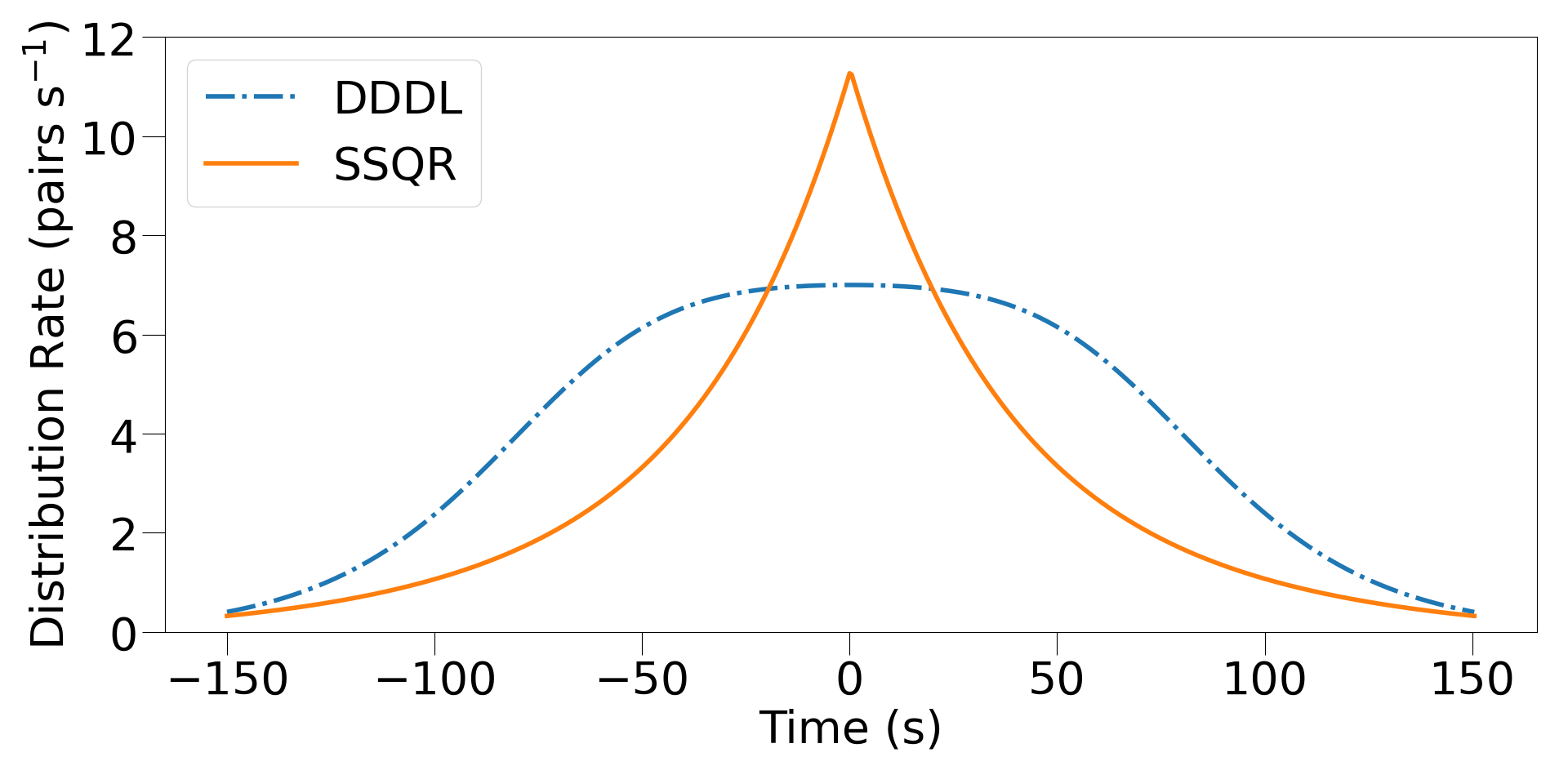}
\caption{Zenith-zenith ($\Delta = 0\ \text{km},\phi = 0^\circ$). Total $\text{PDV}_\text{DDDL, SSQR} = 1230, 911$.}
\end{subfigure}
\hfill
\begin{subfigure}{\figwidth}
\centering
\includegraphics[width=\linewidth]{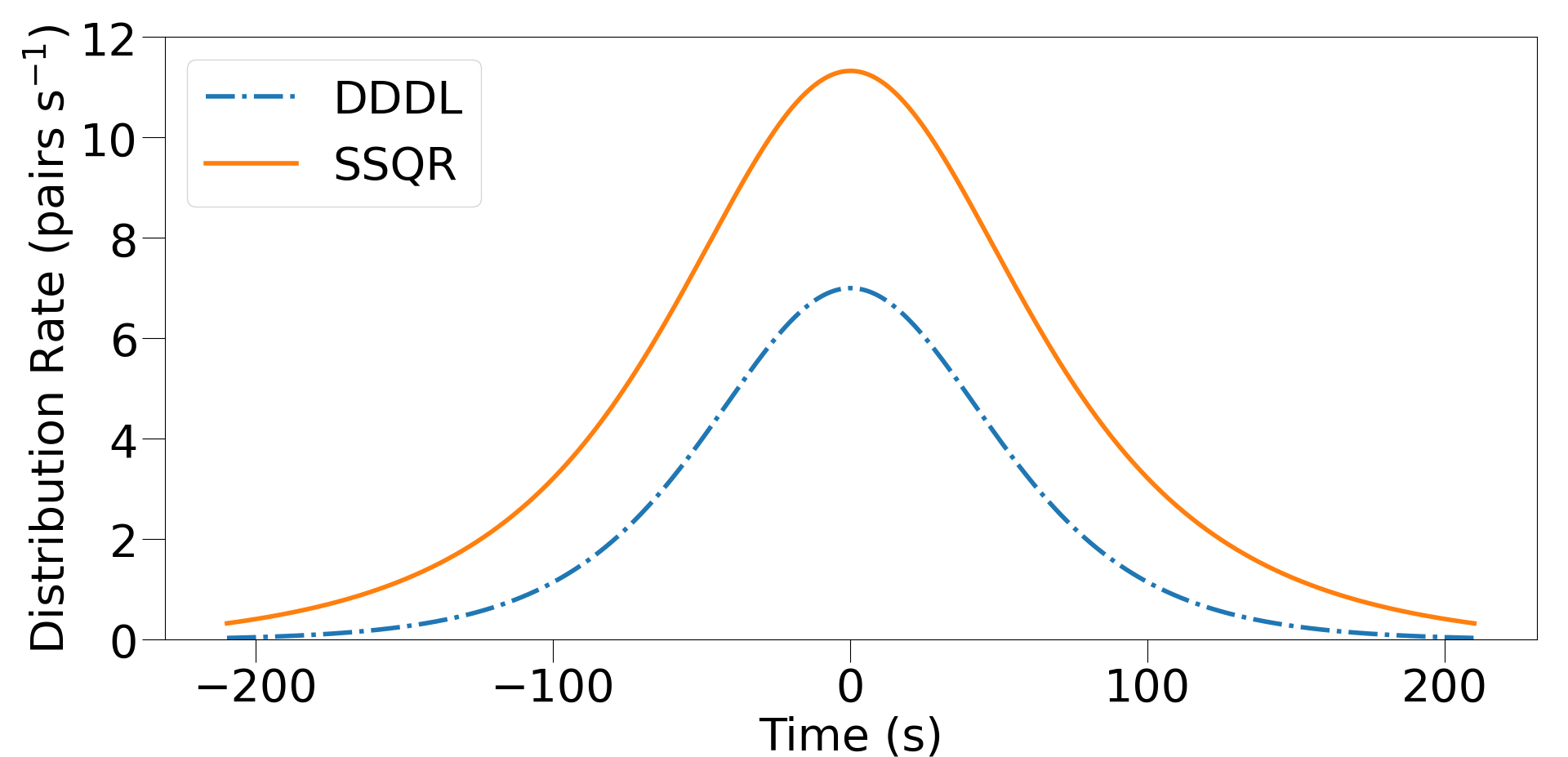}
\caption{Symmetric ($\Delta = 0\ \text{ km}, \phi = 90^\circ$). Total $\text{PDV}_\text{DDDL, SSQR} = 893, 1801$.}
\end{subfigure}
\begin{subfigure}{\figwidth}
\centering
\vspace{12pt}
\includegraphics[width=\linewidth]{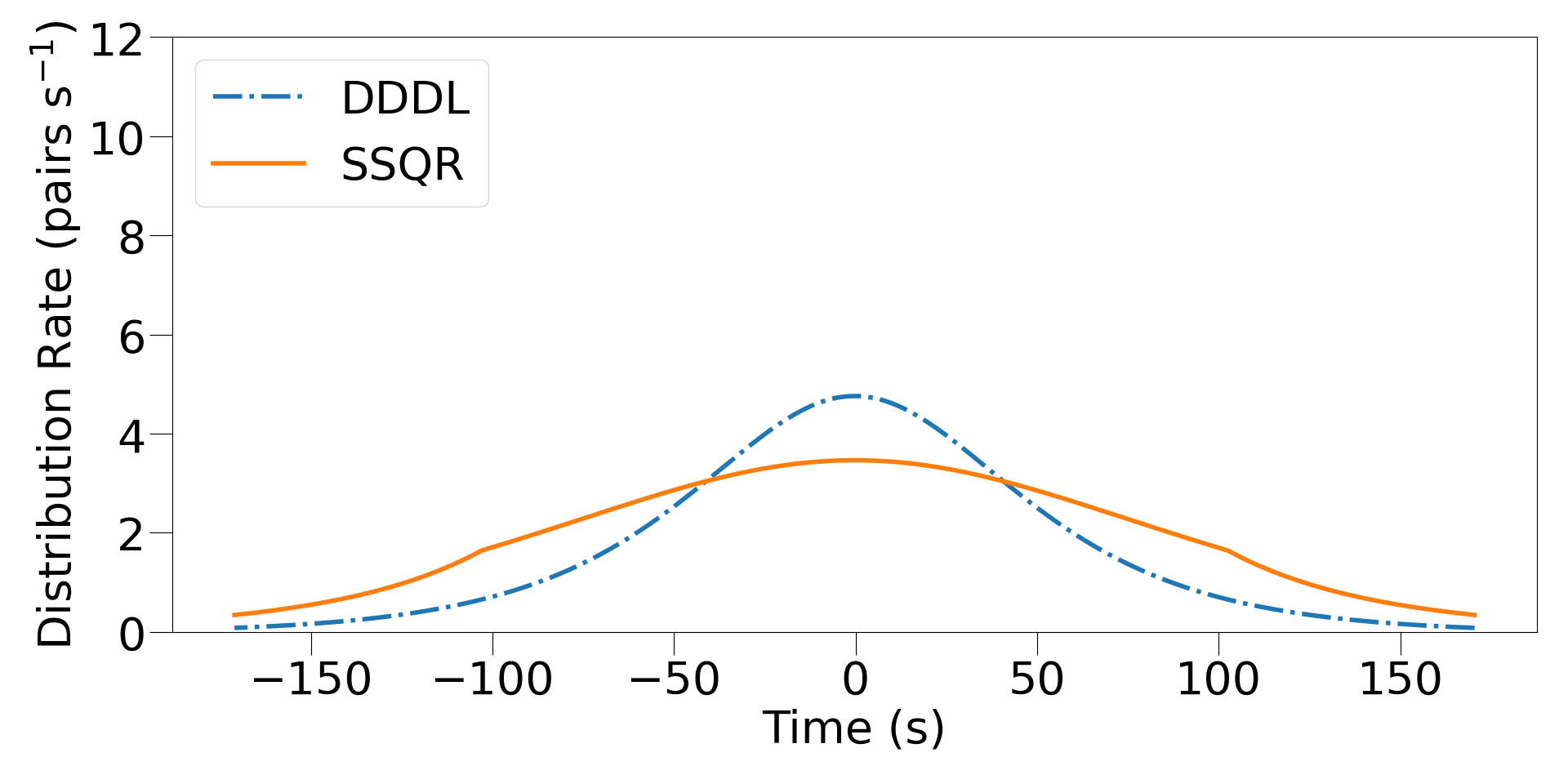}
\caption{Zenith A $90^\circ$ ($\Delta = 500 \text{ km}, \phi = 90^\circ$). Total $\text{PDV}_\text{DDDL, SSQR} = 575, 674$.}
\end{subfigure}
\hfill
\begin{subfigure}{\figwidth}
\centering
\vspace{12pt}
\includegraphics[width=\linewidth]{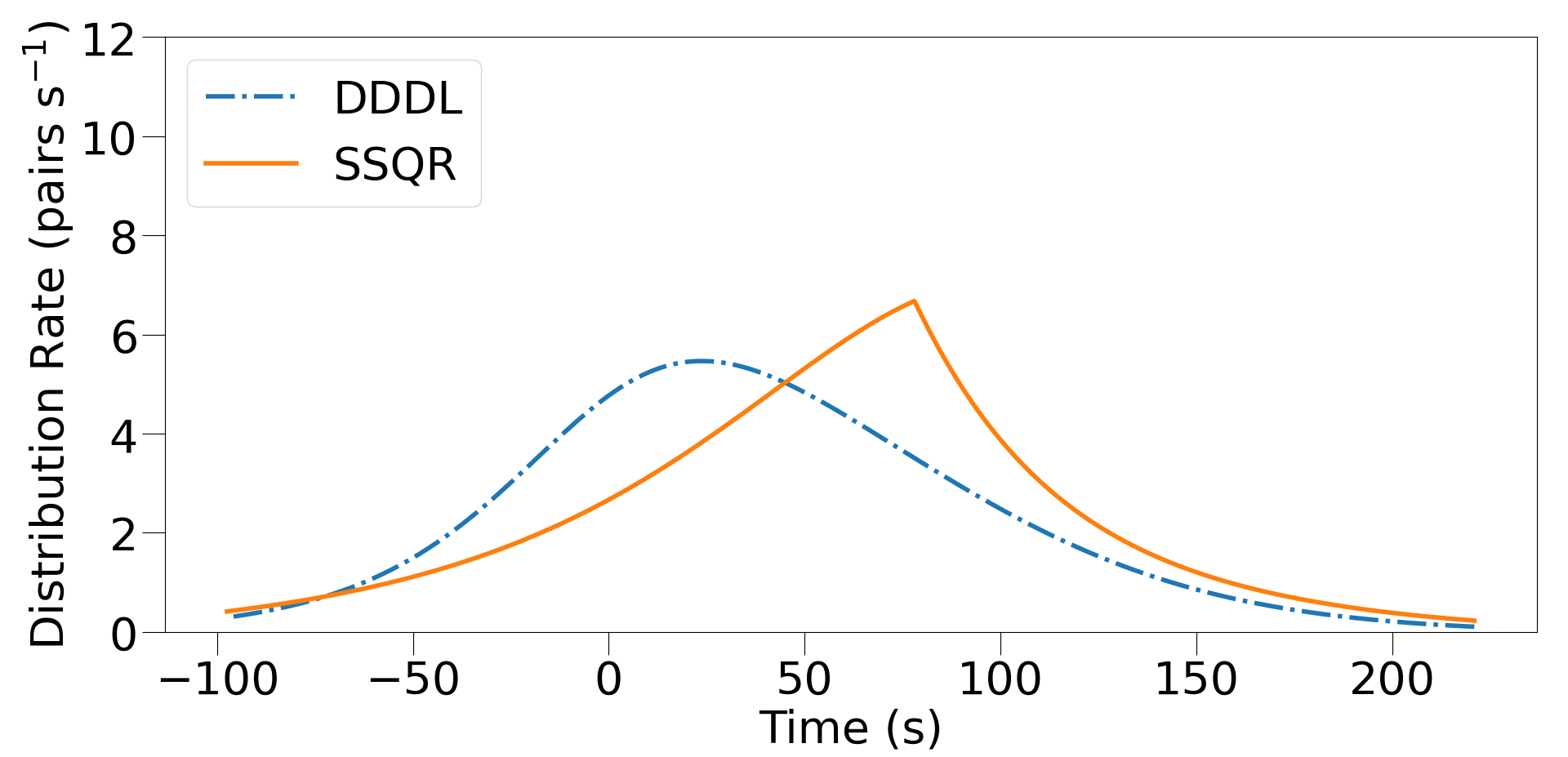}
\caption{Zenith A $45^\circ$ ($\Delta = 500 \text{ km}, \phi = 45^\circ$). Total $\text{PDV}_\text{DDDL, SSQR} = 740, 759$.}
\end{subfigure}
\caption{$R^\text{DDDL,SSQR}$ for (a) Zenith-zenith, (b) Symmetric, (c) Zenith A $90 ^\circ$ and (d) Zenith a $45^\circ$ overpasses. $N_\text{sat}=200$ and $R_\text{EPS}^\text{DDDL}=5.9\times 10^6\ \text{s}^{-1}$. The discontinuities in the gradient of $R^{SSQR}$ correspond to times where the rates to each OGS crossover.}
\label{fig:4PassesPDRs}
\end{figure}

\begin{figure}[!b]
\centering
\includegraphics[width=0.70\linewidth]{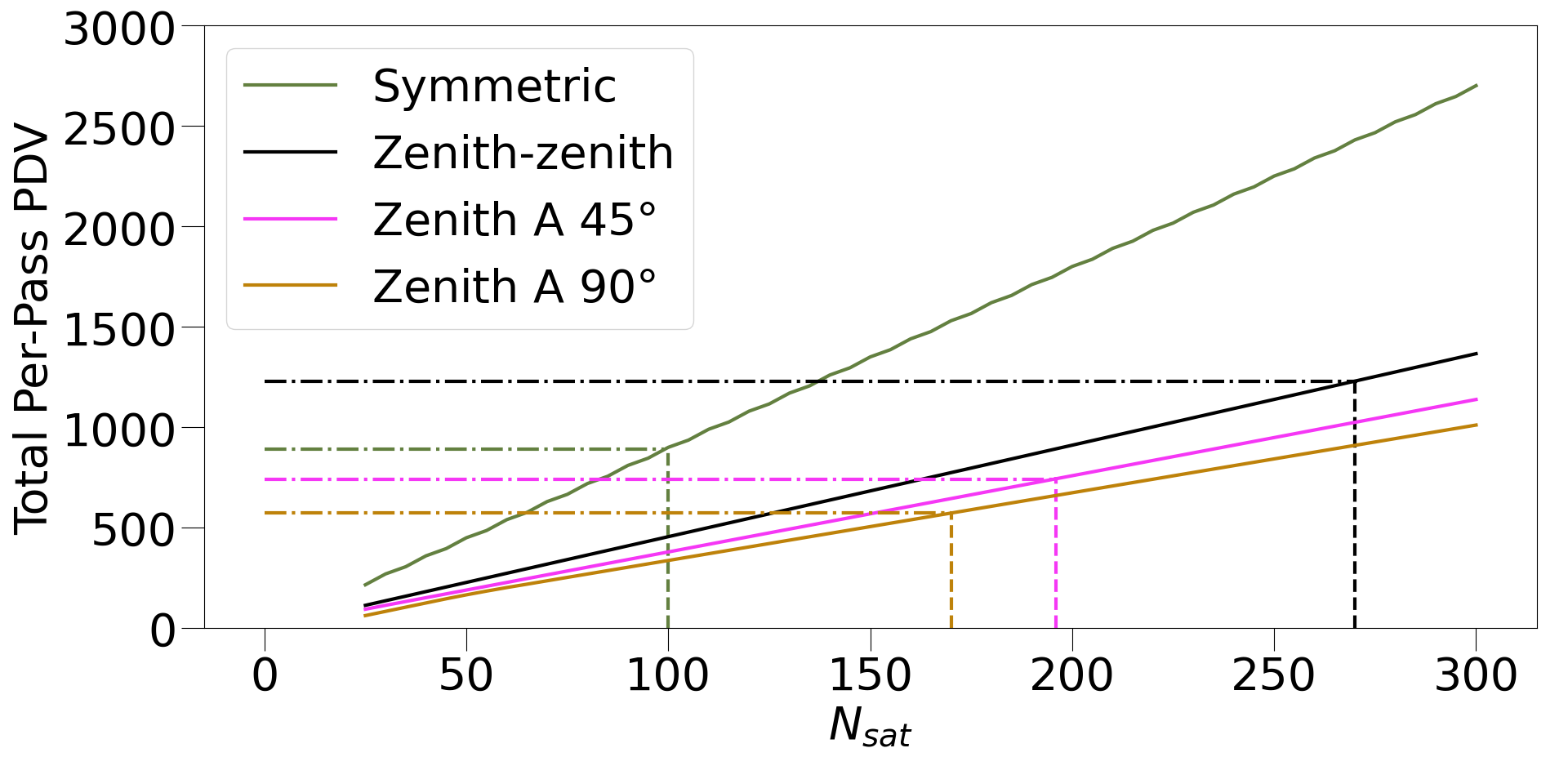}
\caption{Per-overpass PDV for varying $N_\text{sat}$. The horizontal dashed-dotted lines show the DDDL PDV corresponding to the Micius source rate and vertical dashed lines indicated $N_C$ values for each overpass.. The value $N_\text{sat}$ for which the repeater PDV is equal to DDDL defines $N_c$. The DDDL and repeater satellite performed best for the zenith-zenith and symmetric overpasses, respectively.}
\label{fig:4PassenSats}
\end{figure}

We see in Figs.~\ref{fig:4PassesPDRs} and \ref{fig:4PassenSats} a clear dependence of protocol performance on the overpass geometry. A DDDL satellite performs best for zenith-zenith overpass whereas the repeater satellite performs best for the symmetric overpass. Appendix~\ref{sec:overpasscomparison} provides an explanation for this behaviour. This is reflected in the values of $N_c$ for each overpass, which can be seen as $N_\text{sat} = 270 \text{ modes}$ and $N_\text{sat} = 100 \text{ modes}$ for the zenith-zenith and symmetric overpasses, respectively. For the Zenith A $90 ^\circ$ and Zenith A $45^\circ$ overpasses, $N_c$ was found to be $170 \text{ modes}$ and $196 \text{ modes}$, respectively. 

\begin{table}[!tbp]
\centering
\begin{tabular}{l|cc|cc}
\toprule
&\multicolumn{2}{c|}{$N_\text{sat}=200$}&\multicolumn{2}{c}{$N_\text{sat}=2000$}\\
Overpass & $N_A$ & $N_B$ & $N_A$ & $N_B$\\ 
\midrule
Zenith-zenith & $100$ & $100$ & $1000$ & $1000$\\ 
Symmetric & $100$ & $100$ & $1000$ & $1000$\\ 
Zenith A $90^\circ$& $32$&$168$& $323$&$1677$\\ 
Zenith A $45^\circ$ & $71$ & $129$ & $709$ & $1291$\\ 
\bottomrule
\end{tabular}
\caption{Optimal static memory allocations for the representative overpasses for $N_\text{sat} = 200,2000$. $N_A$ ($N_B$) denotes the memory slots assigned to Alice (Bob). In each case the allocations were chosen to maximise the per-overpass PDV and were found via brute force.}
\label{table:staticAllocations200}
\end{table}

These values of $N_c$ were obtained for a satellite adopting an optimised static memory allocation. For a given overpass geometry and $N_\text{sat}$ the optimal allocation was found via brute force. These allocations are given in Table~\ref{table:staticAllocations200} for $N_\text{sat} = 200,2000$. $N_c$ naturally depends on the DDDL source rate, but the linear nature of this scaling means that we can normalise the $N_c$ values to $R_\text{EPS}^\text{DDDL}$. We estimate this by calculating the quantity
\begin{equation}
    \nu_c = \left\lceil \frac{\text{PDV}_\text{DDDL}}{\text{PDV}_\text{Repeater}} \cdot \frac{N_\text{sat}}{R_\text{EPS}^{DDDL}/10^6}\right\rceil
\label{equ:NcCalculation}
\end{equation}
and rounding up to the next even integer, noting that $\nu_c$ has the dimensions of time. For the Zenith-Zenith, Symmetric, Zenith A $90 ^\circ$, and Zenith A $45^\circ$ overpasses, $\nu_c^\text{Z-Z} = 46 \text{ modes}\text{ MHz}^{-1}$, $\nu_c^\text{Sym} = 18 \text{ modes}\text{ MHz}^{-1}$, $\nu_c^{\text{Z-A} 90^\circ} = 30 \text{ modes}\text{ MHz}^{-1}$ and $\nu_c^{\text{Z-A} 45^\circ} = 34 \text{ modes}\text{ MHz}^{-1}$, respectively. $\nu_c$ provides a compact benchmark for the memory capacity of candidate platforms. It also captures the competition between $R_\text{EPS}^\text{DDDL}$ and $N_\text{sat}$ values: increases to $R_\text{EPS}^\text{DDDL}$ means a higher $N_\text{sat}$ is required for PDV parity. 

\begin{figure}[!t]
\centering
\newcommand{\figwidth}{0.45\linewidth}
\newcommand{\figheight}{0.2\textheight}
\begin{subfigure}{\figwidth}
\centering
\includegraphics[width=\linewidth]{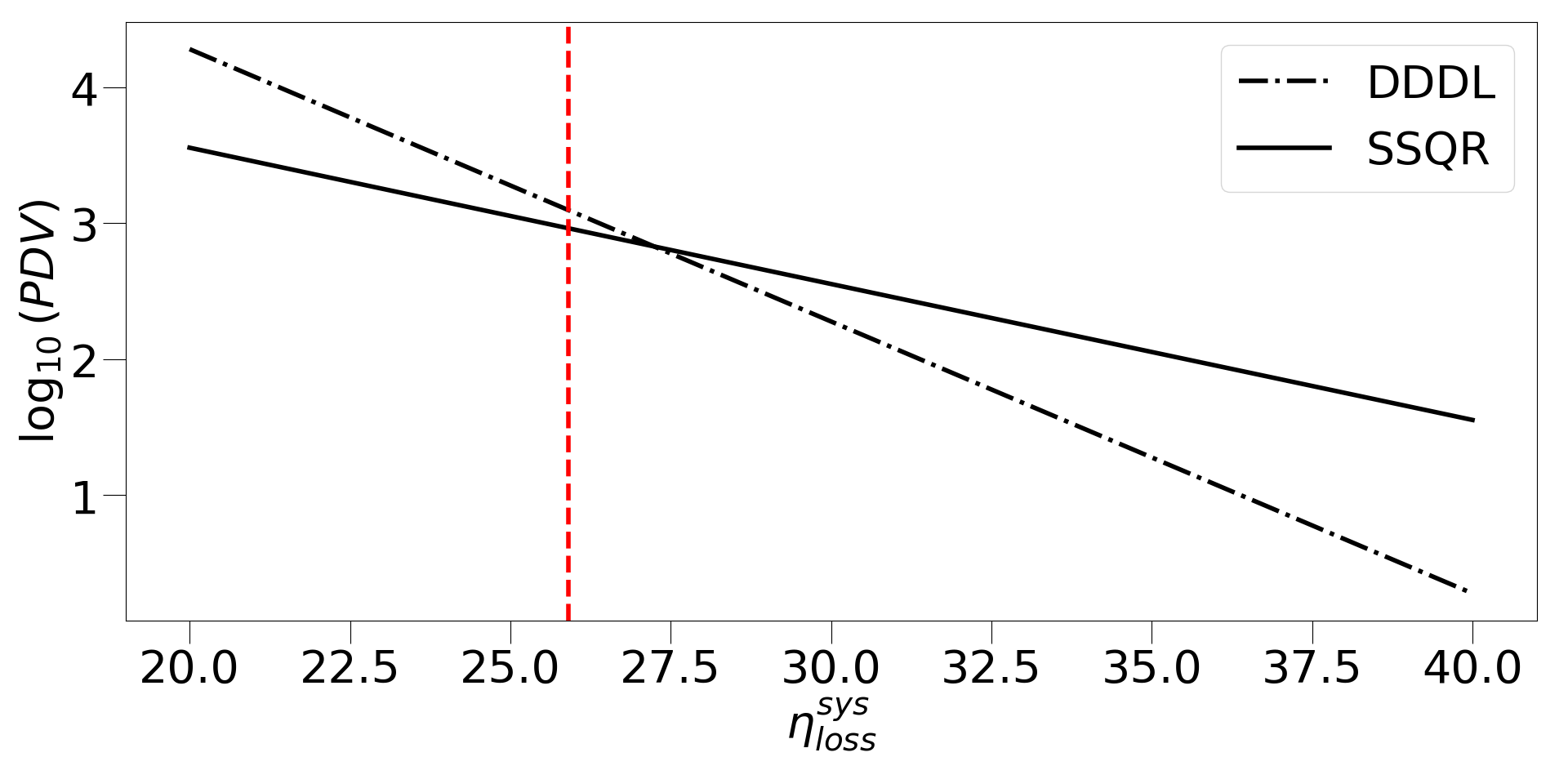}
\caption{Zenith-zenith ($\Delta = 0\ \text{km},\phi = 0^\circ$).}
\end{subfigure}
\hfill
\begin{subfigure}{\figwidth}
\centering
\includegraphics[width=\linewidth]{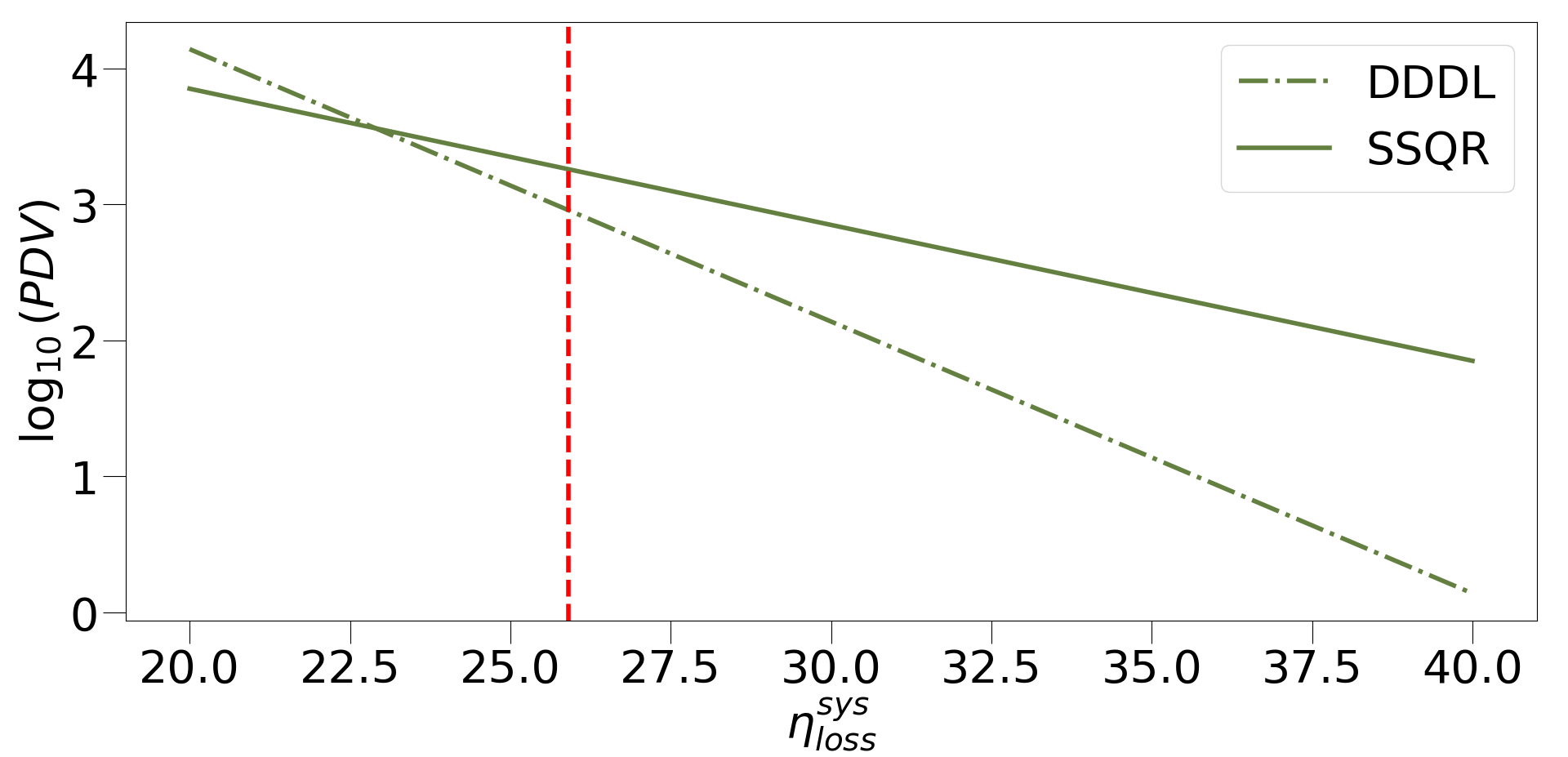}
\caption{Symmetric ($\Delta = 0\ \text{ km}, \phi = 90^\circ$).}
\end{subfigure}
\begin{subfigure}{\figwidth}
\centering
\vspace{12pt}
\includegraphics[width=\linewidth]{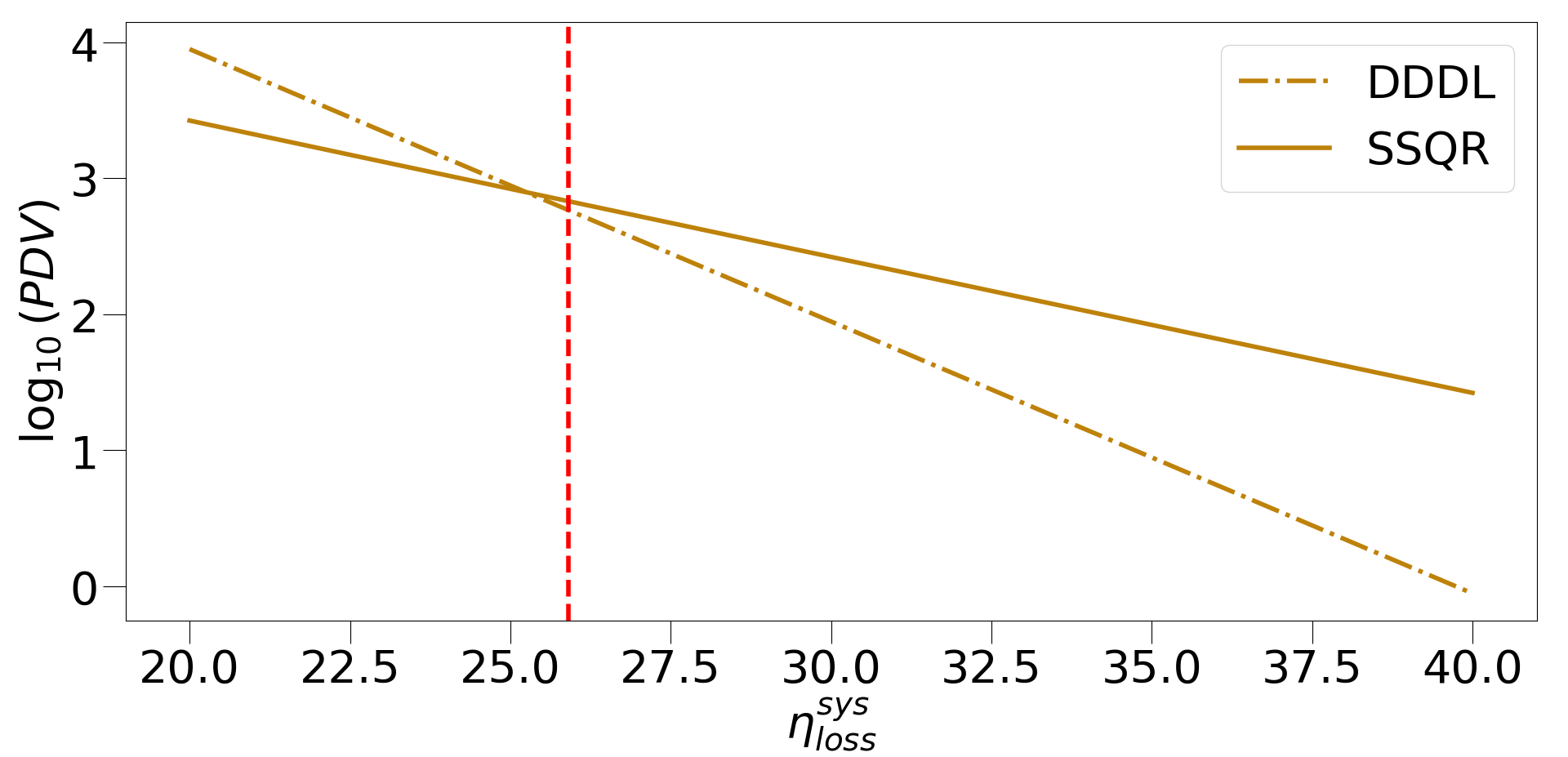}
\caption{Zenith A $90^\circ$ ($\Delta = 500 \text{ km}, \phi = 90^\circ$).}
\end{subfigure}
\hfill
\begin{subfigure}{\figwidth}
\centering
\vspace{12pt}
\includegraphics[width=\linewidth]{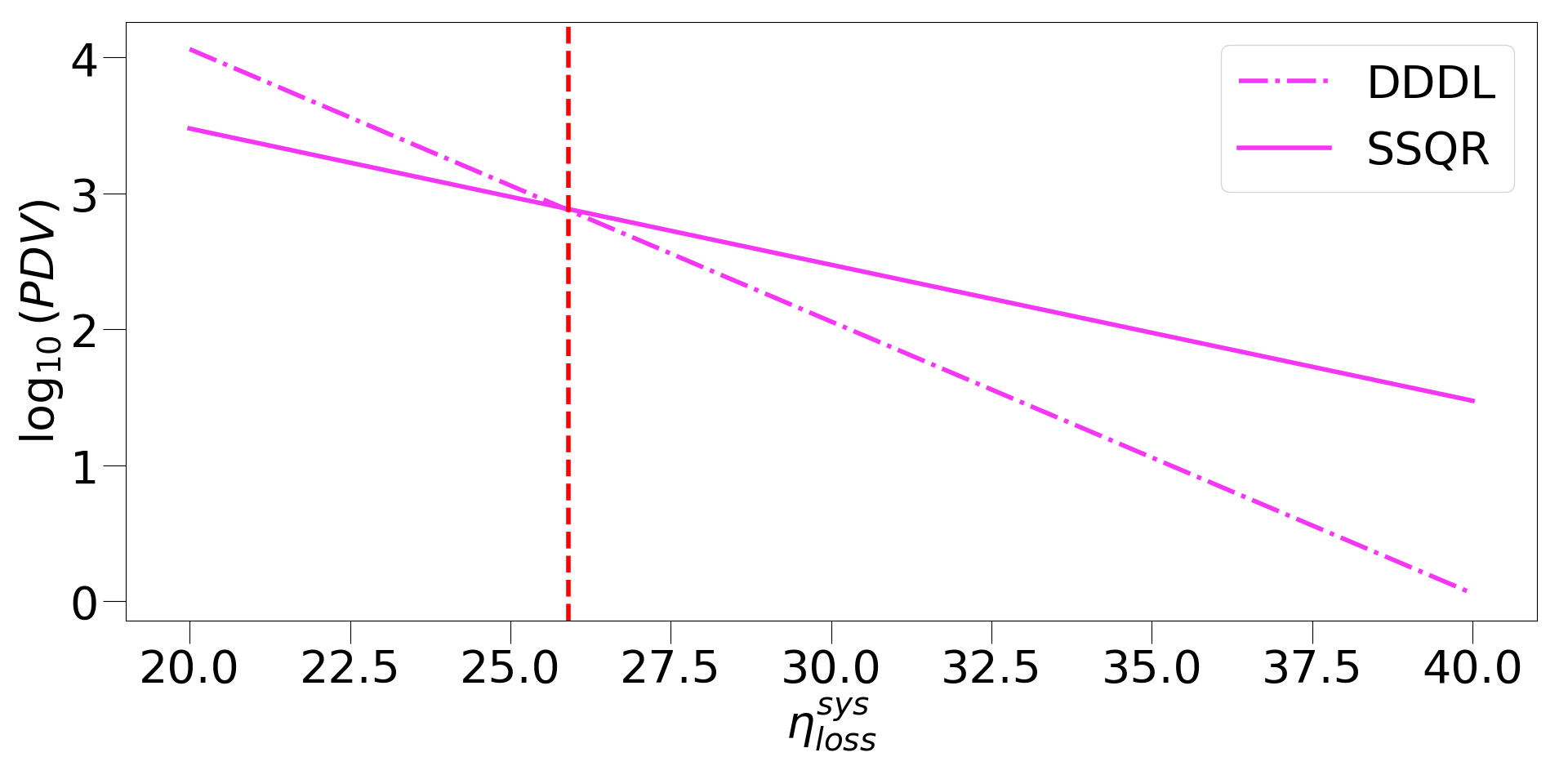}
\caption{Zenith A $45^\circ$ ($\Delta = 500 \text{ km}, \phi = 45^\circ$).}
\end{subfigure}
\caption{Per-overpass PDV against $\eta^\text{sys}_\text{loss}$ for the (a) Zenith-zenith, (b) Symmetric, (c) Zenith A $90^\circ$ and (d) Zenith A $45^\circ$ overpasses. Dashed red line show the baseline $\eta^\text{sys}_\text{loss}=25.9$ dB. $R_\text{EPS}^\text{DDDL}=5.9 \times 10^6 \text{ pairs s}^{-1}$ and the repeater satellite has $N_\text{sat} = 200$. There is a value of $\eta^\text{sys}_\text{loss}$ below which DDDL outperforms the repeater satellite in each case, indicating that the benefit of memories to the PDV may only be realised when channel efficiencies are low.}
\label{fig:baslineLossAnalysis}
\end{figure}

Satellite performance is influenced by factors such as source and memory metrics, as well as system link efficiencies characterised by $\eta^\text{sys}_\text{loss}$. Transmitter and receiver aperture sizes together with internal optical losses are major contributors to link efficiency and these can vary greatly depending on the satellite-OGS system. The improved Micius apparatus at the Delingha ground station has achieved a system loss of $27\ \text{dB}$, for example~\cite{miciusEntQKD2020}, whereas a CubeSat platform and medium sized OGS may have $\eta_\text{loss}^{sys}\sim 40\ \text{dB}$~\cite{sagar2023design}. To incorporate the effects of changing $\eta^\text{sys}_\text{loss}$ in our analysis, we investigate the PDV and $\nu_c$ values for $\eta^\text{sys}_\text{loss}$ values between $20\ \text{dB}$ and $40\ \text{dB}$. This range encapsulates optimistic near-term high-performance OGS and the higher loss regimes associated with small satellite implementations. In Fig.~\ref{fig:baslineLossAnalysis}, for each overpass we observe distinct `crossover' values of $\eta^\text{sys}_\text{loss}$ below which DDDL outperforms the repeater PDV, for a fixed $N_\text{sat}$. For the zenith-zenith and symmetric overpasses, the crossover $\eta^\text{sys}_\text{loss}$ were estimated as $28\ \text{dB}$ and $23\ \text{dB}$, respectively, for $N_\text{sat} = 200$. For the zenith A $90^\circ$ and zenith A $45^\circ$, the estimated value was $26\ \text{dB}$. 

\begin{figure}[!b]
\centering
\includegraphics[width=0.6\linewidth]{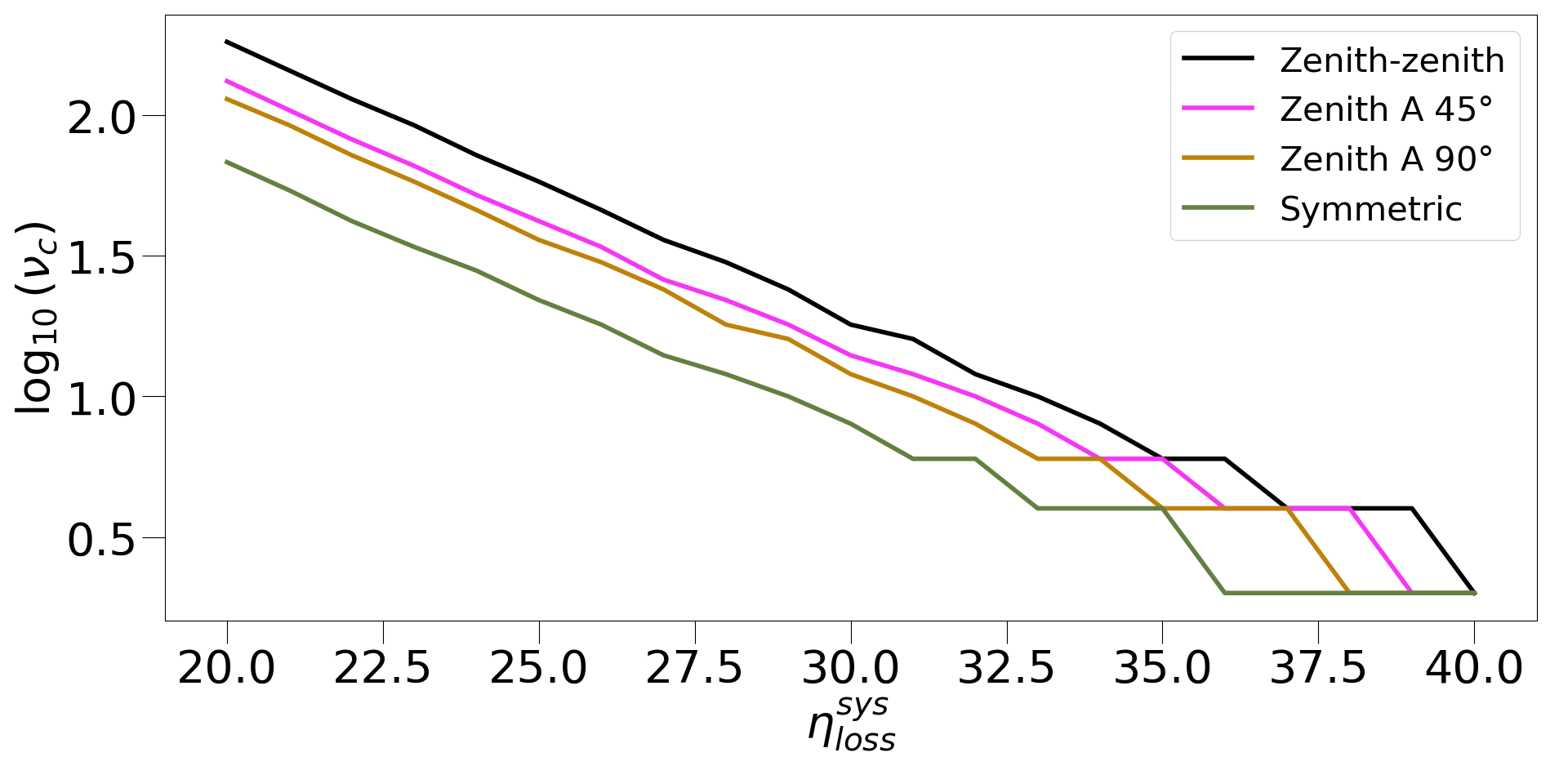}
\caption{Normalised crossover capacity $\nu_c$ dependence on $\eta_\text{sys}^\text{loss}$. We plot the rate-normalised crossover capacity for the representative overpass geometries as a function of increasing system loss. Decreasing $N_c$ with improved $\eta^\text{sys}_\text{loss}$ shows that memories lose their advantage when it comes to the PDV when the $\eta_\text{sys}^\text{loss}$ improves.}
\label{fig:nCVaryDr}
\end{figure}

This dependence on the channel losses of whether memories should be employed  has practical implications. For example, larger receiver apertures on the ground tend to lead to lower a $\eta^\text{sys}_\text{loss}$, so quantum memories on the satellite may not benefit the PDV. On the other, if receiver apertures are small - such as in the case of a portable OGS - then the high $\eta_\text{loss}^\text{sys}$ may result in memories being of substantial advantage to the PDV. The advantage of memories in high loss environments is further reflected in the behaviour of $\nu_c$ with increasing $\eta^\text{sys}_\text{loss}$ (Fig.~\ref{fig:nCVaryDr}).

\begin{figure}[!t]
\centering
\newcommand{\figwidth}{0.45\linewidth}
\begin{subfigure}{\figwidth}
\centering
\includegraphics[width = \linewidth]{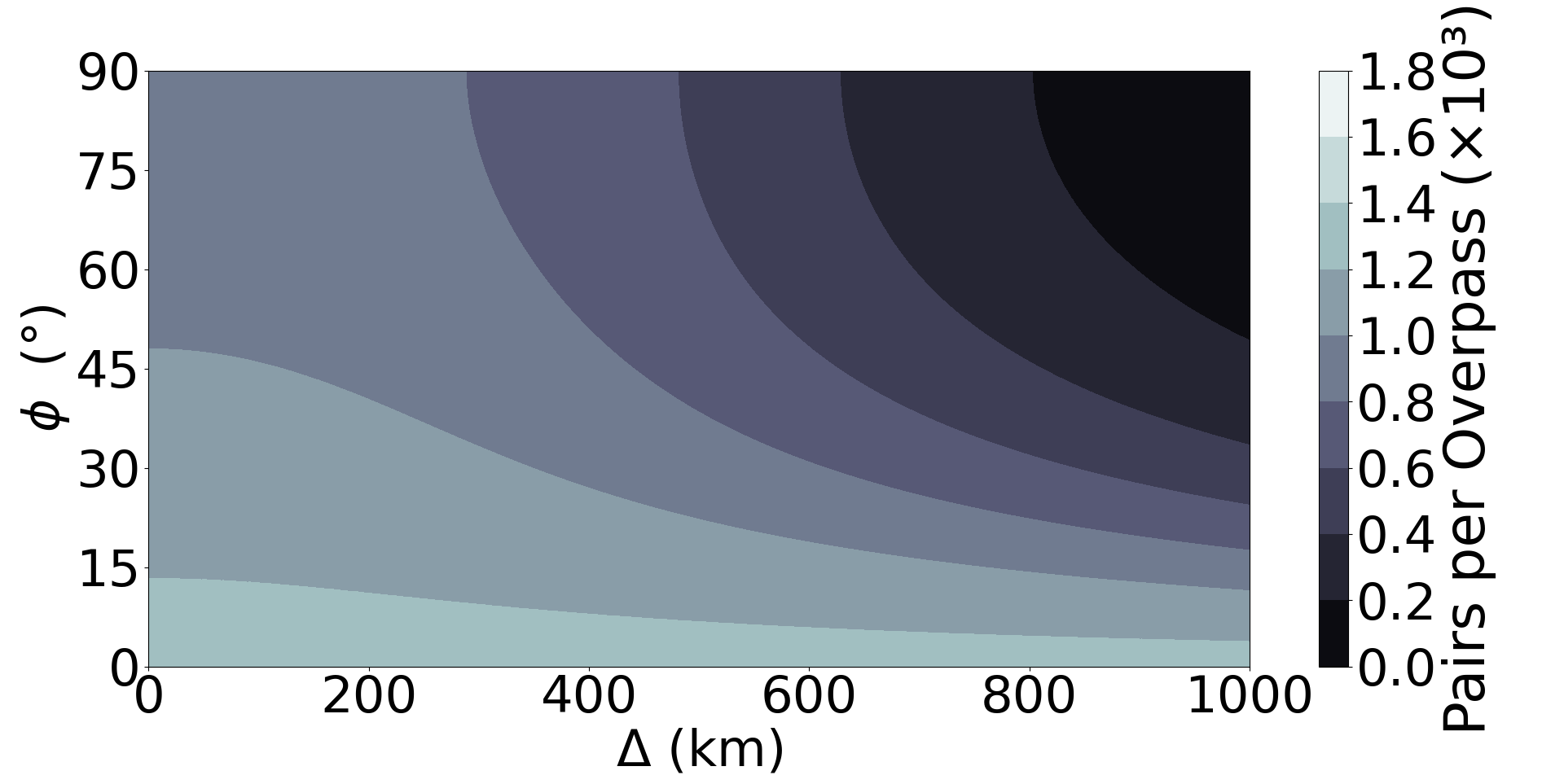}
\caption{DDDL PDV.}
\end{subfigure}
\hfill
\begin{subfigure}{\figwidth}
\centering
\includegraphics[width=\linewidth]{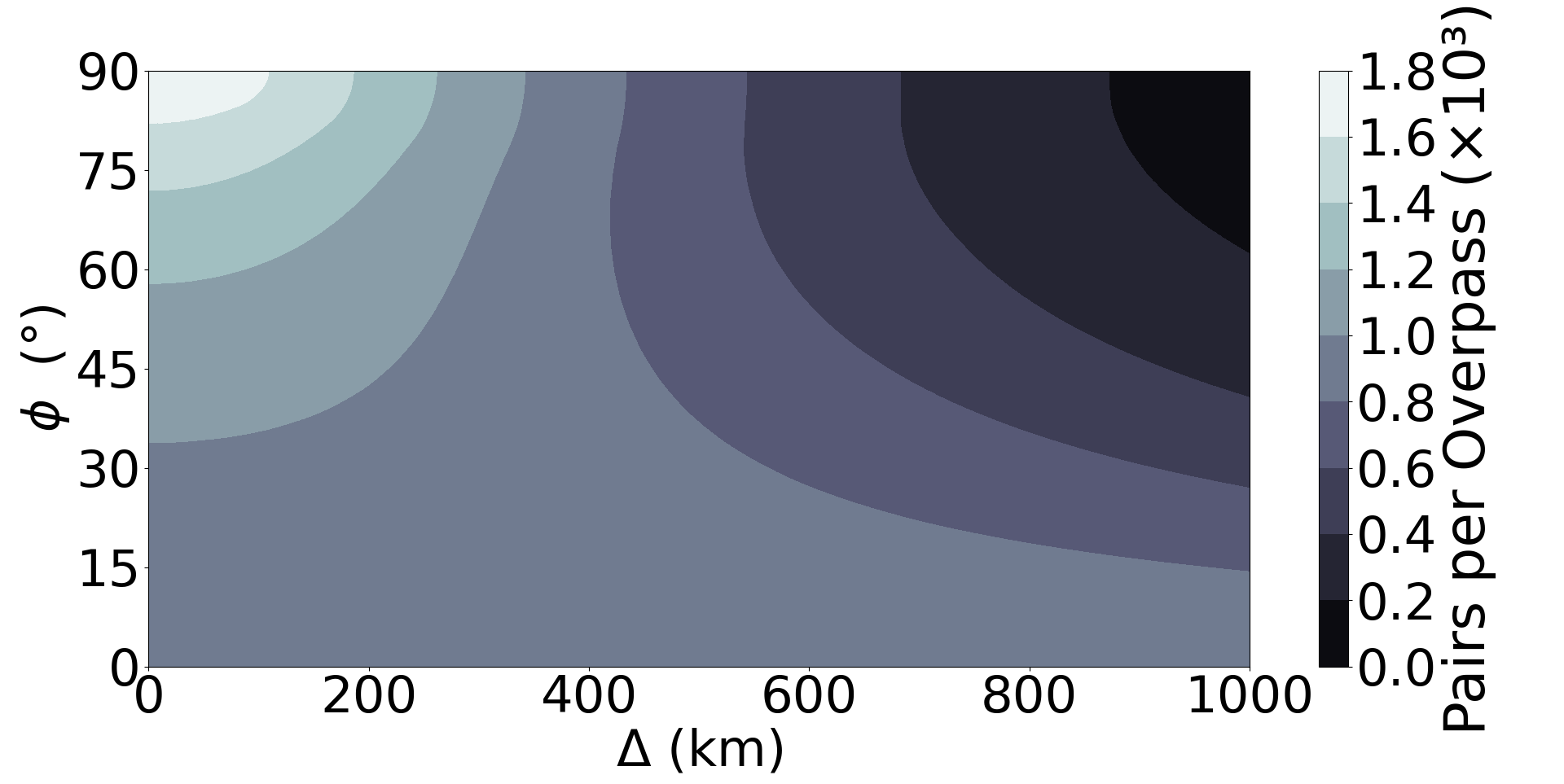}
\caption{SSQR PDV.}
\end{subfigure}
\begin{subfigure}{\figwidth}
\centering
\includegraphics[width=\linewidth]{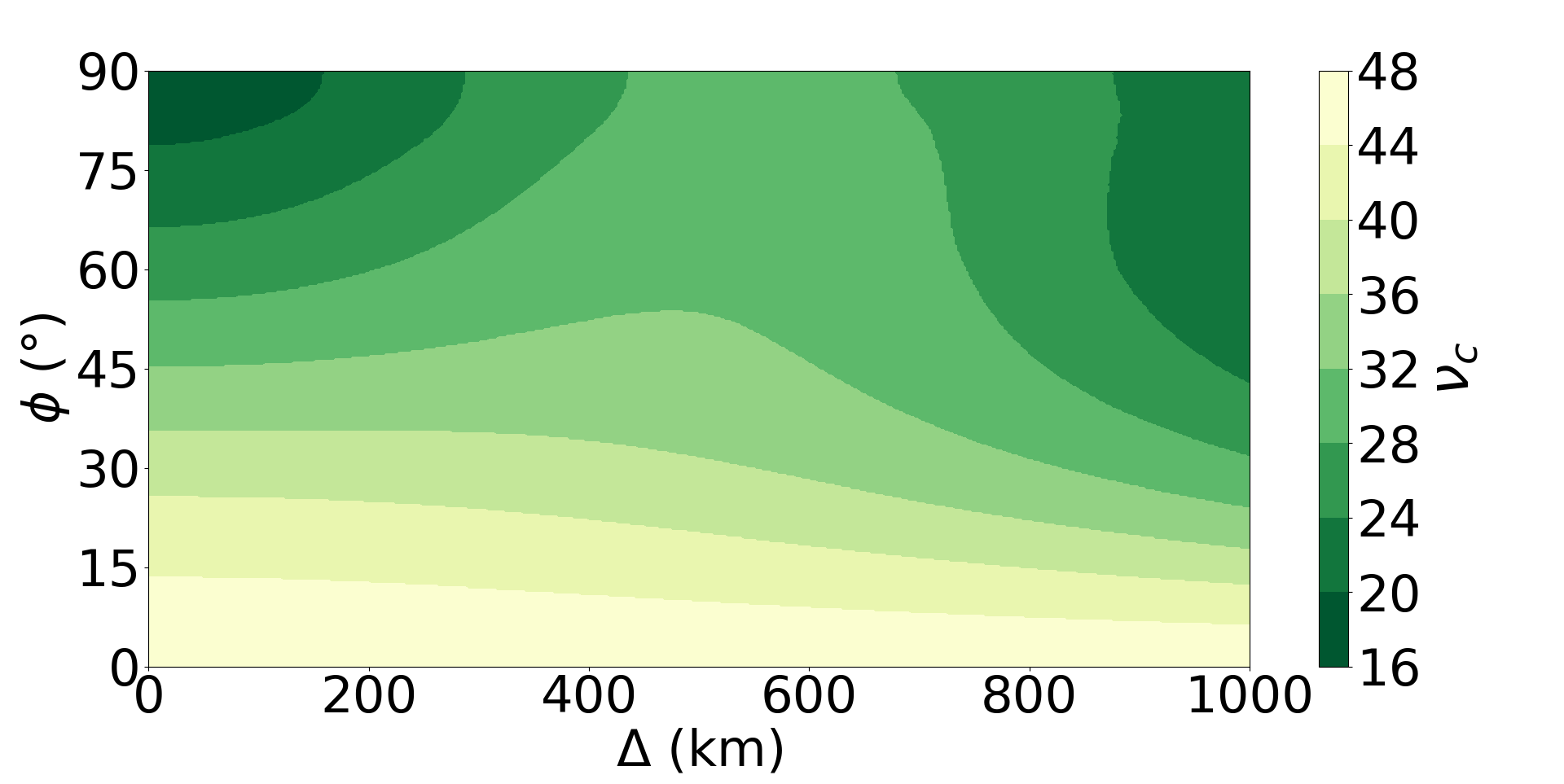}
\caption{Normalised crossover capacity $\nu_c$.\\
}
\end{subfigure}
\hfill
\vspace{0.25cm}
\begin{subfigure}{\figwidth}
\centering
\includegraphics[width=\linewidth]{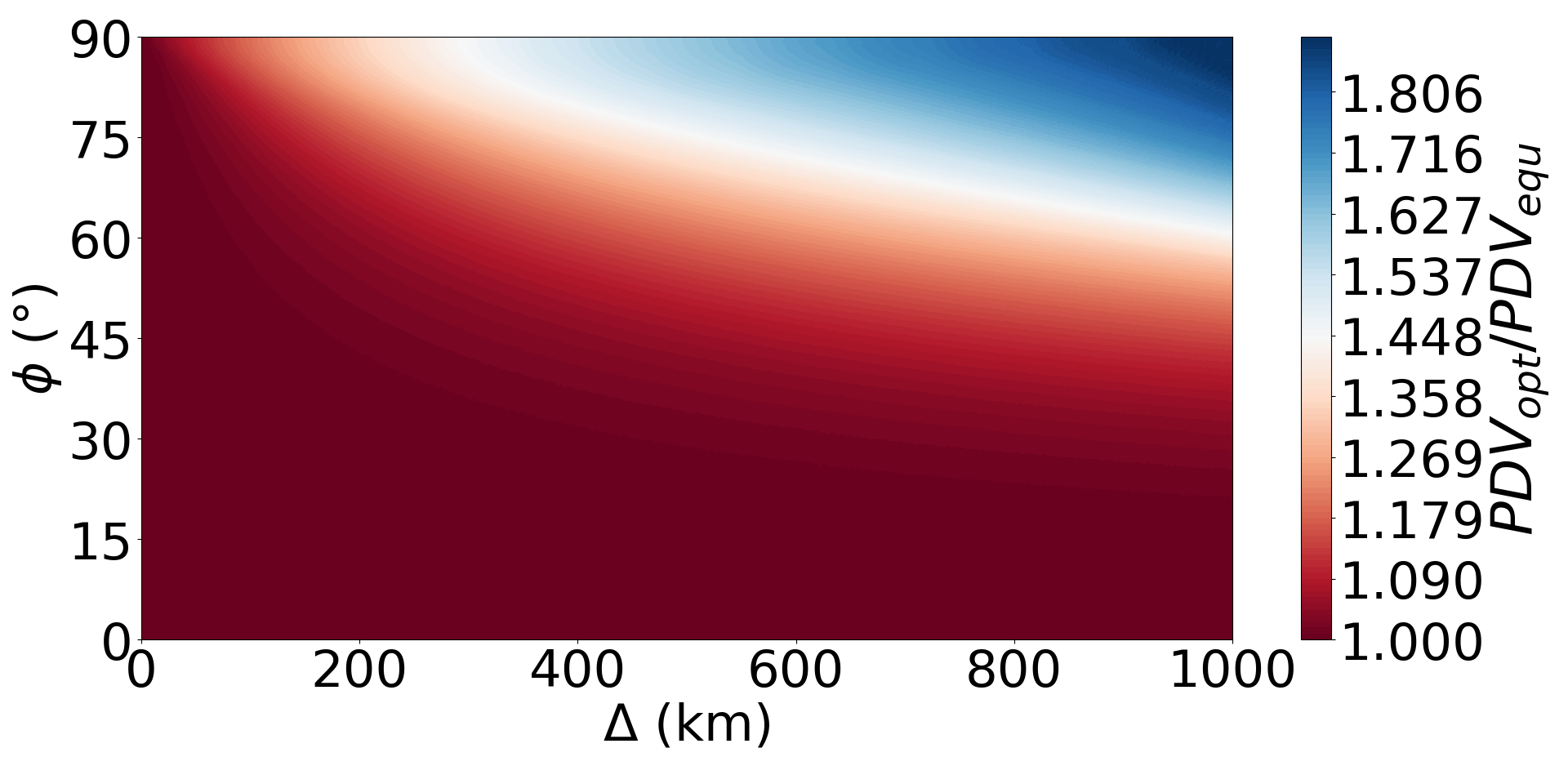}
\caption{PDV ratio of optimatl to static split}
\end{subfigure}
\vspace{0.25cm}
\caption{General Overpass Geometry Performance. Baseline system parameters as per Table~\ref{table:IIIAParameters} . (a) DDDL with $R_\text{ESP}^\text{DDDL} = 5.9 \times 10^6 \text{ pairs s}^{-1}$, (b) SSQR with $N_\text{sat} = 200$ adopting an optimal static memory split. DDDL performs best for overpass geometries with $\phi \sim 0^\circ$ and SSQR performing best when $\Delta \sim 0$ km and $\phi \sim 90^\circ$. These respective optimal regions are in (c), with $\nu_c$ highest for $\phi \sim 0^\circ$ and lower for overpasses with large $\phi$. (d) The PDV ratio for SSQR with optimised vs static memory split shows that for $\phi \gtrsim 45^\circ$ optimising the memory split is beneficial. These correspond to overpasses with highly asymmetric loss profiles, where the benefits of assigning more memory to the downlink with higher losses can greatly improve the PDV. For symmetric link loss efficiencies, the plots are symmetric around $\Delta=0$ km and $\phi=0^\circ$, and periodic around the region $[-90^\circ,90^\circ]$.}
\label{fig:fullLandscapePlots}
\end{figure}

These examples establish that the performance of an entanglement distribution protocol is strongly dependent on the overpass geometry. We now extend the above analysis, calculating the per-overpass PDV's in the full $\Delta-\phi$ space. Different regions of optimality are observed for the DDDL and repeater satellite (Fig.~\ref{fig:fullLandscapePlots}(a),(b)). DDDL performs best for $\phi \sim 0^\circ$ and the repeater satellite for overpasses with $\Delta \sim 0\ \text{km}$ and $\phi \sim 90^\circ$, which is consistent with previous results. This is reflected in the $\nu_c$ values (Fig.~\ref{fig:fullLandscapePlots}(c)) which are highest for overpasses with low $\phi$ values.

The PDVs in Fig.~\ref{fig:fullLandscapePlots}(b) were calculated for optimal static memory splitting. We compare this to the case where the satellite adopts an equal split of the onboard memory, calculating of their PDV ratios in Fig.~\ref{fig:fullLandscapePlots}(d). The benefits of adopting an optimal static split become significant for $\phi \gtrsim 45^\circ$ which correspond to overpass geometries with highly asymmetric loss profiles. The optimal static split allows a more equal link generation rate with the two OGSs throughout the overpass.

\subsection{Average Annual PDV}
\label{sec:annualPDV}

\begin{figure}
    \centering
    \includegraphics[width=\linewidth]{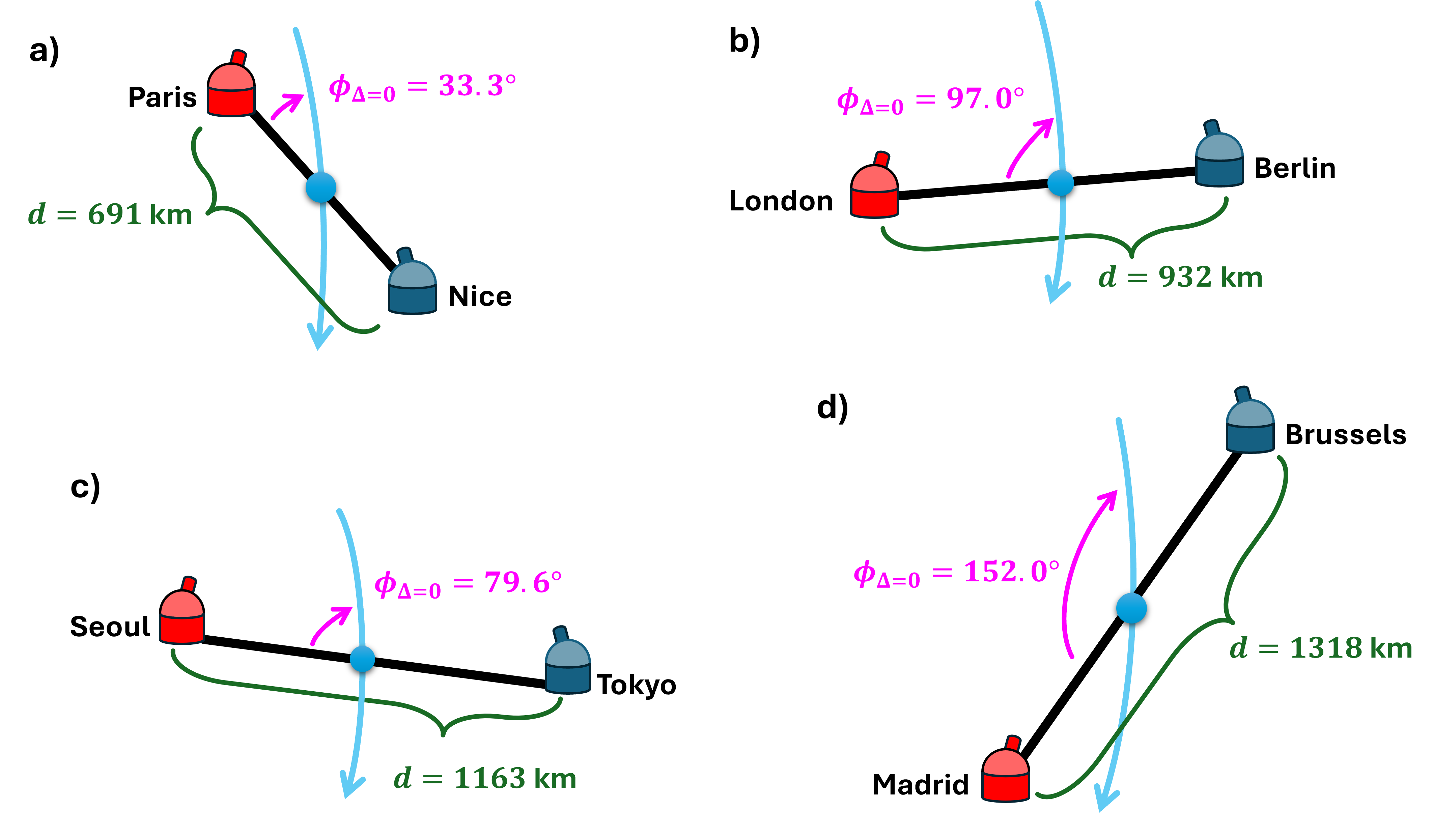}
    \caption{Links considered in the annual PDV analysis: (a) Paris-Nice, (b) London-Berlin, (c) Seoul-Tokyo and (d) Madrid-Brussels. The ground track crossing angles at the OGS baseline midpoint $\phi_{\Delta=0}$ and OGS baseline lengths are indicated. We have used the convention that Alice is to the West of Bob, and that the crossing angle is clockwise from the Alice-to-Bob OGS line  and the North-South ground-track direction.}
    \label{fig:annuaLPDVs4Links}
\end{figure}

Since the PDVs vary significantly with overpass geometry, the long-term average over different overpasses provides a more complete characterisation of the system performance. We choose to determine the expected annual PDV for pairs of OGSs, nominally for a satellite in a Noon-Midnight SSO (downlink only on Midnight overpasses) which we will approximate as a polar orbit. Such an approximation is sufficient for OGS locations situated away from the polar regions.
The annual PDV analysis is performed for a selection of OGS pairs located in major cities (Fig.~~\ref{fig:annuaLPDVs4Links}). Each of these links are characterised by the OGS baseline distance and the $\phi$ value for a polar orbit overpass with $\Delta=0$ km, denoted $\phi_{\Delta = 0}$.
We compare DDDL with SSQR adopting either an equal memory split of an optimal static memory split. This allows for an evaluation of the benefit that optimising the memory allocation in this way could potentially offer over long time scales. 

\begin{figure}[!tbp]
\centering
\newcommand{\figwidth}{0.48\textwidth}
\begin{subfigure}{\figwidth}
\centering
\includegraphics[width=\linewidth]{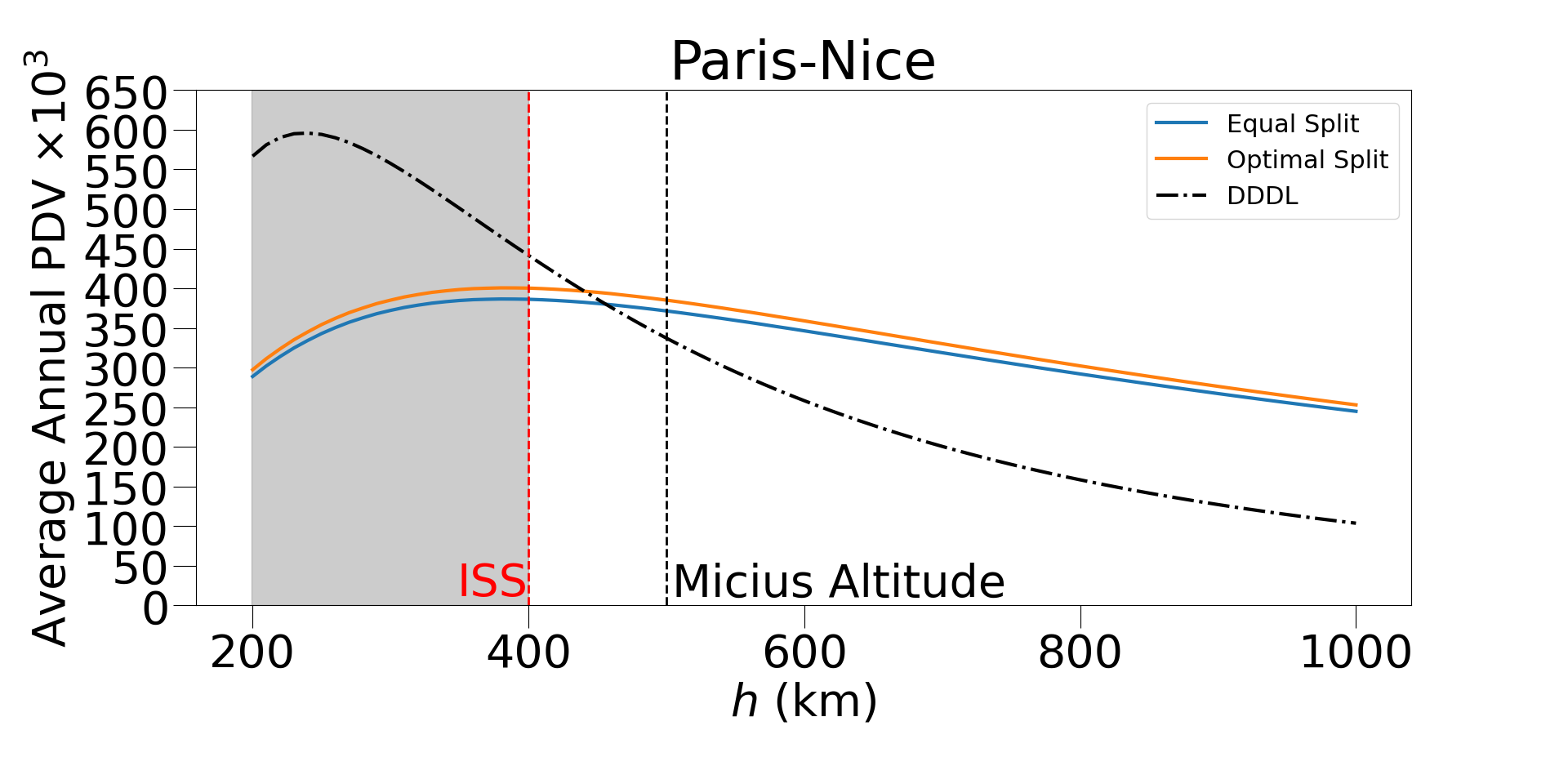}
\caption{Paris-Nice ($d=691$km, $\phi_{\Delta=0}=33.3^\circ$).
}
\end{subfigure}
\hfill
\begin{subfigure}{\figwidth}
\centering
\includegraphics[width=\linewidth]{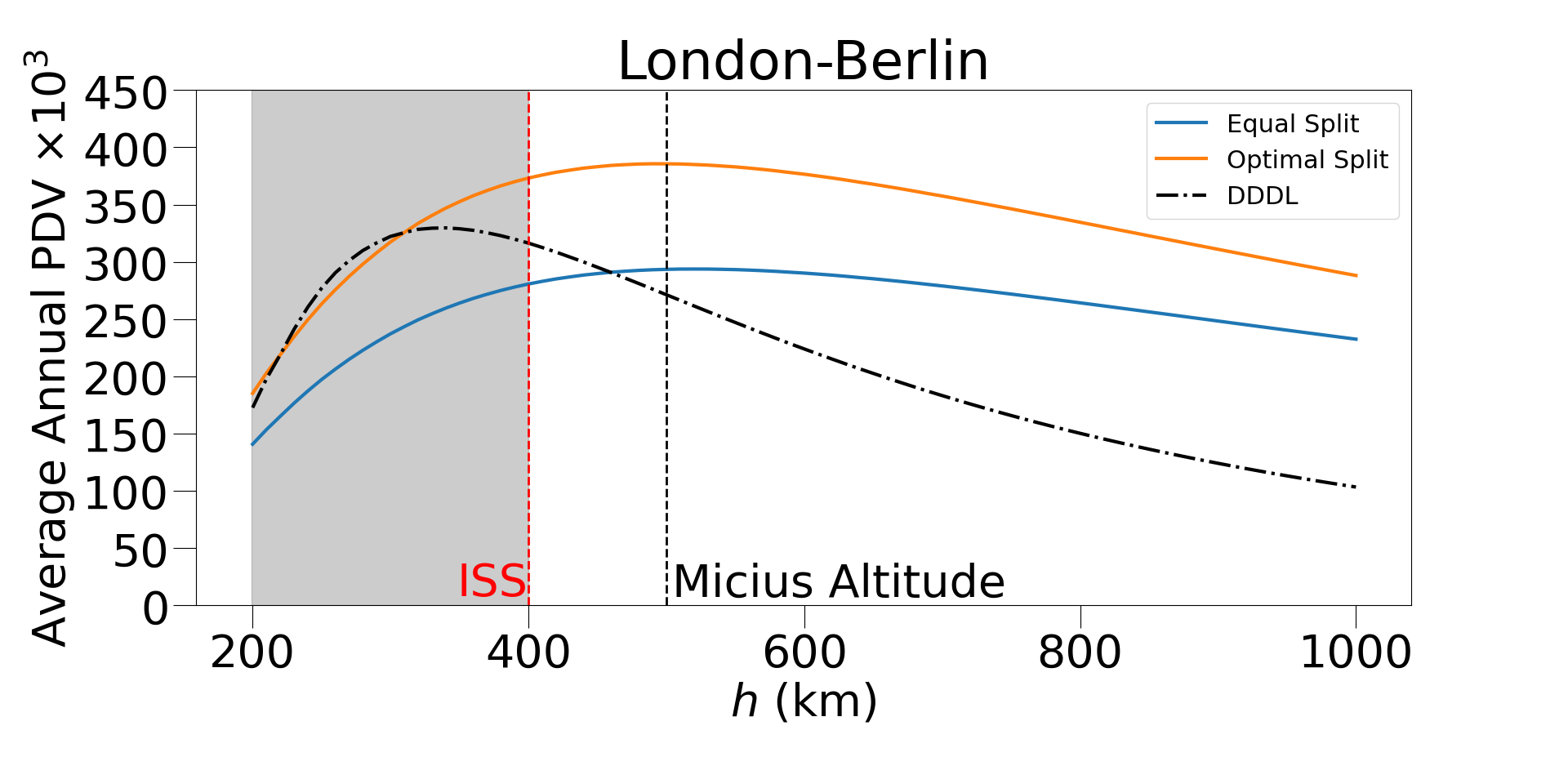}
\caption{London-Berlin ($d=932$km, $\phi_{\Delta=0}=97.0^\circ$).
}
\end{subfigure}
\par\medskip
\begin{subfigure}{\figwidth}
\centering
\includegraphics[width=\linewidth]{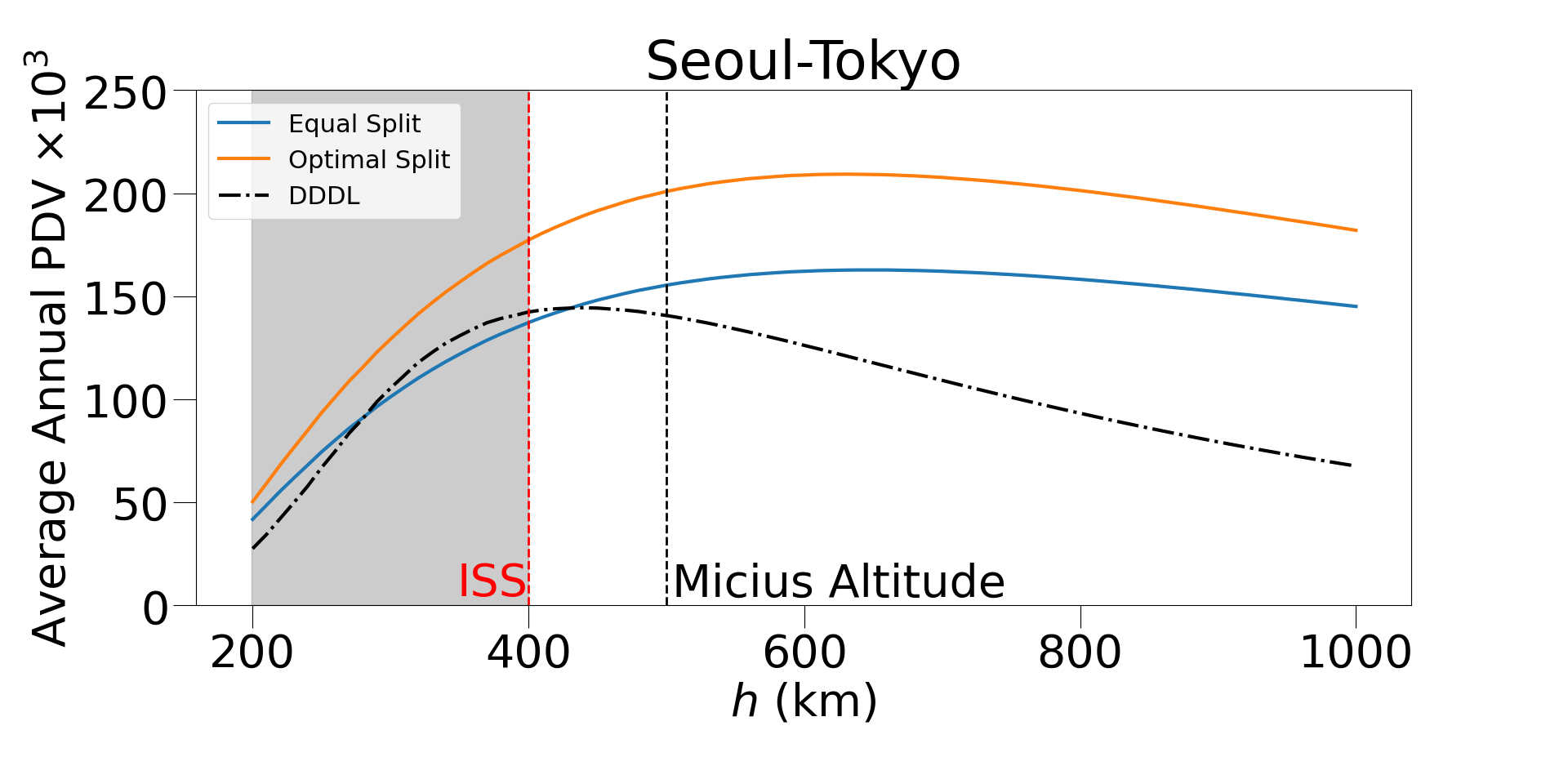}
\caption{Seoul-Tokyo ($d=1163$km, $\phi_{\Delta=0}=79.6^\circ$).
}
\label{fig:seoul}
\end{subfigure}
\hfill
\begin{subfigure}{\figwidth}
\centering
\includegraphics[width=\linewidth]{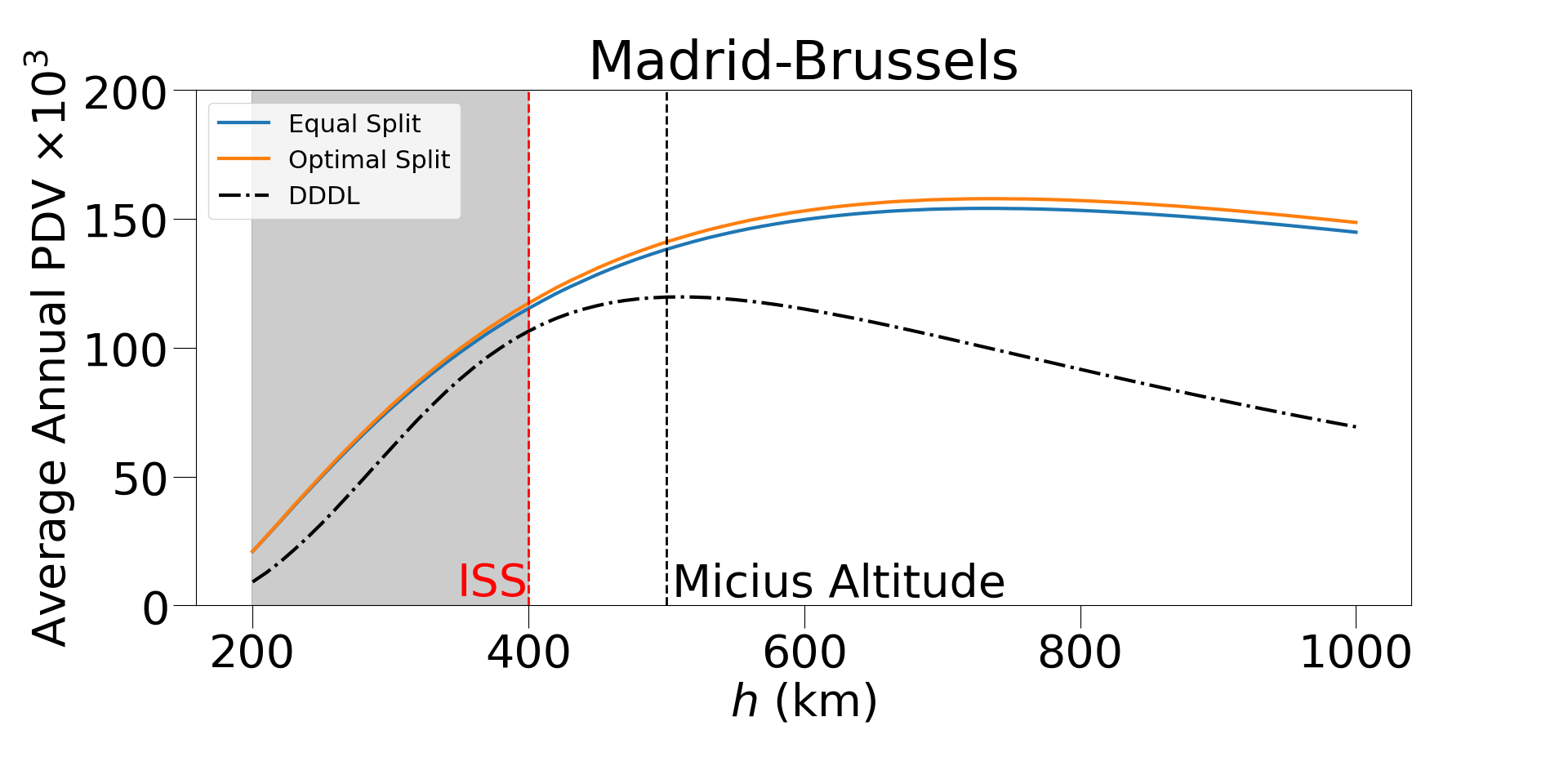}
\caption{Madrid-Brussels ($d =1318$km, $\phi_{\Delta=0}=152.0^\circ$).
}
\end{subfigure}
\caption{Altitudinal PDV dependence. (a) Paris-Nice, (b) London-Berlin, (c) Seoul-Tokyo and (d) Madrid-Brussels links. Assumed system parameters $N_\text{sat}=200$ and $\eta_{loss}^{sys}=25.9$ dB. There are distinct optimal altitudes of operations for DDDL and SSQR, with the latter having higher optimal altitudes. The optimal altitudes are higher for OGS pairs with longer link distances. For orbits $h \gtrsim 500$ km SSQR was superior to DDDL in each case, an important consideration if orbital lifetime at lower altitudes is a concern. The Micius and International Space Station (ISS) orbital altitudes are indicated for comparison.}
\label{fig:avgAnnualPDVs}
\end{figure}

The annual PDV analysis uses the same values of $R_\text{EPS-DDDL}=5.9\times 10^6$ pairs s$^{-1}$ and $\eta^\text{sys}_\text{loss}=25.9\ \text{dB}$ as in Table~\ref{table:IIIAParameters} and $N_\text{sat}=200$. We first examine the altitudinal dependence of the annual PDV (Fig.~\ref{fig:avgAnnualPDVs} and Table~\ref{tab:PDValtitude}) and the optimal altitudes for both protocols.
In each case, SSQR has an optimal altitude higher than that of the DDDL. Increasing altitude leads to longer dual-visibility windows, but at the cost of increasing channel losses. Without memories, the increased losses quickly negate any benefit of longer overpasses. This is evident in Fig.~\ref{fig:avgAnnualPDVs} where the DDDL PDVs decrease sharply when moving beyond the optimal altitude. As the link loss scaling for SSQR is milder, the repeater satellite can therefore better buffer the increasing losses that come with increasing orbital altitudes and exploit the longer overpass times.
For the Paris-Nice and London-Berlin links, SSQR with the optimal static split outperformed DDDL for $h$ greater than $450$~km and $310$~km, respectively. For the Seoul-Tokyo and Madrid-Brussels links, SSQR was superior to DDDL at all altitudes considered. This is consistent with previous observations that the advantage of repeater satellites becomes more pronounced in high loss environments (Fig.~\ref{fig:nCVaryDr}).

A difference in the relative benefit of adopting an optimal static memory split versus an equal split was established. The London-Berlin and Seoul-Tokyo links saw the greatest benefits, with the optimal static memory splits improving the average annual PDV by $31\%$ and $29\%$ compared to the equal split, respectively. This is substantially greater than the benefit seen to the Paris-Nice and Madrid-Brussels links of $4\%$ and $2\%$, respectively. This difference in relative benefit can be explained in terms of the $\phi_{\Delta=0}$. For each OGS pair, the dual-visibility criterion can only be met for a range of $\Delta$ values $[\Delta_-,\Delta_+]$, for a polar satellite this is met for overpass longitudes in an approximately symmetric interval either side of the OGS baseline midpoint longitude~\cite{wang2026scalability}.
For OGSs at moderate latitudes, the overpass $\phi$ values will be similar to $\phi_{\Delta=0}$. 
If $\phi_{\Delta=0}$ is close to $90^\circ$, then the $\phi$ values for visible overpasses will also be close to $90^\circ$ and there is a substantial benefit for adopting an optimal static memory allocation (Fig.~\ref{fig:fullLandscapePlots}d). For $\phi_{\Delta=0}$ close to $0^\circ$ or $180^\circ$, there is little benefit of static memory optimisation since the link-loss balance with each OGS swings between one and the other so that a static split cannot meaningfully improve performance. 

\begin{figure}[!b]
\centering
\newcommand{\figwidth}{0.48\textwidth}
\begin{subfigure}{\figwidth}
\centering
\includegraphics[width=\linewidth]{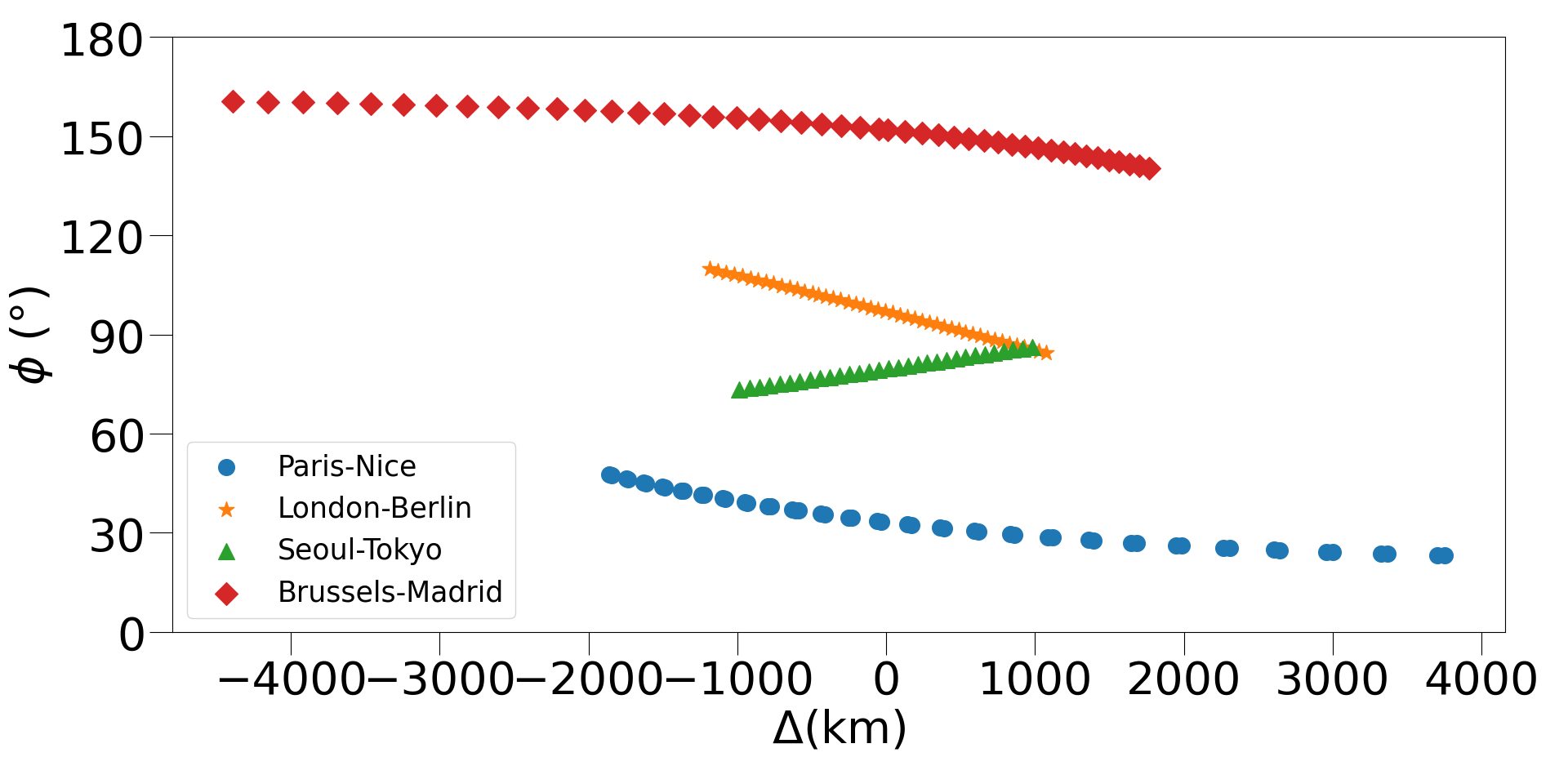}
\caption{Visible $\{\Delta,\phi\}$ for OGS Pairs
}
\end{subfigure}
\hfill
\begin{subfigure}{\figwidth}
\centering
\includegraphics[width=\linewidth]{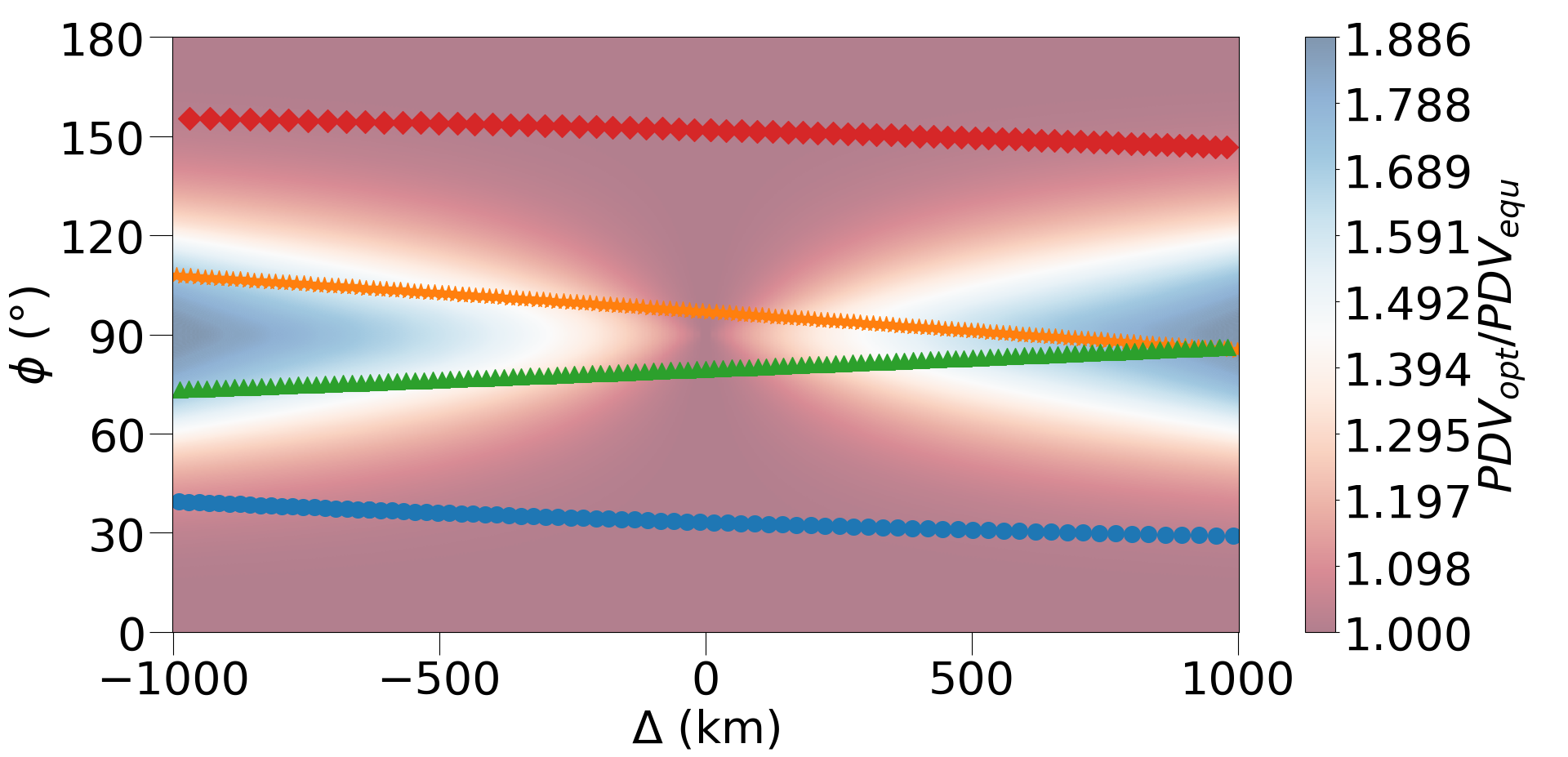}
\caption{$\{\Delta,\phi\}$ values in (a) overlaid on Fig.~\ref{fig:fullLandscapePlots} (d).}
\end{subfigure}
\caption{(a) $\{\Delta,\phi\}$ values whose corresponding overpasses satisfy the dual-visibility criterion. The $\phi$ values depend the OGS baseline inclination and are low for the OGS pairs with a high baseline inclination (Paris-Nice, Madrid-Brussels) and high $\phi$ values for lower inclinations (London-Berlin, Seoul-Tokyo). (b) $\{\Delta,\phi\}$ values in (a) overlaid on Fig.~\ref{fig:fullLandscapePlots} (d), showing the ratio of the optimal static split to equal split PDV. The symmetric of $\{\Delta, \phi\}$ about $\{0, 90^\circ\}$ was used to extend the heat map. For Paris-Nice and Madrid-Brussels, the lower $\phi$ values translate to a small benefit in adopting an optimal static memory split. Conversely, London-Berlin and Seoul-Tokyo receive a significant benefit when optimising memory allocation in this way. Note that Fig. $\ref{fig:fullLandscapePlots}$ (d) assumes $d = 1000\ \text{km}$ - the overlay is illustrative.}
\label{fig:visibleDeltaPhis}
\end{figure}

Fig.~\ref{fig:visibleDeltaPhis} makes this notion concrete for the OGS pairs considered here.
For the Paris-Nice and Madrid-Brussels OGS pairs with low $\phi_{\Delta=0}$ of $33.3 ^\circ$ and $152.0 ^\circ$ respectively, the dual-visibility overpasses are approximately aligned with the OGS baseline, leading to a minimal benefit when the memory allocation is statically optimised. For London-Berlin ($\phi_{\Delta=0}=97.0^\circ$) and Seoul-Tokyo ($\phi_{\Delta=0}=79.6^\circ$), the greater perpendicularity of the overpasses with respect to the OGS baseline translates to a significant benefit when an optimal static memory split is adopted.

\begin{table}[!btp]
    \centering
    \begin{tabular}{l|ccr}
    \toprule
    OGS Pair & Protocol & Max $\bar{N}_\text{year}$ & $h_\text{opt}$\\
    \midrule
    Paris-Nice & DDDL & $596 \times10^3$ &  $240$ km\\
     $d=691$ km, $\phi_{\Delta=0}=33.3^\circ$ & SSQR - Equal Split & $387\times10^3$ &  $380$ km\\
               & SSQR - Optimum Static & $401\times10^3$ &  $380$ km\\
        \hline
    London-Berlin & DDDL & $330 \times10^3$ &  $340$ km\\
    $d=932$ km, $\phi_{\Delta=0}=97.0^\circ$ & SSQR - Equal Split & $294\times10^3$ & $520$ km \\
               & SSQR - Optimum Static & $386\times10^3$ & $490$ km \\
        \hline
    Seoul-Tokyo & DDDL & $144 \times10^3$ & $440$ km \\
    $d=1163$ km, $\phi_{\Delta=0}=79.6^\circ$ & SSQR - Equal Split & $163\times10^3$ &  $650$ km\\
               & SSQR - Optimum Static & $209\times10^3$ &  $630$ km\\
        \hline
    Madrid-Brussels & DDDL & $120 \times10^3$ & $510$ km \\
    $d =1318$ km, $\phi_{\Delta=0}=152.0^\circ$ & SSQR - Equal Split & $154\times10^3$ & $730$ km \\
               & SSQR - Optimum Static & $158\times10^3$ & $740$ km \\   
    \bottomrule
    \end{tabular}
    \caption{Summary of altitudinal dependence. For each of the OGS pairs, the maximum average annual PDVs and respective optimum altitude for the different protocols are listed. As OGS baselines increase, so does the optimum altitude. The benefit of the SSQR per-pass memory split is more noticeable for the lower inclination OGS baselines. Only for Paris-Nice does DDDL outperform SSQR with an optimised split. The SSQR optimum altitudes for the 3 longer baselines are above the ISS altitude and benefit from reduced residual drag and longer orbital durations without reboosting~\cite{Oi_2017CubeSat}.}
    \label{tab:PDValtitude}
\end{table}

\subsection{Fidelity of the Distributed Pairs}
\label{sec:MCFidelity}
The SSQR results in Sections~\ref{sec:perPassPDV} and~\ref{sec:annualPDV} looked at the PDV when the satellite memory was constrained only in its capacity without consideration of the quality of the distributed pairs. We now consider the effect of memory dephasing and its effect on the fidelity of the distributed pairs (see Section~\ref{sec:entFidelityAndMemNoise}). We employ a round-based MC model of the repeater satellite (Section~\ref{sec:MCMethods}) to access qubit waiting time statistics and the resulting fidelities. The waiting-time analysis is performed for the 4 representative overpasses (Fig.~\ref{fig:4passesDiagram}), each repeated 1000 times to obtain statistics for the average and variation.

Due to the dynamic nature of satellite links, the channel parameters will vary with time. This, combined with the stochastic nature of entanglement distribution, results in a spread in qubit waiting times and the resulting fidelities. Understanding how this distribution varies with both overpass geometry and choice of $\tau_\text{cut}$ can provide useful information for future mission design.

A repeater satellite with $N_\text{sat} = 200$ with different choices of cut-off time is considered, with all other parameters kept as in Table~\ref{table:IIIAParameters}. The memory efficiency, i.e. the probability that a transmitted photon is successfully stored and later retrieved, is taken to be $\eta_\text{mem} = 1$. We also assume that memory-retrieval is error free to simplify the analysis and focus on the effects of dephasing. 

\begin{figure}[!btp]
\centering
\newcommand{\figwidth}{0.48\textwidth}
\begin{subfigure}{\figwidth}
\centering
\includegraphics[width=\linewidth]{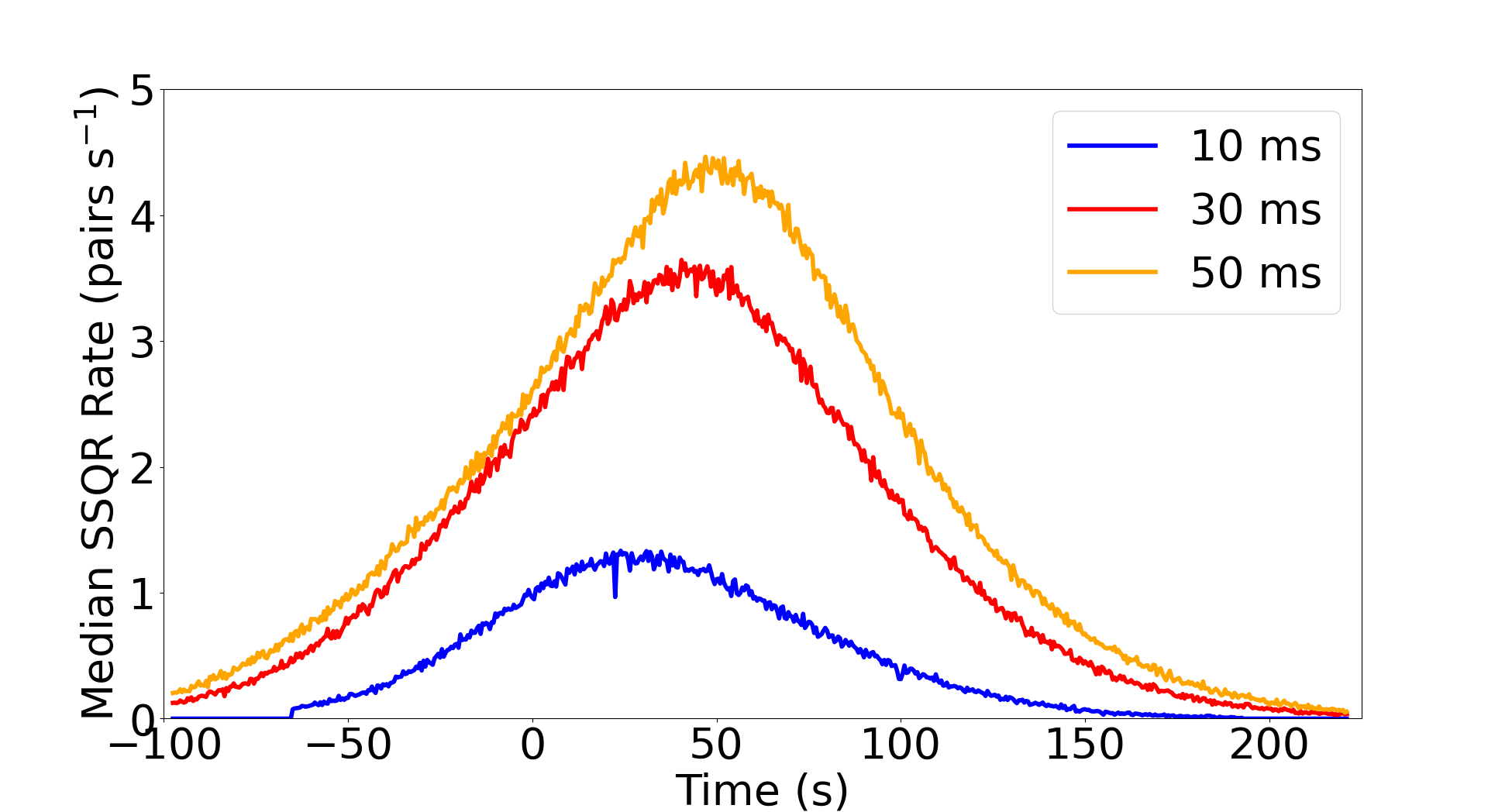}
\caption{Zenith-zenith. Total PDV: $293 \pm 12$, $718 \pm 19$ and $811 \pm 18$, for $\tau_\text{cut} = 10, 30, 50$~ms, respectively.}
\label{fig:zenith200}
\end{subfigure}
\hfill
\begin{subfigure}{\figwidth}
\centering
\includegraphics[width=\linewidth]{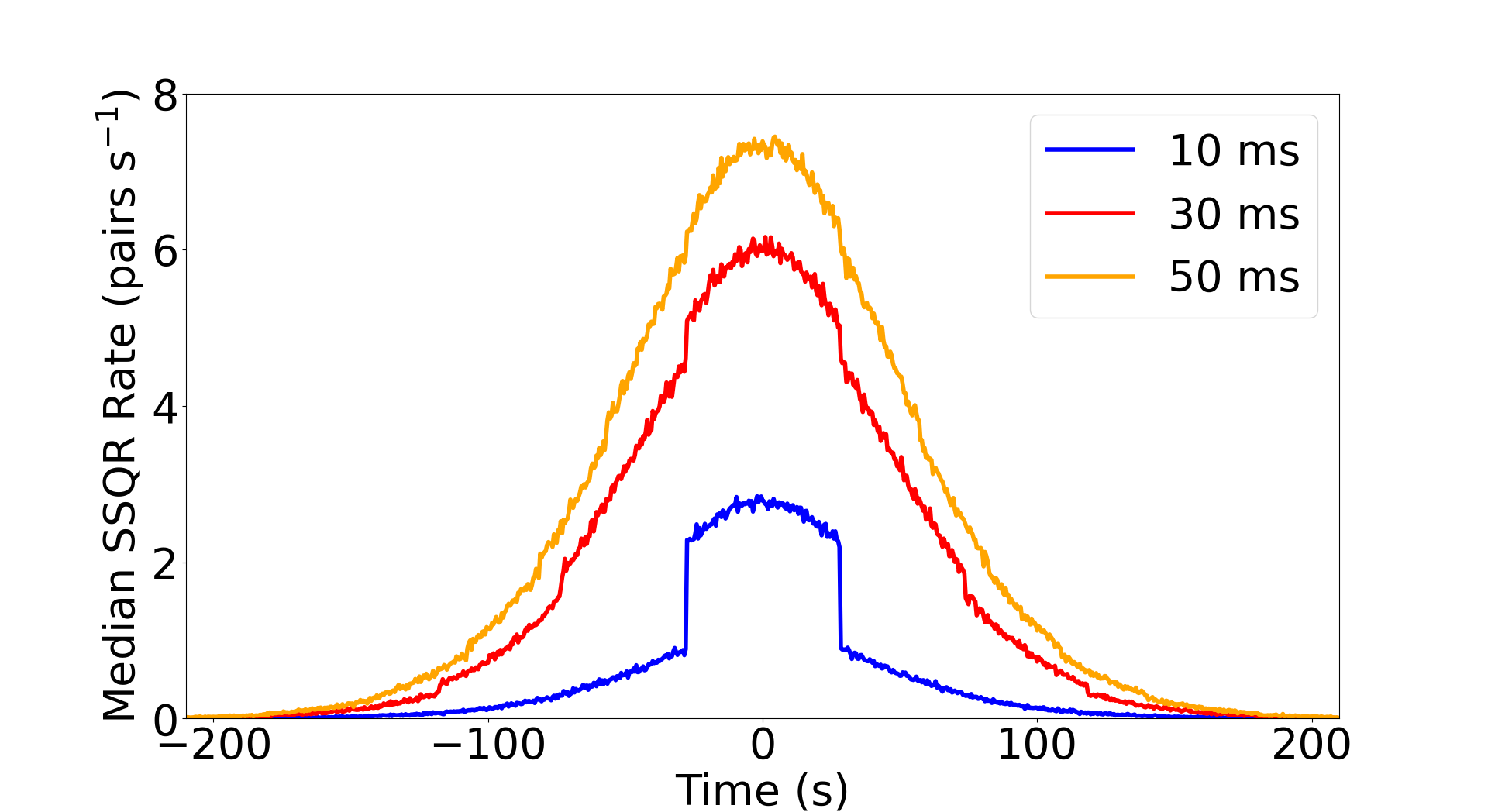}
\caption{Symmetric overpass. Total PDV: $216 \pm 10$, $706 \pm 17$ and $929 \pm 20$, for $\tau_\text{cut} = 10, 30, 50$~ms, respectively.}
\label{fig:symmteric200}
\end{subfigure}
\par\medskip
\begin{subfigure}{\figwidth}
\centering
\includegraphics[width=\linewidth]{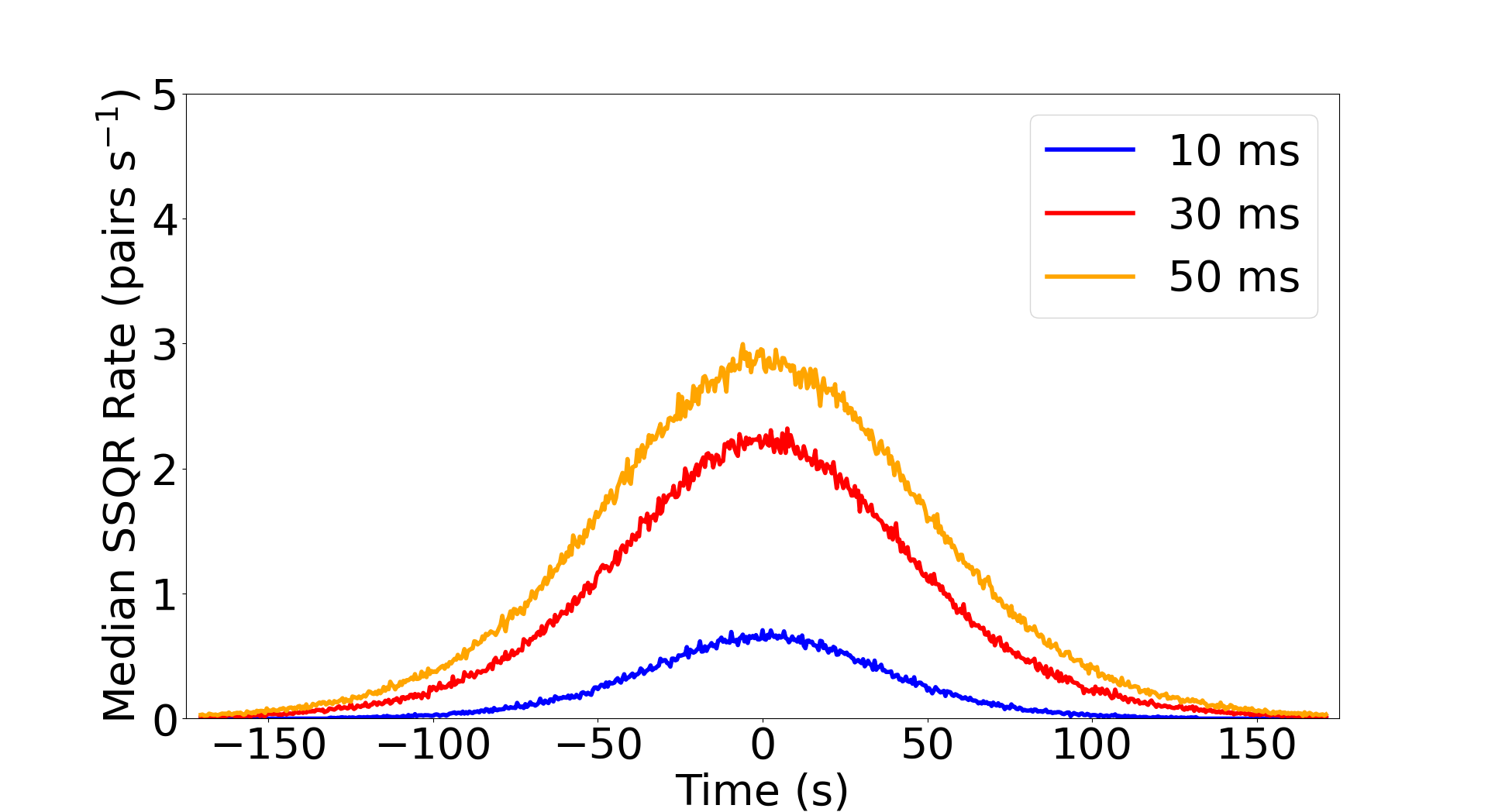}
\caption{Zenith A $90^\circ$. Total PDV: $60 \pm 6$, $251 \pm 11$ and $352 \pm 13$, for $\tau_\text{cut} = 10, 30, 50$~ms, respectively.}
\label{fig:zenitha90200}
\end{subfigure}
\hfill
\begin{subfigure}{\figwidth}
\centering
\includegraphics[width=\linewidth]{Graphics/zenA45_nSat200_mean_rates_different_cutoffs.png}
\caption{Zenith A $45^\circ$. Total PDV: $137 \pm 8$, $447 \pm 15$ and $562 \pm 16$, for $\tau_\text{cut} = 10, 30, 50$~ms, respectively.}
\label{fig:zenitha45200}
\end{subfigure}
\caption{Mean entanglement distribution rates obtained from the Monte Carlo simulation for the (a) Zenith-zenith, (b) Symmetric, (c) Zenith A $90^\circ$ and (d) Zenith A $45 ^\circ$, for $\tau_\text{mem} = 10, 30$ and $50$~ms. The mean total PDV and standard deviation are stated in the sub-figure captions. In each case using a lower $\tau_\text{cut}$ leads to lower rates and a lower total PDV. For $\tau_\text{cut} = 10$~ms there are intervals in each overpass where there is $0$ PDV due to the round-trip time exceeding the cut-off. This suggests that varying $\tau_\text{cut}$ during the overpass would be beneficial. The step-behaviour in the distribution rate around $t=0$~s in the symmetric overpass is an artefact of the round-based approach used in the model.}
\label{fig:different_cutoffs_dist_rates}
\end{figure}

For each of the representative overpasses we simulated the entanglement distribution rate and total PDV for different values of $\tau_\text{cut}$. The results presented are the mean obtained from $1000$ runs of the Monte Carlo model. In each case, (Fig.~\ref{fig:different_cutoffs_dist_rates}) we see that adopting a lower cut-off leads to a reduction in entanglement distribution rates and the final per-overpass PDV.  For the zenith-zenith overpass, for example, using $\tau_\text{cut} = 50$~ms led to an improvement in PDV of a factor $2.8$ compared to $\tau_\text{cut}=10$~ms. For the symmetric, Zenith A $90^\circ$ and Zenith A $45^\circ$ overpasses this improvement factor was greater, at $4.3$, $5.8$ and $5.1$, respectively. This variation suggests that the choice of which $\tau_\text{cut}$ to use depends on the overpass geometry.

By comparing Figs.~\ref{fig:different_cutoffs_dist_rates} and~\ref{fig:4PassesPDRs} it is possible to estimate the reduction in PDV caused by imposing finite cut-off time relative to the case where the memory lifetime is unconstrained. For example, the PDV for the symmetric overpass was found to be $1801$ pairs in the absence of memory lifetime limitations. This is $1585$ pairs more than the average PDV when $\tau_\text{cut} = 10$~ms. For $\tau_\text{cut}=30,50$~ms, the differences are $1095$ and $872$. These quantify the reduction in PDV that results from having finite memory lifetimes, and the accompanying need to discard qubits that have been degraded.

\begin{figure}[!bth]
\centering
\newcommand{\figwidth}{0.48\textwidth}
\begin{subfigure}{\figwidth}
\centering
\includegraphics[width=\linewidth]{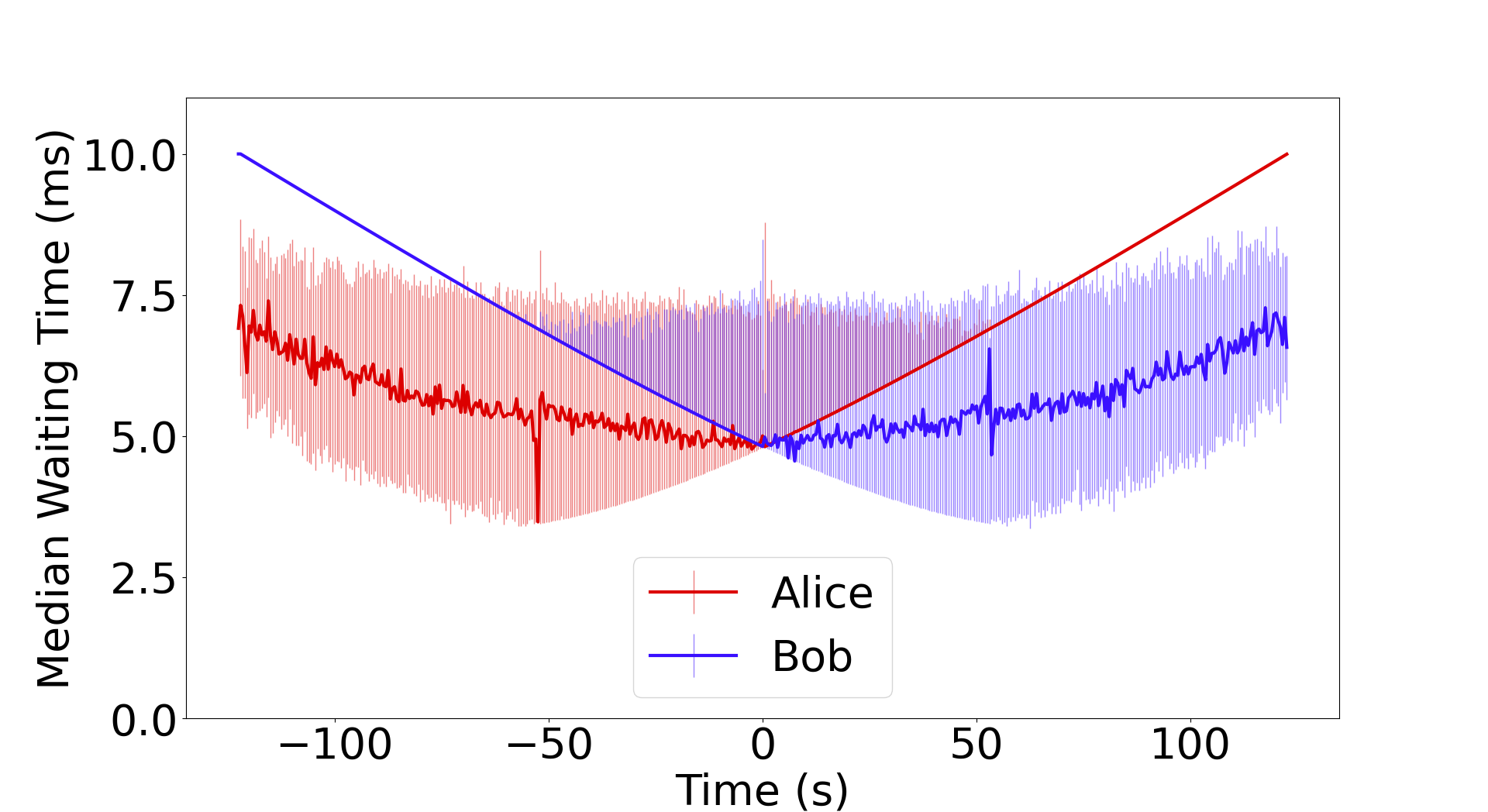}
\caption{Zenith-zenith with $\tau_\text{cut} =10$~ms.}
\end{subfigure}
\hfill
\begin{subfigure}{\figwidth}
\centering
\includegraphics[width=\linewidth]{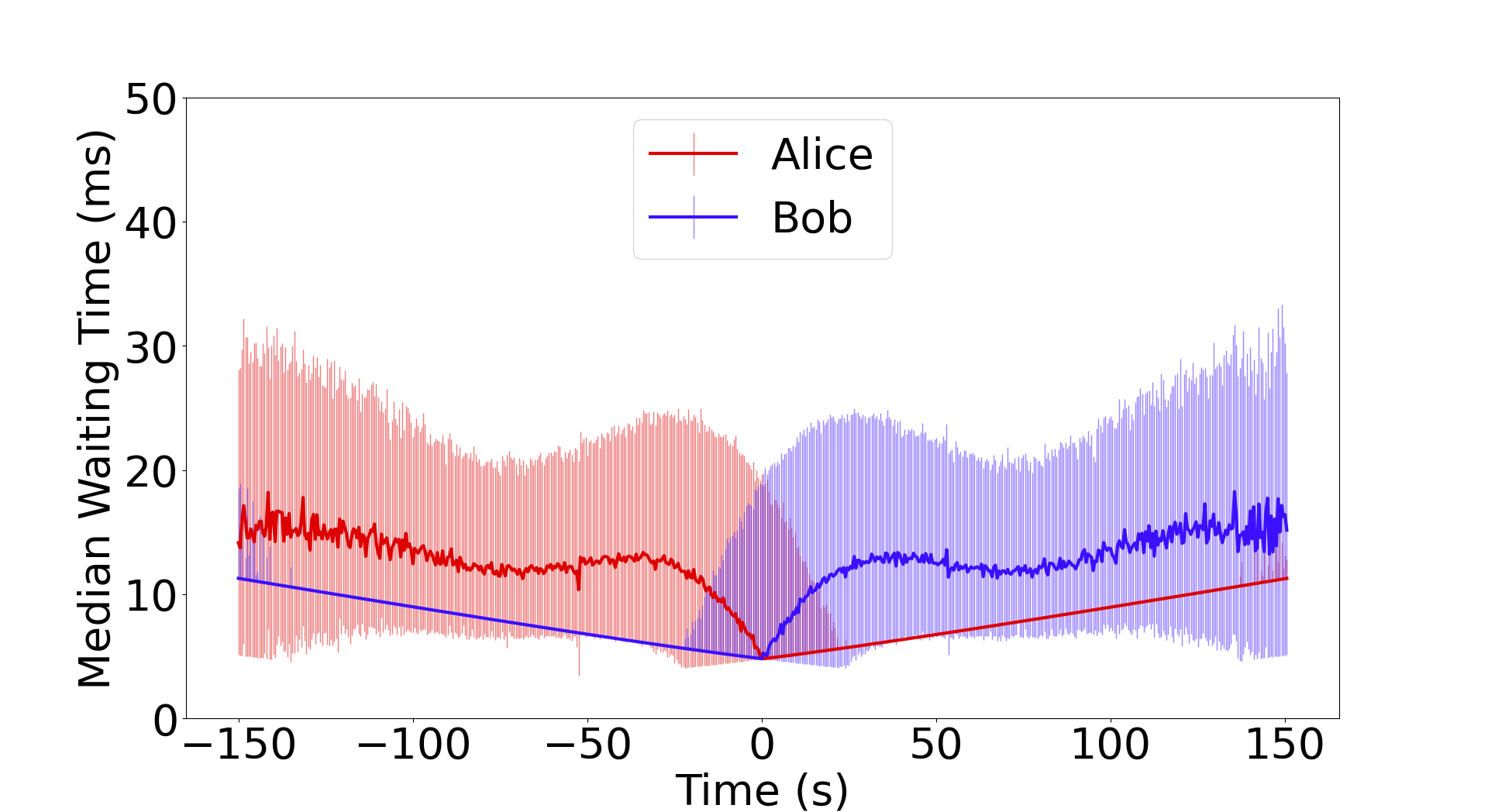}
\caption{Zenith-zenith with $\tau_\text{cut} =50$~ms.}
\end{subfigure}
\par\medskip
\begin{subfigure}{\figwidth}
\centering
\includegraphics[width=\linewidth]{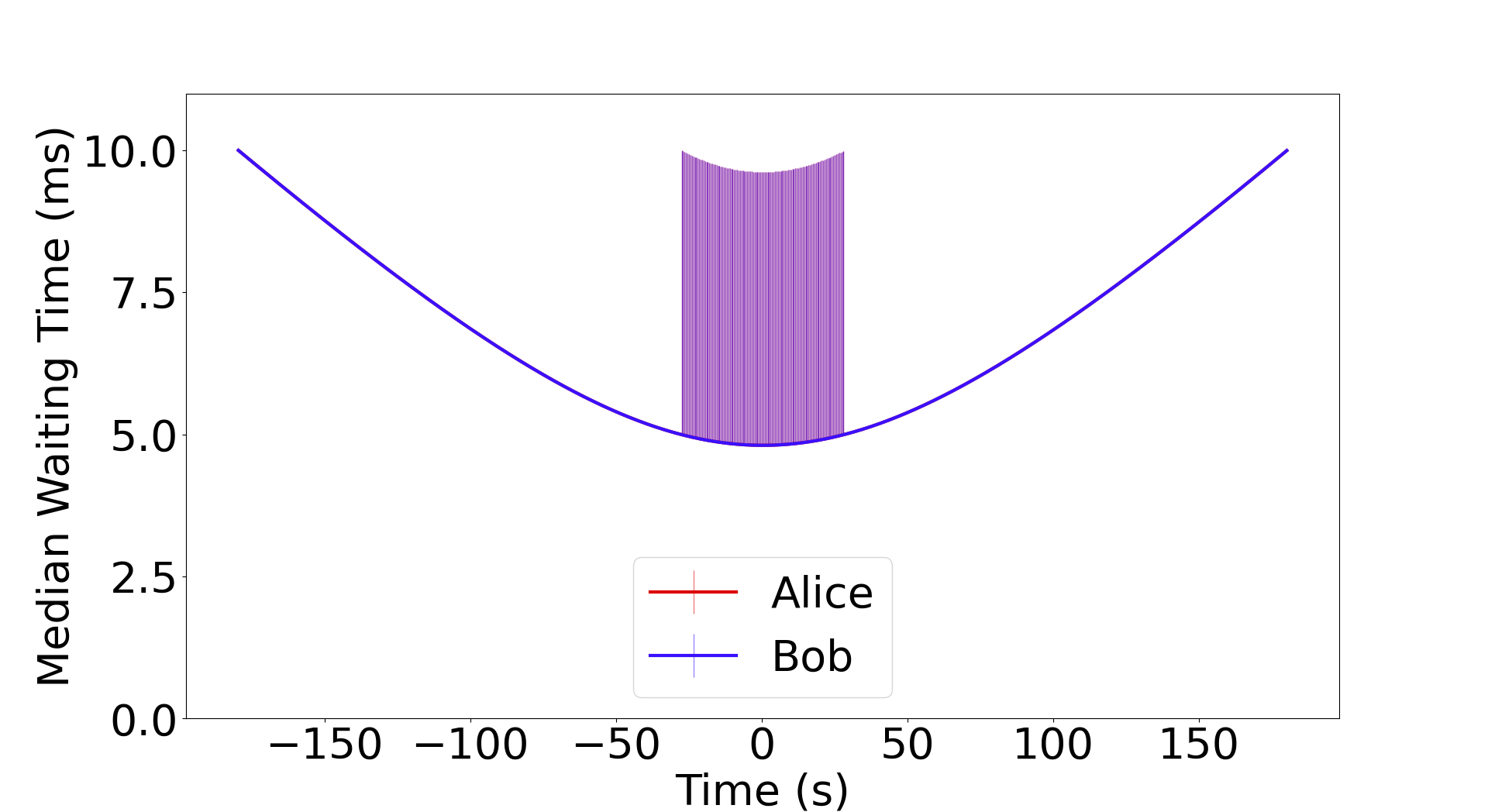}
\caption{Symmetric overpass with $\tau_\text{cut} =10$~ms.}
\end{subfigure}
\hfill
\begin{subfigure}{\figwidth}
\centering
\includegraphics[width=\linewidth]{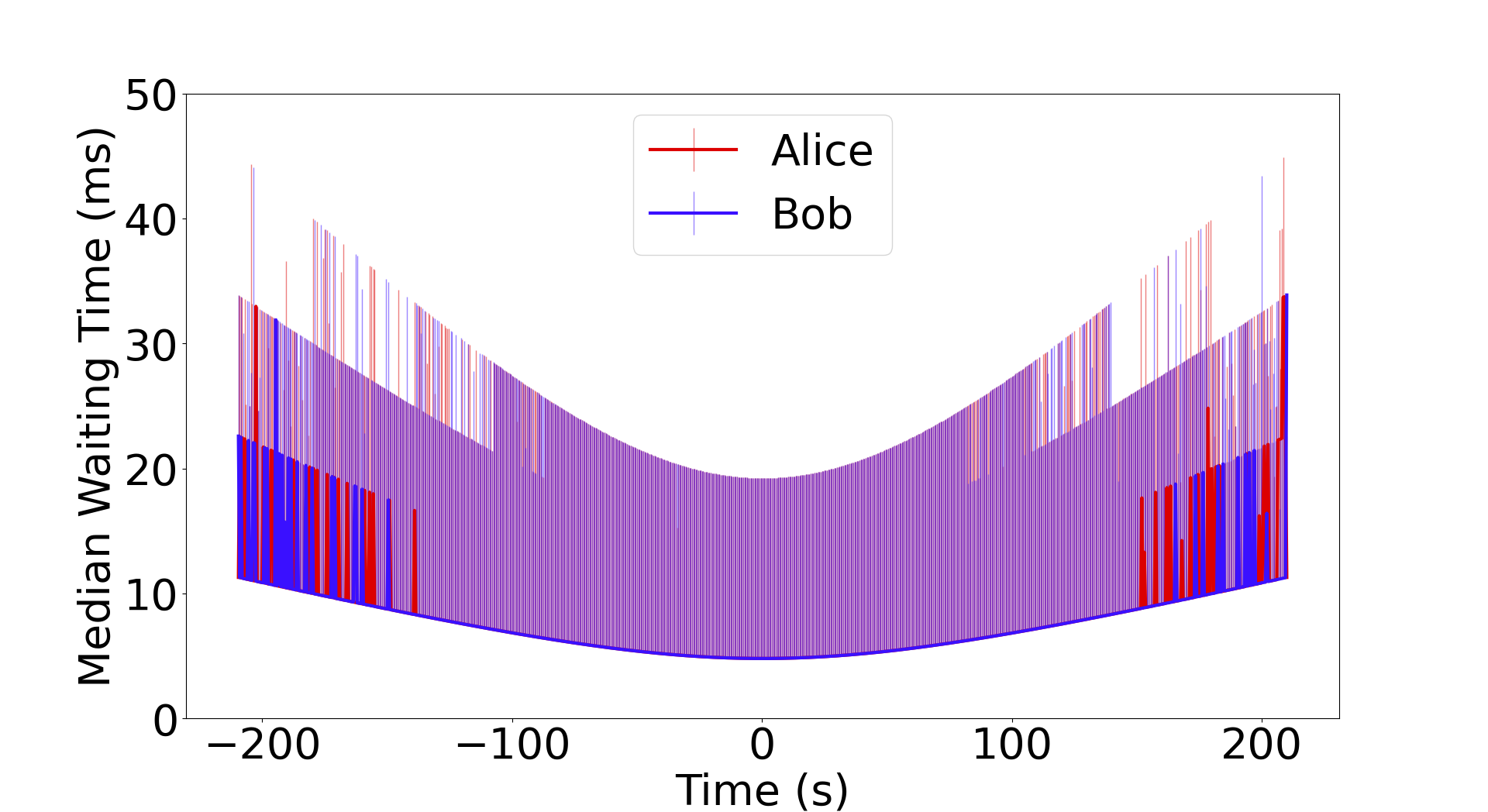}
\caption{Symmetric overpass with $\tau_\text{cut} =50$~ms.}
\end{subfigure}
\par\medskip
\begin{subfigure}{\figwidth}
\centering
\includegraphics[width=\linewidth]{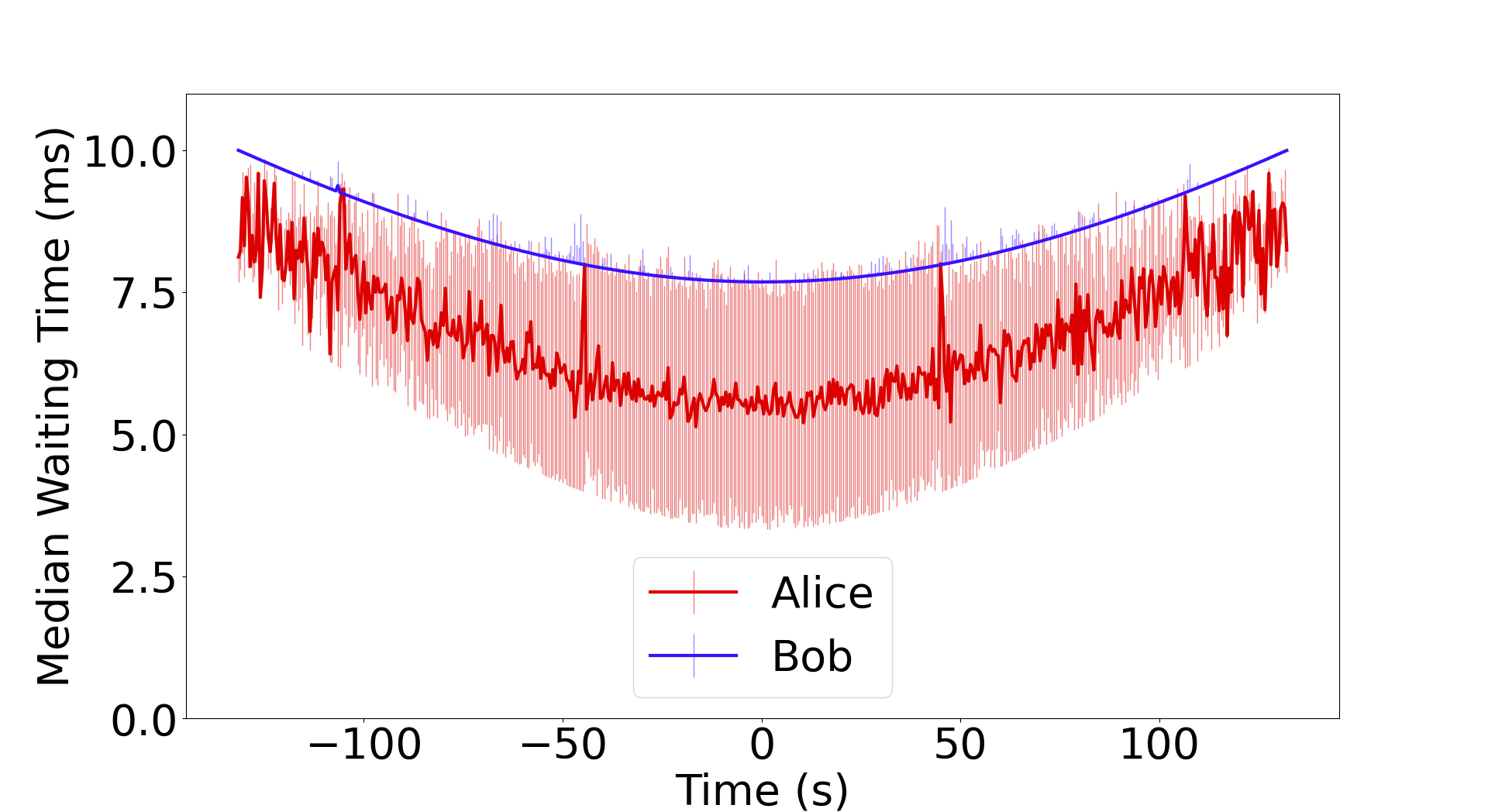}
\caption{Zenith A $90^\circ$ with $\tau_\text{cut} =10$~ms.}
\end{subfigure}
\hfill
\begin{subfigure}{\figwidth}
\centering
\includegraphics[width=\linewidth]{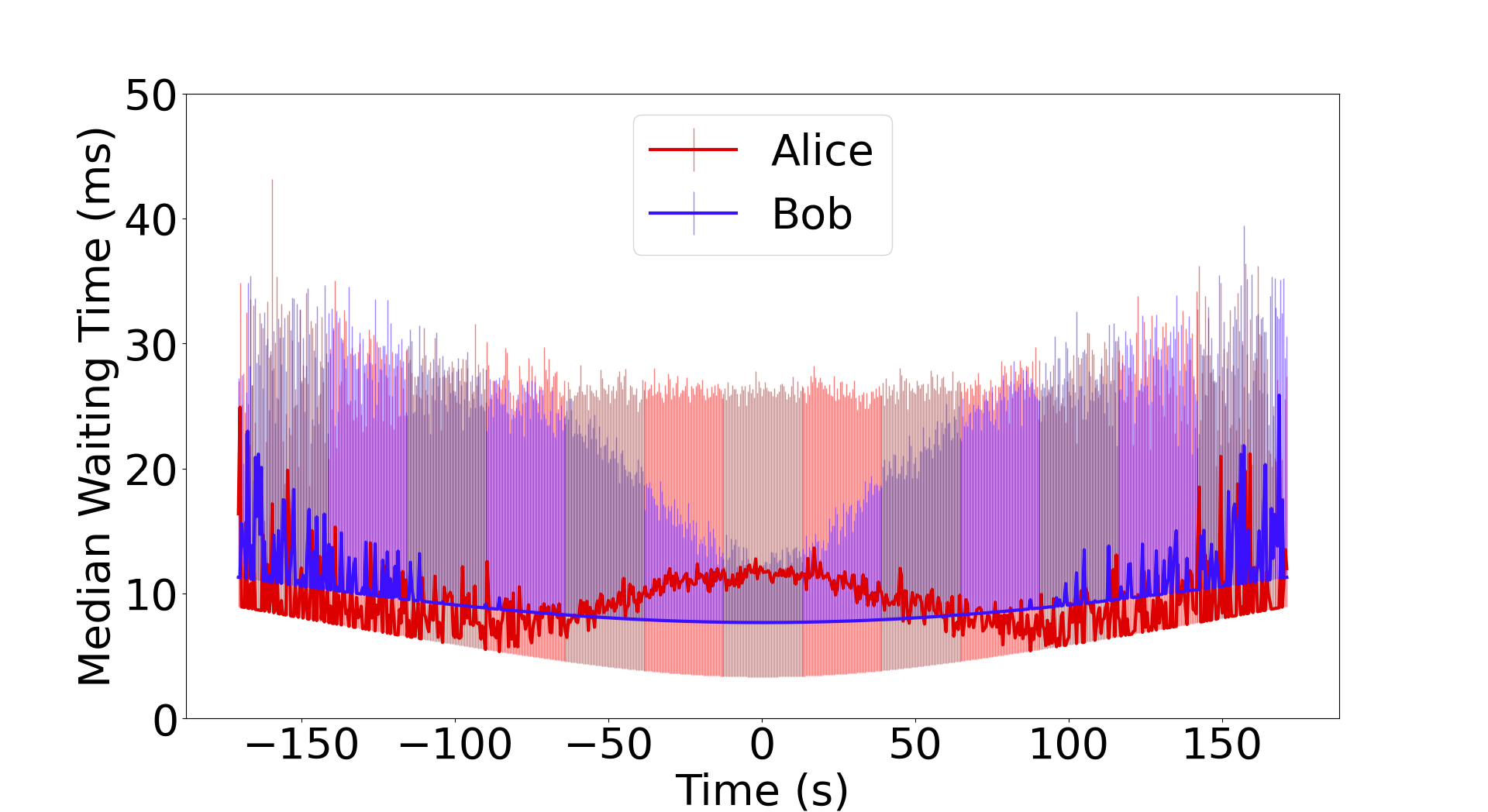}
\caption{Zenith A $90^\circ$ with $\tau_\text{cut} =50$~ms.}
\end{subfigure}
\par\medskip
\begin{subfigure}{\figwidth}
\centering
\includegraphics[width=\linewidth]{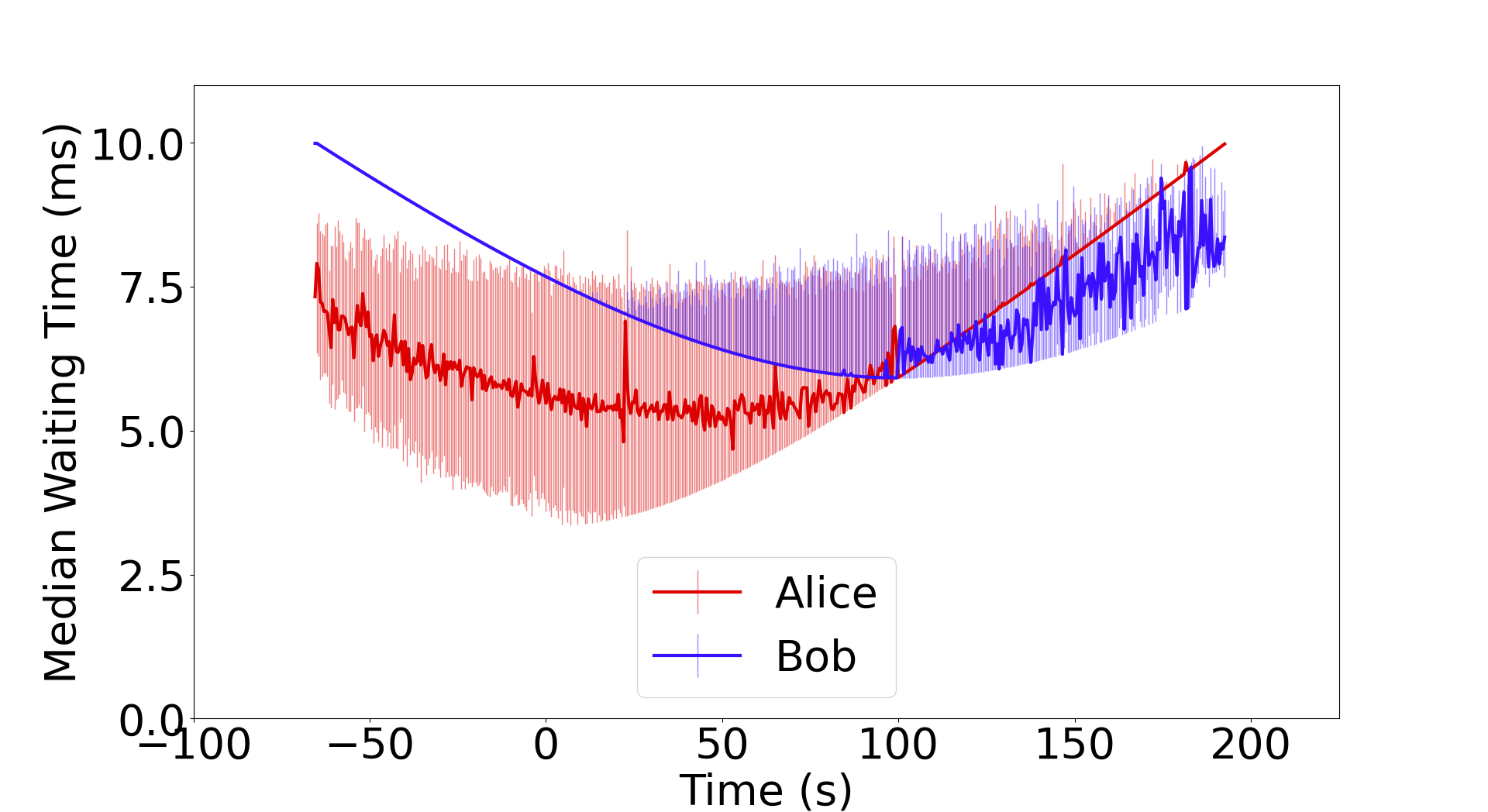}
\caption{Zenith A $45^\circ$ with $\tau_\text{cut} =10$~ms.}
\label{fig:zenitha902000}
\end{subfigure}
\hfill
\begin{subfigure}{\figwidth}
\centering
\includegraphics[width=\linewidth]{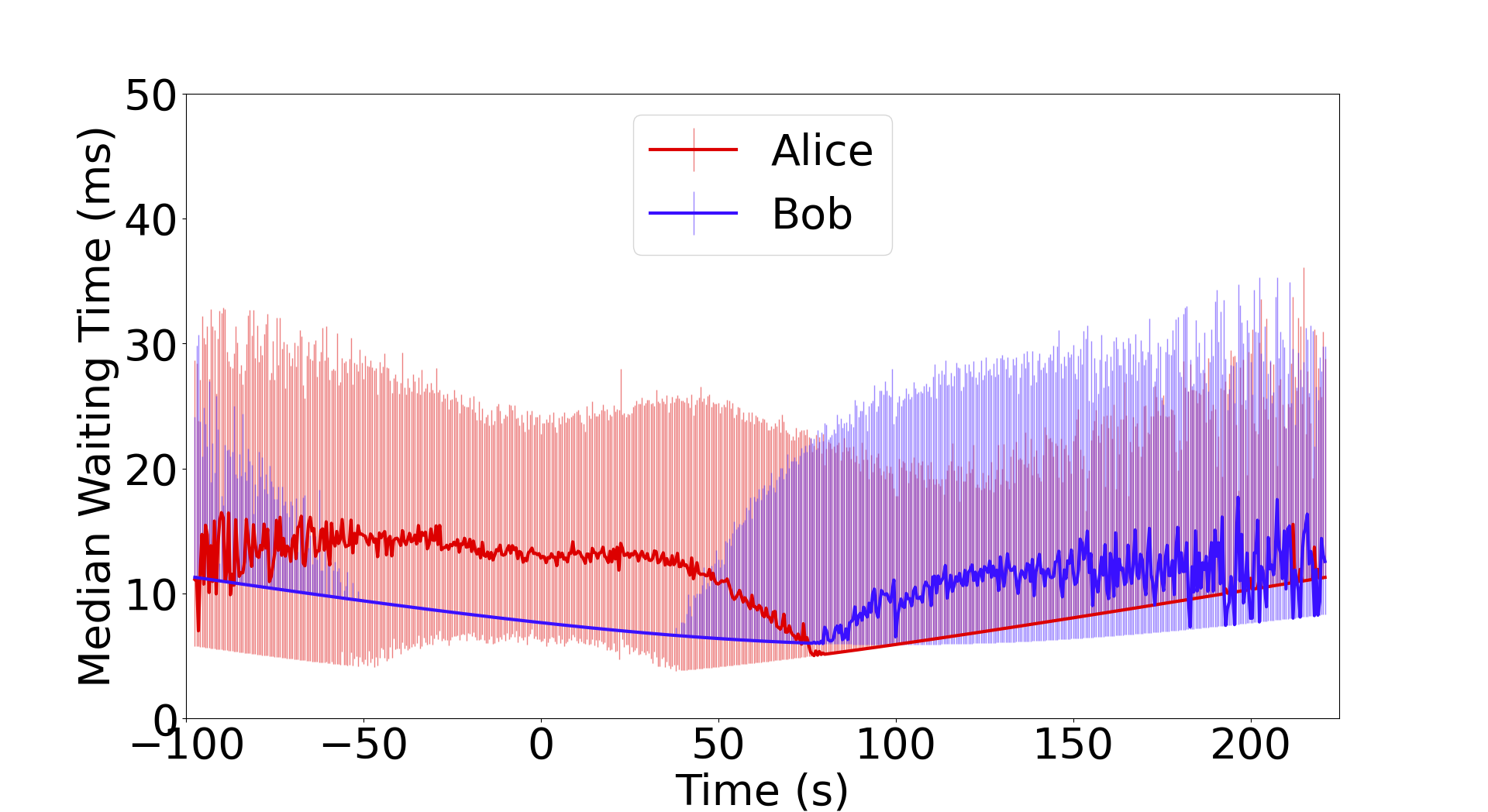}
\caption{Zenith A $45^\circ$ with $\tau_\text{cut} =50$~ms.}
\label{fig:zenitha452000}
\end{subfigure}
\caption{Median qubit waiting times as a function of time for the (a,b) Zenith-zenith, (c,d) Symmetric, (e,f) Zenith A $90^\circ$ and (g,h) Zenith A $45^\circ$ overpasses. The left and right panels correspond to $\tau_\text{cut} = 10$~ms and $50$~ms, respectively. Different choices of $\tau_\text{cut}$ change not only the profile of the waiting time curve (expect for the symmetric overpass) but it also reduces the spread in the distribution of waiting times. We note the abrupt spikes observed in waiting times in e.g. (a) and (e). These are due to the round-based scheduling and occur when the photon round-trip times on the Alice and Bob links approach integer multiples of each other, producing a beating or Moir\'e effect.}
\label{fig:zenZen_zenA90_waiting_times_comparison}
\end{figure}

The choice of $\tau_\text{cut}$ also affects the distribution of the qubit waiting times. This is illustrated in Fig.~\ref{fig:zenZen_zenA90_waiting_times_comparison}, which shows how the median qubit waiting times change during the overpass for $\tau_\text{cut} = 10$~ms and $50$~ms. Imposing a shorter $\tau_\text{cut}$ leads to shorter typical waiting times, as expected. It also leads to a reduction in their variability, suggesting that short $\tau_\text{cut}$ will be required when consistency is an important consideration.

With the exception of the symmetric overpass, changing the cut-off changed the profile of how the qubit waiting times vary with time. For the symmetric overpass, the waiting times are simply random spread about the round-trip time. Note that for $\tau_\text{cut} = 10$~ms there only a small region around $t=0$ for which there is a non-zero IQR in the symmetric overpass waiting times. This an artefact of the round-based Monte Carlo model. During this interval the photon round-trip time falls below $5$~ms. Since trimming of memory qubits is performed before swapping this means that the memory qubits are stored for two entanglement distribution attempts. At all other points in the overpass the round-trip time excess $5$~ms and qubits are only stored for one entanglement distribution attempt, resulting in zero IQR.

We also see that the satellite only accumulates pairs with one OGS at a time, where the accumulation of pairs is associated with waiting times in excess of the round trip time. This stems from the fact that the distribution rates on the two independent downlinks are generally different and the OGS will the higher rate will accumulate surplus pairs on the satellite. The OGS with the lower rate will never accumulate pairs as any qubits successfully transmitted through that link will immediately be used in swapping. 

The accumulation of un-swapped stored qubits and associated long waiting times is an inefficiency. Qubits sitting in memory take up resources that could be used in fresh entanglement generation attempts. The stored qubits also decohere, comprising the quality of the entanglement. Since the accumulation of pairs is fundamentally caused by unequal rates on each downlink, the memory would ideally be allocated to equalise these at all times. Doing so is not possible for the optimal static allocation and would require a dynamic allocation~\cite{fittipaldiEntSwappingInOrbit2024}, which is beyond the scope of this work. 

\begin{figure}[!tbp]
\centering
\newcommand{\figwidth}{0.48\textwidth}
\begin{subfigure}{\figwidth}
\centering
\includegraphics[width=\linewidth]{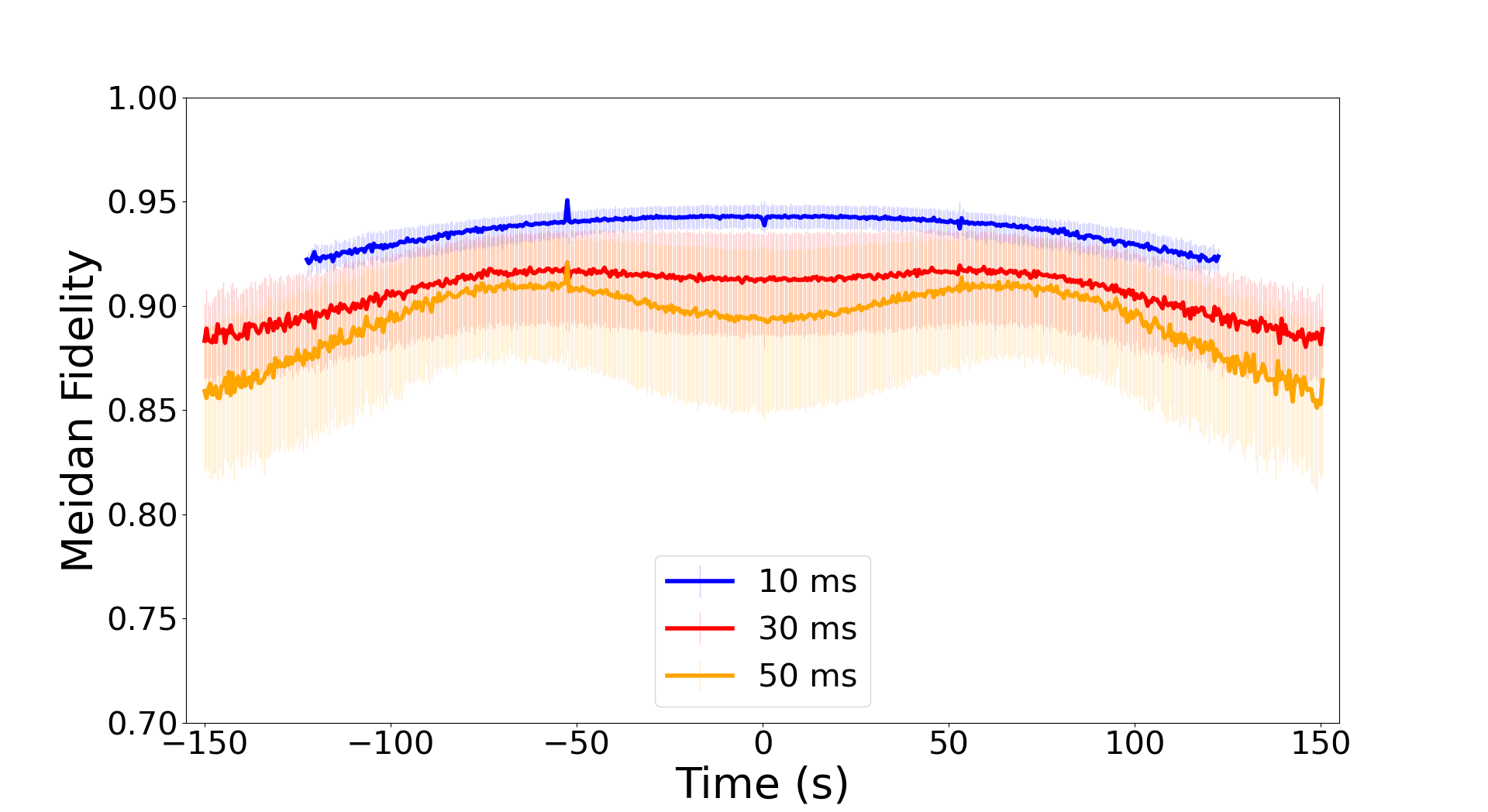}
\caption{Zenith-zenith.}
\end{subfigure}
\hfill
\begin{subfigure}{\figwidth}
\centering
\includegraphics[width=\linewidth]{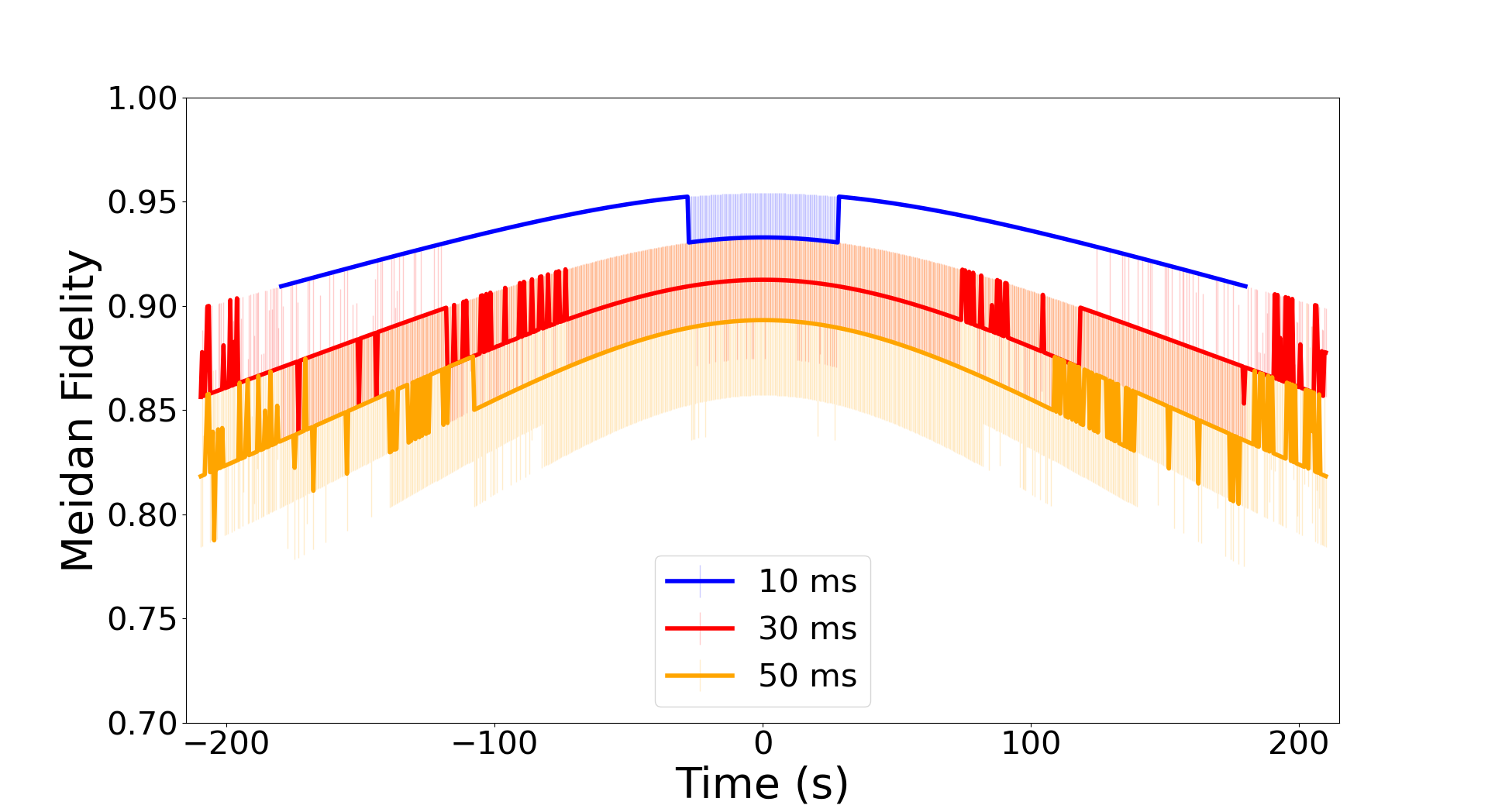}
\caption{Symmetric overpass.}
\end{subfigure}
\par\medskip
\begin{subfigure}{\figwidth}
\centering
\includegraphics[width=\linewidth]{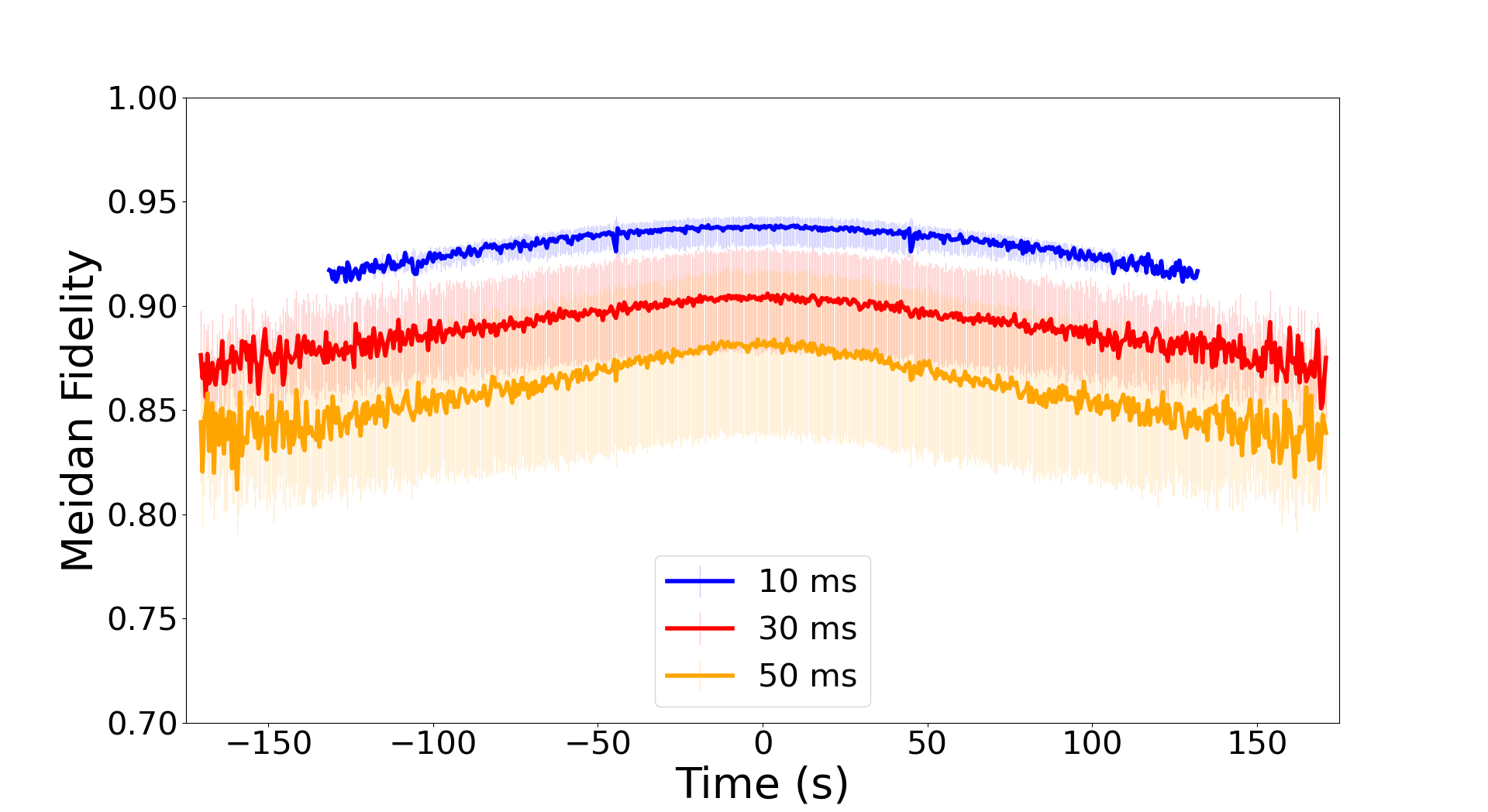}
\caption{Zenith A $90^\circ$.}
\end{subfigure}
\hfill
\begin{subfigure}{\figwidth}
\centering
\includegraphics[width=\linewidth]{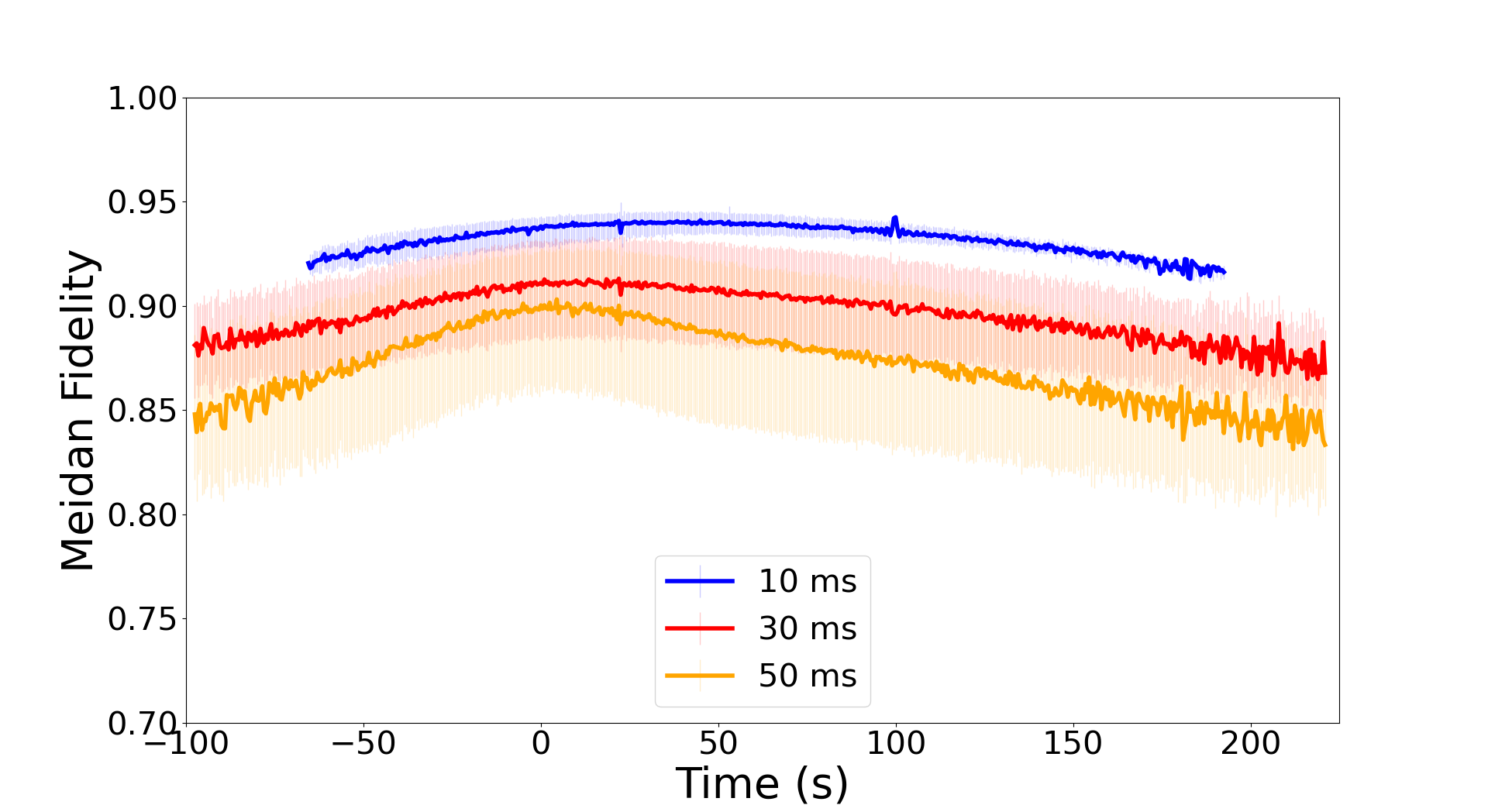}
\caption{Zenith A $45^\circ$.}
\end{subfigure}
\caption{Median fidelities and interquartile ranges as a function of overpass times for (a) Zenith-zenith, (b) Symmetric, (c) Zenith A $90^\circ$ and (d) Zenith A $45^\circ$, for $\tau_\text{cut} = 10, 30$ and $50$~ms. Results in this figure have assumed $N_\text{sat}=2000$ and $\tau_\text{mem}=100$ ms. In each case we see that the median fidelity increases with lower values of $\tau_\text{cut}$, and that there is a reduced spread in the fidelity values. The non-monotonicity in the median fidelity around $t=0$ for the symmetric overpass with $\tau_\text{cut}=10$~ms is a result of the round-based nature of the model and could be suppressed by requiring the satellite to always store qubits for at least one entanglement distribution round. During this interval the entanglement distribution rate continues to increase (Fig.~\ref{fig:different_cutoffs_dist_rates}b) even if the typical pair is of lower fidelity. }
\label{fig:medianFidsVsTime_diffcutoffs}
\end{figure}

\begin{figure}[!tbp]
\centering
\newcommand{\figwidth}{0.48\textwidth}
\begin{subfigure}{\figwidth}
\centering
\includegraphics[width=\linewidth]{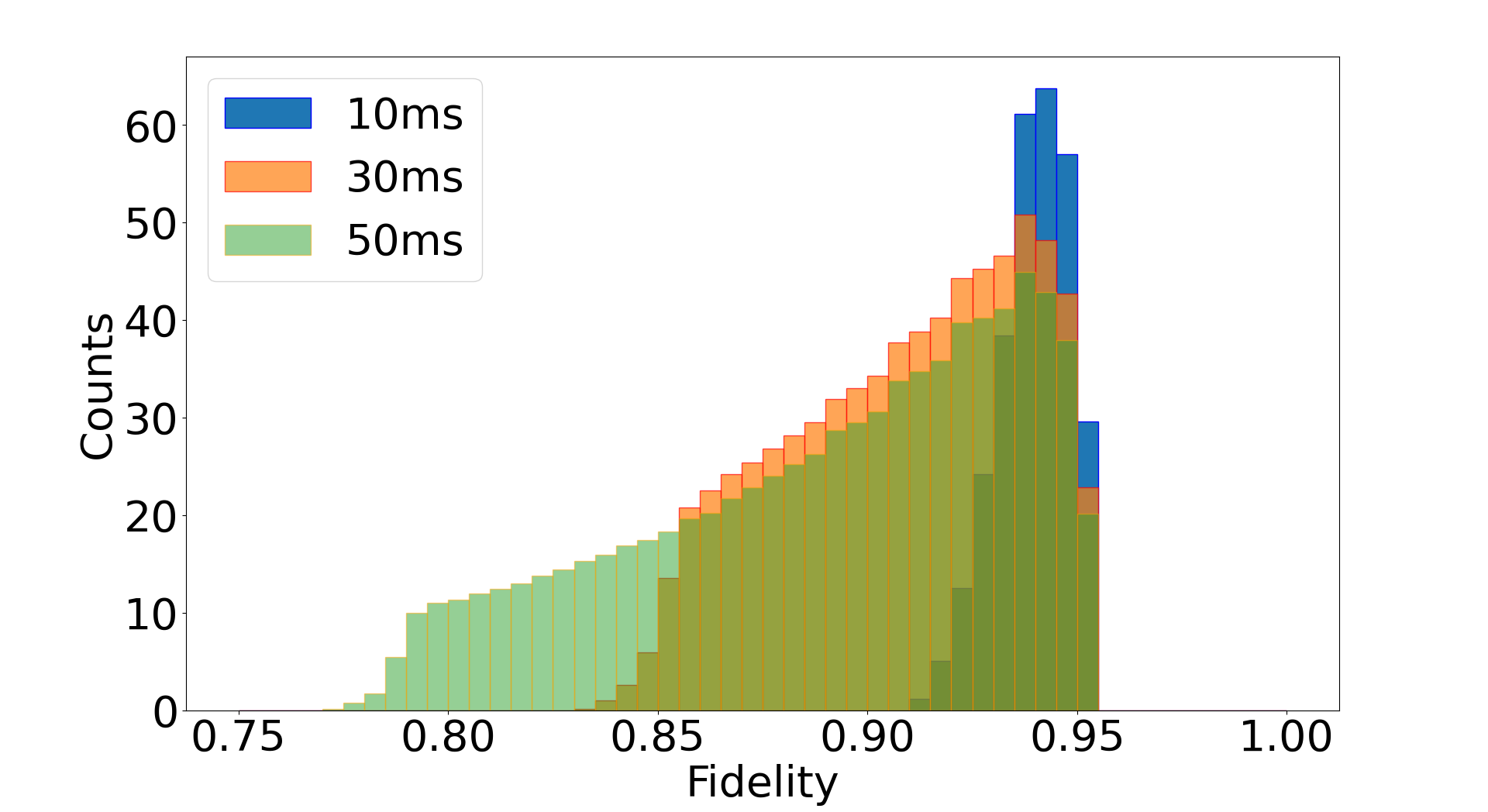}
\caption{Zenith-zenith.}
\end{subfigure}
\hfill
\begin{subfigure}{\figwidth}
\centering
\includegraphics[width=\linewidth]{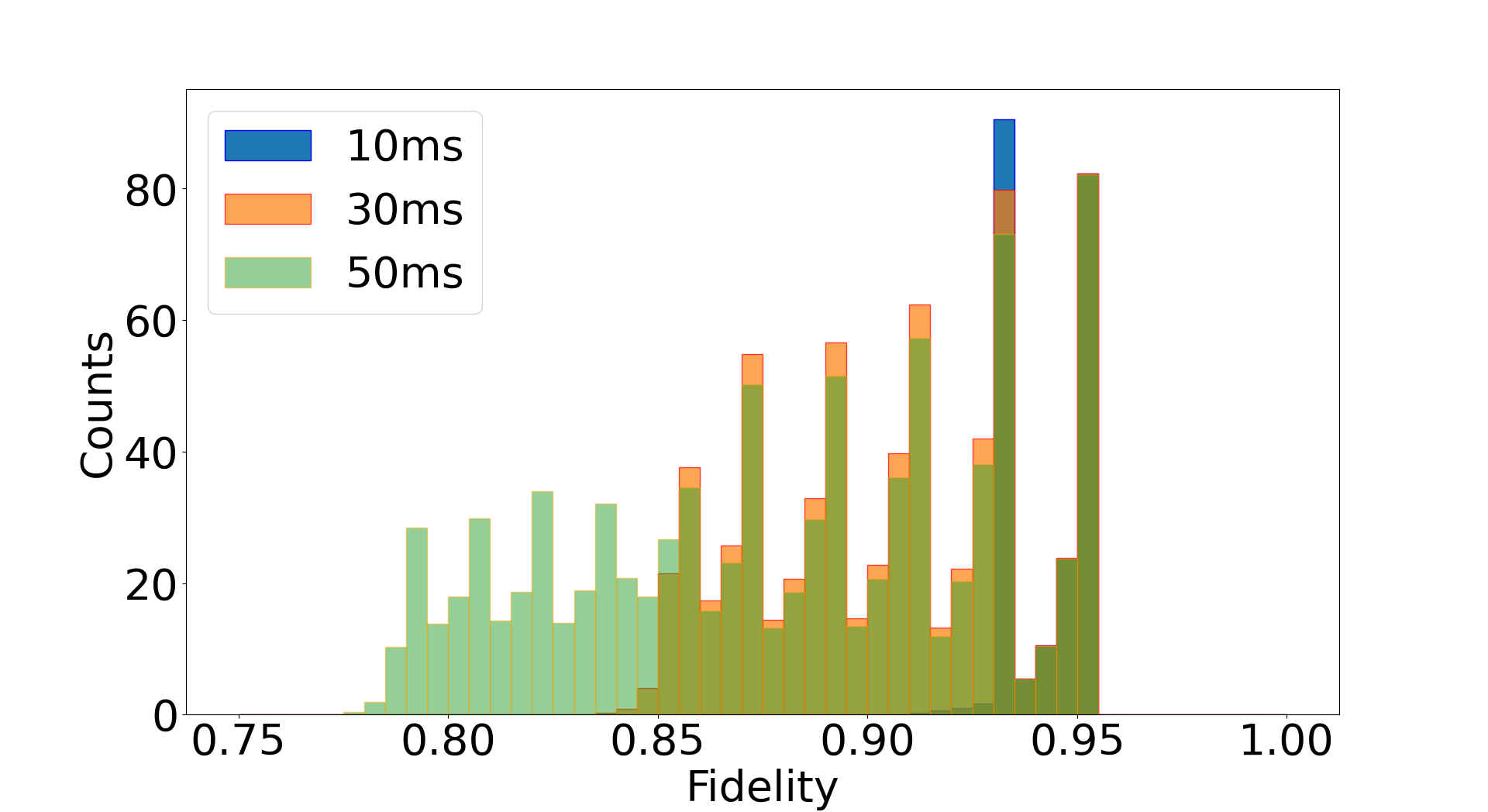}
\caption{Symmetric overpass.}
\end{subfigure}
\par\medskip
\begin{subfigure}{\figwidth}
\centering
\includegraphics[width=\linewidth]{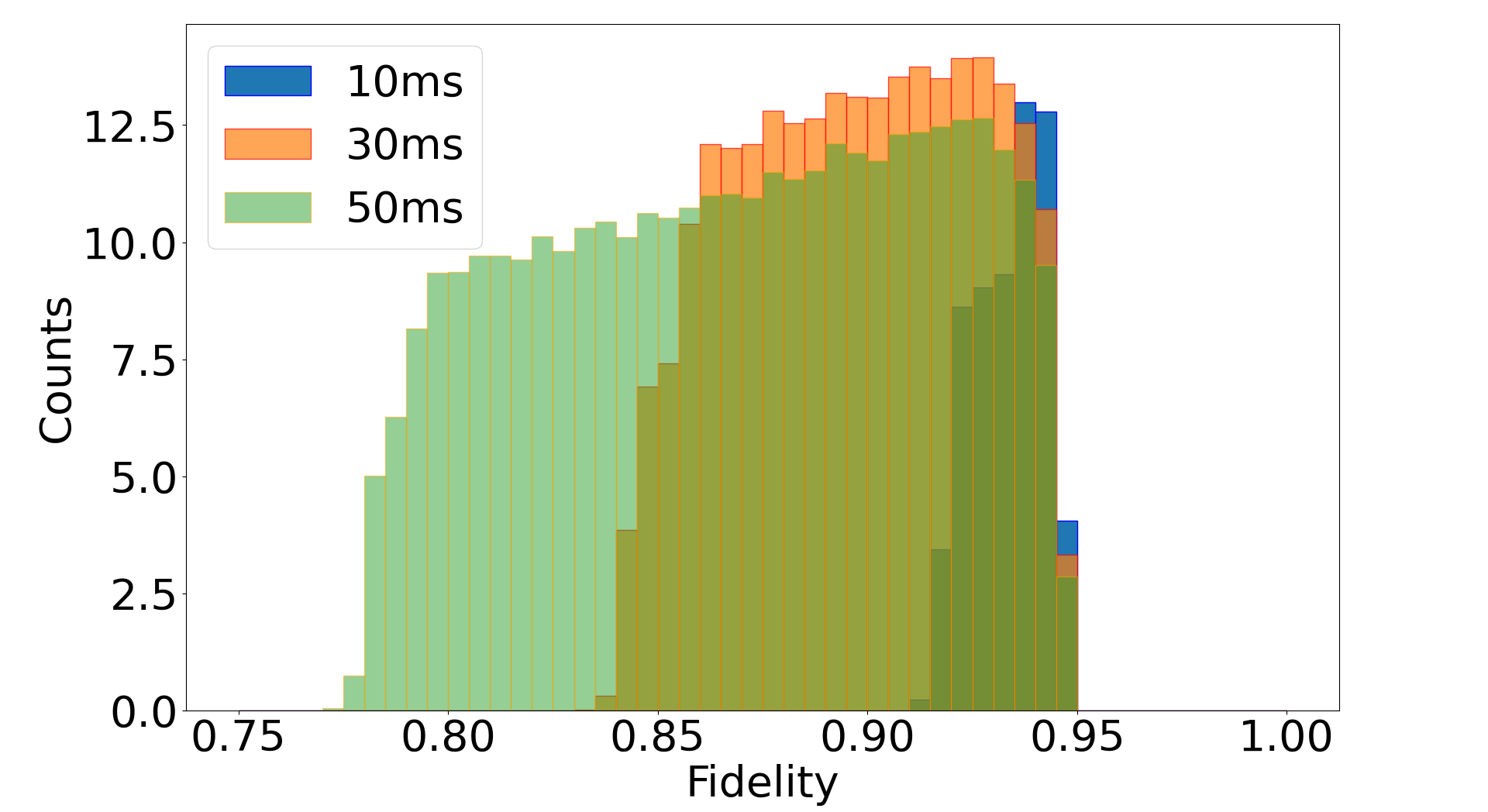}
\caption{Zenith A $90^\circ$.}
\end{subfigure}
\hfill
\begin{subfigure}{\figwidth}
\centering
\includegraphics[width=\linewidth]{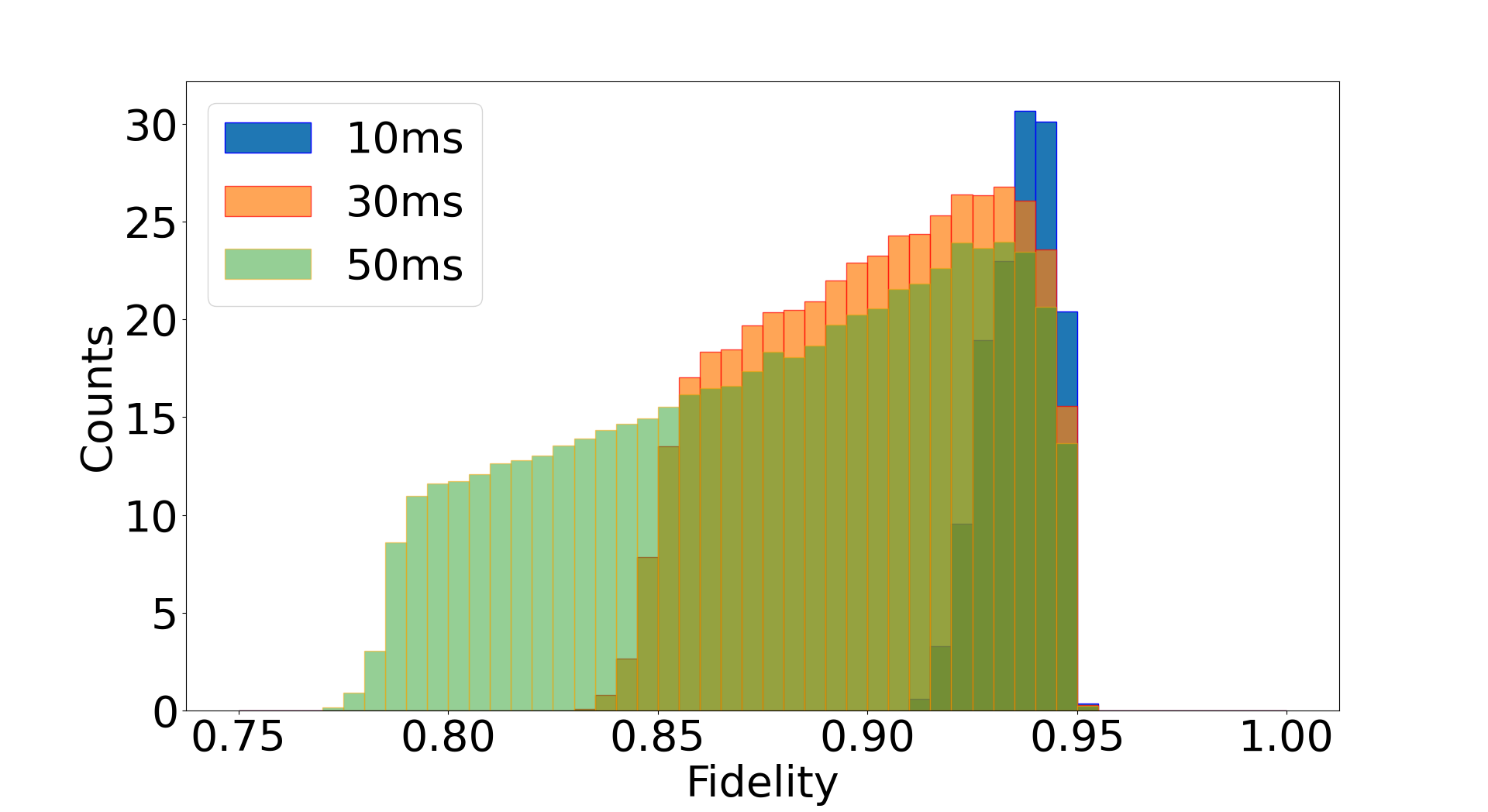}
\caption{Zenith A $45^\circ$.}
\end{subfigure}
\caption{Fidelity histograms (a) Zenith-zenith, (b) Symmetric, (c) Zenith A $90^\circ$ and (d) Zenith A $45^\circ$, for $\tau_\text{cut} = 10, 30$ and $50$~ms, obtained from $1000$ trials. The distribution of fidelities varies with overpass geometry. The dark green sections are points where all $3$ distributions overlap and the lighter green sections are where the distributions for $30$~ms and $50$~ms overlap. Using shorter $\tau_\text{cut}$ has the effect of truncating the low fidelity tail of each distribution, resulting an improvement to the typical fidelity value and its consistency, but reducing the overall pair distribution volume. The jagged appearance for the symmetric overpass in (b) is due to the round-based nature of the MC simulation.}
\label{fig:fidelityHistograms_different_cutoffs}
\end{figure}

The qubit waiting times can be translated into the fidelity of the final pairs. We take a baseline representative memory lifetime $\tau_\text{mem} = 100\ \text{ms}$, potentially achievable in the near-to-midterm in space, and examine how the fidelity varies throughout the overpass and the overall statistical distribution of the fidelity. We see in Fig.~\ref{fig:medianFidsVsTime_diffcutoffs} that fidelity varies throughout the overpass due to different waiting times. Choosing lower values $\tau_\text{cut}$ leads to higher median fidelity. Since having fidelities above a certain threshold is important for a variety of applications, this implies that the choice of cut-off time will depend on the threshold value. The fidelity histograms, shown in Fig.~\ref{fig:fidelityHistograms_different_cutoffs}, display different behaviour for the representative overpass geometries. These results highlight the mechanism by which adopting a lower $\tau_\text{cut}$ improves the typical fidelity and the consistency - truncating the low-fidelity tails of the distribution. Note that a fidelity of $1$ is never achieved due to the unavoidable round-trip time. 

\begin{figure}[!tbp]
\centering
\newcommand{\figwidth}{0.48\textwidth}
\begin{subfigure}{\figwidth}
\centering
\includegraphics[width=\linewidth]{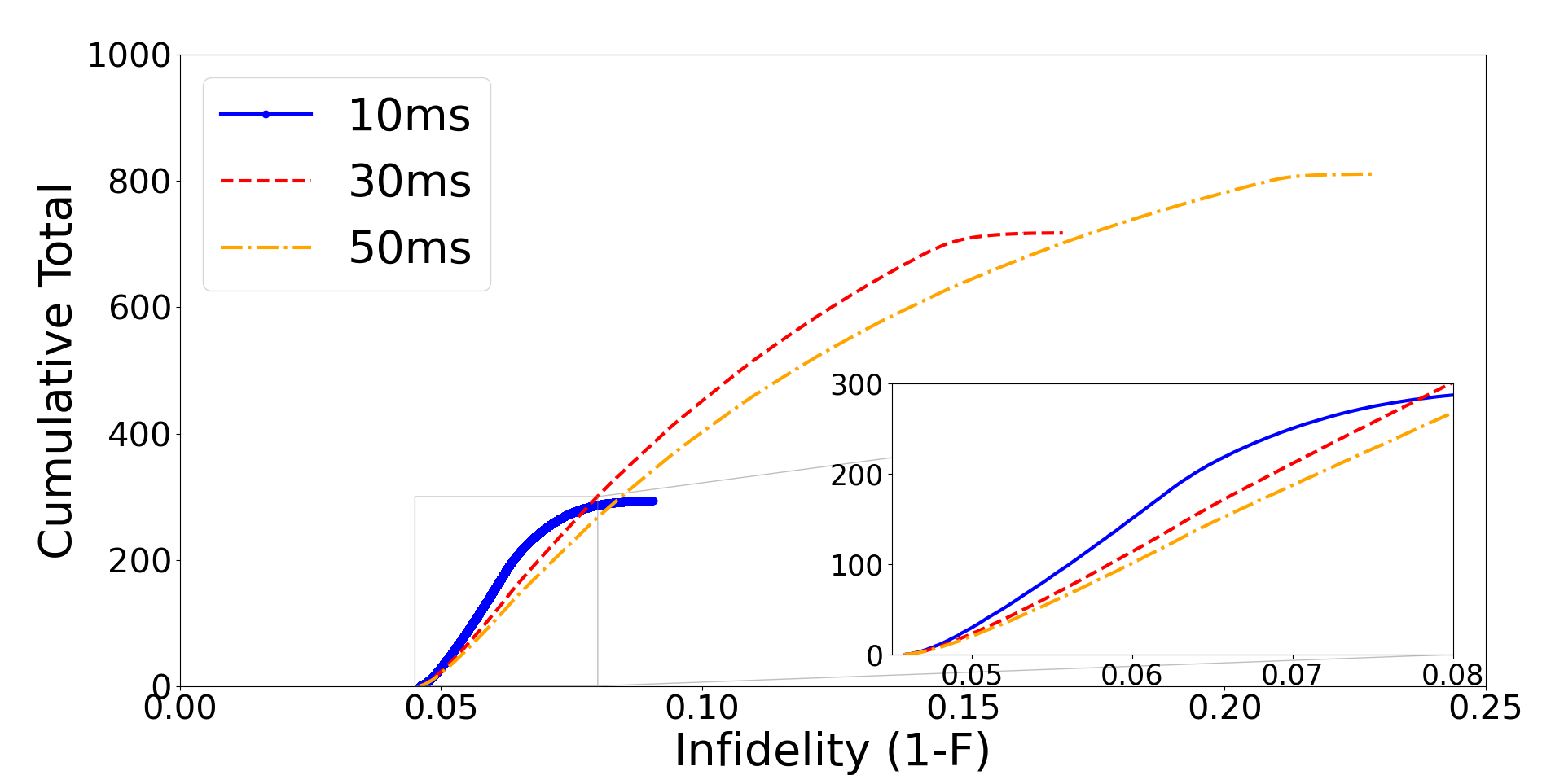}
\caption{Zenith-zenith}
\end{subfigure}
\hfill
\begin{subfigure}{\figwidth}
\centering
\includegraphics[width=\linewidth]{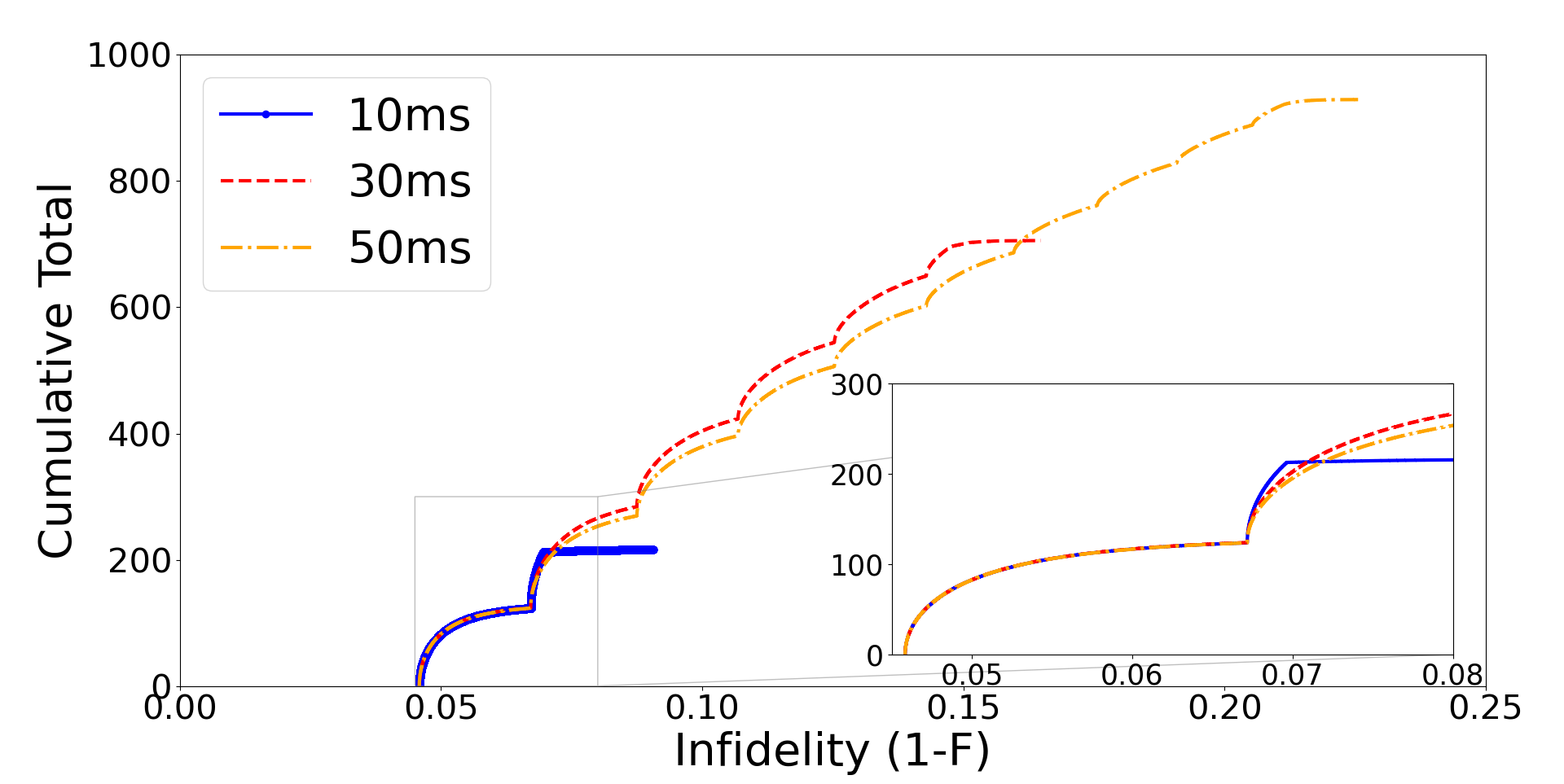}
\caption{Symmetric overpass}
\end{subfigure}
\par\medskip
\begin{subfigure}{\figwidth}
\centering
\includegraphics[width=\linewidth]{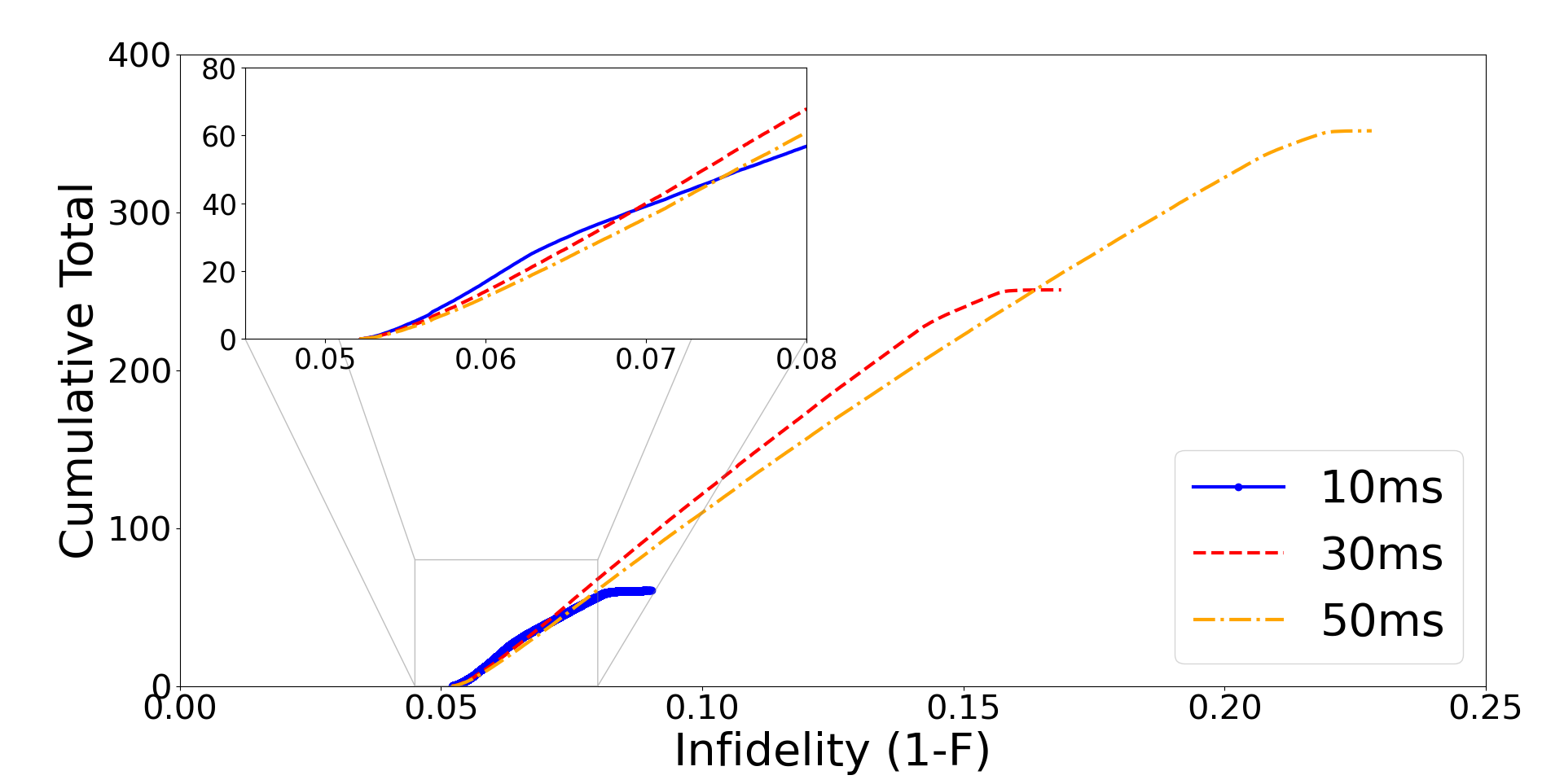}
\caption{Zenith A $90^\circ$.}
\end{subfigure}
\hfill
\begin{subfigure}{\figwidth}
\centering
\includegraphics[width=\linewidth]{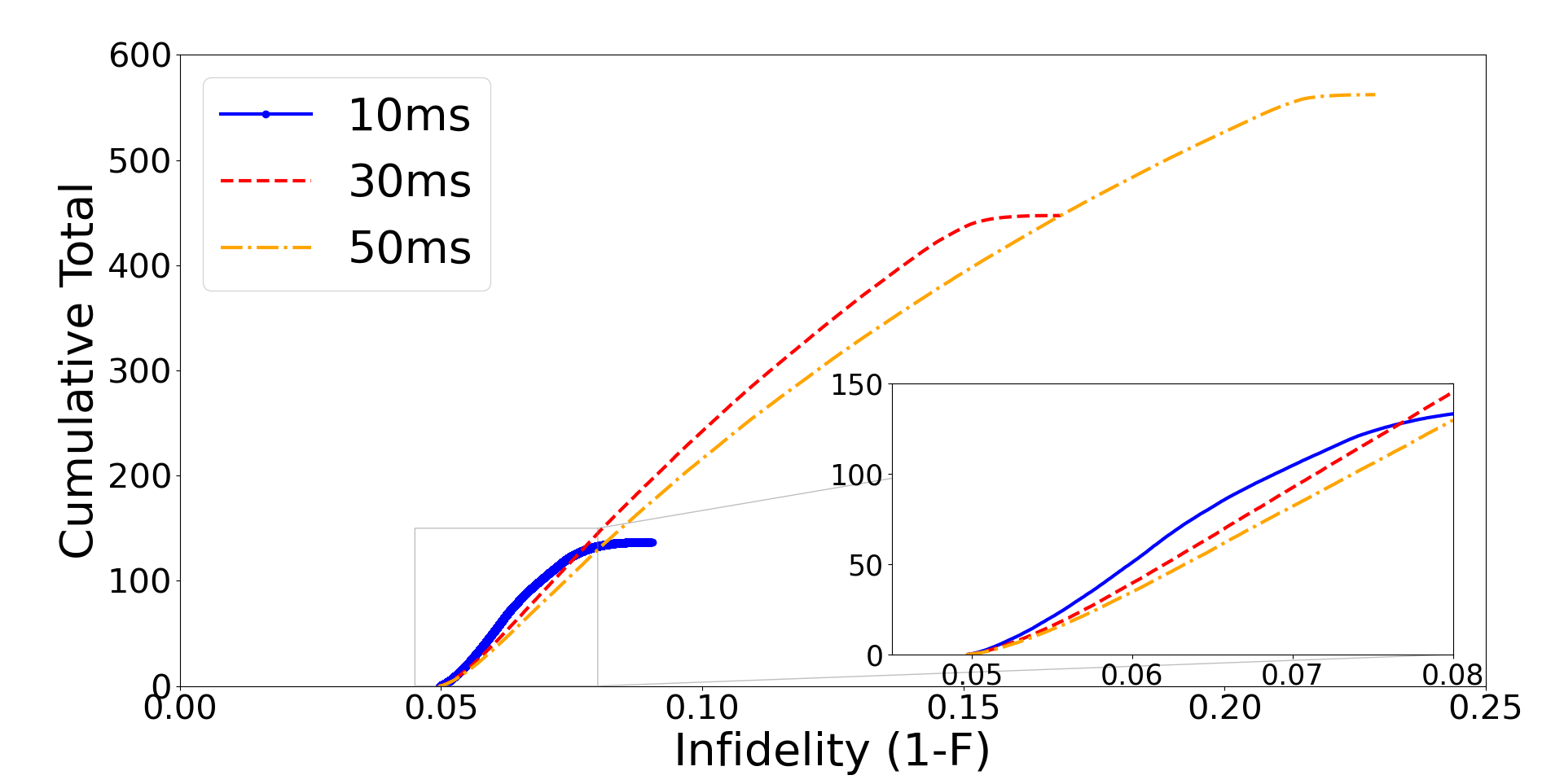}
\caption{Zenith A $45^\circ$.}
\end{subfigure}
\caption{Cumulative infidelities (a) Zenith-zenith, (b) Symmetric, (c) Zenith A $90^\circ$ and (d) Zenith A $45^\circ$, for $\tau_\text{cut} = 10, 30$ and $50$~ms, obtained from $1000$ trials. For the symmetric overpass higher values of $\tau_\text{cut}$ contain the pairs from lower values. In all other cases, adopting a lower value of $\tau_\text{cut}$ leads to more pairs at lower infidelities, but substantially lower overall PDVs. The inset graphs show a zoomed-in view of the low infidelity regions form which it can be seen that $\tau_\text{cut} = 10$~ms leads to a greater number of high-quality pairs. }
\label{fig:cumulativeInfididelity_different_cutoffs}
\end{figure}

 The cumulative distributions in infidelity ($1-F$) is shown in Fig.~\ref{fig:cumulativeInfididelity_different_cutoffs}. Interestingly, in the case of the symmetric overpass, we see an increase in $\tau_\text{cut}$ mainly has the effective of adding pairs to the cumulative distribution, with the distribution for longer values containing the pairs from shorter values. In all other cases, we see that choosing lower values of $\tau_\text{cut}$ leads to an increase in the number of high-quality pairs (low infidelity), at the cost of reducing the total PDV.
 
\begin{figure}[!tbp]
\centering
\newcommand{\figwidth}{0.48\textwidth}
\begin{subfigure}{\figwidth}
\centering
\includegraphics[width=\linewidth]{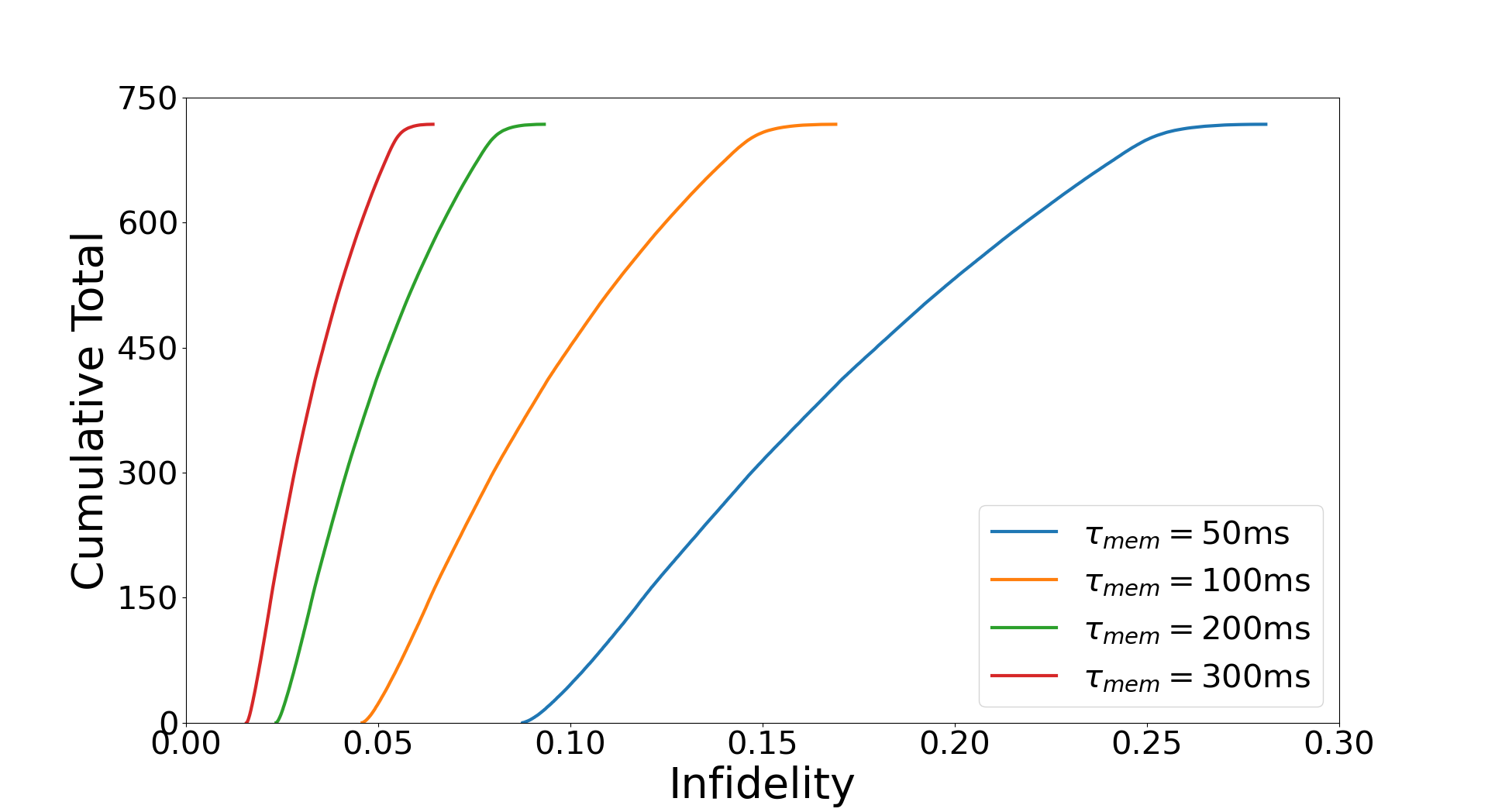}
\caption{Zenith-zenith}
\end{subfigure}
\hfill
\begin{subfigure}{\figwidth}
\centering
\includegraphics[width=\linewidth]{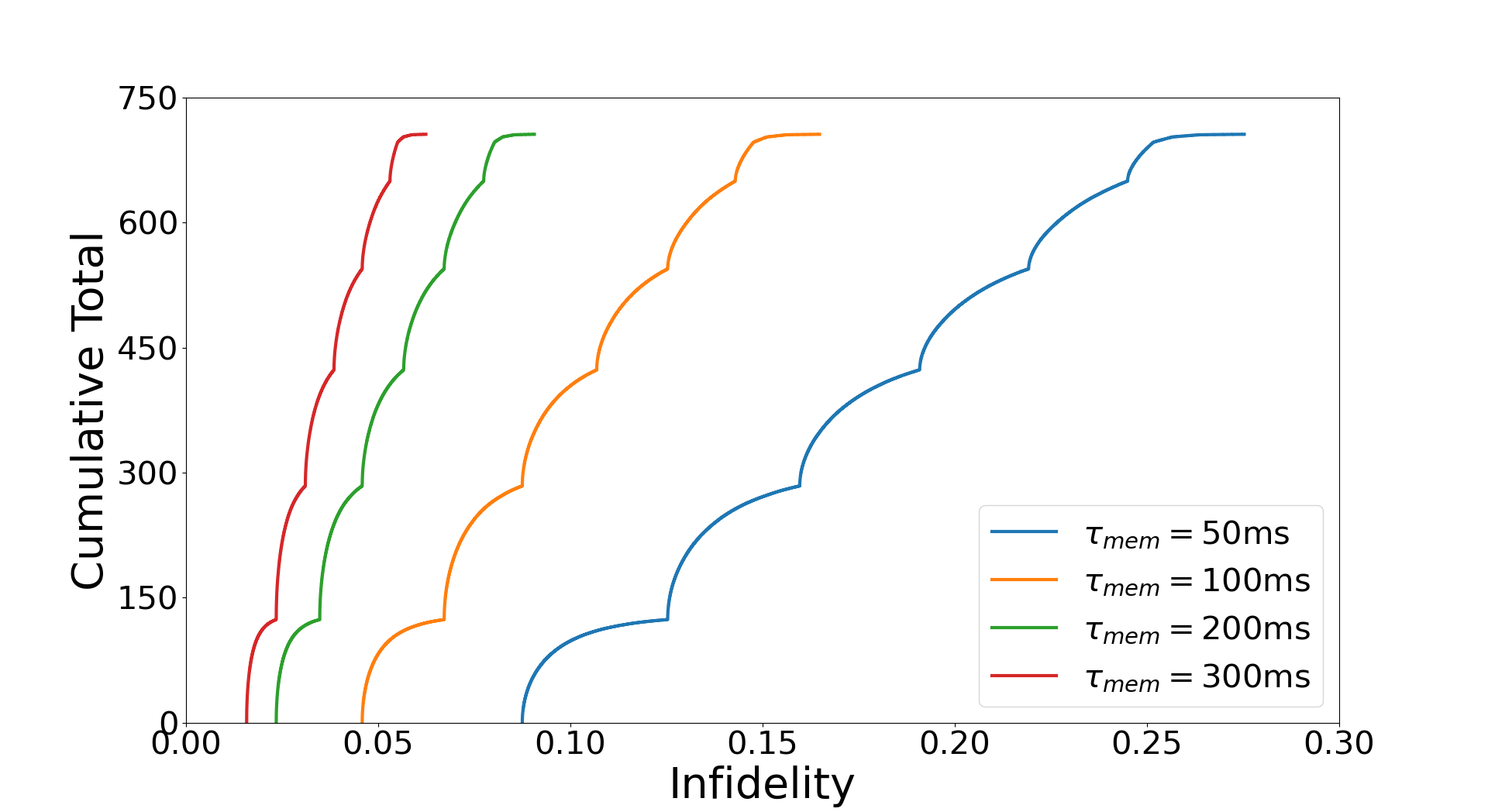}
\caption{Symmetric overpass}
\end{subfigure}
\par\medskip
\begin{subfigure}{\figwidth}
\centering
\includegraphics[width=\linewidth]{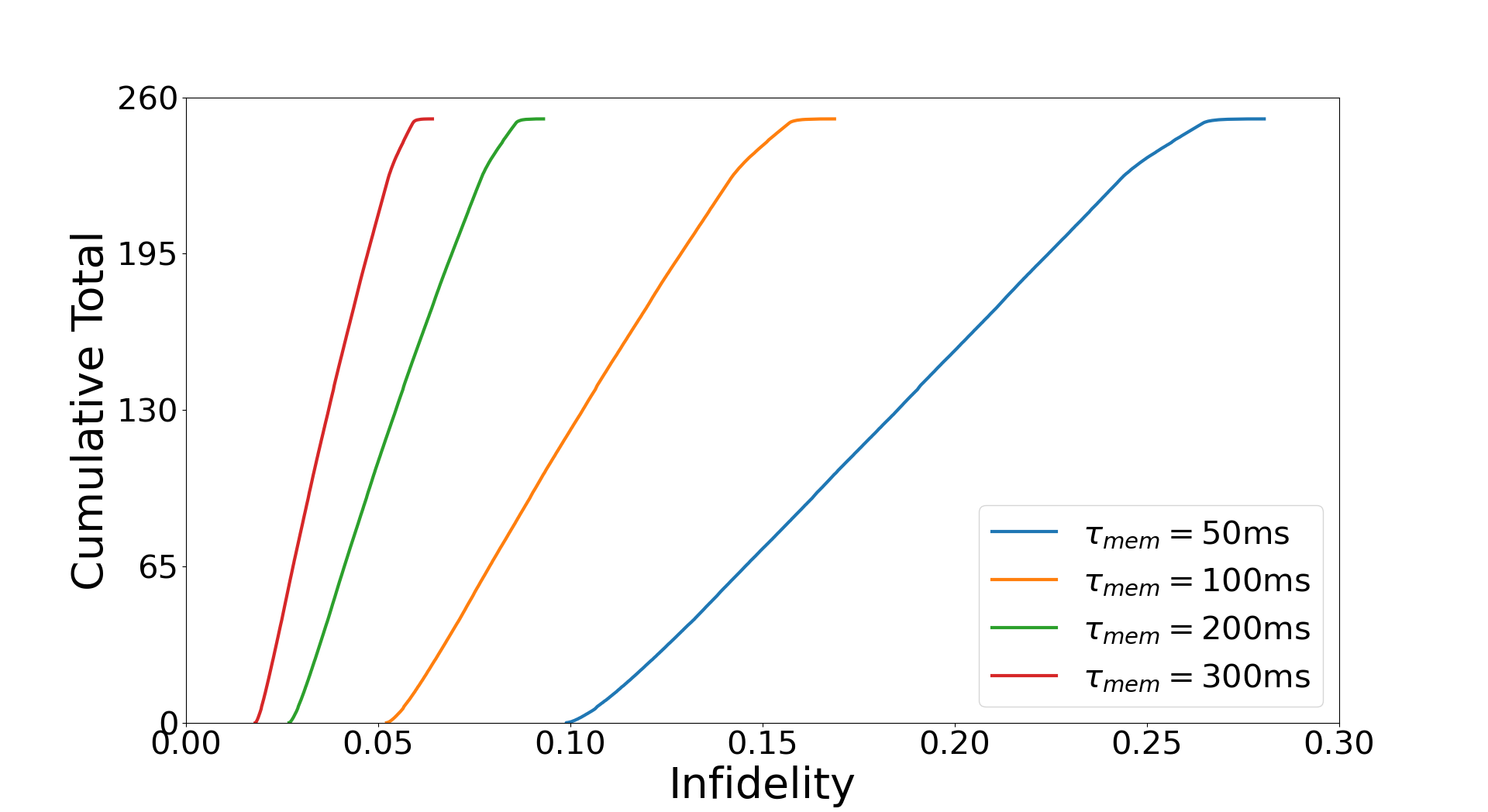}
\caption{Zenith A $90^\circ$.}
\end{subfigure}
\hfill
\begin{subfigure}{\figwidth}
\centering
\includegraphics[width=\linewidth]{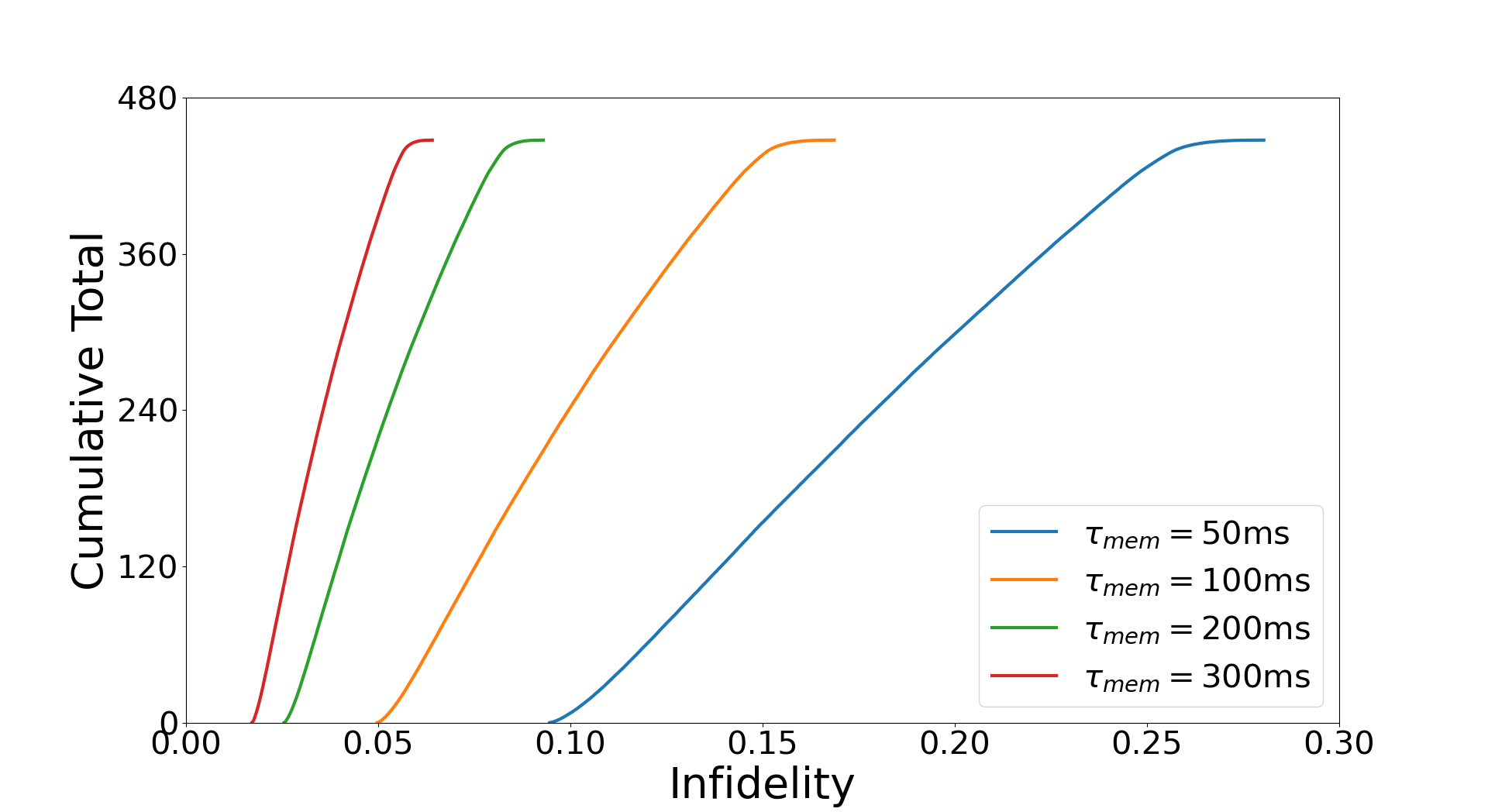}
\caption{Zenith A $45^\circ$.}
\end{subfigure}
\caption{Cumulative infidelities (a) Zenith-zenith, (b) Symmetric, (c) Zenith A $90^\circ$ and (d) Zenith A $45^\circ$ for $N_\text{sat} = 200$ and $\tau_\text{cut} = 30$~ms, for $\tau_\text{mem} = 50, 100, 200$ and $300$~ms. Results obtained form $1000$ trials. A longer $\tau_\text{mem}$ value decreases the minimum infidelity and shifts the cumulative count graphs to the left. It also leads to a steeper slope indicating that there is more consistency in the final fidelity distribution,  with higher $\tau_\text{mem}$ values leading to a greater proportion of high-fidelity pairs.}
\label{fig:varying_tMem_cumulative_infids}
\end{figure}

It is $N_\text{sat}$ and $\tau_\text{cut}$ that determines the qubit waiting time distribution for a given overpass geometry. Once this waiting time distribution is determined the resulting fidelity distribution is determined by $\tau_\text{mem}$. We now investigate the effects of different values of $\tau_\text{mem}$ on the final cumulative infidelity distribution, for $N_\text{sat} = 200$ and $\tau_\text{cut}=30$~ms (Fig.~\ref{fig:varying_tMem_cumulative_infids}). Longer vales of $\tau_\text{mem}$ results in a lower value of minimum infidelity, shifting the cumulative count graph to the right. It also results in a steeper increase from $0$ to the maximum, indicating a more consistent fidelity distribution with a greater proportion of high-fidelity pairs.

The effect of $N_\text{sat}$ on the fidelity distributions for an illustrative example of $N_\text{sat} = 2000$ and $\tau_\text{mem}$ is shown in Appendix~\ref{appen:N2000}. We find that increased memory capacity of the satellite results in a higher typical fidelity of the distributed pairs and eases requirements on memory lifetimes. In Appendix~\ref{appen:bufferBasedMemoryManagement} we consider the case where the satellite employs a buffer-size memory strategy. The overall higher waiting times and greater spread in their distribution lead us to conclude that a buffer-size approach is likely to be unsuitable for near-term memory platforms.

\newpage

\section{Conclusion}
\label{sec:conclusion}

The impact of time-of-flight latencies on SSQR performance necessitates high values of $N_\text{sat}$. The results indicate a modestly-sized quantum memory equipped ($N_\text{sat}=200$) quantum repeater satellite is a viable alternative to direct distribution of entanglement over international scales of $\sim 1000$ km. The advantage of SSQR grows when considering smaller OGSs and higher link losses where the repeater-scaling benefits are more apparent. The scaling of a SSQR memory capacity compared with a given DDDL EPS rate provides a basis of comparison. The different behaviour of the pair distribution volume with overpass geometry has implications for the design of satellite constellations~\cite{goswami2025satellites} based on direct pair downlink versus repeater satellites. To achieve high fidelities of the distributed pairs, memory lifetimes of several $100$ ms will be required for the $N_\text{sat}$ and $\eta_\text{loss}^\text{sys}$ considered. Different memory management~\cite{fittipaldiEntSwappingInOrbit2024} and swapping policies  may provide a relaxation in these requirements, along with trading the quantity of pairs against their quality. A particular feature of satellite quantum repeaters, compared with fibre links, is the influence of the time-dependent losses and range/latencies that complicates the synchronisation and optimisation of the 2 downlinks and their memory subsystems.

We note several limitations and key assumptions of this study. We have simplified the background light and atmospheric channel/weather model to concentrate on the particular characteristics of satellite repeaters. Near-deterministic entanglement generation is a key factor in the operation and efficiency of quantum repeater links, probabilistic sources based on spontaneous parametric downconversion~\cite{anwar2021entangled} are non-ideal in this regard. Practical considerations such as memory bandwidths, preparation, write/read latencies and mode addressability have not been incorporated into this study.

We have not considered storage of entangled states at the OGSs, e.g. they could be measured immediately. Should entanglement need to be preserved, then quantum non-demolition measurement of the received photons, or a heralded memory storage mechanism, will be required and the efficiency of this process will further impact upon the performance of SSQR. A comparison of DDDL and SSQR performance could be made on the basis of the equivalent secure key length that could be generated from the distributed entanglement (e.g. as in~\cite{sidhu2026operational} for DDDL alone). Such an analysis should consider finite key effects. For example, pairs distributed by DDDL will have a higher fidelity than those distributed by SSQR, due to the absence of memory decoherence, but the lower number of distributed pairs  may result in lower secret key compared to SSQR when finite block size effects are taken into account. More general measures of entanglement distribution quantity and quality could be used, depending on the envisaged application.

Future work should incorporate more detailed quantum memory models, taking into account write and read efficiencies, more general decoherence channels, storage-and-retrieval fidelity and system-level latencies. For given pairs of OGS links, optimised orbital and constellation design would augment the investigations of the city pairs considered here. The architecture and operation of the satellite quantum memory is open to extensive exploration, such as going beyond a round-based schedule and towards independent operation of each slot, dynamic allocation (e.g. using a unified pool of memory slots), time-based optimisation of $\tau_\text{cut}$, and swap-pairing optimisation in the long $\tau_\text{mem}$ scenario. More extensive Monte Carlo simulations will provide better estimates of performance variation, especially in the context of service level guarantees for particular application.

We have provided an analysis of the first step of satellite quantum repeater entanglement distribution which forms a starting point for informing quantum repeater design of multi-satellite configurations. With increased numbers of links, the effects of rate asymmetry and variable time-of-flight delays will exacerbate synchronisation, memory management, and swapping policy optimisation. Determining the memory capacity requirements to achieve useful distribution rates for intercontinental and global spanning nested repeater links~\cite{dawar2026feasibility} will be important for platform and payload design, especially the choice of memory platform and the critical supporting technologies that will be required to make them compatible with satellite deployment. In the long-term, satellite relay nodes will need to incorporate entanglement purification and quantum error correction~\cite{pathumsoot2024boosting} to establish the high fidelities required for advanced applications such as networking quantum computers, further increasing the requirements of the quantum memory and repeater subsystem performance.

\funding{We acknowledge: the UK National Quantum Technology Programme EPSRC Quantum Technology Hubs in Quantum Communications (EP/T001011/1) and Integrated Quantum Networks (EP/Z533208/1); EPSRC International Network in Space Quantum Technologies (EP/W027011/1); Horizon Marie Sk{\l}odowska-Curie Fellowship (101211639); UK Space Agency (NSTP3-FT-063, NSTP3-FT2-065, NSIP ROKS Payload Flight Model); Innovate UK projects ReFQ (78161), AirQKD (45364), and ViSatQT (43037); Fraunhofer UK Research Ltd. studentship support; EPSRC Research Excellence Award studentship; and ESA contracts 4000142084/23/UK/AL, 4000148330/25/NL/FGL/lf, and 4000147561/25/NL/FGL/ss.}

\newpage

\bibliographystyle{unsrt}
\bibliography{biblio}

\newpage
\appendix

\renewcommand{\thefigure}{A\arabic{figure}} 
\setcounter{figure}{0}  
\renewcommand{\theequation}{A\arabic{equation}} 
\setcounter{equation}{0}  

\section*{\bf \Large Appendices}

\

\section{Single OGS Links}
\label{appendix:singleOGSLinks}

\begin{figure}[!bp]
    \centering
    \includegraphics[width=0.6\linewidth]{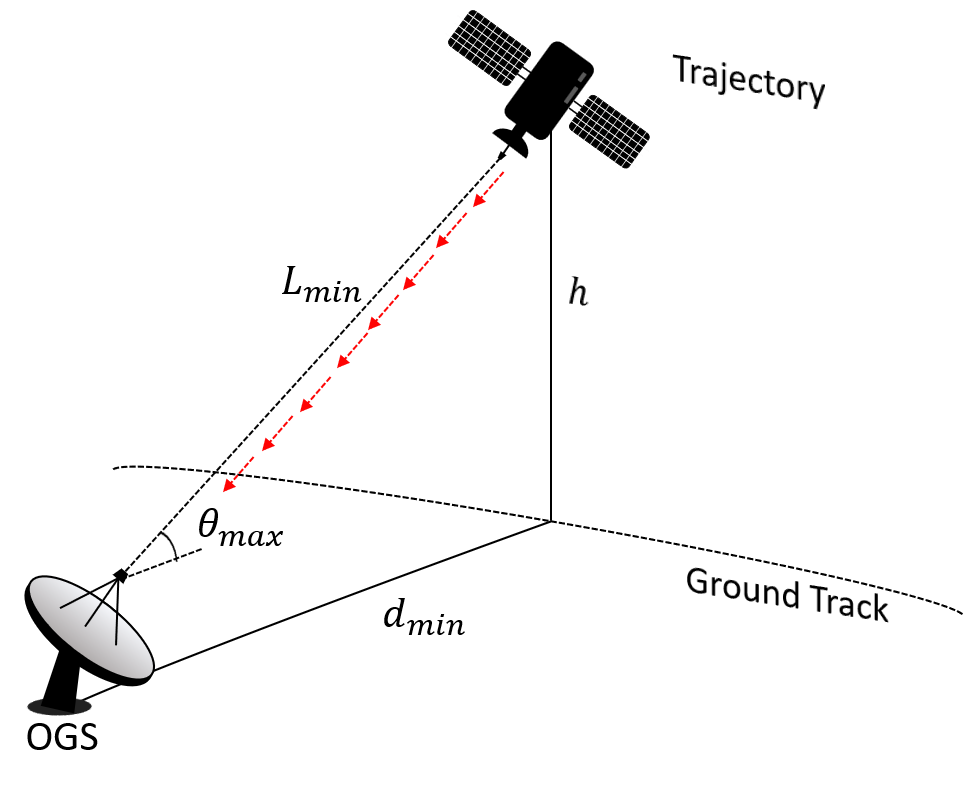}
    \caption{Single OGS links are the simplest kind of space quantum links. We consider satellites in circular orbits of altitude $h$. The projection of the satellite's motion onto the Earth's surface deins the ground track. At any point in time, the link is characterised by the link length, $L(t)$, elevation angle, $\theta(t)$, and transmittance $\eta(t)$. The satellite and OGS only communicate when the satellite is above a minimum $\theta_\text{min}$. The ground track offset, $d_\text{min}$, is defined as the minimum distance from the OGS to the ground track and is the most important parameter for describing the overpass geometry of these links. At this point, the satellite achieves the minimum link length, $L_\text{min}$, and maximum elevation, $\theta_\text{max}$.}
    \label{fig:singleOGSLink}
\end{figure}

The simplest kind of space quantum links are those that consist of a single optical ground station (OGS) communicating with a satellite (Fig.~\ref{fig:singleOGSLink})~\cite{sidhu2021satellite}. A dual-downlink can be considered as the combination of 2 independent single downlink channels. At any time, a channel is described by the link length, $L(t)$ and the elevation of the satellite relative to the OGS, $\theta(t)$. These are used in the calculation of the transmittance of the satellite-ground channel $\eta(t)$, which is the probability that a photon sent from the satellite is successfully transmitted to the OGS. These links are restricted to be above a minimum elevation $\theta_\text{min}$. The times for which this visibility criterion is satisfied defines an overpass.

An important parameter when describing the overpass geometry of a single OGS link is the ground track offset $d_\text{min}$. This is the minimum distance between the OGS and the satellite's ground track. At this point, the satellite achieves a maximum elevation and minimum distance with respect to the ground station. 

\begin{figure}[!tbp]
    \centering
    \raisebox{4.5cm}{
    \begin{subfigure}[height=5m]{0.45\textwidth}
        \centering
        \includegraphics[width=\linewidth]{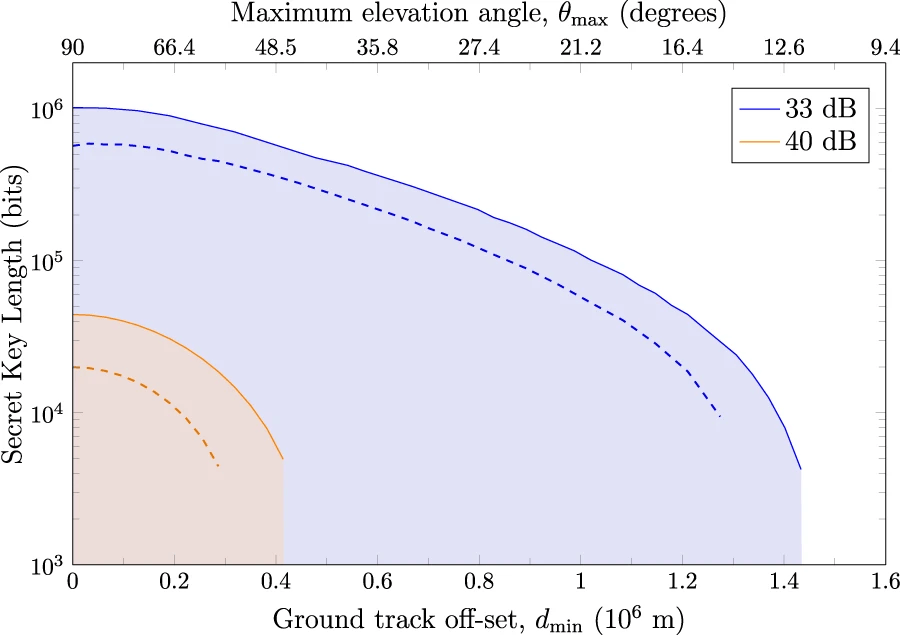}
        \caption{Single-overpass SKL dependence on Ground Track Offset. Efficient (solid) and standard (dashed) BB84 are compared for different values of zenith losses. This analysis can be seen as capturing the effect that a changing overpass geometry has on the system performance.}
        \label{fig:sub1}
    \end{subfigure}
    }
    \hspace{0.2cm}
    \begin{subfigure}{0.45\textwidth}
        \centering
        \includegraphics[trim=0.1cm 0.1cm 0.1cm 0.1cm, clip, height=6cm]{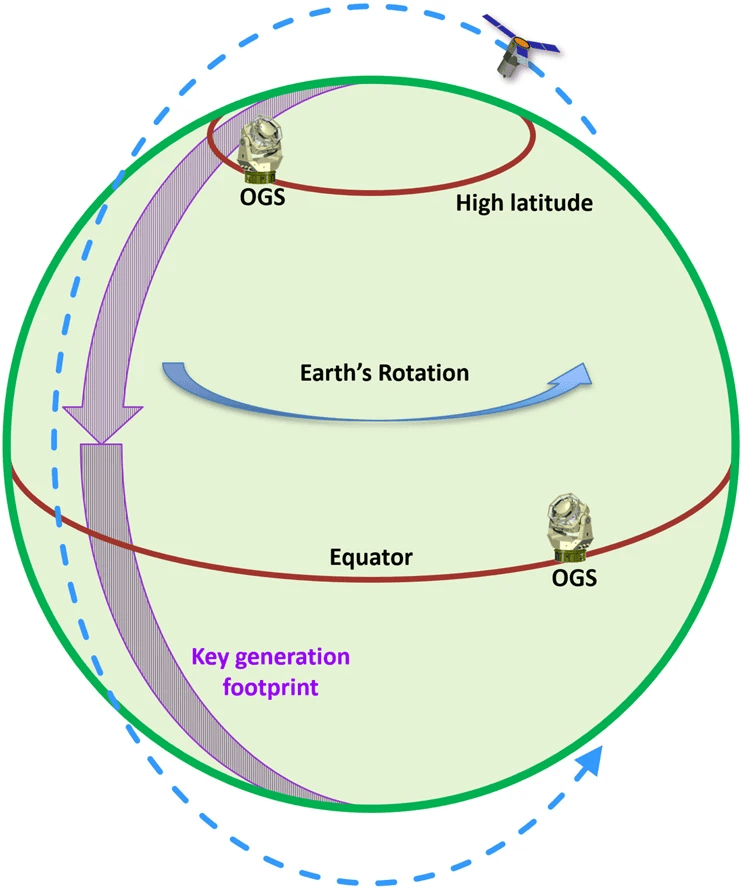}
        \caption{Averaging of Overpass SKL. It is important to characterise satellite QKD performance over long time scales. For SSO/polar orbits, the average annual SKL depends on OGS latitude due to overpasses density.}
        \label{fig:sub2}
    \end{subfigure}
    \caption{Figures taken from~\cite{sidhu2021finitekeyeffectssatellite}, showing examples of previous research on single-OGS links. (a) per-overpass SKL for efficient and standard BB84 under different values of zenith loss as a function of $d_\text{min}$. (b) The long-term SKL was estimated by considering the average annual SKL for a satellite in a SSO. The analysis we present in the main text on two-OGS links can be seen as a generalisation of these results. }
    \label{fig:sidhuResults}
\end{figure}

Single OGS links have been the focus of previous research in the context of QKD~\cite{sidhu2021finitekeyeffectssatellite}. The performance of the satellite have been given in terms of the per-pass secret key length (SKL) (Fig.~\ref{fig:sidhuResults} (a)). This can be seen as quantifying the effect that a changing overpass geometry has on the satellite's performance. 

The per-overpass SKL is highly variable and depends strongly on the specific $d_\text{min}$ value of an overpass. The long-term performance of the satellite is potentially a more meaningful metric of system performance. This has previously been evaluated by the average SKL for a satellite in a sun synchronous orbit (SSO). It was found that the location of the OGS on the Earth had implications for this. OGS's at higher latitudes have higher average annual SKL due to a higher density of overpasses.

\section{Link Budget}

\renewcommand{\thefigure}{B\arabic{figure}} 
\setcounter{figure}{0}  
\renewcommand{\theequation}{B\arabic{equation}} 
\setcounter{equation}{0}  

When calculating the entanglement distribution rates of the satellite it is necessary to calculate the instantaneous transmittances, $\eta$, on each of the individual Alice- Bob- downlinks as a function of time. We denote the loss in decibels as 
\begin{equation}
    \eta^{(\text{dB})} = (-10) \cdot \log_{10} (\eta).
\end{equation}
Our convention is that higher positive dB loss equates to lower transmittance. The channel transmittances are determined by the slant distances $\{L_A(t)$, $L_B(t)\}$ from the satellite to the OGS's and the elevations $\{\theta_A(t), \theta_B(t)\}$. We consider three sources of loss in our modelling due to diffraction, atmospheric scattering \& absorption, and optical inefficiencies in the transmitter, receiver, and pointing.

\subsection{Diffraction Losses}
Unavoidable beam broadening due to diffraction is a dominant loss mechanism in satellite links. A common treatment is to use the Gaussian beam approximation with a beam waist $\omega_0$ at the transmitter aperture and the beam width at distance $L$,
\begin{equation}
    \omega_L^2 = \omega_0^2 \left(1 + \left(\frac{\lambda L}{\pi\omega_0^2} \right)^2 \right),
\end{equation}
where $\lambda$ is the wavelength of the source. The diffraction transmittance is then given by 
\begin{equation}
    \eta_{\text{dif}} = 1 - \exp\left( -\frac{D_r^2}{2\omega_L^2} \right).
\end{equation}

To obtain a more accurate diffraction loss, the effect of a finite aperture and beam clipping (truncation of the wings of the Gaussian beam) can be included. The far field diffraction pattern is given by the Fourier transform of the field distribution across the transmitter aperture, the received intensity is then the integral of the far field pattern at the slant range of the collection aperture. The optimum beam waist can be found and is typically $w_0\approx 0.45 D_t$ for a clear aperture~\cite{bourgoin2013comprehensive}. For a transmission telescope with a secondary obscuration, the optimum beam waist will depend on the ratio of primary to secondary mirror diameters.

\subsection{Atmospheric Losses}
As the beam propagates through the atmosphere, it is attenuated by absorption in aerosols and other atmospheric components. This absorption is elevation and wavelength dependent. We model atmospheric attenuation using a slab atmosphere approximation. If the transmittance at zenith is given by $\eta_\text{zen}$ then the transmittance at elevation $\theta$ is given by 
\begin{equation}
    \eta_\text{atm} = \left(\eta_\text{zen}\right)^{\csc(\theta)}.
\end{equation}
The value of $\eta_\text{zen}$ can be found, for example, using dedicated software such as MODTRAN~\cite{modtran}. For the $\lambda = 780$nm we considered in this work $\eta_\text{zen} = 0.79$ was assumed~\cite{gundoganProposalSpaceMemories2020}.

\subsection{Intrinsic and System Loss Metric}
The final source of loss that we consider is the intrinsic system loss. This is a fixed loss that we introduced to account for several sources of loss that affect satellite links. We list these sources and the conservative corresponding values of loss as 
\begin{itemize}
    \item Internal loss in the transmitter optics ($3$ dB),
    \item Internal loss in the receiver optics ($3$ dB),
    \item Beam wader and excessive broadening ($3$ dB),
    \item Detector inefficiency ($1$ dB),
\end{itemize}
for a total intrinsic loss $\eta^{(\text{dB})}_\text{int} = 10 \ \text{dB}$.

The time-dependent downlink efficiency is, 
\begin{equation}
    \eta^{(\text{dB})}(t) = \eta^{(\text{dB})}_\text{diff}(t) + \eta^{(\text{dB})}_\text{atm}(t) + \eta^{(\text{dB})}_\text{int}.
\end{equation}
The satellite link optoelectronic efficiency is well characterised by the loss at zenith, termed the system loss metric $\eta^\text{sys}_\text{loss}$. For $D_t=100\ \text{mm}$, $D_r=1000\ \text{mm}$, and $w_0=45\ \text{mm}$, there is a clipping loss of $0.4\ \text{dB}$ giving a diffraction loss of $\eta_{diff}^{(dB)}=14.9\ \text{dB}$, leading to a baseline of $\eta_\text{loss}^\text{sys}=25.9\ \text{dB}$ assumed for this work.
\label{appendix:linkBudget}

\section{DDDL and Repeater Satellite Optimal Overpass Geometries}
\label{sec:overpasscomparison}

\renewcommand{\thefigure}{C\arabic{figure}} 
\setcounter{figure}{0}  
\renewcommand{\theequation}{C\arabic{equation}} 
\setcounter{equation}{0}  

In Section~\ref{sec:perPassPDV} the DDDL and repeater satellites performed best for zenith-zenith and symmetric overpasses, respectively. The optimal performance of the different protocols with varying overpass geometry is a key result of this work with important implications for the design of satellite quantum links. Here, we present the geometric factors that underpin this behaviour.

\begin{figure}[!btp]
\centering
\newcommand{\figwidth}{0.48\textwidth}
\begin{subfigure}{\figwidth}
\centering
\includegraphics[width=\linewidth]{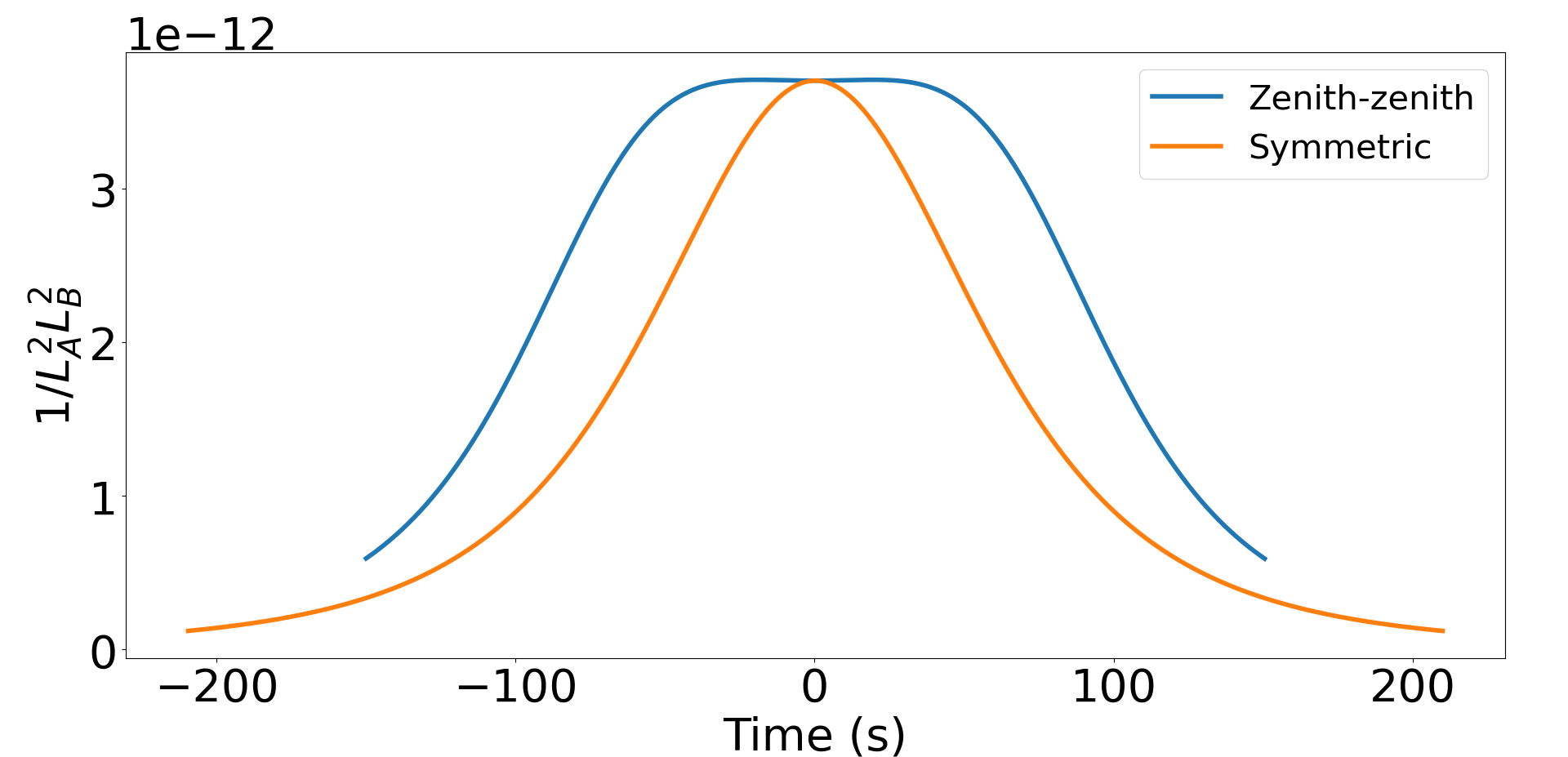}
\caption{DDDL scaling (Eq.~\ref{equ:appendixLSquared})}
\end{subfigure}
\hfill
\begin{subfigure}{\figwidth}
\centering
\includegraphics[width=\linewidth]{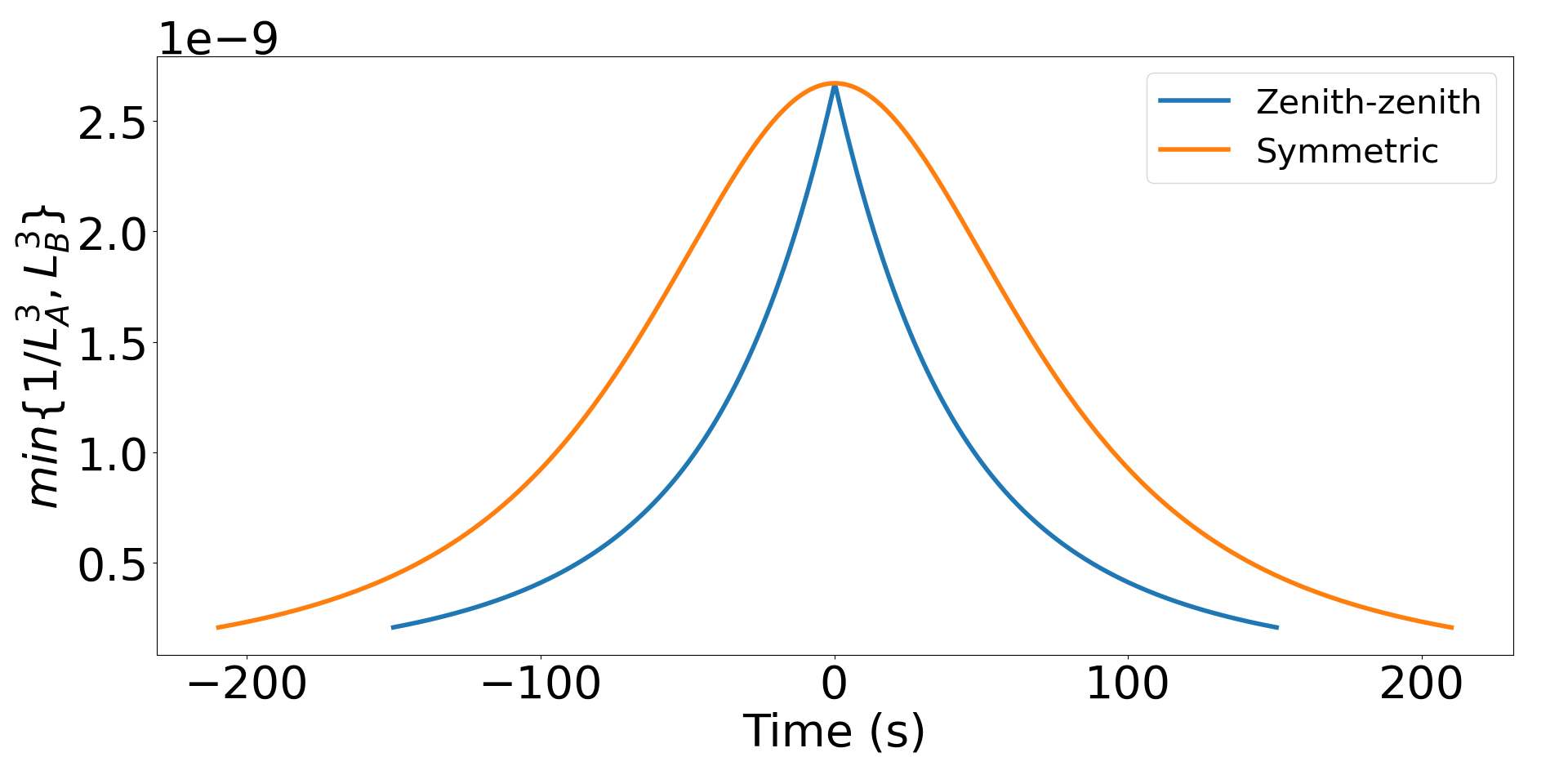}
\caption{SSQR scaling (Eq.~\ref{equ:appendixLCubed})}
\end{subfigure}
\caption{Scaling of the instantaneous pair distributions for the zenith-zenith and symmetric overpasses. The $t=0$ point here corresponds to where the satellite is equidistant from Alice and Bob ($\sim 720$km for $d = 1000$km and $h=500$km). Since the per-overpass PDV is found by integrating the instantaneous pair distribution rates, a broader distribution leads to superior performance. Thus zenith-zenith is optimal for DDDL and symmetric is optimal for SSQR.}
\label{fig:appendixScalingGraphs}
\end{figure}

From Eq.~\ref{equ:DDDLentDistRate} in the main text, it can be seen that the instantaneous DDDL pair distribution rate behaves as 
\begin{equation}
    R^\text{DDDL} \sim \frac{1}{L^2_A L^2_B},
\label{equ:appendixLSquared}
\end{equation}
with the inverse square scaling arising from diffraction-dominated in the far-field. This scaling is illustrated in Fig.~\ref{fig:appendixScalingGraphs}(a) for the zenith-zenith and symmetric overpasses. The comparatively flatter distribution in the zenith-zenith case leads to a greater per-overpass PDV (found by integration). Furthermore, any deviation from the zenith-zenith would lead to a sharper distribution and a reduced PDV. 

For the repeater satellite, the instantaneous PDR behaves as,(Eq.~\ref{eq:latencylimitedrate})
\begin{equation}
    R^\text{SSQR} \sim \min \{1/L^3_A, 1/L^3_B \},
    \label{equ:appendixLCubed}
\end{equation}
due to channel loss $\sim \frac{1}{L_{A,B}^2}$ and the attempt rate $\sim \frac{1}{L_{A,B}}$. Fig.~\ref{fig:appendixScalingGraphs}(b) illustrates this scaling for the zenith-zenith and symmetric overpasses. In this case the symmetric overpass yields a flatter rate distribution than zenith-zenith, with a corresponding higher per-overpass PDV after integration. The higher PDV in this case can also be seen directly from Eq.~\ref{equ:appendixLCubed}, where the rate is maximised when $L_A = L_B$, which is satisfied for all points in a symmetric overpass.

\section{Memory Noise Model}
\label{appendix:MemoryNoiseModel}

\renewcommand{\thefigure}{D\arabic{figure}} 
\setcounter{figure}{0}  
\renewcommand{\theequation}{D\arabic{equation}} 
\setcounter{equation}{0}  

Here, we detailed our calculations for the fidelity of the pair that Alice and Bob having in hand after the satellite performs entanglement swapping. We assume that the satellite produces perfect entanglement pairs of fidelity $1$. Under this assumptions, the only fidelity losses in the system are due to memory imperfections. We denote the memory lifetimes $\tau_\text{mem}$, which is assumed to be equal across all memory slots on board the satellite. The pairs that the satellite generates on board of the form 
\begin{equation}
    \rho_{AS_1BS_2} = \ket{\Phi^+}\bra{\Phi^+}_{AS_1} \otimes \ket{\Phi^+}\bra{\Phi^+}_{BS_2,}
\end{equation}
where A and Bob are Alice and Bob, respectively, and $S_1$, $S_2$ are internal states of the satellite memory. While the states are held in memory waiting for confirmation of successful transmission, they are subject to dephasing noise, a common model of memory noise that is present in the literature \cite{Walln_fer_2022}. The qubit held in memory undergoes dephasing noise
\begin{equation}
    D(t) \rho: \rho \mapsto (1-\lambda(t))\rho \otimes \lambda(t) Z \rho Z,
\end{equation}
where
\begin{equation}
    \lambda(t) = \frac{1 - \exp(-t/\tau_\text{mem})}{2},
\end{equation}
and $t$ is the time the qubit has been waiting in memory. The qubits held in ground station memories are not subjected to any noise in our model. One can easily see that 
\begin{equation}
    (I \otimes D(t))\ket{\Phi^+}\bra{\Phi^+} = (1- \lambda(t)) \ket{\Phi^+}\bra{\Phi^+} + \lambda(t) \ket{\Phi^-}\bra{\Phi^-}.
\end{equation}
The state of the two qubit pairs on-board the satellite just performed swapping is performed is given by 
\begin{equation}
     \rho_{AS_1BS_2}(t_A,t_B) = (I_A \otimes D(t_A)_{S_1}) \otimes  (I_B \otimes D(t_B)_{S_2}) \rho_{AS_1BS_2},
\end{equation}
where $t_A$ and $t_B$ are the total times that each pair has been sitting in memory.
The satellite then performs entanglement swapping on this pair, in order to establish direct entanglement between Alice and Bob. This consists of performing a Bell state measurement on the $S_1S_2$ pair, and classically communicating the results to the ground stations. One can show that under the above memory noise model, the state that the ground stations have in their possession is given by 
\begin{align}
    \rho_{AB} = &\left[\lambda\left(t_A\right)\lambda\left(t_B\right)+\left(1-\lambda\left(t_A\right)\right)\left(1-\lambda\left(t_B\right)\right)\right]\ket{\Phi^+}\bra{\Phi^+}\\&
    +\left[\lambda\left(t_A\right)\left(1-\lambda\left(t_B\right)\right)+\lambda\left(t_A\right)(1-\lambda\left(t_B\right))\right]\ket{\Phi^-}\bra{\Phi^-}.
\end{align}
The fidelity can be easily read off the above equation as the coefficient in front of $\ket{\Phi^+}\bra{\Phi^+}$.

\section{Waiting Times and Fidelities for $N_\text{sat}=2000$}
\label{appen:N2000}

For a given overpass the waiting time and subsequent fidelity distributions are determined primarily by $N_\text{sat}$ and $\tau_\text{cut}$. In the main text we focused on the effect of the latter. To explore the effect of the former, we present an illustrative example in which $N_\text{sat}= 2000$. We present the median fidelity as a function of time and the final cumulative infidelity counts for the representative overpasses, comparing the distributions obtained for $N_\text{sat}=2000$ to  $N_\text{sat} = 200$. 

\begin{figure}[!tbp]
\centering
\newcommand{\figwidth}{0.48\textwidth}
\begin{subfigure}{\figwidth}
\centering
\includegraphics[width=\linewidth]{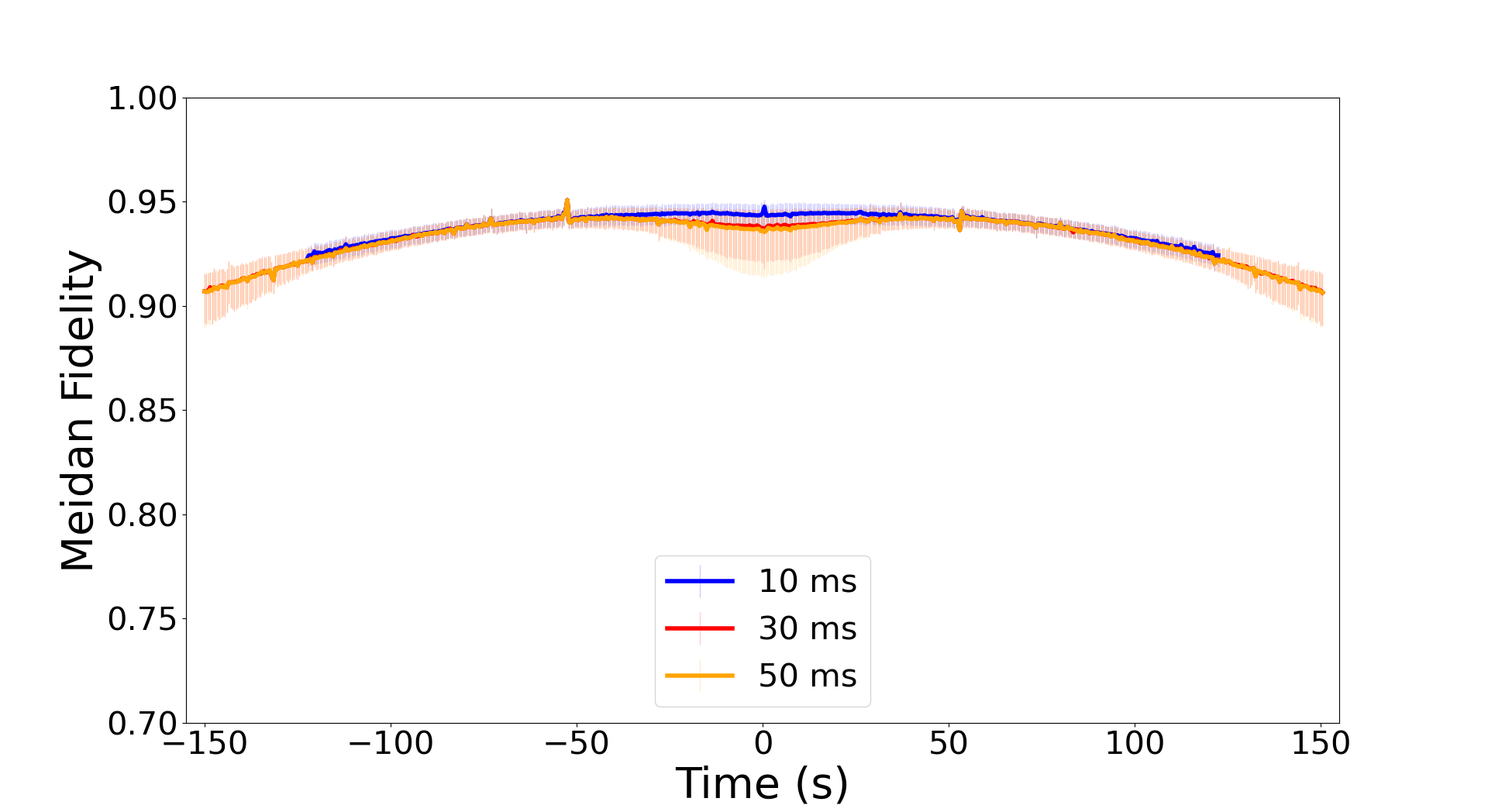}
\caption{Zenith-zenith. Total PDV: $7951 \pm 62$, $8936 \pm 62 $, $8994 \pm 65 $, for $\tau_\text{cut} = 10, 30$ and $50$~ms, respectively.}
\end{subfigure}
\hfill
\begin{subfigure}{\figwidth}
\centering
\includegraphics[width=\linewidth]{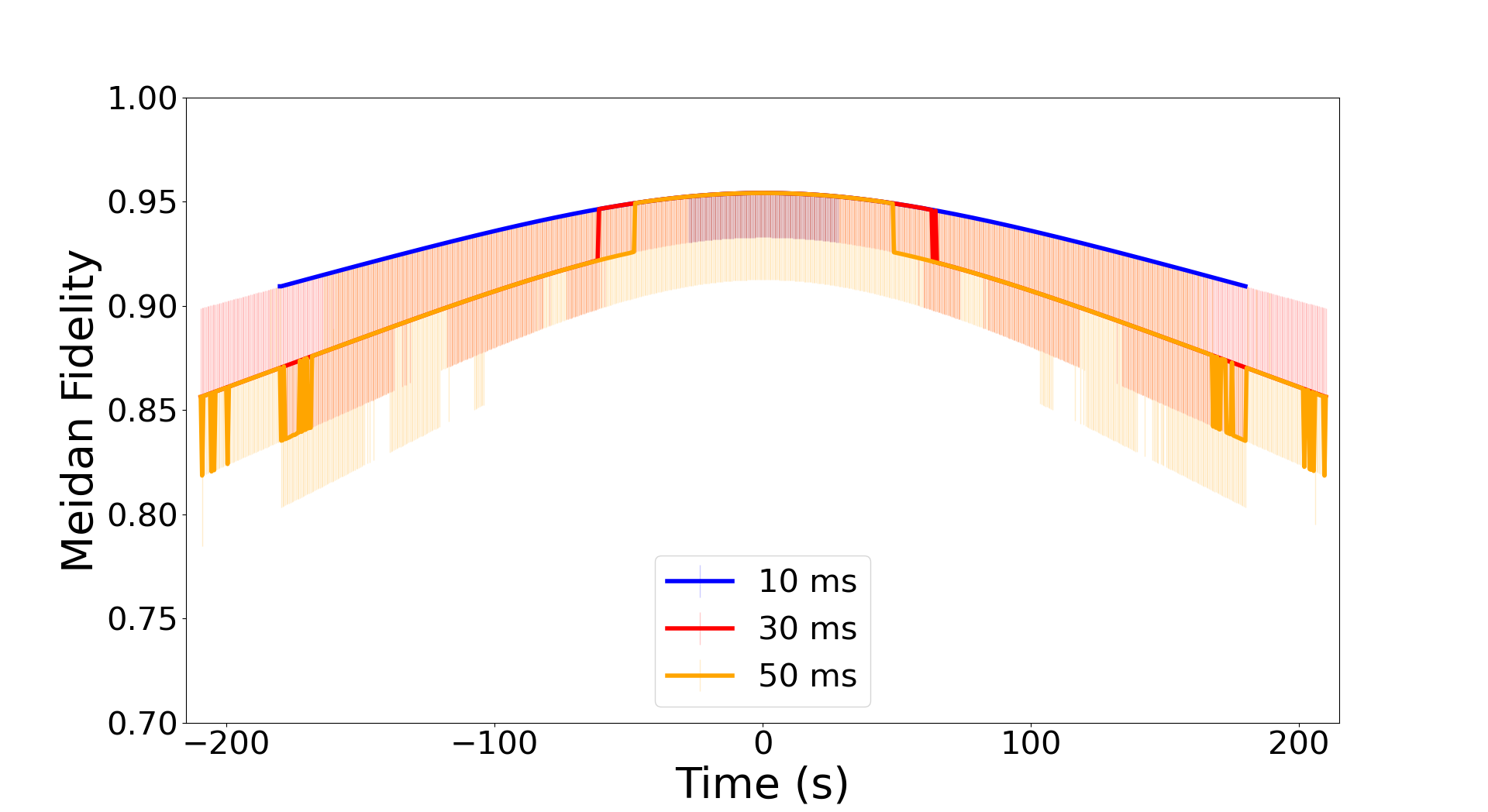}
\caption{Symmetric overpass.  Total PDV: $8587 \pm 59$, $14058 \pm 71$, $15171 \pm 73$, for $\tau_\text{cut} = 10, 30$ and $50$~ms, respectively.}
\end{subfigure}
\par\medskip
\begin{subfigure}{\figwidth}
\centering
\includegraphics[width=\linewidth]{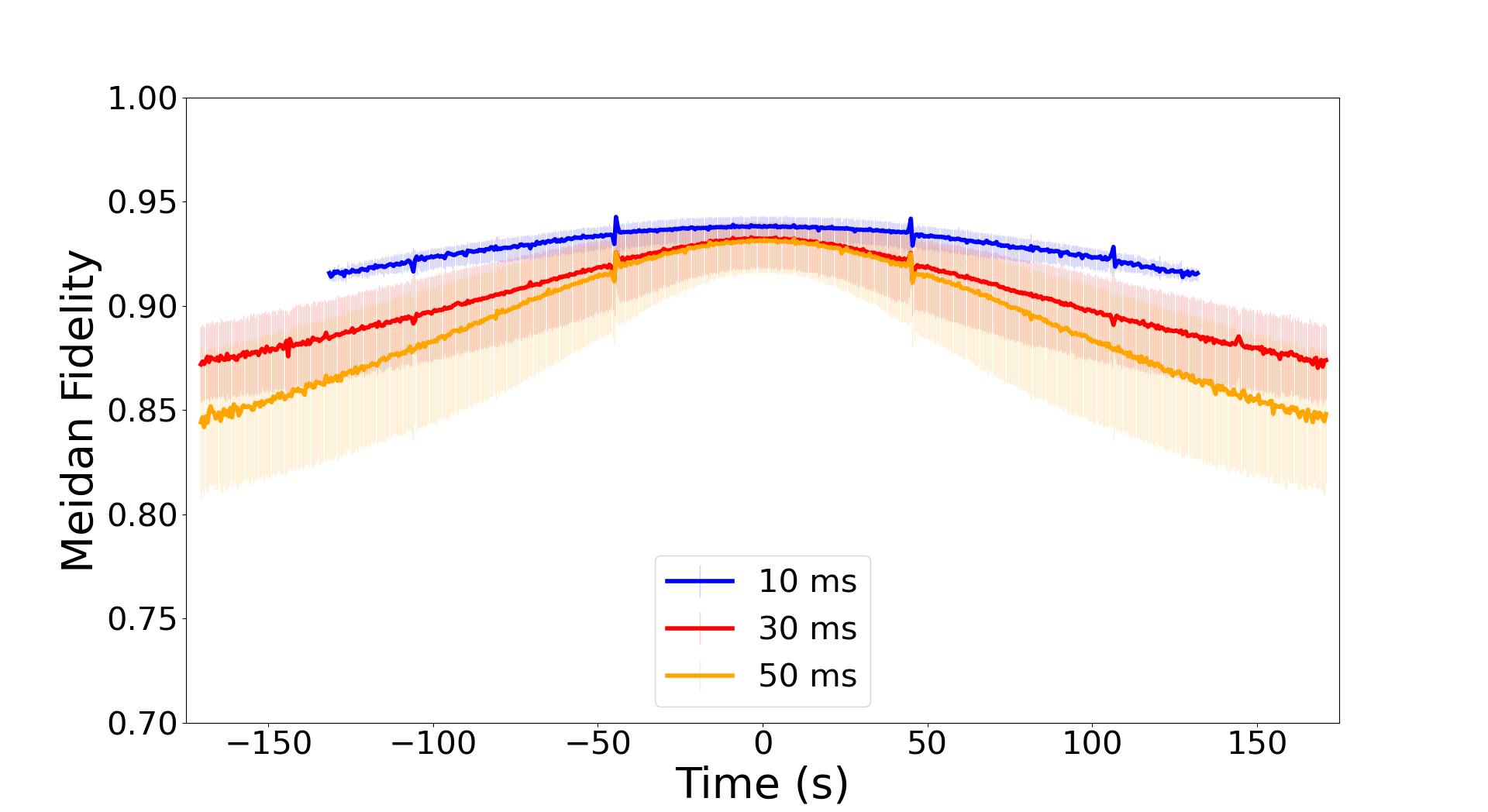}
\caption{Zenith A $90^\circ$. Total PDV: $3127 \pm 38$, $5647 \pm 53 $, $6055 \pm 52$, for $\tau_\text{cut} = 10, 30$ and $50$~ms, respectively.}
\end{subfigure}
\hfill
\begin{subfigure}{\figwidth}
\centering
\includegraphics[width=\linewidth]{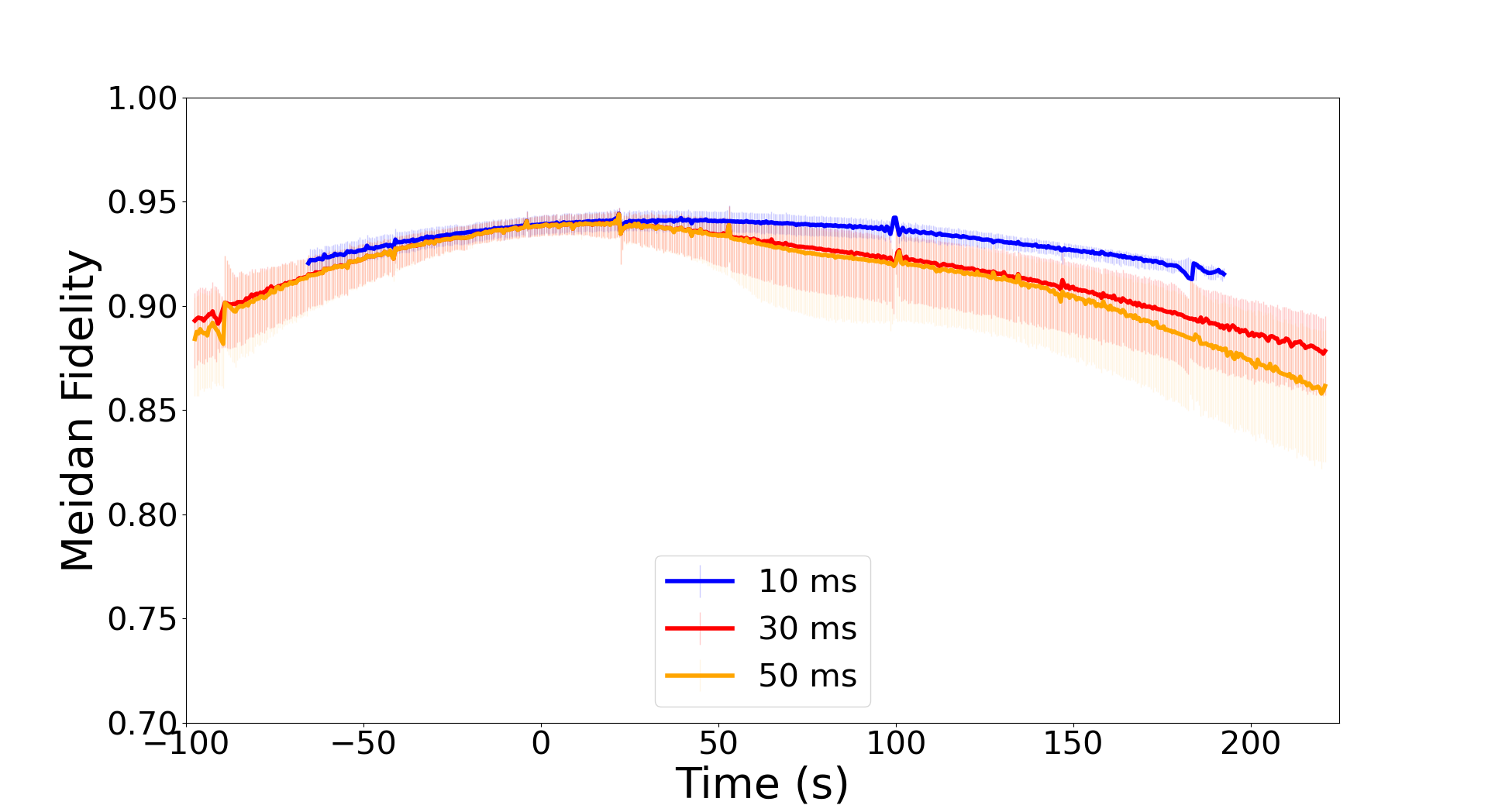}
\caption{Zenith A $45^\circ$.  Total PDV: $5171 \pm 48$, $7267 \pm 58$, $7422 \pm 58$, for $\tau_\text{cut} = 10, 30$ and $50$~ms, respectively.}
\end{subfigure}
\caption{Median fidelities as a function of time for  (a) Zenith-zenith, (b) Symmetric, (c) Zenith A $90^\circ$ and (d) Zenith A $45^\circ$, for $N_\text{sat} = 2000$ and $\tau_\text{cut} = 10, 30$ and $50$~ms. Results were obtained from $1000$ trials and the error bars show the interquartile ranges. The total overpass PDVs in the figure captures are stated as the mean and standard deviation. Compared to $N_\text{sat} = 200$ (Fig~\ref{fig:medianFidsVsTime_diffcutoffs} the typical fidelity for $N_\text{sat}=2000$ is higher and there is less spread. The benefit to PDV with the increase to $N_\text{sat}$ (Fig.~\ref{fig:different_cutoffs_dist_rates}) depends on $\tau_\text{cut}$, with $\tau_\text{cut} = 10$~ms having a much greater increase to PDV than $\tau_\text{cut}= 30$ and $50$~ms. The non-monotonicity in median fidelity fidelity for the symmetric overpass that was observed for $N_\text{sat}=200$ was suppressed when $N_\text{sat}=2000$.}
\label{fig:N2000_median_fids_times}
\end{figure}

The median fidelities as a function of time for $N_\text{sat} = 2000$ are shown in Fig.~\ref{fig:N2000_median_fids_times}. A comparison to the $N_\text{sat} = 200$ ((Fig.~\ref{fig:medianFidsVsTime_diffcutoffs} and Fig.~\ref{fig:different_cutoffs_dist_rates}) shows two notable features. First, $N_\text{sat}=2000$ results in higher typical fidelities with less spread in the distribution. Second, we see that the increase in total PDV with greater $N_\text{sat}$ depends on $\tau_\text{cut}$. For the Zenith-zenith overpass, e.g., the increase from $N_\text{sat}=200$ to $2000$ led to a $\sim27 \times$ increase to the PDV for $\tau_\text{cut}=10$~ms. This is much greater than the $\sim12\times$ and $\sim11\times$ increases obtained for $\tau_\text{cut} = 30$ and $50$~ms, respectively. Both of these observations can be explained by the simple fact greater $N_\text{sat}$ results in more successes within each entanglement distribution attempt, resulting in lower typical waiting times and fewer discarded pairs.

\begin{figure}[!tbp]
\centering
\newcommand{\figwidth}{0.48\textwidth}
\begin{subfigure}{\figwidth}
\centering
\includegraphics[width=\linewidth]{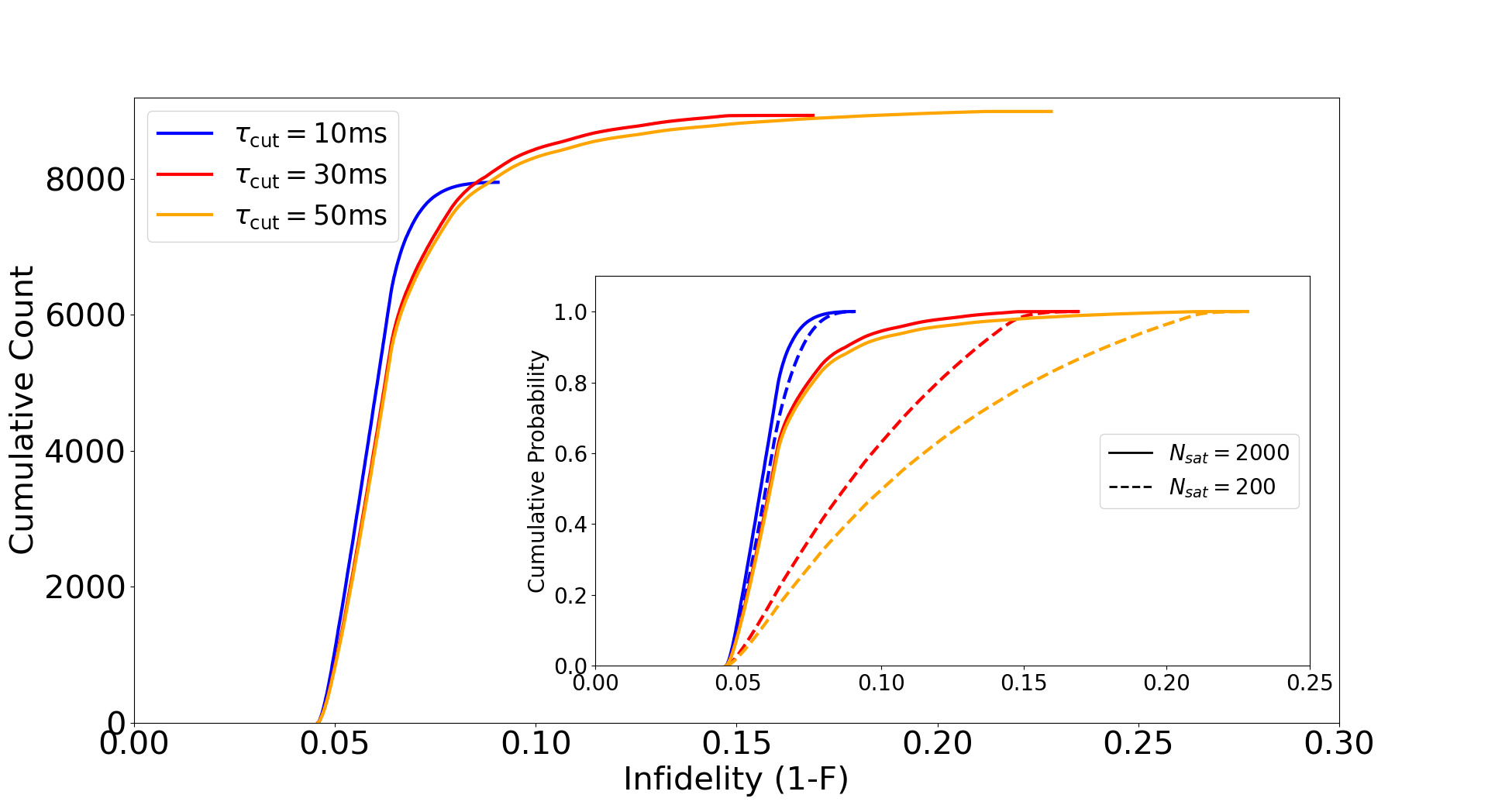}
\caption{Zenith-zenith.}
\end{subfigure}
\hfill
\begin{subfigure}{\figwidth}
\centering
\includegraphics[width=\linewidth]{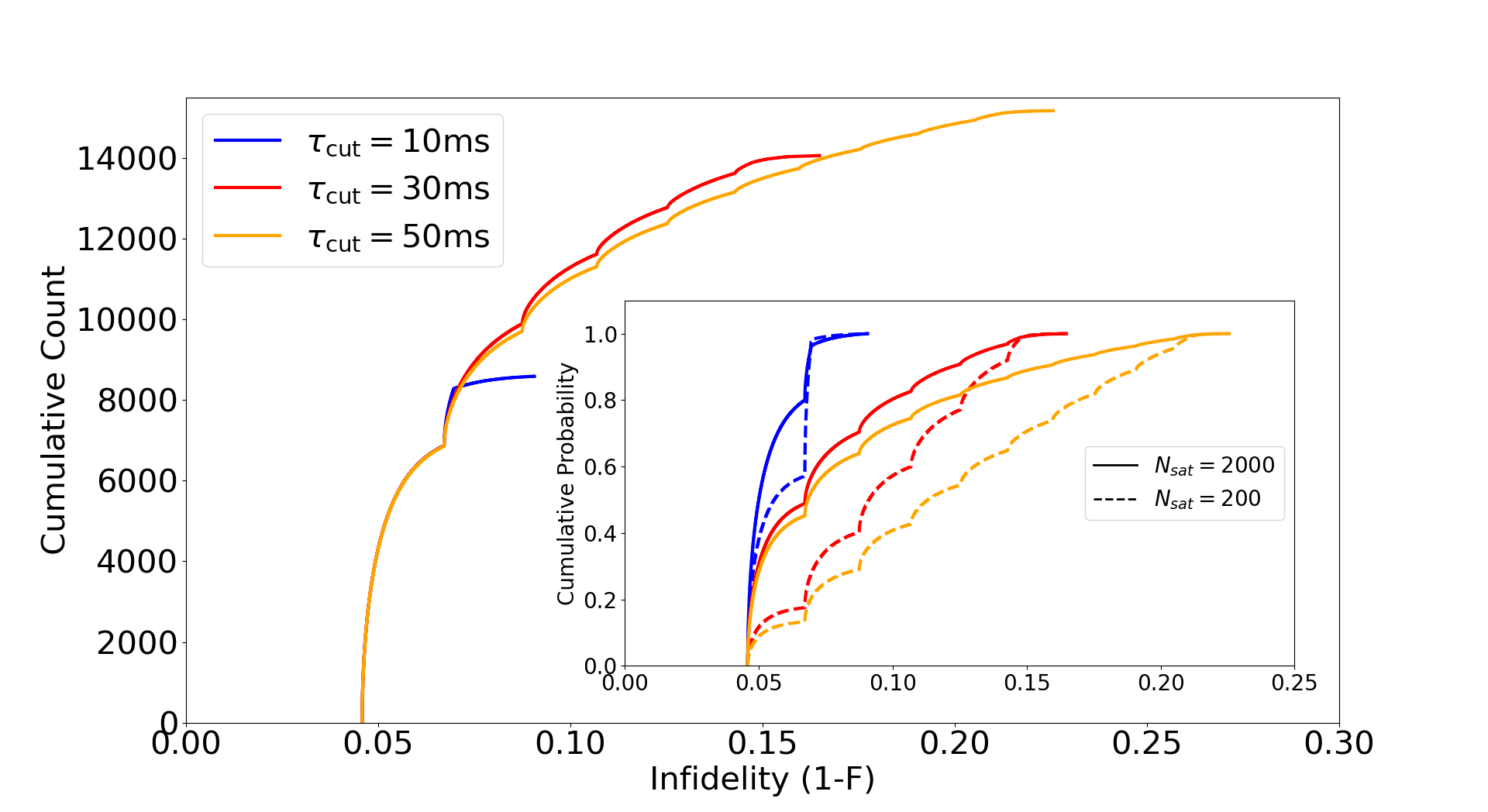}
\caption{Symmetric overpass.}
\end{subfigure}
\par\medskip
\begin{subfigure}{\figwidth}
\centering
\includegraphics[width=\linewidth]{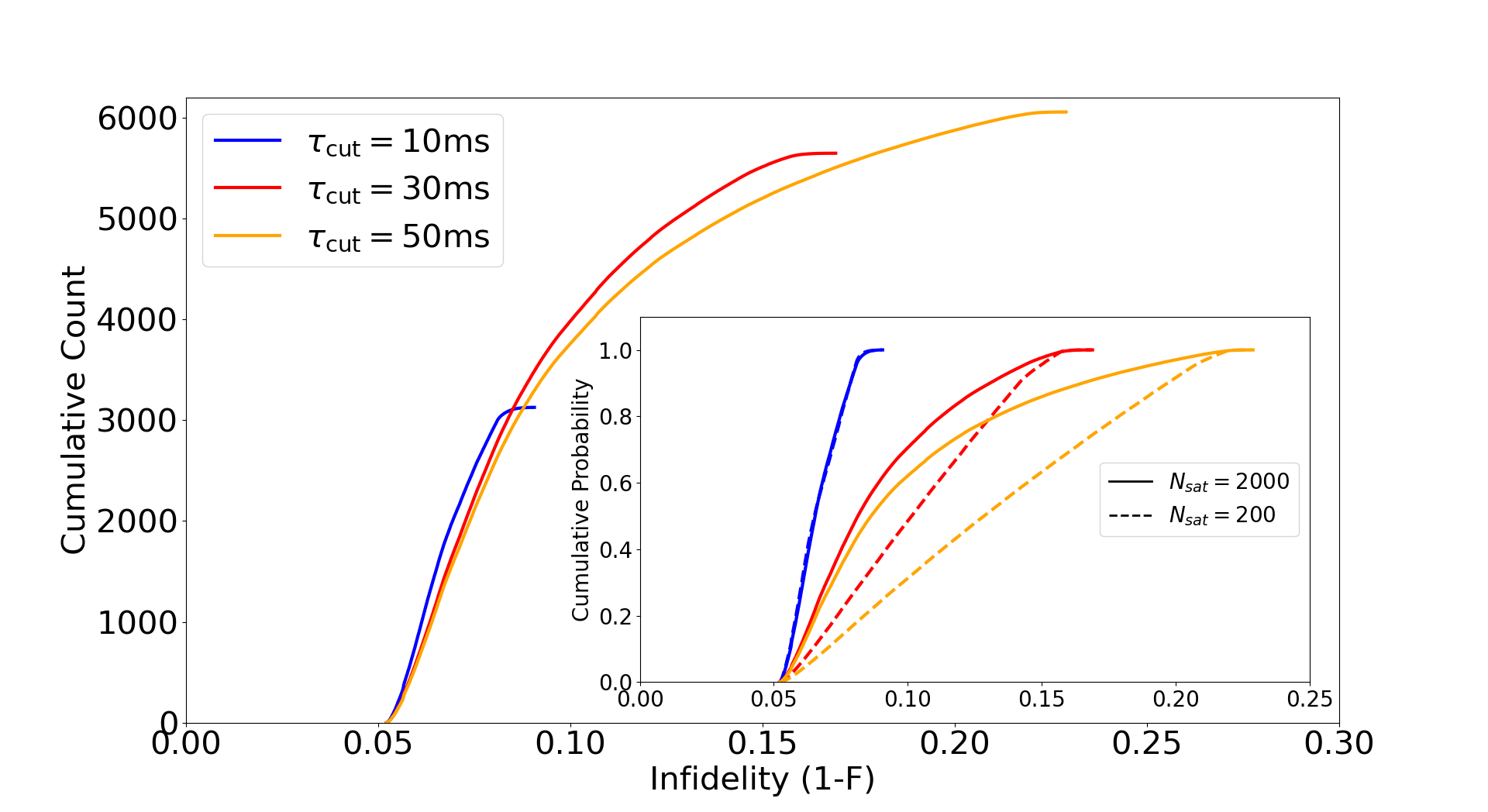}
\caption{Zenith A $90^\circ$.}
\end{subfigure}
\hfill
\begin{subfigure}{\figwidth}
\centering
\includegraphics[width=\linewidth]{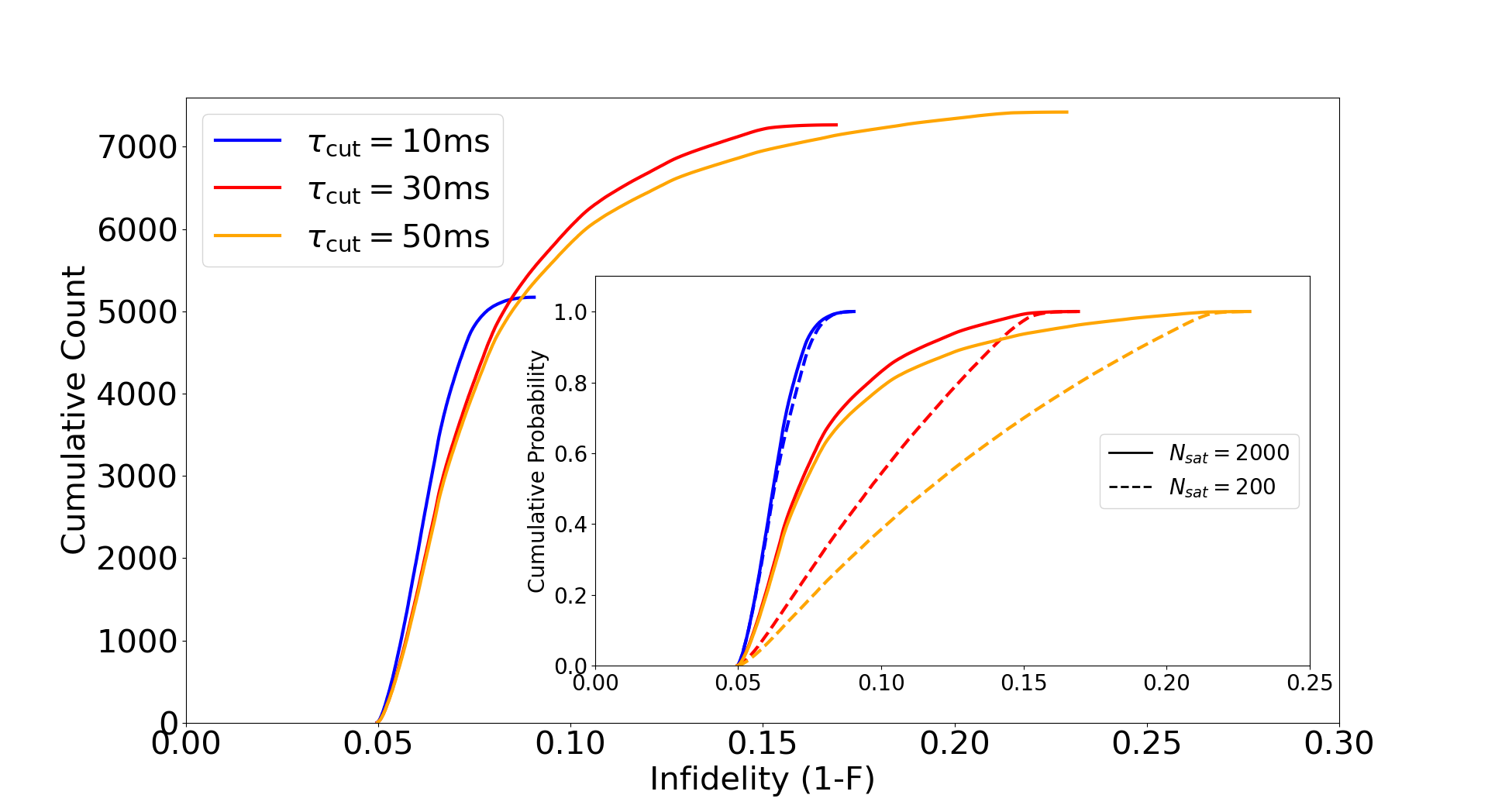}
\caption{Zenith A $45^\circ$.}
\end{subfigure}
\caption{Cumulative infidelity counts for (a) Zenith-zenith, (b) Symmetric, (c) Zenith A $90^\circ$ and (d) Zenith A $45^\circ$, for $N_\text{sat} = 2000$ and $\tau_\text{cut} = 10, 30$ and $50$~ms. Results were obtained from $1000$ trials. The inset graphs present the cumulative infidelity distributions for $N_\text{sat} = 200$ (dashed lines) and  $2000$ (solid lines). Increasing $N_\text{sat}$ results in not only a greater PDV but also increases the typical fidelity. This is reflected in the stepper slopes of the cumulative probability graphs for $N_\text{sat}=2000$ compared to $N_\text{sat}=200$. }
\label{fig:n2000_cumulative_infids_with_inset}
\end{figure}

The cumulative infidelity counts are shown in Fig.~\ref{fig:n2000_cumulative_infids_with_inset}. Inset graphs show the normalised cumulative infidelity probability distributions for $N_\text{sat} = 200$ and $2000$. The steeper slopes of the cumulative infidelities for $N_\text{sat}=2000$ compared to $N_\text{sat}=200$ indicate a narrower spread, with the former having a greater typical fidelity also. Together with Fig.~\ref{fig:N2000_median_fids_times} this shows that increasing the memory capacity of the satellite eases requirements on its memory lifetimes, in the sense that the typical fidelity improves for a fixed $\tau_\text{mem}$.

\section{Waiting Times and Fidelities for Buffer Size Memory Management}
\label{appen:bufferBasedMemoryManagement}
In the main text, we employed a memory management policy based on a maximum storage time, to ensure that the entanglement fidelity on each sublink remained sufficiently high. Alternatively, we can choose to keep only a predetermined maximum number of successfully distributed qubits, $B$, in memory, and reuse the other slots for fresh entanglement distribution attempts. We term this a `buffer size' memory management policy. Here, we compare the performance of these two policies to each other, in terms of PDV and obtained waiting time statistics.

\begin{figure}[!tbp]
\centering
\newcommand{\figwidth}{0.48\textwidth}
\begin{subfigure}{\figwidth}
\centering
\includegraphics[width=\linewidth]{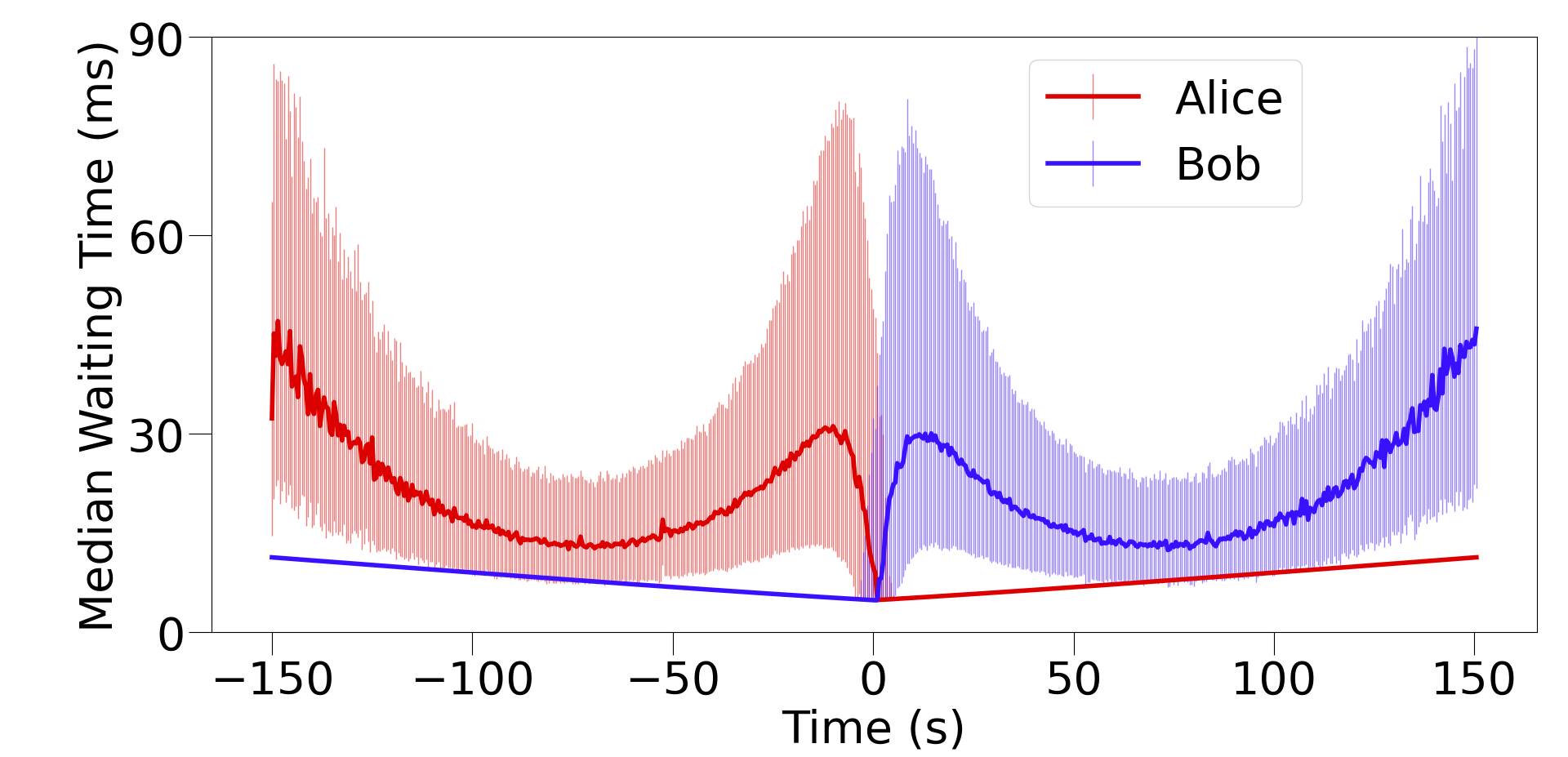}
\caption{Zenith-zenith. Total PDV: $900 \pm 20$}
\end{subfigure}
\hfill
\begin{subfigure}{\figwidth}
\centering
\includegraphics[width=\linewidth]{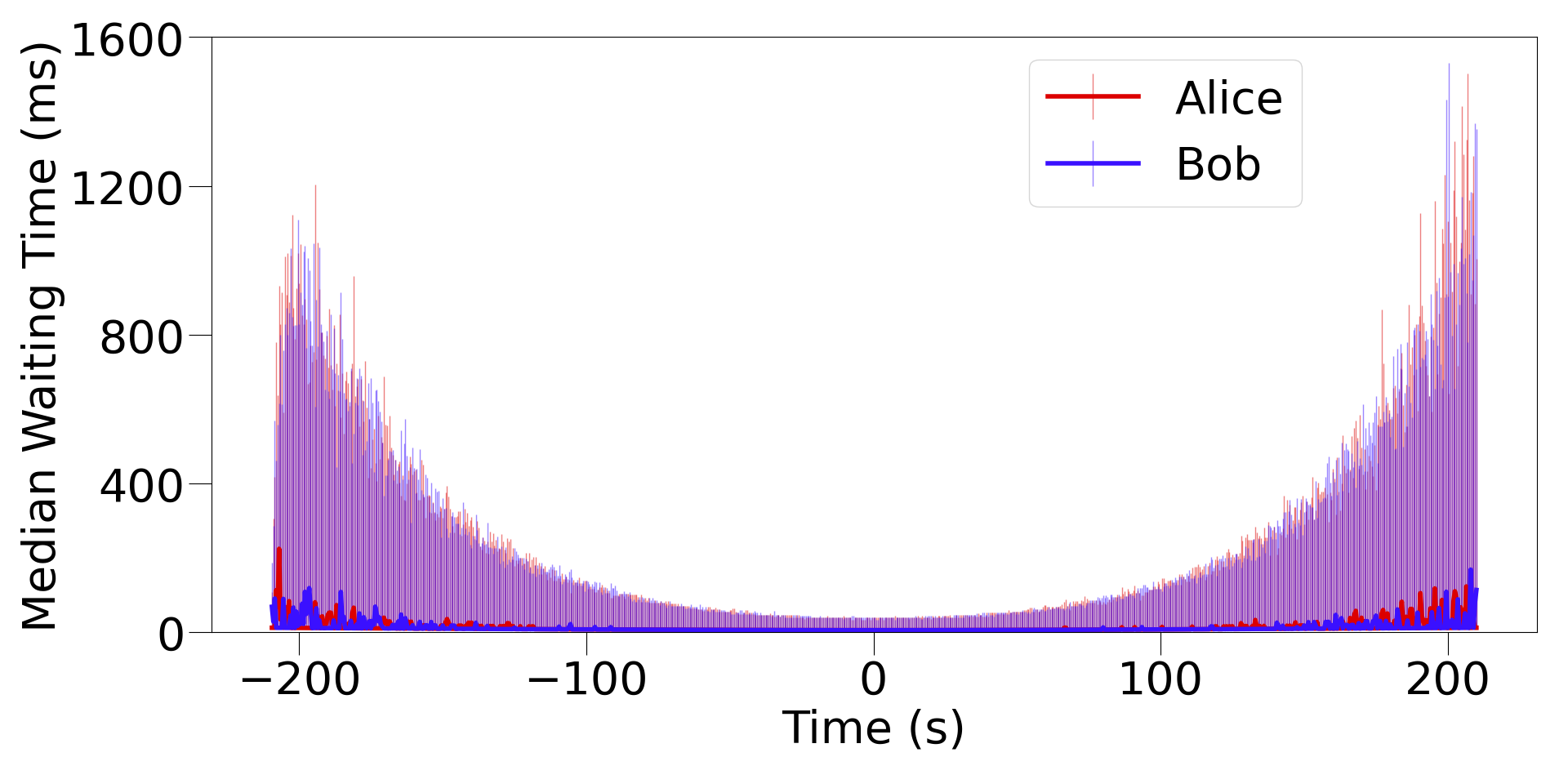}
\caption{Symmetric overpass. Total PDV: $1632 \pm 22$.}
\end{subfigure}
\par\medskip
\begin{subfigure}{\figwidth}
\centering
\includegraphics[width=\linewidth]{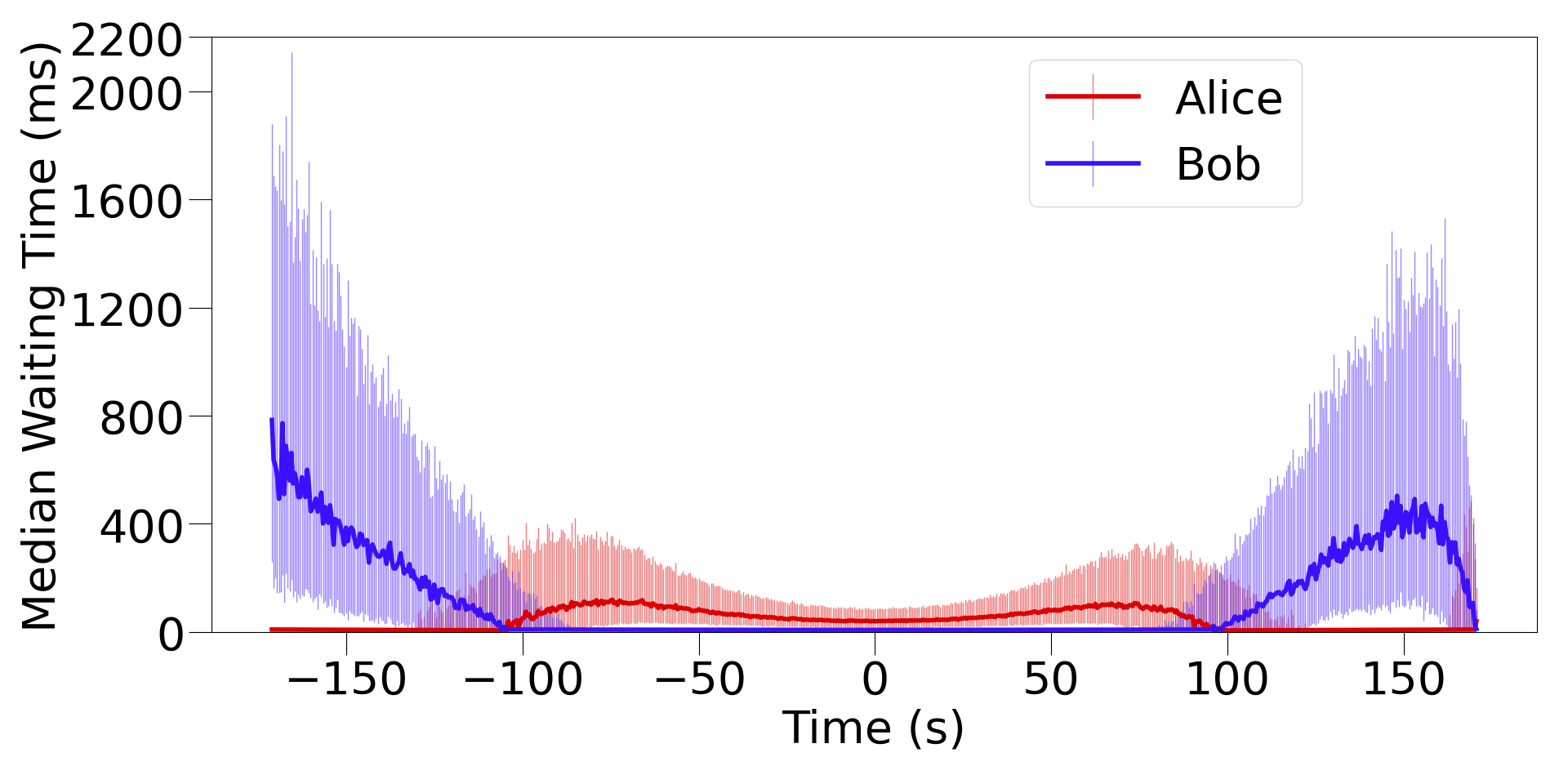}
\caption{Zenith A $90^\circ$. Total PDV: $661 \pm 18$.}
\end{subfigure}
\hfill
\begin{subfigure}{\figwidth}
\centering
\includegraphics[width=\linewidth]{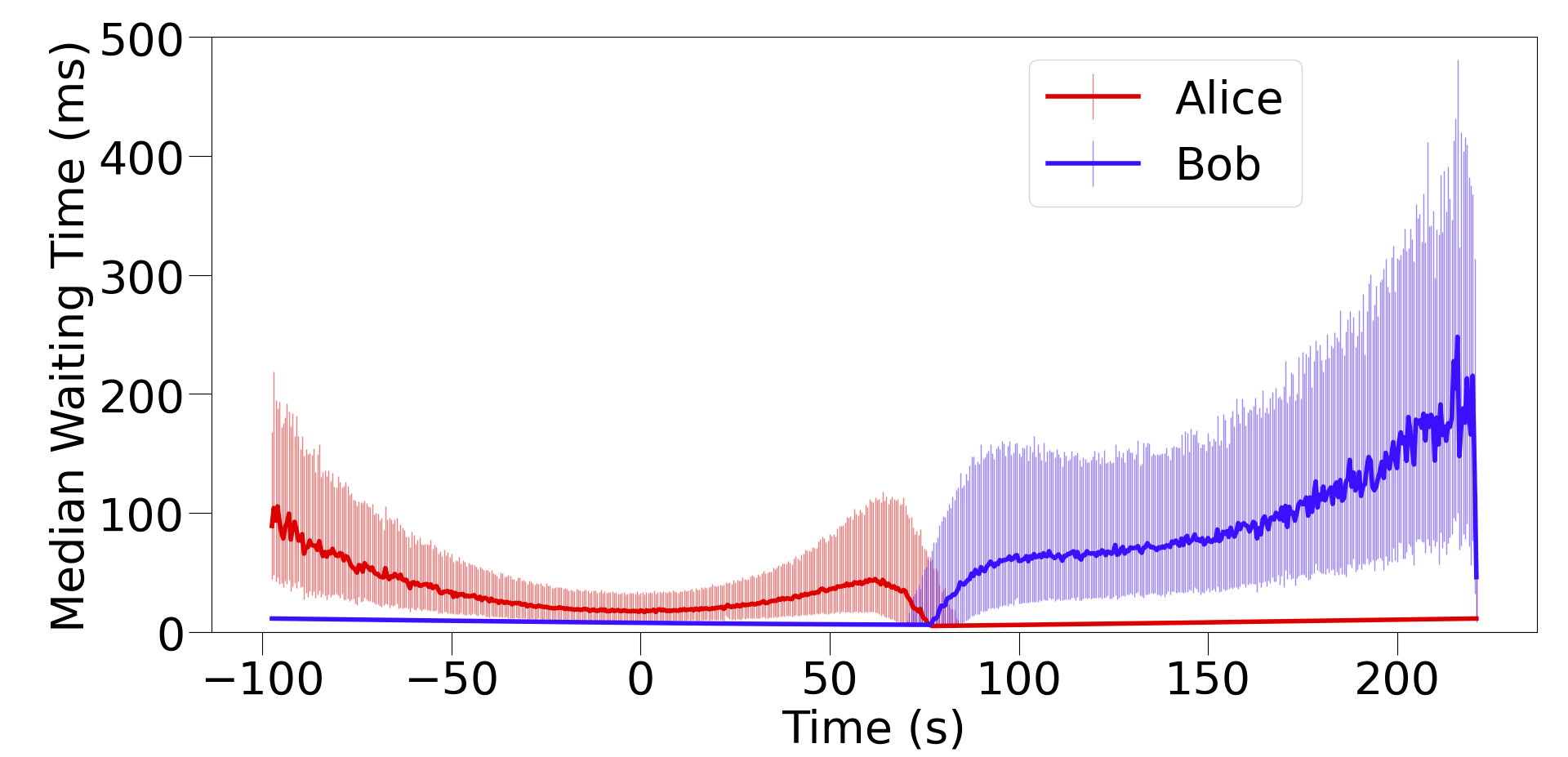}
\caption{Zenith A $45^\circ$. Total PDV: $749 \pm 18$.}
\end{subfigure}
\caption{Median waiting times for  (a) Zenith-zenith, (b) Symmetric, (c) Zenith A $90^\circ$ and (d) Zenith A $45^\circ$, for $N_\text{sat} = 200$ and $B=5$, obtained from $1000$ trials. Error bars show the interquartile ranges. The total overpass PDVs are stated as the mean and standard deviation. The buffer strategy resulted in more total PDV than the cut-off time strategy (Fig~\ref{fig:different_cutoffs_dist_rates}), especially when compared to $\tau_\text{cut} = 10$~ms. However, the typical waiting times and their variability were much greater in comparison. This implies that the buffer-size approach will be unsuitable for platforms where $\tau_\text{mem}$ is short or consistency in waiting times is desired.}
\label{fig:buffer5_waitingTimes}
\end{figure}

We obtained qubit waiting time and fidelity distributions for the representative overpasses (Fig.~\ref{fig:4passesDiagram}, for $N_\text{sat} = 200$ and $B=5$. It is immediately clear that this buffer-strategy results in substantially higher typical waiting times than the cut-off-based approach (~\ref{fig:buffer5_waitingTimes}). The higher waiting times indicate that the buffer-strategy will likely be unsuitable for platforms with short $\tau_\text{mem}$. Furthermore, the higher variability in the waiting times as indicated by the large IQRs implies that the buffer-size strategy is also unsuitable for applications where consistency is desired.

\begin{figure}[!tbp]
\centering
\newcommand{\figwidth}{0.48\textwidth}
\begin{subfigure}{\figwidth}
\centering
\includegraphics[width=\linewidth]{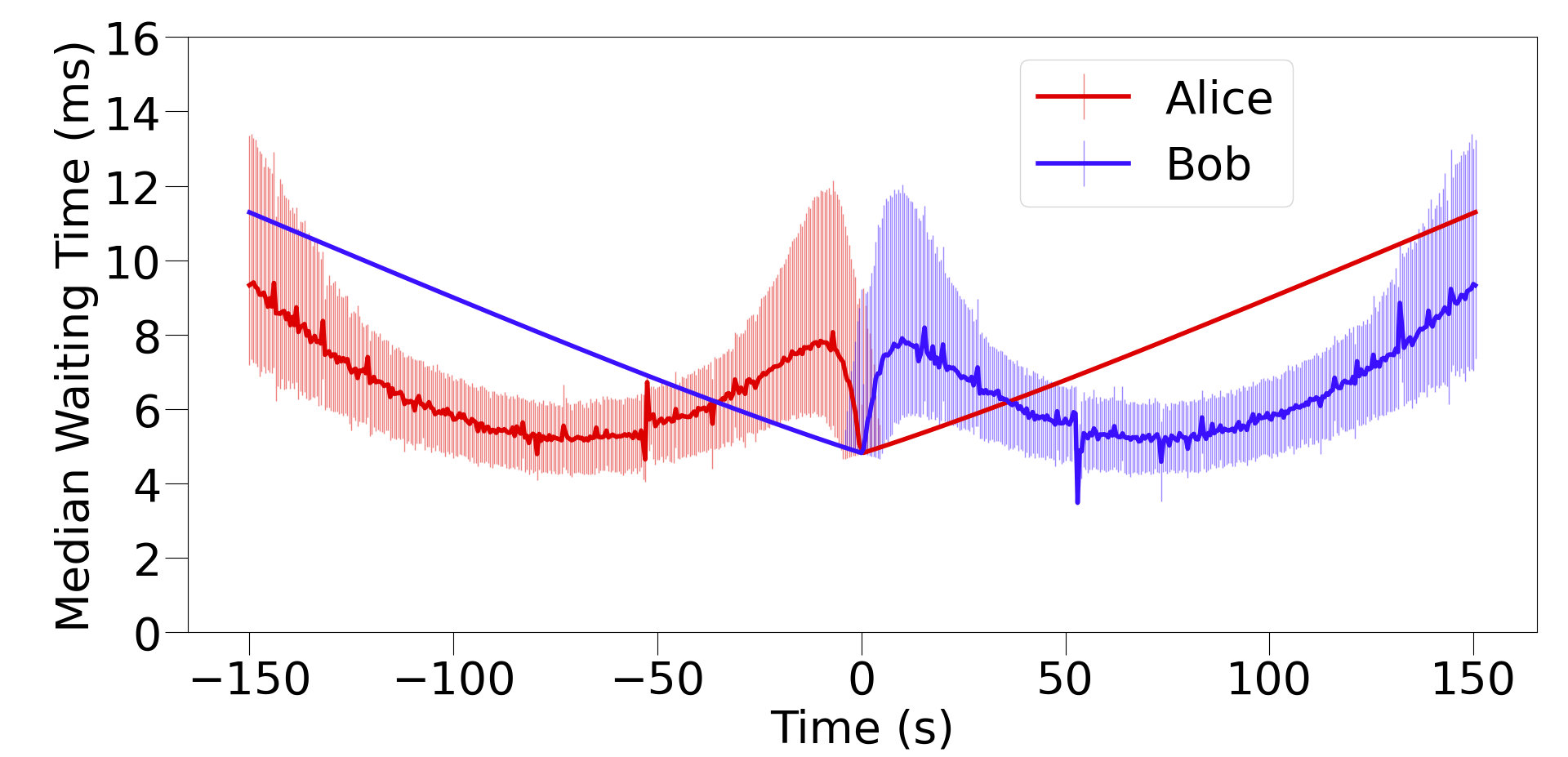}
\caption{Zenith-zenith. Total PDV: $8869 \pm 68$}
\end{subfigure}
\hfill
\begin{subfigure}{\figwidth}
\centering
\includegraphics[width=\linewidth]{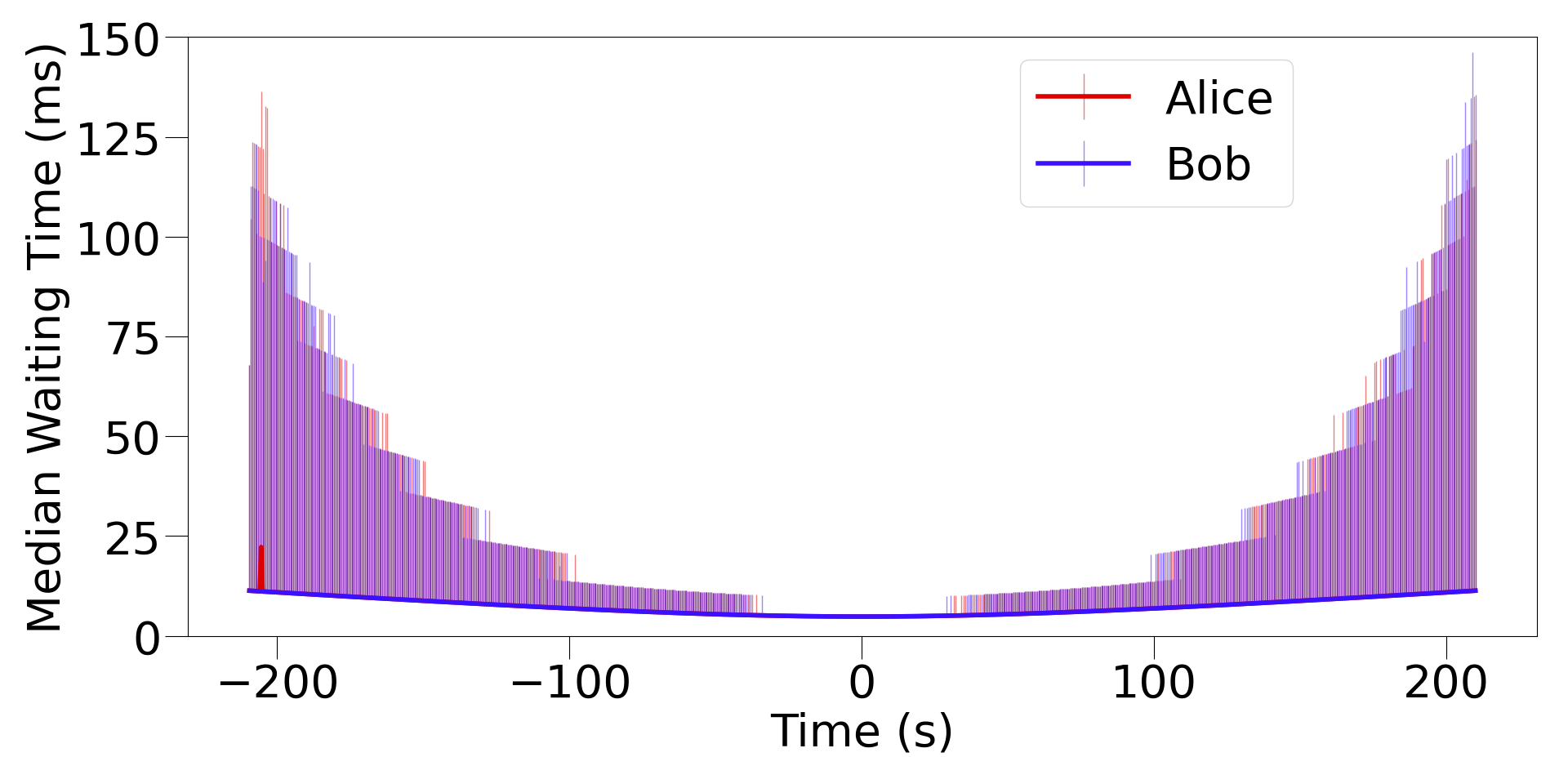}
\caption{Symmetric overpass. Total PDV: $16459 \pm 70$.}
\end{subfigure}
\par\medskip
\begin{subfigure}{\figwidth}
\centering
\includegraphics[width=\linewidth]{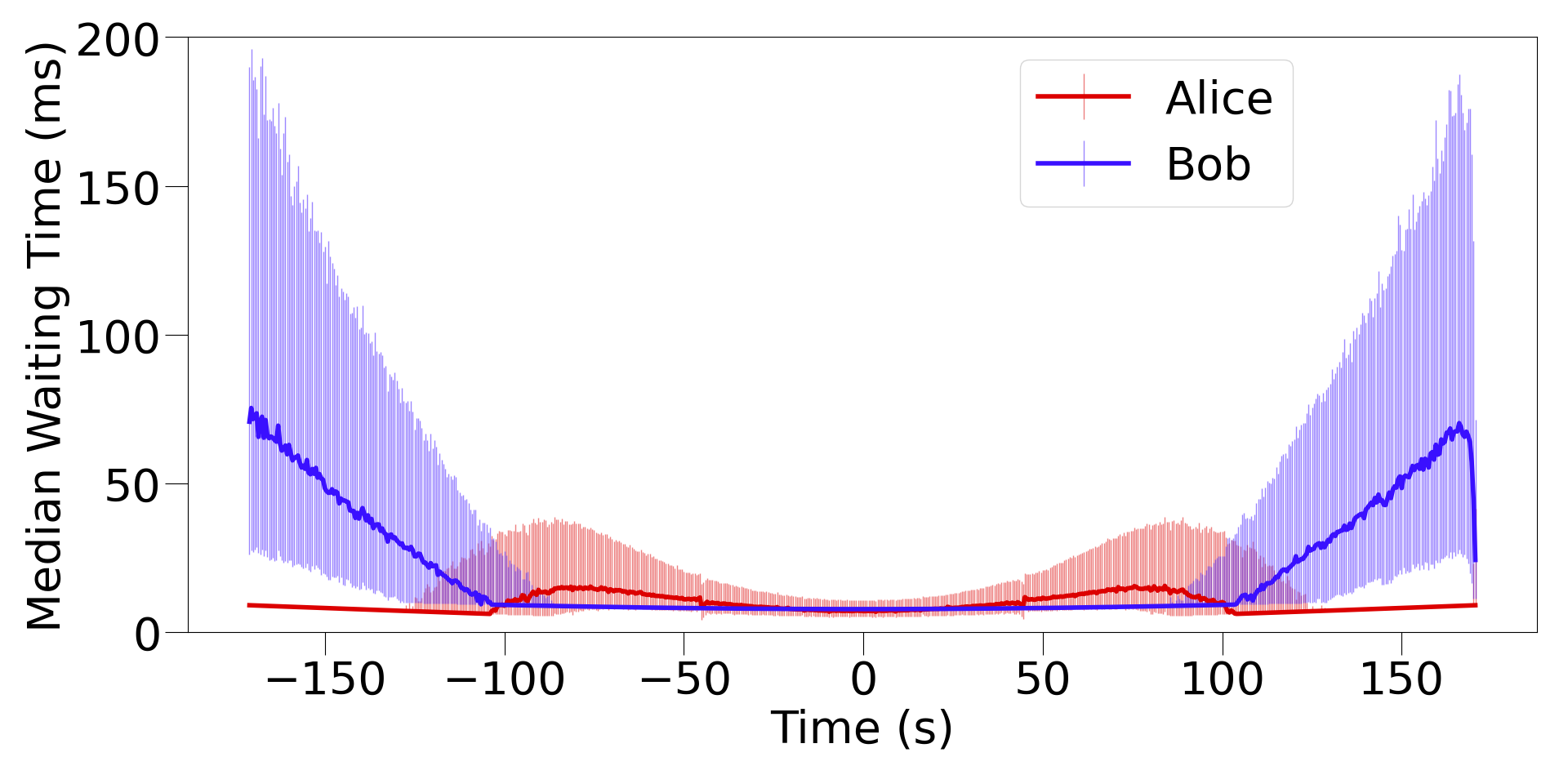}
\caption{Zenith A $90^\circ$. Total PDV: $6619 \pm 55$.}
\end{subfigure}
\hfill
\begin{subfigure}{\figwidth}
\centering
\includegraphics[width=\linewidth]{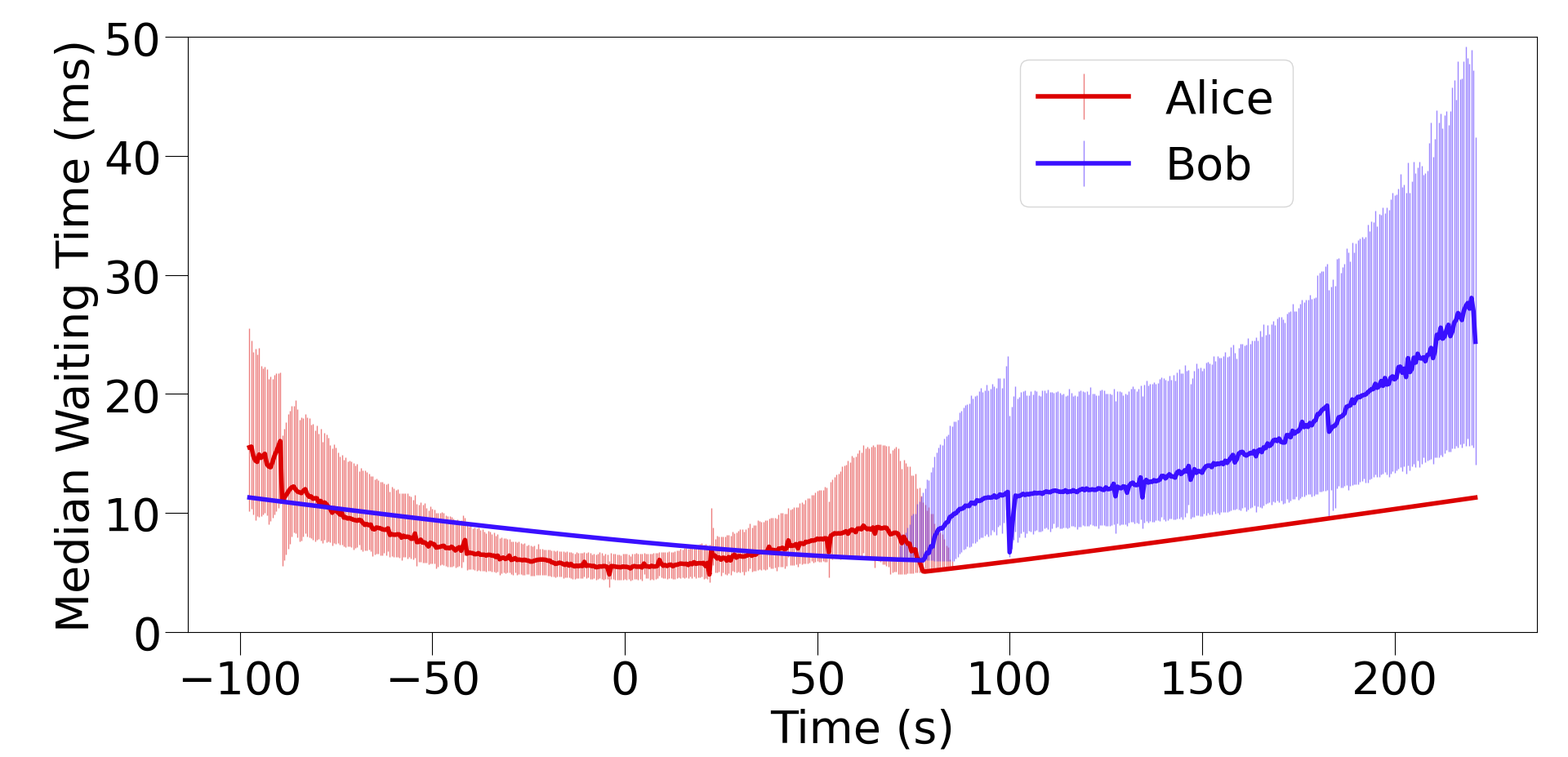}
\caption{Zenith A $45^\circ$. Total PDV: $7485 \pm 59$.}
\end{subfigure}
\caption{Median waiting times for  (a) Zenith-zenith, (b) Symmetric, (c) Zenith A $90^\circ$ and (d) Zenith A $45^\circ$ for $N_\text{sat} = 2000$ and $B=5$, obtained from $1000$ trials. Error-bars show the interquartile ranges. The total overpass PDVs are stated as the mean and standard deviation. The waiting times are lower compared to to the $N_\text{sat}=200$ case (Fig.~\ref{fig:buffer5_waitingTimes} and show smaller variability, although the Symmetric and Zenith A $90^\circ$ still exhibit high IQRs. The achievable improvement in waiting times is limited by the unavoidable round-trip time. These results indicate that increasing the memory capacity can ease requirements on $\tau_\text{mem}$.}
\label{fig:buffer5_waitingTimes_2000}
\end{figure}

Increasing the satellite's memory capacity decreases the waiting times of the buffer strategy stemming from the greater number of successful photon transmission that an increased memory capacity gives. Increasing $N_\text{sat}$ to $2000$ leads to lower waiting times and less spread than for $N_\text{sat} = 200$ and a linear increase in total PDV (Fig.~\ref{fig:buffer5_waitingTimes_2000}). This indicates that increasing the capacity of a memory platform will ease the requirements on its $\tau_\text{mem}$. However, it remains important to improve both in tandem to optimize entanglement distribution rates and fidelities. 

Reducing the buffer size has the effect of reducing the typical waiting time and associated spread. We performed this analysis for the symmetric overpass with $B = 1,2,3,4$ and $N_\text{sat} = 2000$. For each value of $B$ the median waiting times correspond to the round trip times, with decreases in buffer size reducing the interquartile range. This improved consistency comes at the cost of reduced PDV, with $B=1$ resulting in $3055$ fewer total pairs than $B=4$. 

Overall, the buffer-size memory management strategy results in higher waiting time and greater variability than the cut-off based approach. We therefore conclude that buffer-size strategies are not likely to be suitable for near-term quantum memory platforms.

\begin{figure}[!tbp]
\centering
\newcommand{\figwidth}{0.48\textwidth}
\begin{subfigure}{\figwidth}
\centering
\includegraphics[width=\linewidth]{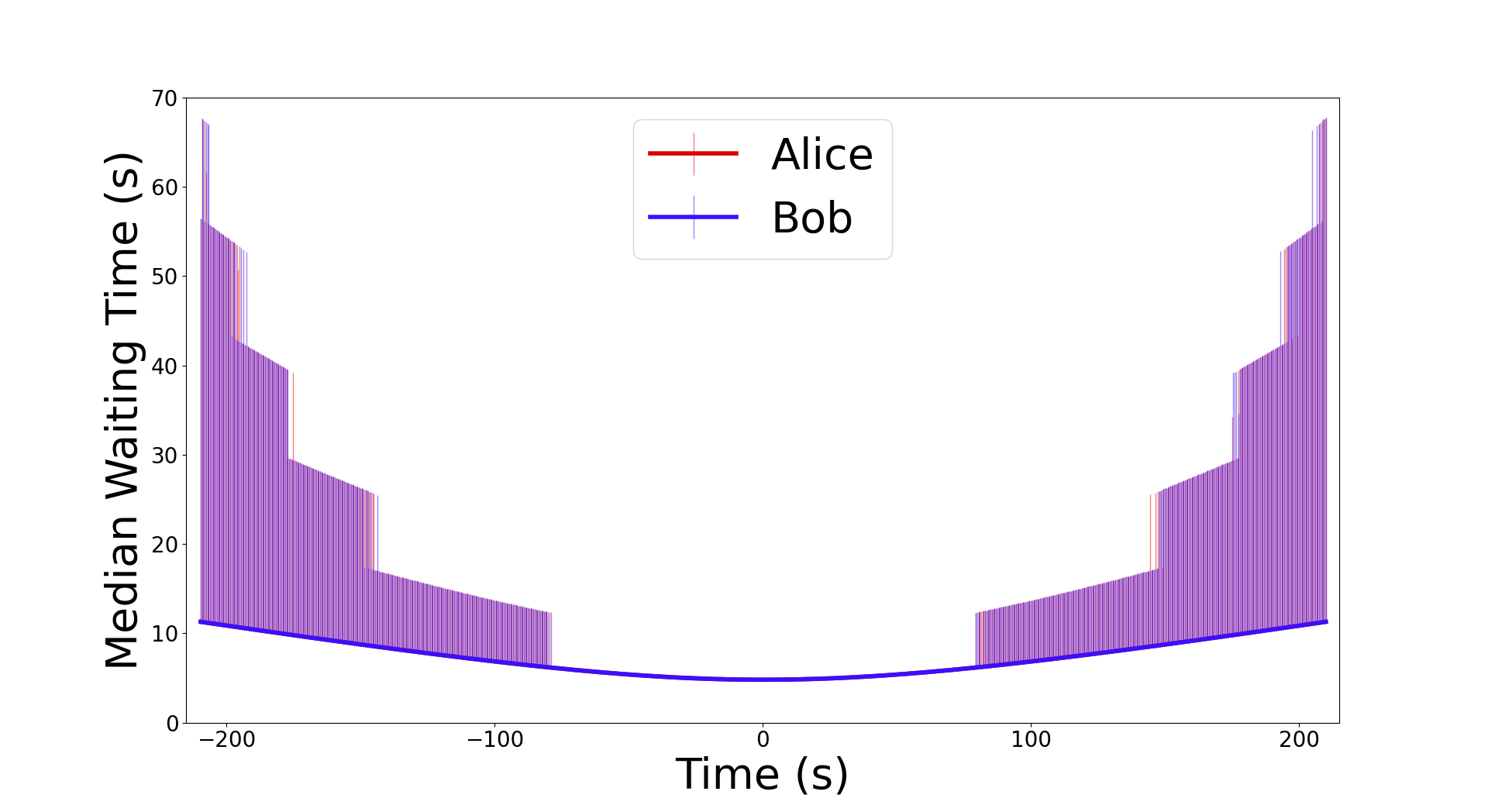}
\caption{$B=1$. Total PDV: $13089 \pm 62$.}
\end{subfigure}
\hfill
\begin{subfigure}{\figwidth}
\centering
\includegraphics[width=\linewidth]{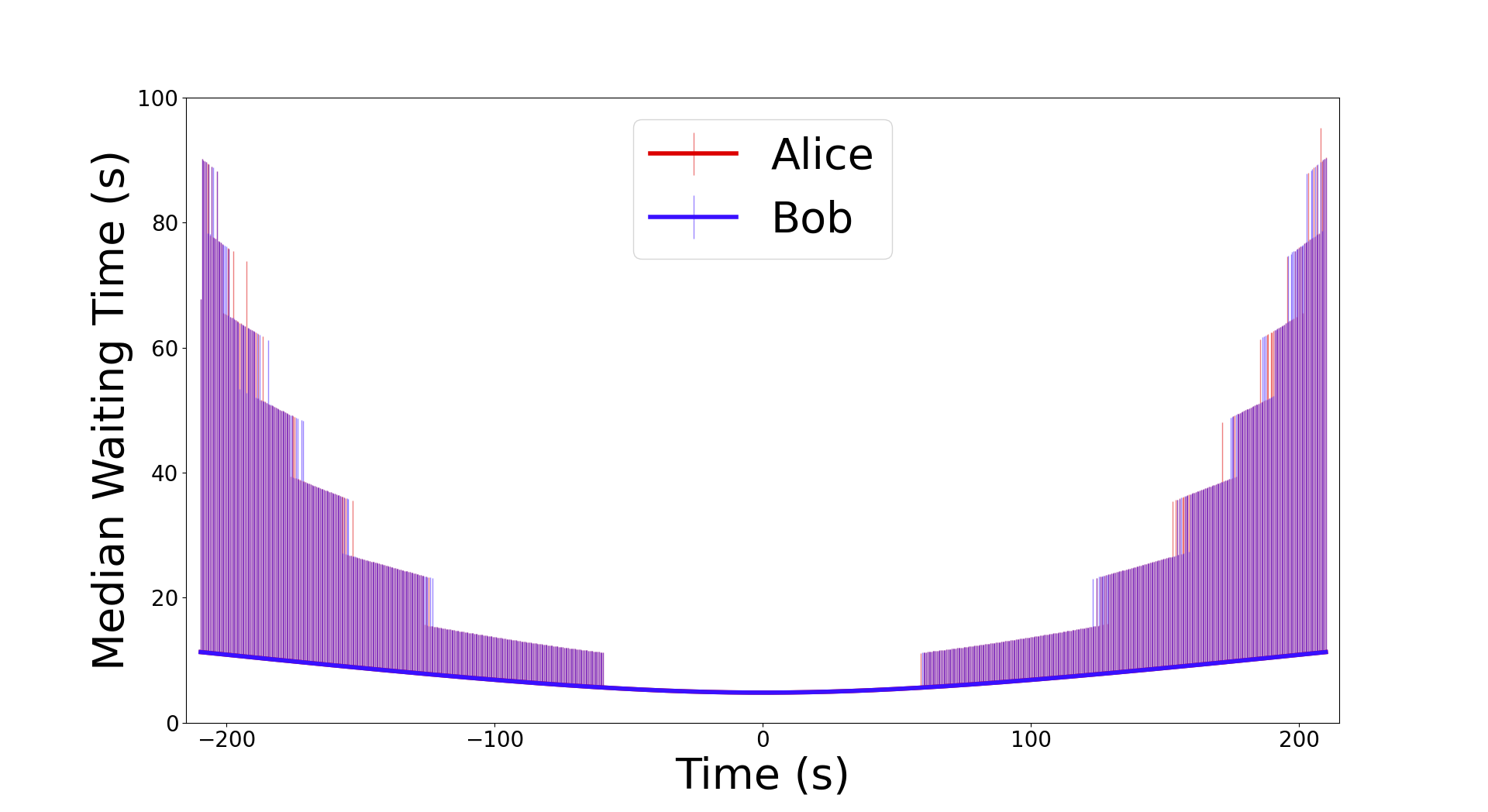}
\caption{$B=2$. Total PDV: $14827 \pm 67$.}
\end{subfigure}
\par\medskip
\begin{subfigure}{\figwidth}
\centering
\includegraphics[width=\linewidth]{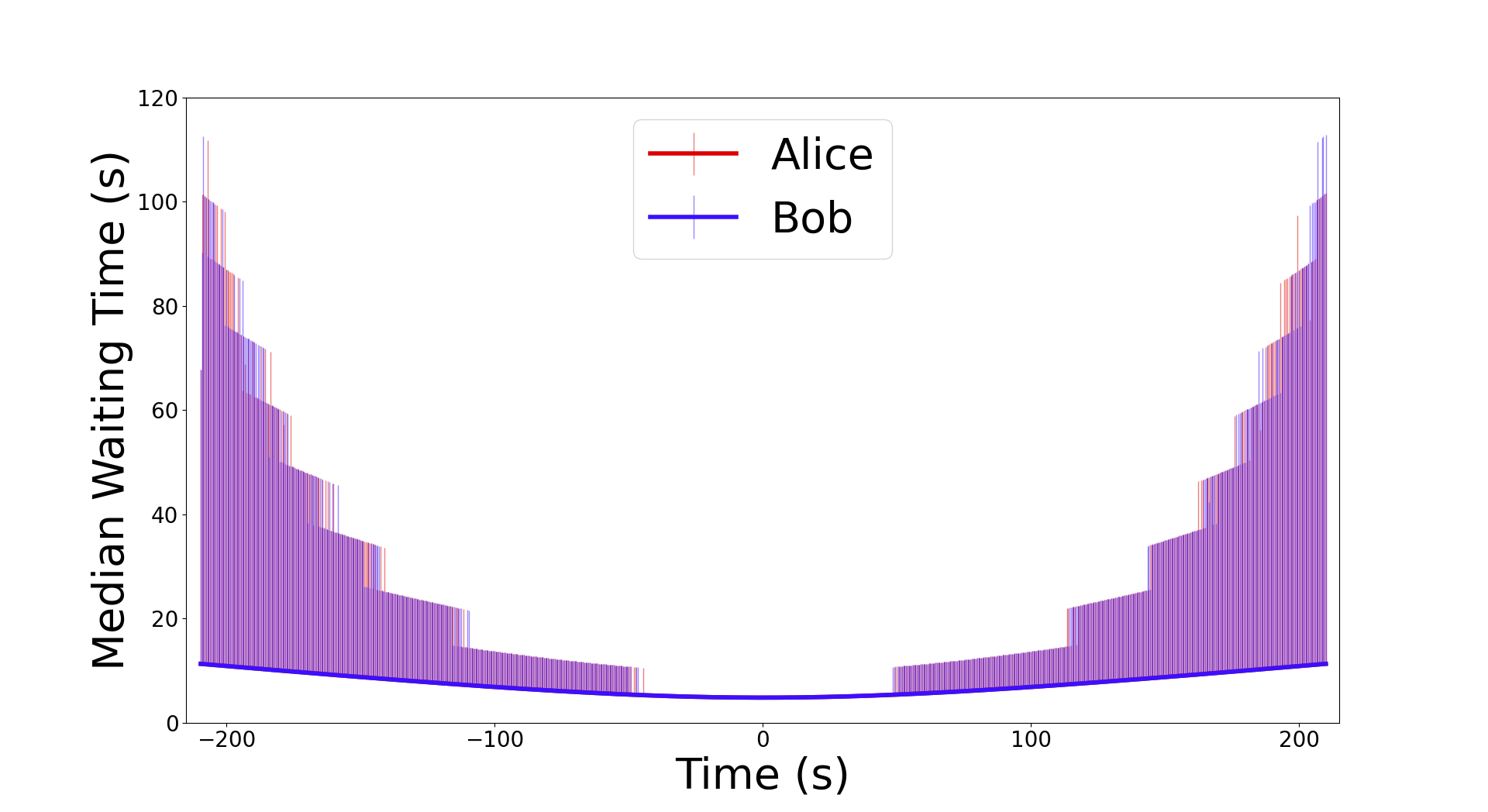}
\caption{$B=3$. Total PDV: $15656 \pm 70$.}
\end{subfigure}
\hfill
\begin{subfigure}{\figwidth}
\centering
\includegraphics[width=\linewidth]{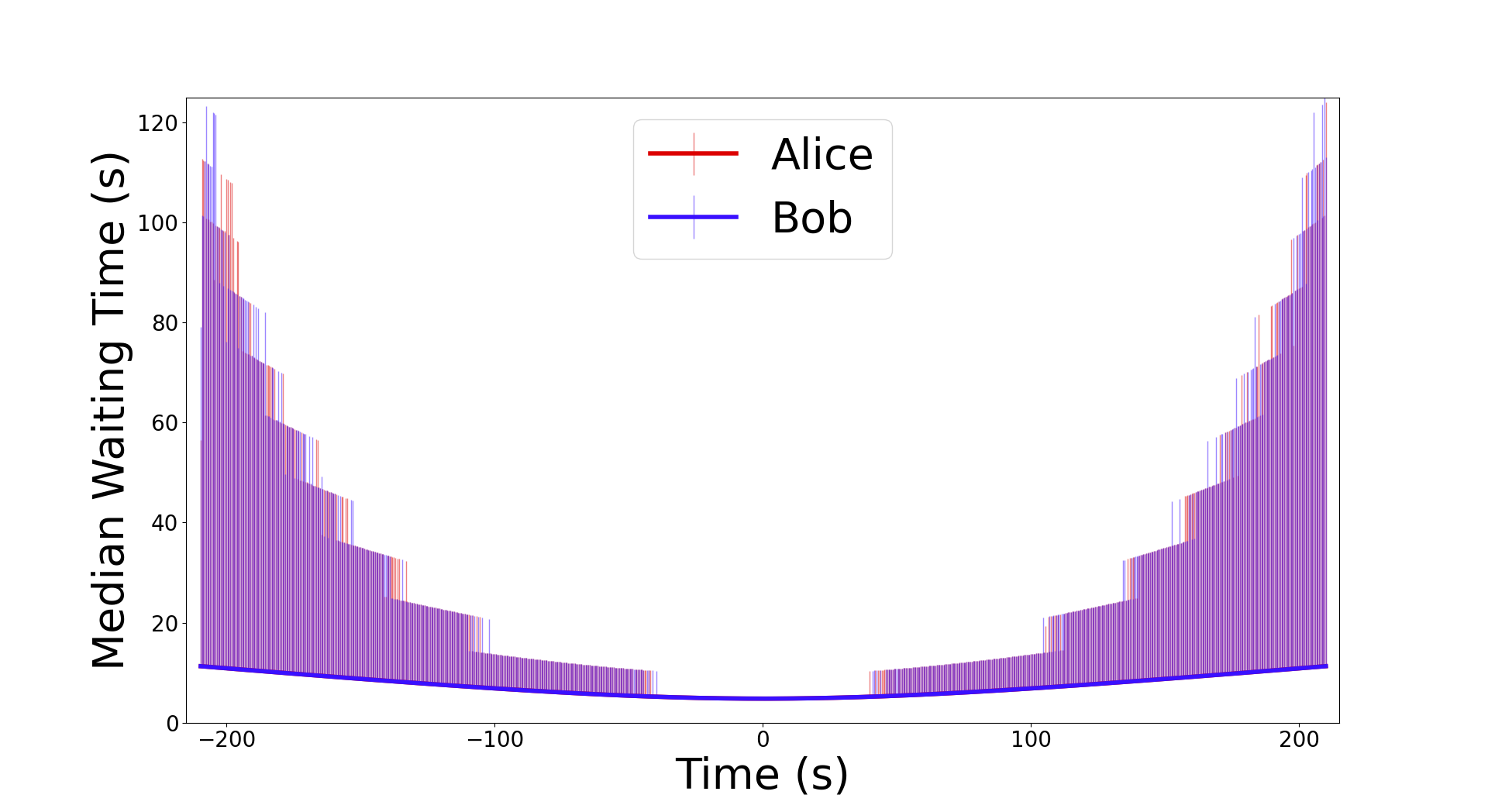}
\caption{$B=4$. Total PDV: $16144 \pm 70$.}
\end{subfigure}
\caption{Median waiting times for the symmetric overpass for (a) $B=1$, (b) $B=2$, (c) $B=3$ and (d) $B=4$ for $N-\text{sat} = 2000$, obtained from $1000$ trials. The error bars indicate the inter-quartile ranges. Total PDVs stated in the sub figure captions correspond to the mean and standard deviation. The observed median waiting times correspond to the round-trip time in each case, but reducing $B$ has the effect of reducing the variability of the waiting times. This improved consistence does come at a cost to the total PDV.}
\label{fig:symm_changing_buffer_size}
\end{figure}

\clearpage

\appendixpageoff
\appendixtitleoff
\renewcommand{\appendixtocname}{Supplementary material}
\begin{appendices}
\setcounter{section}{19}
\crefalias{section}{supp}
\section*{Supplementary Information}
\subsection{Monte Carlo Methods}

In the main text, we evaluated the fidelity of the pairs distributed by a single satellite quantum repeater (SSQR) to two ground stations. In order to do this, it is necessary to access the waiting time statistic of memory qubits, for which we employed a Monte Carlo (MC) simulation of SSQR operation. In this supplementary information, we give pseudocode for the MC methods underlying the main text.

A list of variables used in the MC processes is presented in Table~\ref{table:MCVariables}. The satellite is propagated in time steps of $\text{dt}$ (set to $1/2$~s in the main work). The channel efficiencies and photon transit times are updated at the beginning of each time step. During each time step, with starting time $t_0$, the satellite performs the sequence of operations described in Algo.~\ref{algor:MCProcessMain} until its internal time reaches $t_0 + dt$:
\begin{enumerate}
    \item The satellite trims the memory (Algor.~\ref{algor:trimMemory}). This involves discarding any memory qubits whose current storage time is greater than $\tau_\text{cut}$. This is implemented by updating arrays that store the current state and waiting time of memory qubits to reset expired pairs.
    \item The satellite performs entanglement swapping between qubits stored simultaneously in both the Alice and Bob memory registers, using the youngest available qubits (Algor.~\ref{algor:entSwapping}). The waiting times of qubits used in entanglement swapping are recorded. In the code, this is done by identifying which memory qubits correspond to the youngest pairs and appending their waiting times to a list. The corresponding memories and waiting times are then reset, and the total number of swapping events in incremented.
    \item The satellite attempts to generate entanglement with the ground stations (Algor.~\ref{algor:entGeneration}). Based on the success and failure notifications received from the ground stations, available memory resources are allocated to subsequent transmission attempts.
\end{enumerate}
This process is repeated over the full overpass. Repeating this for many overpasses allows for meaningful statistics to be obtained from the MC model. 

In the Appendix of the main text we also consider the case where the satellite trims its memory based on the number of stored pairs and not a cut-off time. In this case, the satellite discards memory qubits in line $8$ of Algor.~\ref{algor:MCProcessMain} oldest first, until a predetermined number remain in storage. This is described in Algor.~\ref{algor:trimBuffer}.

\begin{table*}[t]
\centering
\begin{tabular}{llp{8cm}}
\toprule
Variable & Description\\ 
\midrule
$M_{A(B)}$ & Arrays tracking the state of the Alice (Bob) memory register: $1 (0)$ indicates (un)avaialable for entanglement swapping  \\ 
$T_{A(B)}$ & Arrays recording memory qubit loading times: updated upon confirmation of transmission, $0$ otherwise.\\  
timeA(B) & Arrival time of the next confirmation signal from Alice (Bob).  \\
$W_{A(B)}$ & Lists recording waiting times of Alice (Bob) memory qubits, updated when qubit is swapped.\\
$\tau_{A(B)}$ & One-way photon transit time to Alice (Bob). \\ 
$\eta_{A(B)}$ & Alice (Bob) channel efficiencies.\\ 
$t_0 $& Time at beginning of current time step.  \\ 
$t_\text{internal}$ & Internal time for tacking time elapsedin current time step. \\
$\tau_\text{cut}$ & Memory qubit cut-off time.\\  
$N_\text{SSQR}$& Number of successfully swapped pairs in this time steps. \\
\bottomrule
\end{tabular}
\caption{Description of key variables used in the MC process. }
\label{table:MCVariables}
\end{table*}

\begin{algorithm}[!htb]
    \SetKwInOut{Input}{Input}
    \SetKwInOut{Output}{Output}

    \underline{SSQR\_MonteCarlo}$\big(M_A, M_B, T_A , T_B, \tau_A,\tau_B, \eta_A, \eta_B, t_0, \tau_\text{cut}\big)$\\;
    \Input{memory arrays, storage time arrays, photon transit times, channel transmissivities, time-step beginning time and cut-off time.}
    \Output{updated memory arrays, updated time arrays, waiting times, SSQR counts, Alice counts, Bob counts, final time.}
    $N_\text{SSQR} \leftarrow 0$ \\
    $W_A \leftarrow []$, $W_B \leftarrow []$\\
    $t_\text{internal} \leftarrow 0$ \\
    timeA $\leftarrow$ $2\tau_A$, timeB $\leftarrow 2 \tau_B$ \\
    \While{$t_\text{internal} \leq $dt }{
    $M_A$, $T_A$ $\leftarrow$ \underline{TrimMemory}($M_A$, $T_A$, $t_0$, $t_\text{internal}$, $\tau_\text{cut}$) \\
    $M_B$, $T_B$ $\leftarrow$ \underline{TrimMemory}($M_B$, $T_B$, $t_0$, $t_\text{internal}$, $\tau_\text{cut}$) \\
    $M_A$, $M_B$, $T_A$,$T_B \leftarrow $\underline{entanglementSwapping}($M_A$, $M_B$, $T_A$, $T_B$) \\
    $M_A, M_B, T_A, T_B, t_\text{internal} \leftarrow$ \underline{EntanglementGeneration}($M_A, M_B, T_A, T_B$)
    }
    \Return{$M_A, M_B, T_A, T_B, W_A, W_B, N_\text{SSQR},  N_\text{Alice},  N_\text{Bob}$, $t_0 + t_\text{internal}$}
    \caption{Pseudocode for the MC process performed within each time step. The satellite performs the process of trimming the memory registers, performing entanglement swapping and attempting entanglement generation. The $M_{A(B)}$ and $T_{A(B)}$ arrays are outputted to initialise the next time step. The SSQR counts $N_\text{SSQR}$ and waiting times lists $W_A(B)$ are recorded to built statistics for the full overpass.}
    \label{algor:MCProcessMain}
\end{algorithm} 

\begin{algorithm}
    \SetKwInOut{Input}{Input}
    \SetKwInOut{Output}{Output}
    \underline{TrimMemory}$\big(M, T, t_0, t_\text{internal},  \tau_\text{cut}\big)$\\;
    \Input{memory array, storage time array, time-step beginning time , internal time, cut-off time.}
    \Output{updated memory and storage time arrays.}

    \For{$i \leftarrow 1$ \KwTo $\mathrm{length}(M)$}{
    \If{$T[i] - (t_0 + t_\text{internal})> \tau_{\text{cut}}$}{
    $T[i] \leftarrow 0$ \\ 
    $M[i] \leftarrow 0$
    }
    }

    \Return{M, T}
    \caption{Algorithm for trimming the satellite memory. Any memory qubits whose current waiting time is in excess of $\tau_\text{cut}$ is discarded. The corresponding positions in the $M_{A,B}$ and $T_{A, B}$ arrays are reset. }
    \label{algor:trimMemory}
\end{algorithm}

\begin{algorithm}
    \SetKwInOut{Input}{Input}
    \SetKwInOut{Output}{Output}
    \underline{EntanglementSwapping}$\big(M_A,  M_B, T_A, T_B, t_0, t_\text{internal}\big)$\\;
    \Input{memory arrays, storage time arrays.}
    \Output{updated memory arrays, storage time arrays, time step beginning time, internal time.}
    GLOBAL $N_\text{SSQR}, W_A, W_B$ \\
    \While{$\mathrm{sum}(M_A) > 0$  AND $\mathrm{sum}(M_B) > 0$}{
    $N_\text{SSQR}$ += 1 \\
    readyAlice $\leftarrow$ [], readyBob $\leftarrow$ [] \\
    readyAliceTimes $\leftarrow []$, readyBobTimes $\leftarrow []$ \\
    \For{$i \leftarrow 1$ \KwTo $\mathrm{length}(M_A)$}{
    \If{$M_A[i] >0$}{
    append i to readyAlice\\
    append $T_A$[i] to readyAliceTimes\\
    }
    }
    \For{$i \leftarrow 1$ \KwTo $\mathrm{length}(M_B)$}{
    \If{$M_B[i] >0$}{
    append i to readyBob\\
    append $T_B$[i] to readyBobTimes\\
    }
    }
    \tcp{choose youngest pairs to swap}
    aliceSlot $\leftarrow \mathrm{argmax}(readyAliceTimes)$ \\
    bobSlot $\leftarrow \mathrm{argmax}(readyBobTimes)$\\
    append ($t_0 + t_\text{internal}$ - $T_A$[aliceSlot]) to  $W_A$\\
    append  ($t_0 + t_\text{internal}$ - $T_B$[bobSlot]) to  $W_B$\\
    $M_A$[aliceSlot] $\leftarrow 0$, $T_A$[aliceSlot] $\leftarrow 0$ \\
    $M_B$[bobSlot] $\leftarrow 0$,  $T_B$[bobSlot] $\leftarrow 0$ \\
    }

    \Return{$M_A$, $M_B$, $T_A$, $T_B$}
    \caption{Entanglement swapping. When there is at least one memory qubit in each of the Alice and Bob memory registers the satellite performs entanglement swapping. The youngest qubits are selected for use in swapping (chosen by argmax). $N_\text{SSQR}$ is incremented and the $M_{A,B}$ and $T_{A,B}$ arrays are updated to reflected which qubits were used. The waiting times of the qubits when they were in swapping are recorded to built overpass statistics.
    }
    \label{algor:entSwapping}
\end{algorithm}

\begin{algorithm}
    \SetKwInOut{Input}{Input}
    \SetKwInOut{Output}{Output}
    \underline{EntanglementGeneration}$\big(M_A,M_B, T_A, T_B\big)$\\;
    \Input{memory arrays, storage time arrays.}
    \Output{updated memory and storage time arrays.}
    \If{$timeA = timeB$}{
    event $\leftarrow ['Alice', 'Bob']$ \\
    $t_\text{internal} \leftarrow timeA$ \\
    }
    \ElseIf{$timeA < timeB$}{
    event $\leftarrow ['Alice']$  \\
    $t_\text{internal} \leftarrow timeA$  \\
    }
    \ElseIf{$timeB < timeA$}{
    event $\leftarrow ['Bob']$ \\
    $t_\text{internal} \leftarrow timeB$
    }
    \For{x in event}{
    \If{x = Alice}{
    counter $\leftarrow $ 0 \\
    sample $\leftarrow$ SampleBinomial(length($M_A$) - sum($M_A$), $\eta_A$) \\
    \For{$i \leftarrow 1$ \KwTo $\mathrm{length}(M_A)$}{
        \If{$M_A[i] = 0$ AND counter $<$ sample}{
            $M_A[i] \leftarrow 1$\\
            $T_A[i] \leftarrow t_0 + t_\text{interal}  - 2\tau_A$ \\
            counter += 1
        }
    }
    timeA += 2$\tau_A$
    }
  \If{x = Bob}{
    counter $\leftarrow $ 0 \\
    sample $\leftarrow$ SampleBinomial(length($M_B$) - sum($M_B$), $\eta_B$) \\
    \For{$i \leftarrow 1$ \KwTo $\mathrm{length}(M_B)$}{
        \If{$M_B[i] = 0$ AND counter $<$ sample}{
            $M_B[i] \leftarrow 1$\\
            $T_B[i] \leftarrow t_0 + t_\text{interal}  - 2\tau_B$ \\
            counter += 1
        }
    }
    timeB += 2$\tau_B$
    }
    }
    \Return{$M_A, M_B, T_A, T_B, t_\text{internal}$ }
    \caption{Entanglement generation algorithm. The OGS from which the next confirmation signal is expected is first identified. The number of successfully transmitted photons are then determined by sampling from a binomial distribution (~\ref{alg:SampleBinomial}) and the internal time is updated. The memory and loading time arrays are updated conditioned on successful transmission, with the $-2\tau$ terms (e.g. line 24) reflecting the fact that the qubits were loaded into memory in the previous entanglement distribution attempt}
    \label{algor:entGeneration}
\end{algorithm} 

\begin{algorithm}
    \SetKwInOut{Input}{Input}
    \SetKwInOut{Output}{Output}
    \underline{SampleBinomial}$\big(n ,p)$\\;
    \Input{number of samples, success probability}
    \Output{number of successes.}
    $nSucceed \leftarrow 0$ \\
    \For{$i \leftarrow 1$ \KwTo $n$}{
    \If{$\mathrm{rand}() < p$}{
    nSucceed += 1     
    }
    }
    \Return{$nSucceed$}
    \caption{Algorithm used for sampling $n$ times from a binomial distribution with success probability $p$, used in Algor.~\ref{algor:entGeneration}. This algorithm makes use of rand(), the computer's built in random number generator that returns a random number betwween $0$ and $1$.}
    \label{alg:SampleBinomial}
\end{algorithm} 

\begin{algorithm}
    \SetKwInOut{Input}{Input}
    \SetKwInOut{Output}{Output}
    \underline{TrimToBuffer}$\big(M, T,  b\big)$\\;
    \Input{memory array, storage time array, buffer size.}
    \Output{updated memory and storage time arrays }
    \While{$sum(M) > b$}{
     listMems $\leftarrow$ [] \\
     timesList $\leftarrow$ []\\
     \For{$i \leftarrow 1$ to $\text{length}(M)$}{
            \If{$M[i] > 0$}{
                append i to listMems \\
                append $T[i]$ to timesList\\
            }
      }
      \tcp{discard oldest pairs} 
      $M[listMems[argmin(listTimes)]] \leftarrow 0$ \\ 
      $T[listMems[argmin(listTimes)]] \leftarrow 0 $
      }

    \Return{M, T}
    \caption{Trim to buffer size. If the current number of held (successfully transmitted) pairs is greater than the buffer size then qubits are discarded until $b$ remain. The oldest pairs, chosen by argmin, are chosen for to be discarded. }
    \label{algor:trimBuffer}
\end{algorithm} 

\clearpage

\subsection{Code}
This section presents the code used to generate the results in the main text. The code is split into two parts: the main Python script that runs the Monte Carlo simulation and saves the outputs, and a Cython file that contains all the functions used in the main script.

\subsubsection{Python code for running and saving Monte Carlo outputs}
Here, we present the main Python script used to run the satellite Monte Carlo simulation and saves the resulting outputs. The outputs are the waiting times of the memory qubits and the number of entangled pairs established by the repeater satellite. The number of trials, overpass geometry and memory allocations to each of Alice and Bob are set at the top of the script.
\begin{lstlisting}[breaklines=true]
import numpy as np
import os
from datetime import datetime
from cython_MC_functions_submission import atmosphericTransmissivity
from cython_MC_functions_submission import diffractionTransmissivity
from cython_MC_functions_submission import pointingLosses
from cython_MC_functions_submission import generateOrbitGeomtery
from cython_MC_functions_submission import SSQR_MonteCarlo
from cython_MC_functions_submission import save2DList

def generateLosses():
    for t in range(0, len(bothVisibleTimes)):
        aliceAtmdB  = (-10)*np.log10(atmosphericTransmissivity(zenithTransmissivity, elevationsAlice[t]))
        bobAtmdB  = (-10)*np.log10(atmosphericTransmissivity(zenithTransmissivity, elevationsBob[t]))
       
        aliceDiffdB = (-10)*np.log10(diffractionTransmissivity(distancesAlice[t]*1000, wavelength, Dr, deltaTheta))
        bobDiffdB = (-10)*np.log10(diffractionTransmissivity(distancesBob[t]*1000, wavelength, Dr, deltaTheta))
        
        alicePointdB = (-10)*np.log10(pointingLosses(distancesAlice[t]*1000, pointError, wavelength, deltaTheta))
        bobPointdB = (-10)*np.log10(pointingLosses(distancesBob[t]*1000, pointError, wavelength, deltaTheta))
        
        alicedBs[t] = aliceAtmdB + aliceDiffdB + alicePointdB + detectionEfficiencydB
        bobdBs[t]= bobAtmdB + bobDiffdB + bobPointdB + detectionEfficiencydB
    
    return 0 

saveToggle = True
pi = np.pi
rE = 6378
zenithTransmissivity = 0.79
detectionEfficiency = 0.7
detectionEfficiencydB = (-10)*np.log10(detectionEfficiency)
wavelength = 780e-9
deltaTheta = 10e-6
Dr = 50e-2
pointError = 1e-6
c = 2.9979e5
distance = 1000 
satAlt = 500 
mu = 3.9860043543609598e5
temp1 = mu/(rE+satAlt)
omega = np.sqrt(temp1/(rE+satAlt)/(rE+satAlt))
T = 2*pi/omega
integerT = round(T)
memCap = 200 
trials = 1000
here = os.getcwd()
pBSM = 1/2
dt = 1/2
here = os.getcwd()
times = np.arange(0, integerT + dt, dt, dtype = float)
satelliteCartesians  = np.zeros((len(times),3))
for t in range(0, len(times)):
    satelliteCartesians[t][0] = (rE+satAlt)*np.cos(omega*times[t])
    satelliteCartesians[t][1] = 0 
    satelliteCartesians[t][2] = (rE+satAlt)*np.sin(omega*times[t]) 
    

tCutoff = 50e-3
deltas = np.array([0])
alphas = np.array([0])
bothVisibleTimes, distancesAlice, distancesBob, elevationsAlice, elevationsBob = generateOrbitGeomtery(deltas[0], alphas[0], distance, satAlt, times, satelliteCartesians)

alicedBs = np.zeros(len(bothVisibleTimes))
bobdBs = np.zeros(len(bothVisibleTimes))
avgRates = np.zeros(len(bothVisibleTimes))

generateLosses()

aliceTaus = distancesAlice/c
bobTaus = distancesBob/c


aliceCap = int(100)
bobCap =  int(memCap - aliceCap)


if (saveToggle == True):
    outpath = str(round(tCutoff*1000))+'ms'
    if not os.path.exists(outpath):
       print('Making output directory:', outpath)
       os.makedirs(outpath)
    os.chdir(outpath)
    out = datetime.now()
    temp = str(out.day) + '.'  + str(out.month) + '.'  + str(out.year)
    outpathParent = 'dt=' + str(round(dt,2)) + temp + 'static_ACap=' + str(aliceCap) + '_BCap=' +str(bobCap) + '_h=' + str(satAlt) + '_trials=' + str(trials) + 'D=' + str(deltas[0]) + 'phi=' + str(round(alphas[0],2)) +'_tCut=' + str(round(tCutoff,2))
    if not os.path.exists(outpathParent):
        print('Making output directory:', outpathParent)
        os.makedirs(outpathParent)
    os.chdir(outpathParent)
    np.savetxt('bothVisibleTimes.csv', bothVisibleTimes)
    np.savetxt('alicedBs.csv', alicedBs)
    np.savetxt('bobdBs.csv', bobdBs)
    np.savetxt('distancesAlice.csv',distancesAlice)
    np.savetxt('distancesBob.csv', distancesBob)
    
   
for j in range(trials):
    
    aliceWaitGrid = [[] for _ in range(len(bothVisibleTimes))]
    bobWaitGrid = [[] for _ in range(len(bothVisibleTimes))]
    SRCounts = np.zeros(len(bothVisibleTimes))
    
    globalTime = 0 

    aliceMemories = np.zeros(aliceCap, dtype = int)
    bobMemories = np.zeros(bobCap, dtype = int) 
    aliceTimes = np.zeros(aliceCap)
    bobTimes = np.zeros(bobCap) 
    
    for t in range(len(bothVisibleTimes)):
        aliceMemories, bobMemories, aliceTimes, bobTimes, SRCounts[t], aW, bW, globalTime = SSQR_MonteCarlo(aliceMemories, bobMemories, aliceTimes, bobTimes, aliceTaus[t], bobTaus[t], alicedBs[t], bobdBs[t], globalTime, (t+1)*dt, tCutoff)
        aliceWaitGrid[t].extend(aW)
        bobWaitGrid[t].extend(bW)
        
    if (saveToggle == True):
        outpath = 'Overpass ' + str(j)
        if not os.path.exists(outpath):
            print('Making output directory:', outpath)
            os.makedirs(outpath)
        os.chdir(outpath)
        np.savetxt('SRCounts.csv', SRCounts)
        save2DList(aliceWaitGrid, 'aliceWaitGrid.csv')
        save2DList(bobWaitGrid, 'bobWaitGrid.csv')
        os.chdir('..')
\end{lstlisting}

\clearpage
\subsubsection{Cython code containing functions used in main script}
This section presents the Cython code that contains the functions called in the main Python script.
\begin{lstlisting}[breaklines=true]
import numpy as np
cimport numpy as np
import cython
cimport cython
from libc.math cimport sin
from libc.math cimport cos
from libc.math cimport exp
from libc.math cimport acos
from libc.math cimport log10
from libc.stdlib cimport rand, RAND_MAX

cpdef atmosphericTransmissivity(double zenithTransmission, double satelliteElevation):
    return zenithTransmission**(1/(sin(satelliteElevation)))

cpdef diffractionTransmissivity(double L, double Lambda, double receieverAperture, double divergence):
    cdef double omega0, temp1, temp2, omegaLSq, temp3
    cdef pi = 3.141592653589793
    omega0 = Lambda/(divergence*pi)
    
    temp1 = (Lambda*L)/(pi*omega0*omega0)
    temp2 = 1 + ((temp1*temp1))
    
    omegaLsq = omega0*omega0*temp2
    
    temp3 = (receieverAperture*receieverAperture)/(2*omegaLsq)
    return 1 - exp(-temp3)

cpdef save2DList(list data, str filename):
    cdef int i, j
    cdef int rows = len(data)
    cdef int cols
    cdef list row

    with open(filename, "w", encoding="utf-8") as f:
        for i in range(rows):
            row = data[i]
            cols = len(row)

            for j in range(cols):
                f.write(str(row[j]))
                if j < cols - 1:
                    f.write(",")

            f.write("\n")
            
cpdef pointingLosses(double z, double pointingError, double Lambda, double divergence):
    cdef pi = 3.141592653589793
    cdef double omega0, temp1, temp2, omegazsq, temp3
    omega0 = Lambda/(divergence*pi)
    
    temp1 = (Lambda*z)/(pi*omega0*omega0)
    temp2 = 1 + temp1*temp1
    
    omegazsq = omega0*omega0*temp2
    
    temp2 = pointingError*z
    
    temp3 = (-8)*temp2*temp2/omegazsq
    
    return exp(temp3)


cpdef cartesian(double r, double lat, double longitude):  
    return np.array([r*cos(lat)*cos(longitude), r*sin(longitude)*cos(lat), r*sin(lat)], dtype = np.double)



cpdef elevationAngle(double distance, double satAlt, double rE):
    cdef double temp1, temp2
    cdef pi = 3.141592653589793
    temp1 = 2*distance*rE
    temp2 = satAlt**2 + 2*satAlt*rE - distance**2
    
    return pi/2 - acos(temp2/temp1)

cpdef SampleBinomial(int n, double p):
    cdef int  j
    cdef int count = 0 
    count  =  0
    for j in range(n):
        if rand() / RAND_MAX < p:
            count += 1
    
    return count

cpdef generateOrbitGeomtery(double displacement, double intersectionAngle, double distance, double satAlt, double[:] times, double[:,:] satelliteCartesians):
    cdef pi = 3.141592653589793
    cdef double temp, angularSeparation, temp1, temp2, temp3, elevOGS1, elevOGS2
    cdef list bothVisibleTimes, bothVisibleDistancesOGS1, bothVisibleDistancesOGS2, bothVisibleElevationsOGS1, bothVisibleElevationsOGS2
    cdef np.ndarray[np.double_t, ndim = 1] latlongOGS1 = np.empty(2, dtype = np.double)
    cdef np.ndarray[np.double_t, ndim = 1] latLongOGS2 = np.empty(2, dtype = np.double)
    cdef np.ndarray[np.double_t, ndim = 1] cartesianOGS1 = np.empty(3, dtype = np.double)
    cdef np.ndarray[np.double_t, ndim = 1] cartesianOGS2 = np.empty(3, dtype = np.double)
    cdef np.ndarray[np.double_t, ndim = 1] coordOGS1 = np.empty(3, dtype = np.double)
    cdef np.ndarray[np.double_t, ndim = 1] coordOGS2 = np.empty(3, dtype = np.double)
    cdef np.ndarray[np.double_t, ndim = 1] distancesOGS1 = np.empty(len(times), dtype = np.double)
    cdef np.ndarray[np.double_t, ndim = 1] distancesOGS2 = np.empty(len(times), dtype = np.double)
    cdef np.ndarray[np.double_t, ndim = 1] elevationsOGS1 = np.empty(len(times), dtype = np.double)
    cdef np.ndarray[np.double_t, ndim = 1] elevationsOGS2 = np.empty(len(times), dtype = np.double)
    cdef int y, x                              

    cdef double minimumElevation = 10*pi/180
    cdef rE = 6378
    temp = displacement/(rE) 
    angularSeparation = distance/rE
    latLongOGS1  = np.array([pi/2 + angularSeparation/2 - temp, 0]) 
    latLongOGS2  = np.array([pi/2 - angularSeparation/2 - temp, 0])
   
    cartesianOGS1  = cartesian(rE, latLongOGS1[0], latLongOGS1[1])
    cartesianOGS2 = cartesian(rE, latLongOGS2[0], latLongOGS2[1]) 
    
    coordOGS1= np.array([cartesianOGS1[0]*np.cos(intersectionAngle), cartesianOGS1[0]*np.sin(intersectionAngle), cartesianOGS1[2]])
    coordOGS2 = np.array([cartesianOGS2[0]*np.cos(intersectionAngle), cartesianOGS2[0]*np.sin(intersectionAngle),cartesianOGS2[2]])
   
    distancesOGS1= np.zeros(np.size(times))
    distancesOGS2 = np.zeros(np.size(times))
   
    elevationsOGS1 = np.zeros(np.size(times))
    elevationsOGS2 = np.zeros(np.size(times))
    for y in range(0, np.size(times)):
        temp1 = satelliteCartesians[y][0] - coordOGS1[0]
        temp2 = satelliteCartesians[y][1] - coordOGS1[1]
        temp3 = satelliteCartesians[y][2] - coordOGS1[2]
       
        distancesOGS1[y] = (temp1*temp1+temp2*temp2+temp3*temp3)**(1/2)
        elevationsOGS1[y] = elevationAngle(distancesOGS1[y],satAlt,rE)  
       
        temp1 = satelliteCartesians[y][0] - coordOGS2[0]
        temp2 = satelliteCartesians[y][1] - coordOGS2[1]
        temp3 = satelliteCartesians[y][2] - coordOGS2[2]
       
        distancesOGS2[y] = (temp1*temp1+temp2*temp2+temp3*temp3)**(1/2)
        elevationsOGS2[y] = elevationAngle(distancesOGS2[y],satAlt,rE)
       
       
    bothVisibleTimes = []
    bothVisibleDistancesOGS1 = []
    bothVisibleDistancesOGS2 = []
    bothVisibleElevationsOGS1 = []
    bothVisibleElevationsOGS2 = [] 
    for x in range(0,np.size(times)):
        elevOGS1 = elevationsOGS1[x]
        elevOGS2 = elevationsOGS2[x]

        if ((elevOGS1 >= minimumElevation) and (elevOGS2 >= minimumElevation)):
            bothVisibleTimes.append(times[x])
            bothVisibleDistancesOGS1.append(distancesOGS1[x])
            bothVisibleDistancesOGS2.append(distancesOGS2[x])
            bothVisibleElevationsOGS1.append(elevationsOGS1[x])
            bothVisibleElevationsOGS2.append(elevationsOGS2[x])
           
           
    cdef double[:] bothVisibleTimesArr = np.array(bothVisibleTimes, dtype = np.double)
    cdef double[:] bothVisibleDistancesOGS1Arr = np.array(bothVisibleDistancesOGS1, dtype = np.double)
    cdef double[:] bothVisibleDistancesOGS2Arr = np.array(bothVisibleDistancesOGS2, dtype = np.double)
    cdef double[:] bothVisibleElevationsOGS1Arr = np.array(bothVisibleElevationsOGS1, dtype = np.double)
    cdef double[:] bothVisibleElevationsOGS2Arr = np.array(bothVisibleElevationsOGS2, dtype = np.double)
    
    return np.asarray(bothVisibleTimesArr), np.asarray(bothVisibleDistancesOGS1Arr),np.asarray(bothVisibleDistancesOGS2Arr), np.asarray(bothVisibleElevationsOGS1Arr), np.asarray(bothVisibleElevationsOGS2Arr)

cpdef TrimMemory(np.ndarray[np.int64_t, ndim = 1] memoryArray,  np.ndarray[np.double_t, ndim = 1] timesArray, double t0, double tInternal, double tCut):
    for i in np.where(((t0 + tInternal) - timesArray)>=tCut)[0]:
        memoryArray[i] = 0 
        timesArray[i] = 0 
    return np.asarray(memoryArray), np.asarray(timesArray)
        
cpdef EntanglementSwapping(np.ndarray[np.int64_t, ndim = 1] aliceMemories, np.ndarray[np.int64_t, ndim = 1] bobMemories, np.ndarray[np.double_t, ndim = 1] aliceTimes, np.ndarray[np.double_t, ndim = 1] bobTimes, double t0, double tInternal, list aliceWaits, list bobWaits, int SRCounts):
    cdef int currentAlicePairs = aliceMemories.sum()
    cdef int currentBobPairs = bobMemories.sum()
    cdef int aSlot, bSlot
    cdef list alive_a, alive_b, TimesA, TimesB
    
    while(currentAlicePairs> 0 and currentBobPairs > 0):
        SRCounts += 1
        alive_a = [i for i, v in enumerate(aliceMemories) if v>0]
        alive_b =[i for i, v in enumerate(bobMemories) if v>0]
        TimesA = [aliceTimes[i] for i in alive_a]
        TimesB = [bobTimes[i] for i in alive_b] 
        
        aSlot = alive_a[np.array(TimesA).argmax()]
        bSlot = alive_b[np.array(TimesB).argmax()]
       
        aliceWaits.append((t0 + tInternal) - aliceTimes[aSlot])
        bobWaits.append((t0 + tInternal) - bobTimes[bSlot])
        
        aliceMemories[aSlot] = 0 
        bobMemories[bSlot] = 0
        
        aliceTimes[aSlot] = 0 
        bobTimes[bSlot] = 0
        
        currentAlicePairs = aliceMemories.sum()
        currentBobPairs = bobMemories.sum()   
        
    return np.asarray(aliceMemories), np.asarray(bobMemories), np.asarray(aliceTimes), np.asarray(bobTimes), aliceWaits, bobWaits, SRCounts
    
cpdef EntanglementGeneration(np.ndarray[np.int64_t, ndim = 1] aliceMemories, np.ndarray[np.int64_t, ndim = 1] bobMemories, np.ndarray[np.double_t, ndim = 1] aliceTimes, np.ndarray[np.double_t, ndim = 1] bobTimes,
                             double t0, double tInternal, double aliceTau, double bobTau, 
                             double etaAlice, double etaBob, double timeA, double timeB, int aliceCap, int bobCap):
    cdef list Events
    cdef int sampleAlice, sampleBob
    cdef np.ndarray[np.int64_t, ndim = 1] slotsToFill
    
    if (abs(timeA - timeB) < 1e-8):
        internalTime = timeA
        Events = ['Alice', 'Bob']
    elif (timeA <timeB): #alice
        Events = ['Alice']
        internalTime = timeA
    elif (timeB < timeA):
        Events = ['Bob']
        internalTime = timeB
        
    for event in Events: 
        if (event == 'Alice'):
            sampleAlice = np.random.binomial(aliceCap - aliceMemories.sum(), etaAlice)
            if (sampleAlice > 0):
                slotsToFill = np.where(aliceMemories == 0)[0][:sampleAlice]
                aliceMemories[slotsToFill]= 1
                aliceTimes[slotsToFill] = t0 + tInternal  - 2*aliceTau
            timeA += 2*aliceTau
        if (event == 'Bob'):
            sampleBob = np.random.binomial(bobCap - bobMemories.sum(), etaBob)
            if (sampleBob > 0):
                slotsToFill = np.where(bobMemories == 0)[0][:sampleBob]
                bobMemories[slotsToFill]= 1
                bobTimes[slotsToFill] = t0 + tInternal - 2*bobTau
            timeB += 2*bobTau
            
    return np.asarray(aliceMemories), np.asarray(bobMemories), np.asarray(aliceTimes), np.asarray(bobTimes), timeA, timeB, internalTime
        
        
        
cpdef SSQR_MonteCarlo(np.ndarray[np.int64_t, ndim = 1] aliceMem, np.ndarray[np.int64_t, ndim = 1] bobMem, np.ndarray[np.double_t, ndim = 1] timesAlice, np.ndarray[np.double_t, ndim = 1] timesBob, double aliceTau, double bobTau, double alicedB, double bobdB, double globalTime, double goalTime, double tCut):
    cdef int SRCounts = 0
    cdef list aliceWaits = []
    cdef list bobWaits = []
    cdef np.ndarray[np.int64_t, ndim = 1] aliceMemories =  aliceMem
    cdef np.ndarray[np.int64_t, ndim = 1] bobMemories = bobMem
    cdef np.ndarray[np.double_t, ndim = 1] aliceTimes = timesAlice
    cdef np.ndarray[np.double_t, ndim = 1] bobTimes = timesBob
    cdef double etaAlice = 10**(alicedB/-10)
    cdef double etaBob = 10**(bobdB/-10)
    cdef double timeA = 2*aliceTau
    cdef double timeB = 2*bobTau
    cdef int aliceCap = len(aliceMem)
    cdef int bobCap = len(bobMem)   
            
    
    cdef double internalTime = 0
    while(globalTime + internalTime < goalTime):
        
        aliceMemories, aliceTimes = TrimMemory(aliceMemories, aliceTimes, globalTime, internalTime, tCut)
        bobMemories, bobTimes = TrimMemory(bobMemories, bobTimes, globalTime, internalTime, tCut)
       
        aliceMemories, bobMemories, aliceTimes, bobTimes, aliceWaits, bobWaits, SRCounts = EntanglementSwapping(aliceMemories, bobMemories, aliceTimes, bobTimes,  globalTime, internalTime,aliceWaits, bobWaits, SRCounts)
        
        aliceMemories, bobMemories, aliceTimes, bobTimes, timeA, timeB, internalTime = EntanglementGeneration(aliceMemories, bobMemories, aliceTimes, bobTimes,globalTime,  internalTime,  aliceTau,  bobTau, etaAlice,  etaBob,  timeA,  timeB, aliceCap, bobCap)

    return np.asarray(aliceMemories), np.asarray(bobMemories), np.asarray(timesAlice), np.asarray(timesBob), SRCounts, aliceWaits, bobWaits, (globalTime + internalTime)
\end{lstlisting}

\end{appendices}

\end{document}